%% file: thesis.tex
\label{gen_header}
\documentclass[12pt,a4paper]{book}
\usepackage[bindingoffset=0.5cm,margin=3cm]{geometry}
\usepackage[T1]{fontenc} 
\usepackage[utf8]{inputenc}
\usepackage{amsmath}
\usepackage{amsfonts}
\usepackage{amssymb}
\usepackage{tikz}
\usetikzlibrary{patterns}
\usepackage{graphicx}
\graphicspath{ {figures/} }
\usepackage{color}
\usepackage{hyperref} 
\hypersetup{
    colorlinks=false,
    linkcolor=blue,
    filecolor=magenta,      
    urlcolor=red,
    linktocpage=true
    }

\usepackage{appendix}
\usepackage{chngcntr}
\usepackage{etoolbox}
\usepackage{layout}
\AtBeginEnvironment{subappendices}{%
\chapter*{Appendix}
\addcontentsline{toc}{section}{Appendices}
\counterwithin{figure}{section}
\counterwithin{table}{section}
}

\AtEndEnvironment{subappendices}{%
\counterwithout{figure}{section}
\counterwithout{table}{section}
}

\newcommand{\zb}[2][]{#2_{#1}, \bar{#2}_{#1}}

\newcommand{\blankpage}{
\newpage
\thispagestyle{empty}
\mbox{}
\newpage
}

\label{covercmds}

\usepackage{xcolor}
\definecolor{bordeau}{rgb}{0.3515625,0,0.234375}
\usepackage[absolute,overlay]{textpos}
\usepackage{lipsum}
\usepackage{array}
\usepackage{caption}
\usepackage{multicol}
\newcommand{\PhDTitle}{Conformal bootstrap in two-dimensional conformal field theories with non-diagonal spectrums}
\newcommand{\PhDname}{Santiago Migliaccio} 															
\newcommand{\NNT}{2018SACLS362} 															

\newcommand{\ecodoctitle}{Physique en Île-de-France} 													
\newcommand{\ecodocacro}{EDPIF}																
\newcommand{\ecodocnum}{564} 																
\newcommand{\PhDspeciality}{Physique} 										
\newcommand{\PhDworkingplace}{l'Université Paris-Sud.\\  Institut de physique théorique, Université Paris Saclay, CEA, CNRS.} 										
\newcommand{\defenseplace}{Saint-Aubin} 											
\newcommand{\defensedate}{10 octobre 2018} 															

\newcommand{\logoEt}{logo-ipht} 																
\newcommand{\vpos}{0.1}																	
\newcommand{\hpos}{11}																		
\newcommand{\logoEtt}{UPSUD}  																
\newcommand{\vpostt}{0.1} 																	
\newcommand{\hpostt}{6}																	

\label{juryinfo}

\newcommand{\jurynameA}{Pascal Baseilhac}
\newcommand{\juryadressA}{Directeur de Recherche, Laboratoire de Mathématiques et Physique Théorique Université de Tours, CNRS.}
\newcommand{\juryroleA}{Président}

\newcommand{\jurynameB}{Valentina Petkova}
\newcommand{\juryadressB}{Professeur Emérite, Laboratory of Theory of Elementary Particles, Institute of Nuclear Research and Nuclear Energy, Bulgarian Academy of Sciences.}
\newcommand{\juryroleB}{Rapporteur}

\newcommand{\jurynameC}{Oleg Lisovyy}
\newcommand{\juryadressC}{Professeur des Universités, Laboratoire de Mathématiques et Physique Théorique Université de Tours, CNRS. }
\newcommand{\juryroleC}{Rapporteur}

\newcommand{\jurynameD}{Raoul Santachiara}
\newcommand{\juryadressD}{Chargé de Recherche, Laboratoire de Physique Theorique et Modeles Statistiques, Université Paris-Sud, CNRS. }
\newcommand{\juryroleD}{Examinateur}

\newcommand{\jurynameE}{Yacine Ikhlef}
\newcommand{\juryadressE}{Chargé de Recherche, 
Laboratoire de Physique Théorique et Hautes Énergies, Sorbonne Université, CNRS.}
\newcommand{\juryroleE}{Examinateur}

\newcommand{\jurynameF}{Sylvain Ribault}
\newcommand{\juryadressF}{Chargé de Recherche, Institut de Physique Théorique, Université Paris Saclay, CEA, CNRS. }
\newcommand{\juryroleF}{Directeur de thèse}

\label{backcovercmds}
\newcommand{\logoEd}{EDPIF}																		
\newcommand{\PhDTitleFR}{ Bootstrap conforme en théorie conforme bidimensionnelle avec spectre non-diagonal.}													
\newcommand{\keywordsFR}{Théorie conforme, Bootstrap conforme, spectre non-diagonal}														
\newcommand{\abstractFR}{La symétrie conforme impose de très fortes contraintes sur les théories quantiques des champs. En deux dimensions, l’algèbre des symétries conformes est infinie, et les théories conformes bidimensionnelles peuvent être complètement résolubles, dans le sens où toutes leurs fonctions de corrélation peuvent être calculées.  Ces théories ont un grand domaine d'application, de la théorie des cordes jusqu'aux systèmes critiques en physique statistique, et elles ont été largement étudiées pendant les dernières décennies. 

Dans cette thèse nous étudions les théories conformes bidimensionnelles dont l’algèbre de symétrie est celle de Virasoro, en suivant l'approche connue sous le nom de bootstrap conforme.  Sous l'hypothèse de l'existence de champs dégénérés, nous généralisons le bootstrap conforme analytique aux théories avec des spectres non-diagonaux. Nous écrivons les équations qui déterminent les constantes de structure, et nous trouvons des solutions explicites en termes de fonctions spéciales. Nous validons ces résultats en faisant des calculs numériques des fonctions de corrélation à quatre points dans des modèles minimaux diagonaux et non-diagonaux, et en vérifiant que la symétrie de croisement est respectée. 

En outre, nous construisons une proposition pour une famille de théories conformes non-diagonales et non-rationnelles pour toute charge centrale telle que $\Re{c} < 13$. Cette proposition est motivée par les limites des spectres des modèles minimaux de la série D. Nous réalisons des calculs numériques des fonctions à quatre points dans ces théories, et nous trouvons qu'elles obéissent à la symétrie de croisement. Ces théories peuvent être interprétées comme des extensions non-diagonales de la théorie de Liouville.  }

\newcommand{\PhDTitleEN}{Conformal bootstrap in two-dimensional conformal field theories with with non-diagonal spectrums.}													
\newcommand{\keywordsEN}{Conformal field theory, Conformal bootstrap, non-diagonal spectrum}														
\newcommand{\abstractEN}{Conformal symmetry imposes very strong constraints on quantum field theories. In two dimensions, the conformal symmetry algebra is infinite-dimensional, and two-dimensional conformal field theories can be completely solvable, in the sense that all their correlation functions may be computed. These theories have an ample range of applications, from string theory to critical phenomena in statistical physics, and they have been widely studied during the last decades.  

In this thesis we study two-dimensional conformal field theories with Virasoro algebra symmetry, following the conformal bootstrap approach. Under the assumption that degenerate fields exist, we provide an extension of the analytic conformal bootstrap method to theories with non-diagonal spectrums. We write the equations that determine structure constants, and find explicit solutions in terms of special functions. We validate this results by numerically computing four-point functions in diagonal and non-diagonal minimal models, and verifying that crossing symmetry is satisfied.

In addition, we build a proposal for a family of non-diagonal, non-rational conformal field theories for any central charges such that $\Re{c} < 13$. This proposal is motivated by taking limits of the spectrum of D-series minimal models. We perform numerical computations of four-point functions in these theories, and find that they satisfy crossing symmetry. These theories may be understood as non-diagonal extensions of Liouville theory.  } 

\label{docstart}
\begin{document}

\pdfbookmark{Cover}{cover}
\input{cover}\newpage\cleardoublepage

\pdfbookmark{Abstract}{abstract}
\input{chapters/abstract}\newpage\cleardoublepage
\input{chapters/acknowledgements}\newpage\cleardoublepage
\pdfbookmark{\contentsname}{toc}
\tableofcontents 


\input{chapters/intro}\newpage\cleardoublepage
\input{chapters/confsym}\newpage\cleardoublepage
\input{chapters/bootstrap}\newpage\cleardoublepage
\input{chapters/mmlim}\newpage\cleardoublepage
\input{chapters/conclusions}\newpage\cleardoublepage
\input{chapters/resume}\newpage\cleardoublepage

\input{refs.tex}

\cleardoublepage
\blankpage
\pdfbookmark{Back cover}{backcover}

\input{backcover}

\end{document}

%% file: cover.tex
\begin{titlepage}
\thispagestyle{empty}
\phantom{Including text for fixing overlapping page issue}
\begin{textblock}{5.1}(0,0)
	\textblockcolour{bordeau}
	\includegraphics [scale=1]{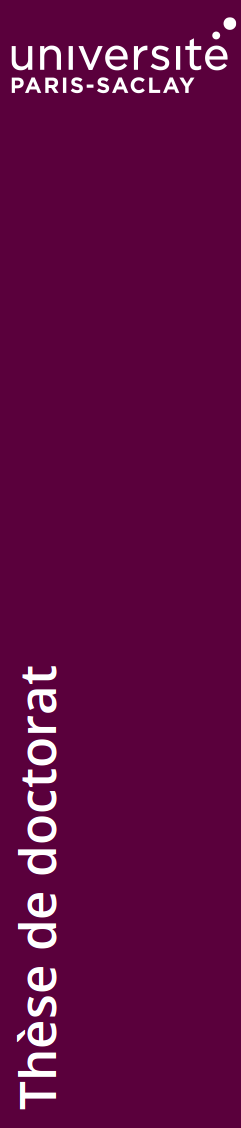}
	\vspace{300mm}
\end{textblock}

\begin{textblock}{1}(0.3,3)
	\Large{\rotatebox{90}{\color{white}{NNT : \NNT}}}
\end{textblock}

\begin{textblock}{1}(\hpostt,\vpostt)
	\textblockcolour{white}
	\includegraphics[scale=1]{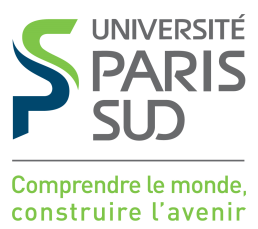}
\end{textblock}

\begin{textblock}{1}(\hpos,\vpos)
	\textblockcolour{white}
		\includegraphics[scale=.6]{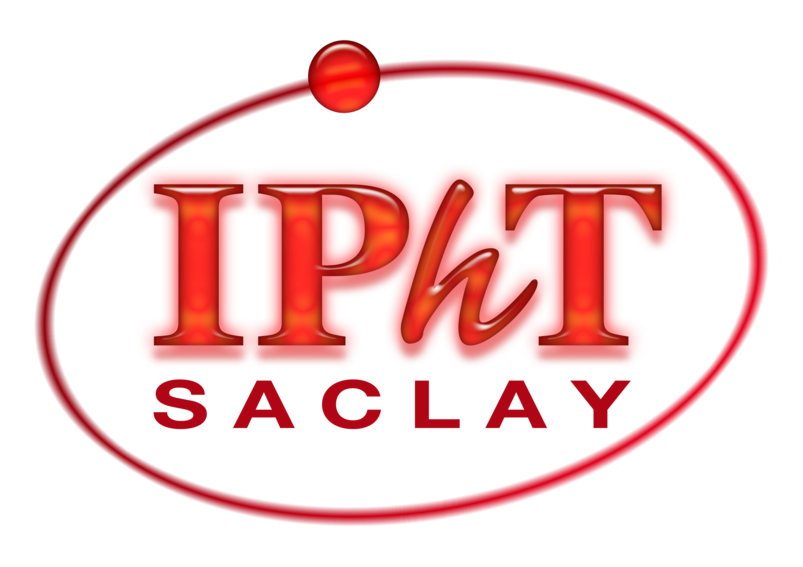}
\end{textblock}

\begin{textblock}{10.1}(5.5,3)
	\textblockcolour{white}
	
	\color{bordeau}
	\begin{flushright}
		\huge{\PhDTitle} \bigskip 
		\vfill
		\color{black} 
		\normalsize {Thèse de doctorat de l'Université Paris-Saclay} \\
		préparée à \PhDworkingplace \\ \bigskip
		\vfill
		Ecole doctorale n$^{\circ}$\ecodocnum ~\ecodoctitle ~(\ecodocacro)  \\
		
		\small{Spécialité de doctorat: \PhDspeciality} \bigskip 
		\vfill  
		\footnotesize{Thèse présentée et soutenue à \defenseplace, le \defensedate, par} \bigskip
		\vfill
		\Large{\textbf{\textsc{\PhDname}}} 
		\vfill
	\end{flushright}
	
	\color{black}
	\begin{flushleft}
		
		\small Composition du Jury :
	\end{flushleft}

	\small
	\newcolumntype{L}[1]{>{\raggedright\let\newline\\\arraybackslash\hspace{0pt}}m{#1}}
	\newcolumntype{R}[1]{>{\raggedleft\let\newline\\\arraybackslash\hspace{0pt}}lm{#1}}
	
	\label{jury} 																				
	\begin{flushleft}
	\begin{tabular}{@{} L{10cm} R{4cm}}
		\jurynameA  \\ \juryadressA & \juryroleA \\
		\jurynameB  \\ \juryadressB & \juryroleB \\
		\jurynameC  \\ \juryadressC & \juryroleC \\
		\jurynameD  \\ \juryadressD & \juryroleD \\
		\jurynameE  \\ \juryadressE & \juryroleE \\
		\jurynameF  \\ \juryadressF & \juryroleF \\
	\end{tabular} 
	\end{flushleft}   
\end{textblock}
\vfill
\end{titlepage}


%% file: chapters/abstract.tex
\chapter*{Abstract}  \label{sec:abstract}


In this thesis we study two-dimensional conformal field theories with Virasoro algebra symmetry, following the conformal bootstrap approach. Under the assumption that degenerate fields exist, we provide an extension of the analytic conformal bootstrap method to theories with non-diagonal spectrums. We write the equations that determine structure constants, and find explicit solutions in terms of special functions. We validate this results by numerically computing four-point functions in diagonal and non-diagonal minimal models, and verifying that crossing symmetry is satisfied.

In addition, we build a proposal for a family of non-diagonal, non-rational conformal field theories for any central charges such that $\Re{c} < 13$. This proposal is motivated by taking limits of the spectrum of D-series minimal models. We perform numerical computations of four-point functions in these theories, and find that they satisfy crossing symmetry. These theories may be understood as non-diagonal extensions of Liouville theory. 

%% file: chapters/acknowledgements.tex
\chapter*{Acknowledgements, Remerciements, Agradecimientos}

I would like to begin by thanking the two Rapporteurs, Valentina Petkova and Oleg Lisovyy, for the time devoted to carefully reading this manuscript and for the valuable feedback they provided. Furthermore, I would like to thank the jury members Pascal Baseilhac, Raoul Santachiara and Yacine Ikhlef, for accepting to evaluate this work.

I am very grateful to Jesper Jacobsen and Mariana Graña, who did not hesitate to take the roles of scientific tutor and `godmother', respectively, and were part of my follow-up committee providing advice and support. In particular, Mariana was my first contact with IPhT, and her encouragement was very important to me while I was looking for a position. 

Pendant ces années de doctorat, j'ai rencontré des collègues qui m'ont aidé à progresser. Je voudrais remercier en particulier Raoul et Nina pour des discussions enrichissantes, et Benoît Estienne et Yacine Ikhlef, pour ses commentaires sur nos travaux et pour accepter que je fasse un séminaire au LPTHE. Je remercie aussi les participants et organisateurs des écoles d'été à Cargèse de 2016 et 2017.

Bien évidemment, les conseils et soutien de Sylvain ont été fondamentaux pendant le déroulement de la thèse. J'apprécie sa prédisposition à l'écoute et à la discussion, et aussi ses critiques toujours constructives et bien fondées. Je suis très reconnaissant de son soutien continu, et j'admire son engagement pour une science plus ouvert et accessible.  

Plusieurs institutions et organismes ont été impliqués dans la réalisation de cette thèse : le CEA, l'Université Paris-Sud/Paris Saclay, et l'École Doctorale de Physique en Île de France. Ces institutions ont veillé pour le bon déroulement des travaux thèses des nombreux étudiants, ce qui implique l'effort de plusieurs personnes. En particulier, je voudrais remercier le CEA pour le financement de ma formation, le bureau d'accueil international du CEA pour m'avoir aidé avec les nombreuses démarches nécessaires pour venir en France, et M. Claude Pasquier, pour son travail comme directeur adjoint EDPIF pour Paris-Saclay. 

À l'IPhT j'ai trouvé un laboratoire très vivant et accueillante, et ceci est grâce au travail de ses équipes scientifiques, administratives et informatiques. Je voudrais remercier tous les membres du labo en générale, et en particulier Sylvie Zafanella, Anne Angles, Laure Sauboy, Patrick Berthelot, Laurent Sengmanivanh, Riccardo Guida, Ivan Kostov, François Gelis, Vincent Pasquier et Didina Serban. J'oublie sûrement des noms importants, mais, malgré cela, je suis très reconnaissant d'avoir fait parti de ce laboratoire.   

I want to thank the Argentinian State for providing me and all its citizens with high quality free public education at all levels. I believe that the importance of public education in our country cannot be overstated, and that the system should be protected and continuously improved, as it has been and still is one of the most important elements for promoting the development of the country and ensuring that a majority of the population has opportunities of social mobility and improving their quality of life. I am grateful to the institutions I have been a part of, the IES Juan B. Justo and the University of Buenos Aires, where I've had many great teachers. Among them, I would like to extend these acknowledgements to Martín Schvellinger, who was my advisor during my Licenciatura.       

When I arrived in Paris in 2015, I settled in the Argentinian House (Maison de l'Argentine) of the Cité Internationale Universitaire de Paris. There I found a welcoming and warm environment, which served as a home and helped me and other students and artists that were living there take our first steps in an unknown city. Sadly, since the year 2017, the House has been undergoing a rapid and steady decline, and the welcoming environment I had first found became authoritarian, oppressive, and outright hostile towards the residents. The Argentinian House can play an important and positive role in the careers of the students, artists, and academics pursuing their studies and research in this city, I am grateful for the place I had there. I sincerely wish the House will soon recover the nurturing atmosphere it needs in order to fulfill its mission.   

During these years I encountered numerous other PHD students and postdocs, for whose friendship and company I am grateful. Thanks to Séverin, for welcoming me when I first arrive, for his support, and for frequently bringing food and plants into the office we shared. To the Chouffe team, Pierre, Luca, Christian and Niall, for pushing me to reach up to the highest shelves of scientific excellence, and also to Niall for being a great RER B partner. Many others helped make lunches, coffee breaks and summer schools memorable: Raphael, Kemal, Benoît, Jonathan, Steven, Nico, Romain, Etienne, Thiago, Hannah, Linnea, Laïs, Vardan, Guillaume, Thibault, Michal, Samya, Valerio, Louise, Riccardo, Ana, Lucía, Juan Miguel, Lilian, Orazio, Stefano, Alex, Anna, Jules, Long, Federico, Corentin, Debmalya, Ruben, Sebastian, and all the people I am surely but unintentionally forgetting.

De los amigos que hice en la Casa Argentina aprendí muchas cosas, y su apoyo fue fundamental para poder completar mi tesis. Una lista incompleta: Leo y Nati, Lu, Florence, Mirjana, Ceci, Christian, Luchi, El Chino, Fer, Carla, Jime, Guido, Mati, Juan Martín, Santi Q, K2, Joanna, Estefi, Clari, Gabi, Facu, Ouma y muchos otros. Valoro su amistad y admiro su valentía. 

A mis amigos del grupo L.C.G., tanto en Argentina como en Europa, les agradezco su constante apoyo. Compartir mi experiencia con ustedes me ayuda a seguir avanzando. 

A los pibes, con quienes pareciera que ni la distancia ni las diferencias horaria existen, gracias por más de 20 años de amistad.

A mi familia quiero agradecerle su apoyo incondicional y constante, y su cariño a la distancia. No podría estar acá sin todo el trabajo que hicieron por mi, y el amor que siempre dieron. Gracias. 

Me gustaría poder compartir esta tesis con mi tío Ricardo. Sé que estaría muy contento de poder verla. Me hará mucha falta. 

Por último, quiero agradecerle a Cami por acompañarme, por su apoyo, paciencia, compresión y amor. Sos tan responsable de que haya escrito esta tesis como yo, tal vez más. Me hace feliz poder compartir mi vida con vos.

%% file: chapters/intro.tex
\chapter{Introduction}

Symmetry is an immensely powerful tool to understand physical systems, as it can help to simplify or extract information about potentially complex problems. The most important example of this is perhaps Noether's theorem, which states that for any continuous symmetry there is an associated conserved quantity. It is often the case that symmetry considerations give rise to results with a wide range of applications, as there may be many different physical systems sharing the same symmetries. 

In this work we focus on conformal symmetry, and its consequences on two dimensional quantum field theories. Conformal transformations are symmetry transformations that leave angles invariant, and they form a larger class of transformations than those of the Poincaré group. Then, conformal field theories have an enhanced symmetry that makes them more tractable than a generic QFT. Sometimes conformal field theories can become completely solvable, in the sense that all their correlation functions can in principle be computed. 
Among conformal transformations there are dilatations, or scale changes. That these transformations are a symmetry of a system might seem unphysical, because it is known that physical laws are usually strongly dependent on scale. The exceptions come when the characteristic distances of a system become either $0$ of $\infty$, and there are many examples where a conformal field theory description is possible.  In the next section we list some of them.

\section{The various applications of CFTs}

Conformal field theories (CFTs) provide a very interesting example where the many constraints coming from conformal symmetry can make complicated quantum field theories more tractable. In some cases, this is enough to render interacting theories completely solvable. In this sense, conformal field theories may be seen as a stepping stone in understanding more complex theories. 

Conformal field theories are one of the key elements of the AdS/CFT correspondence. Perhaps the most studied example is the duality between type $IIB$ string theory in AdS$_{5} \times S^5$ and the super-conformal gauge theory $\mathcal{N} = 4$ Super Yang-Mills in four dimensions\cite{DHoker:2002nbb}. As another application, conformal field theories play an important role in the world-sheet description of string theory. 

The AGT correspondence, named after Alday, Gaiotto and Tachikawa, \cite{Alday:2009aq}, provides one further example where conformal field theories play an important role. This correspondence proposes a relation between a four dimensional $SU(2)$ gauge theory and Liouville field theory, a two dimensional conformal field theory we will discuss more in detail in section \ref{sec:liouville}. 

Moreover, conformal field theories have important applications in the domain of critical phenomena. In this case, systems that undergo a second order phase transition see their correlation length diverge as they approach the critical point. Then, their continuum limit can be described by a conformal field theory, identified by the critical exponents of the theory. In this context there is a striking phenomenon know as \textbf{critical universality} which refers to the fact that, near the critical point, the continuum limit becomes independent of the microscopic details of the underlying models, and many different systems are described by the same conformal field theory. The canonical example of a system with this behaviour is the Ising model. In three dimensions, the CFT that describes its continuum limit is also common to other systems, such as water at the critical point \cite{Cardy:1996xt}. In recent years there has been an important effort to better understand this theory, for example \cite{PhysRevD.86.025022}. 

The case of two dimensional conformal field theories is particularly interesting, because the symmetry algebra becomes infinite dimensional. In this work we focus on the simplest theories, where the symmetry algebra is the Virasoro algebra. There are  other theories whose symmetry algebras contain the Virasoro algebra, such as theories with $\mathcal{W}$-algebra symmetry, which will be left out of the scope of this work. 

The Virasoro algebra is characterized by a parameter called the central charge, $c$, whose value will be crucial to determine many properties of a CFT:  from the very existence of the theory to the structure of its spectrum and properties such as unitarity. In this sense, different values of $c$ correspond to different physical systems, but there may be different systems that are described by conformal field theories with the same value of $c$. 
  
In this work we do not focus on any specific application of two dimensional CFTs. 
Instead, we will base our analysis on the constraints arising solely symmetry and self-consistency, following an approach known as the Conformal bootstrap.

\section{The conformal bootstrap approach}

In this thesis we study two dimensional conformal field theories via the conformal bootstrap approach. This approach attempts to build, classify and solve conformal field theories by studying the constraints imposed by symmetry and self-consistency only. The idea is to start from very general principles, shared by large classes of theories, hoping to solve simultaneously many different problems. This approach has been applied to conformal field theories such as Liouville theory and the generalized minimal models, proving its effectiveness, see \cite{Ribault:2014hia, Teschner:2017del} for reviews. Furthermore, analysis in the conformal bootstrap approach have allowed to extend known results, giving rise for example to a description of Liouville theory with a central charge $c < 1$ \cite{rs15}.

Two dimensional CFTs are particularly well suited for the bootstrap approach, as shown by the pioneering work in \cite{Belavin:1984vu}. The reason is that the infinite dimensional symmetry algebra gives rise to infinitely many constraints on the correlation functions. These constraints are known as the local Ward identities, and in section \ref{sec:ward} we will define $N$-point correlation functions as solutions to these identities, satisfying a few other conditions. This contrasts with the more usual Lagrangian approach, where correlation functions are defined using path integrals. In this sense, the bootstrap approach can  be used to study systems where a Lagrangian description may be ill defined, or not even available.  
Furthermore, relying on constraints originating from the general structure of conformal field theories gives the conformal bootstrap approach the power to produce non-perturbative results. 

A negative side of taking the bootstrap point of view may be that once a theory has been built and solved through this method, it may not be straightforward to determine which physical systems -if any- it represents. We depend, then, on external information in order to make this identification, as opposed to a case where taking a continuous limit of a discrete system can give rise to a Lagrangian describing the limit system. 

In the two dimensional conformal bootstrap approach, a theory is identified by specifying its symmetry algebra, in our case the Virasoro algebra,  its spectrum, i.e the set of fields whose correlation functions it describes, and a set of rules that controls how products of nearby fields behave, called the fusion rules. 
One of the consequences of having Virasoro algebra symmetry is that the spectrum can be organized into conformal \textit{families}: each family is identified by a particular field, called a \textbf{primary field}, and it contains many other fields which are obtained by acting with the symmetry generators on the primary, and are called the descendants of the primary field.  We will see how the Ward identities determine that correlation functions can be expressed as a combination of universal functions, called the conformal blocks, and structure constants, which depend on the theory. In this context, we will say that a theory is solved once we determine all the elements necessary to compute, at least in principle, its correlation functions. In practice this amounts to computing four-point correlation functions, which are the first ones not to be completely determined by symmetry.  

The conformal bootstrap approach requires assumptions to be made explicitly. This can help one identify which assumptions are fundamental, and which can be lifted to give rise to a more general case. For example, a common assumption is the \textbf{diagonality} of the theories, which means that all the primary fields in the spectrum are scalars. However, conformal field theories need not be diagonal in general, and  in this work we follow the conformal bootstrap approach without assuming diagonality. 

In the following section we mention some examples of conformal field theories whose solutions are known, and where the conformal bootstrap approach has been followed through successfully.

\section{Overview of known models}\label{sec:cmap}

There are many different two dimensional conformal field theories that are accessible trough the conformal bootstrap approach. Here we give an overview of certain two-dimensional CFTs that are relevant to the ideas in this work, and offer some discussion about their classification.

We have mentioned that the value of the central charge $c$ is crucial to determine many properties of a theory, and in particular its existence. Let us then show a map of two dimensional CFTs in the complex plane of the central charge $c$, where unitary theories are distinguished by solid colors. This figure  is obtained from a similar one in \cite{Ribault:2014hia}:
\begin{align}
\begin{tikzpicture}[scale = 4.5, baseline=(current  bounding  box.center)]
 \filldraw [red!50, opacity = .15] (2.15, -.3) -- (2.15, .75) -- (-.6, .75) -- (-.6, -.3) -- cycle;
 \filldraw [blue!50, opacity = .15] (2.15, -.3) -- (2.15, .75) -- (-.6, .75) -- (-.6, -.3) -- cycle;
 \node [draw = red, fill = red!70, right] at (-.5, .65) {Liouville theory};
\node [draw = blue, fill = blue!15, right] at (-.5, .49) {Generalized minimal models};
 \node[draw = green!70!black, fill = green!16!black!7, right] at (-.5, .12) {Minimal models};
  \filldraw [red, opacity = .5] (.98, -.02) -- (.98, .02) -- (2.15, .02) -- (2.15, -.02) -- cycle;
  \filldraw [green!70!black, opacity = .3] (-.6, -.02) -- (-.6, .02) -- (.99, .02) -- (.99, -.02) -- cycle;
 \foreach \p in {2,...,20}
  {
  \draw [green!70!black, thick, opacity = 1] ({1-6/(\p*(\p+1))}, -.03) -- ({1-6/(\p*(\p+1))}, .03);
  }
  \node [below] at (0, -.02) {$0$};  \node [below] at (.99, -.02) {$1$};
  \node[draw,circle,inner sep=2pt,yellow] at (.99, 0) {} ; \node [above] at (.99, .02) {Ashkin-Teller};
  \draw[-latex] (-.6, 0) -- (2.15, 0) node [below left] {$c$};
\end{tikzpicture} \label{fig:cftmap}
\end{align}
The red color in figure \ref{fig:cftmap} represents Liouville field theory, a diagonal theory with a continuous, infinite spectrum, that exists for every value of $c \in \mathbb{C}$. Liouville theory is unitary for $c \in \mathbb{R}_{\geq 1}$, and from this line it can be extended to all values $c \notin \mathbb{R}_{<1}$, albeit losing unitarity. For $c \in \mathbb{R}_{<1}$ there is another version of Liouville theory, which has the same diagonal, continuous spectrum but cannot be obtained as a continuation from the other regions \cite{rs15}. In blue we show Generalized minimal models, theories that exist also for every value of $c \in \mathbb{C}$, and whose spectrum is diagonal and infinite, but discrete, as opposed to the continuous spectrum of Liouville theory. In green we have marked the Minimal Models, theories that exist only for an infinite, discrete set of values of $c$, which are however dense in the line $c \in \mathbb{R}_{<1}$. Minimal models are characterized by their discrete, finite spectrums, which can exist due to the special values of the central charge. Minimal models obey an A-D-E classification \cite{Cappelli:2010}, and different models have different spectrums: The A-series minimal models have diagonal spectrums, while the D-and-E-series minimal models spectrums contain a non-diagonal sector. The green lines identify values of $c$ for which there are unitary Minimal models, and we can see that these values accumulate near $c = 1$. Finally, the yellow circle at $c = 1$ signals the CFT corresponding to the Ashkin-Teller model, an example of a theory with an infinite, discrete non-diagonal spectrum.  

The theories represented in figure \ref{fig:cftmap} serve as an illustration of the different types of theories accessible through the conformal bootstrap approach. However, the examples of non-diagonal theories discussed above are restricted to special values of the central charge, and in the case of D-series minimal models, to finite spectrums. In particular, there is no example of a theory with a non-diagonal sector that is defined for every value of $c$ (or at least continuously many), as happens with Liouville theory or the Generalized minimal models, and the only example of an infinite non-diagonal spectrum is confined to the very particular case $c =1$. Non-diagonal CFTs can however be related to interesting physical systems, like the loop models discussed in \cite{ei15} or the Potts model at criticality. Building generic, consistent non-diagonal theories could represent an important step towards extending the current understanding of conformal field theories, and contribute to the description of the aforementioned systems. This leads us to stating the objectives of the present work. 

\section{Goals of this thesis}

As described in \cite{Ribault:2014hia}, the conformal bootstrap approach has been successfully applied to diagonal conformal field theories, giving rise to explicit results for their structure constants. The key to obtaining these results is the use of degenerate fields, fields whose fusion rules are particularly simple. This method was applied to Liouville theory in \cite{Teschner:1995yf}, and we refer to it as the analytic conformal bootstrap.

This thesis has two main objectives, related to the generalization of the analytic conformal bootstrap results to theories containing non-diagonal fields:

Our first objective is to extend the results of the analytic conformal bootstrap by lifting the assumption of diagonality, and obtain a way to compute structure constants for generic non-diagonal conformal field theories on the Riemann sphere. In order to pursue this analysis we will make three basic assumptions, namely: that the central charge can take any value $c \in \mathbb{C}$, and the correlation functions depend analytically on it, that the correlation functions are single-valued, and that two independent degenerate fields exist, in the sense that we may study their correlation functions with fields in our theory although they need not be part of the theory's spectrum. These assumptions will be further discussed in section \ref{sec:assumptions}, where we will study how they condition the types of non-diagonal theories we can study. An important remark is that in principle we will leave aside logarithmic CFTs, and an extension to this case may very well be non-trivial. 

The first difficulty in extending the analytic bootstrap results to the non-diagonal case is to generalize the degenerate fusion rules in a way that remains consistent. This will lead us, in section \ref{sec:dnd}, to give a definition of diagonal and non-diagonal fields that takes degenerate fusion rules into account. This definition will allow us to perform the desired extension, and we will validate our result by arguing that they give rise to consistent solutions of the theories appearing on the map \ref{fig:cftmap}. 

Important work in this direction was performed in \cite{ei15}, where the authors studied non-diagonal theories related to loop models and derived constraints arising from the existence of only one degenerate field. Although the method discussed here can be described as a continuation of the analysis of \cite{ei15}, our results may not be directly applicable to the systems discussed in that article. The reason is that assuming the existence of two degenerate fields, instead of only one, results in more restrictive constraints on the spectrum of the theories we can study. In exchange, having two independent degenerate fields is enough to completely determine the structure constants, which enables us to compute four-point correlation functions to verify the consistency of these results. 

Our second objective is to build a non-diagonal theory, or family of theories, that exist for a wide range of values of $c$, by making use of the non-diagonal analytic conformal bootstrap results. Such a theory would cover the missing examples of the map \ref{fig:cftmap} and provide a way to validate the analytic conformal bootstrap results in a previously unknown context. 

In order to build these families of theories, we will seek for a proposal of a non-diagonal spectrum and fusion rules by taking limits of D-series minimal models where the central charge approaches irrational values. Then, we will combine these proposals with the structure constants from the non-diagonal conformal bootstrap, and numerically verify that their combination gives rise to a consistent class of correlation functions. The existence of these correlation functions can be seen as evidence that the looked-after theories indeed exist, and can be added to the map \ref{fig:cftmap}. In chapter \ref{ch:concl} we will discuss the idea of interpreting these theories as a non-diagonal extension of Liouville theory.

This thesis is organised as follows: In chapter \ref{ch:csym} we give a review of conformal symmetry in two dimensions. We  discuss Ward identities, and give definitions of conformal blocks, structure constants, operator product expansions, fusion rules and degenerate fields. In chapter \ref{ch:cboots} we perform the analytic conformal bootstrap analysis. We describe crossing symmetry, both general and degenerate, and define diagonal and non-diagonal fields that have consistent fusion rules with degenerate fields. Then, we write the constraints on  the structure constants coming from degenerate crossing symmetry, and find some explicit solutions. In addition, we discuss the relationship between the diagonal and non-diagonal solutions. The first part of chapter \ref{ch:crsymfun} discusses how the conformal bootstrap solutions hold for the theories included in the map of figure \ref{fig:cftmap}, and shows numerical examples of crossing-symmetric four-point functions in the minimal models. The second part focuses on the construction of a family of non-diagonal, non-rational CFTs, whose existence is suggested by taking limits of non-diagonal minimal models. We provide evidence for the existence of these theories by numerically testing crossing symmetry of different four-point functions. The main results of chapters \ref{ch:cboots} and \ref{ch:crsymfun} have been published in \cite{Migliaccio:2017dch}, but here we go further in building explicit solutions, include a validation of the bootstrap results in non-diagonal minimal models, and extend the discussion about building the limit theories.  Finally, chapter \ref{ch:concl} offers a summary of our results, and some suggestions for future work. 

%% file: chapters/confsym.tex
\chapter{General notions of CFT}\label{ch:csym}

This chapter presents conformal symmetry and its implications on the structure and observables of a quantum field theory. We begin by reviewing conformal transformations in $d$ dimensions, and then specify to the 2-dimensional case. We describe the algebra of the generators and its central extension, the Virasoro algebra. Then we study the irreducible representations of the Virasoro algebra, which  are used to build the spectrum of CFTs. Finally, we study the constraints on correlation functions imposed by conformal symmetry, the Ward identities, and set the basis for the conformal bootstrap method, discussed in the next chapter. 

\section{Conformal transformations}

In this section we review conformal transformations, and present the generators of infinitesimal transformations that form the symmetry algebra. The content of this section is fairly standard and can be found in many textbooks and review articles. Our main references are the classical book \cite{fms97}, together with \cite{Ketov:1995yd} and the latest version of \cite{Schellekens:1996tg}.

\subsection{Conformal symmetry in \texorpdfstring{$d$}{d} dimensions}
In Euclidean $d$-dimensional space, conformal transformations are coordinate transformations that preserve angles locally. Equivalently, they can be defined as the transformations leaving the metric $g_{\mu \nu}$ invariant, up to a scale factor. Under a change of coordinates $x^{\mu} \to \tilde{x}^{\mu}(x)$ this condition can be expressed as
\begin{align}
g_{\mu \nu}(x) \to \tilde{g}_{\mu \nu}(\tilde{x}) &= \frac{\partial x^{\rho}}{\partial \tilde{x}^{\mu}}\frac{\partial x^{\sigma}}{\partial \tilde{x}^{\nu}} g_{\rho \sigma}(x(\tilde{x})) \label{eq:gtilde}\\ & = \Omega^2(x)g_{\mu \nu}(x)\, . 
\end{align}
where $\Omega^2(x)$ is a scaling factor. 

From this definition we see that the transformations of the Poincaré group are conformal, since they leave the metric invariant ($\Omega^2(x) = 1$). In order to find the rest of the transformations of the conformal group we consider the infinitesimal transformation given by $x^{\mu} \to \tilde{x}^{\mu} = x^{\mu} + \epsilon^{\mu}(x)$, and compute the new metric up to first order in $\epsilon(x)$. Then, equation  \eqref{eq:gtilde} gives
\begin{align}
\tilde{g}_{\mu \nu} (\tilde{x}) = g_{\mu \nu}(x) - (g_{\mu \sigma} \partial_{\nu} \epsilon^{\sigma} + g_{\nu \sigma} \partial_{\mu} \epsilon^{\sigma}) \ , \label{eq:gtildeeps}
\end{align}
where we have used $g_{\mu \nu} = g_{\nu \mu}$. In order for $\epsilon^{\mu}(x)$ to define a conformal transformation we need the term between brackets in \eqref{eq:gtildeeps} to be proportional to $g_{\mu \nu}$. Contracting with $g^{\mu \nu}$ we can find the proportionality factor, and we arrive at the condition
\begin{align}
\boxed{g_{\nu \sigma} \partial_{\mu} \epsilon^{\sigma} + g_{\mu \sigma} \partial_{\nu} \epsilon^{\sigma} = \frac{2}{d} (\partial\cdot \epsilon) g_{\mu\nu}}\,  .\label{eq:confeps}
\end{align}
Notice that here the space-time dimension $d$ appears explicitly in the proportionality factor, as a result of the contraction $g^{\mu \nu} g_{\mu \nu}$. 

For $d=1$, the condition \eqref{eq:confeps} is identically true, and any transformation is conformal. The case $d = 2$ is special, and will be the focus of the rest of this work. 
Let us for now discuss the case $d \geq 2$ without mentioning the special features of the two-dimensional case:

Infinitesimal translations and rotations satisfy conditions \eqref{eq:confeps}, since they are given by
\begin{align}
\text{Translations: }\quad \epsilon^{\mu}(x) &= \epsilon^{\mu} = \text{constant} \\
\text{Rotations: }\quad \epsilon^{\mu}(x) &= \omega^{\mu}_{\nu} x^{\nu}\, ,\ \omega_{\mu \nu} = -\omega_{\nu \mu} \, .
\end{align}

In addition to these transformations we have 
\begin{align}
\text{Dilatations: } \quad \epsilon^{\mu}(x) &= \lambda x^{\mu}\, , \lambda > 0
\end{align}
generating changes of scale, and special conformal transformations (SCTs), 
\begin{align}
\text{SCTs: }\quad \epsilon^{\mu}(x) &= b^{\mu} x_{\rho}x^{\rho} - 2x^{\mu} b_{\rho}x^{\rho}\,  .
\end{align}

The finite versions of these infinitesimal transformations form the global conformal group, which in $d$-dimensional Euclidean space is $SO(d+1,1)$\cite{fms97}. 
The generators of the conformal transformations are 
\begin{equation}
\begin{aligned}
\text{Translations:}&\quad P_{\mu} = -i \partial_{\mu}\, , \\
\text{Rotations:}&\quad M_{\mu \nu} = i (x_{\mu}\partial_{\nu} -x_{\nu}\partial_{\mu} )\, ,\\
\text{Dilations:}&\quad D = -i x^{\mu}\partial_{\mu}\, , \\
\text{SCTs:}&\quad K_{\mu} = i (x^2\partial_{\mu} - 2x_{\mu}x^{\rho}\partial_{\rho})\,\ ,  \label{eq:confgend}
\end{aligned}
\end{equation}
and the algebra defined by their commutation relations is isomorphic to $\mathfrak{so}(d+1,1)$. 
It's worth mentioning that inversions, given by $x^{\mu} \to \tilde{x}^{\mu} = \frac{x^{\mu}}{x^2}$, are also conformal transformations in the sense \eqref{eq:gtilde}. However, they are not connected to the identity transformation (there is no infinitesimal inversion), and thus they cannot be found through the analysis above. Nonetheless, special conformal transformations are obtained by performing two inversions, with one translation in between. 

\subsection{Conformal symmetry in 2 dimensions}\label{sec:2dsym}

We now study the particular case of two dimensions. Of course, we will again find the group $SO(3,1)$ describing global conformal transformations, but we will see that there is a larger class of transformations satisfying the conformal symmetry constraints \eqref{eq:confeps}. Let us write these equations explicitly, 
\begin{align}
\partial_1 \epsilon^1 &= \partial_2 \epsilon^2\, ,\\
\partial_1 \epsilon^2 &= -\partial_2 \epsilon^1\, . \label{eq:2dconf}
\end{align}
In this case, the constraints of conformal symmetry take the form of the Cauchy-Riemann equations, which encourages us to give a description in terms of the complex variables $z, \bar{z}$,
\begin{align}
z = x^1 + i x^2\ , \quad \bar{z} = x^1 - i x^2\, , \label{eq:zbzx}
\end{align} 
with the derivatives with respect to these variables taking the form
\begin{align}
\partial_{z} = \frac{1}{2} (\partial_1 -i \partial_2)\ , \quad\partial_{\bar{z}} = \frac{1}{2} (\partial_1 +i \partial_2)\ . \label{eq:cderv}
\end{align}
The infinitesimal transformation is then given by
\begin{align}
z \to z + \epsilon(z, \bar{z})\, ,\quad \epsilon(z, \bar{z}) = \epsilon^1(x) + i \epsilon^2(x), \\
\bar{z} \to \bar{z} + \bar{\epsilon}(z, \bar{z}) \, , \quad \bar{\epsilon}(z, \bar{z}) = \epsilon^1(x) - i \epsilon^2(x) \, .
\end{align}
Then, equations \eqref{eq:2dconf} become
\begin{align}
\partial_{z} \bar{\epsilon}(z, \bar{z}) = \partial_{\bar{z}} \epsilon(z, \bar{z}) = 0 \, .
\end{align}
Thus, $\epsilon$ is a function only of $z$, while $\bar{\epsilon}$ depends only on $\bar{z}$, meaning that any meromorphic function $\epsilon(z)$ defines a conformal transformation. These transformations can map points to the point at infinity, so we  choose to work in the Riemann sphere $\mathbb{C} \cup \{\infty\}$, rather than the complex plane. 

Any meromorphic transformation may be expanded as
\begin{align}
\epsilon(z) = \sum_{n \in \mathbb{Z}} \epsilon_n z^{n+1}\, ,  \label{eq:expeps}
\end{align}
and the corresponding generators are
\begin{align}
l_n = -z^{n+1}\partial_{z}\, , \quad \bar{l}_n =- \bar{z}^{n+1}\partial_{\bar{z}}\, , \quad n \in \mathbb{Z}\, .
\end{align}

Thus, in the two dimensional case, conformal symmetry is described by the infinite-dimensional Witt algebra, whose commutation relations are
\begin{align}
[l_n, l_m] = (n-m) l_{m+n}\, , \label{eq:Witt}
\end{align}
and similarly for the antiholomorphic generators $\bar{l}_{n}$.
This constitutes an important symmetry enhancement with respect to the case $d \geq 3$, where the conformal symmetry algebra has $\frac{(d+1)(d+2)}{2}$ generators \eqref{eq:confgend}. However, most of the transformations given by \eqref{eq:expeps} will have poles, and are only locally conformal. On the Riemann sphere, the subset of transformations that are globally well defined is still finite dimensional, and corresponds to the values $n = -1, 0, 1$; these transformations are at most quadratic in $z$, as in the higher dimensional case. Restricting the algebra to the global generators we find
\begin{align}
[l_{\pm 1}, l_0] &=\pm l_{\pm 1}\, , \\
[l_{1}, l_{-1} ] &= 2 l_0\, ,
\end{align}
which are the commutation relations of the algebra $\mathfrak{sl}(2)$.
The generators of the global conformal transformations are associated, as expected, with the transformations \eqref{eq:confgend}. We can express these generators in terms of the the coordinates $(z, \bar{z})$ by using equations \eqref{eq:zbzx} and \eqref{eq:cderv}. For example, the generators of Dilatations and rotations are
\begin{align}
\text{Dilations:}\ & i(l_0+\bar{l}_0) \ ,\\
\text{Rotations:}\ & (l_0-\bar{l}_0)\ .
\end{align}

The symmetry group associated to the global conformal transformations is $PSL(2, \mathbb{C})$, given by the Möbius transformations
\begin{align}
z \to \frac{az+b}{cz+d}\, , \quad ad - bc = 1\, . \label{eq:mob}
\end{align}
This group is isomorphic to $SO(3,1)$ and thus global transformations are given by the same group as in higher dimension (i.e $SO(d+1,1$)). 

Finite local conformal transformations are given by
\begin{align}
z \to f(z)\ , 
\end{align}
with $f(z)$ a meromorphic function such that its derivative is non-zero in a certain domain. We can see that such a transformation is conformal by writing the line element $ds^2 = dzd\bar{z}$, which transforms as, 
\begin{align}
d\tilde{s}^2 = \left|\frac{df(z)}{dz}\right|^2 ds^2 \, ,
\end{align}
where the scale factor is $|\frac{df(z)}{dz}|^2$. 

The holomorphic and antiholomorphic generators $l_n$ and $\bar{l}_n$, satisfy 
\begin{align}
[l_n, \bar{l}_m] = 0\, ,
\end{align}
and we can treat $z$ and $\bar{z}$ as independent coordinates. Calculations become easier in this way, and at the end of the day we may recover the original setup by setting $\bar{z} = z^{*}$.

\section{Virasoro algebra}

In order to build a quantum theory, it is necessary to consider the  projective action of the symmetry algebra \cite{Weinberg:1995mt}. This is equivalent to the action of a central extension of said algebra, which leads us to consider central extensions of the Witt algebra \eqref{eq:Witt}. This algebra  allows for a unique central extension \cite{fms97}, called the Virasoro algebra and denoted by $\mathfrak{V}$. Its generators are denoted by $L_n$, $n \in \mathbb{Z}$, and their commutations relations are 
\begin{align}
[L_n, L_m] = (n-m)L_{m+n} + \frac{c}{12}\delta_{m+n,0} (n-1)n(n+1)\, , \label{eq:Virasoro}
\end{align}
where the parameter $c$ is called the central charge, and it should be interpreted as the parameter accompanying an operator that commutes with every $L_n$. Throughout this work we generally consider $c \in \mathbb{C}$. Notice that although the relations \eqref{eq:Virasoro} contain a central term that is absent in Witt's algebra, the commutators of $L_{-1}$, $L_{0}$ and $L_{1}$, associated with global conformal transformations, are unaffected by it, and the global conformal subalgebra remains $\mathfrak{sl}(2)$. 

A two-dimensional conformal field theory can be defined as a quantum field theory whose spacetime symmetry algebra is (or contains) the Virasoro algebra \cite{Teschner:2017del}. Furthermore, since we will follow a description in terms of two complex coordinates $(\zb{z})$, the full symmetry algebra is actually $\mathfrak{V} \otimes \bar{\mathfrak{V}}$, where $\mathfrak{V}$ corresponds to the left-moving or holomorphic sector, related to the variable $z$, and $\bar{\mathfrak{V}}$ corresponds to the right-moving or antiholomorphic one, related to $\bar{z}$. Even though we consider the variables $(\zb{z})$ as independent, certain constraints will be included in section \ref{sec:aboots} in order to make observables single valued in the case $\bar{z} = z^*$. 

Quantizing a field theory requires the choice of a \textit{time} direction, which in Euclidean space is somewhat arbitrary. Here and in the following we take the approach known as radial quantization, where the radial direction is signaled as time, and time evolution is generated by dilatations. In this approach the dilatation operator $L_0$ plays the role of the Hamiltonian, and we will focus our attention on theories in which $L_0$ is diagonalizable. This takes into account many important examples, like minimal models and Liouville theory, but leaves aside logarithmic CFT. 

In what follows, we explore the consequences of Virasoro symmetry on the observables of a quantum field theory. We begin by discussing some of the representations of the symmetry algebra, which will be used in section \ref{ssec:FW} to define the fields of the CFT. For simplicity, we will discus the left-moving sector, but the right-moving one is completely analogous.  

\subsubsection{Highest weight representations}
Highest weight representations of the Virasoro algebra can be constructed in much the same way as the well known example of $\mathfrak{su}(2)$ representations for spin. Suppose there exists an eigenstate $|\Delta\rangle$ of the dilatation operator $L_0$,  such that 
\begin{align}
L_0 |\Delta\rangle  = \Delta|\Delta\rangle\, ,
\end{align}
where $\Delta $ is called the conformal dimension or conformal weight of the state. 
From the commutation relations \eqref{eq:Virasoro} we can see that any state $|\chi\rangle = \prod_{n_i} L_{n_i} |\Delta \rangle $, for any set of integers $\{n_i\}$, is also an eigenstate of $L_0$:
\begin{align}
L_0  |\chi \rangle = (\Delta - \sum_{ \{n_i\} } n_i)  |\chi\rangle \label{eq:deseigen}
\end{align}
In particular, the operators $L_{n<0}$ act as raising operators, increasing the value of $\Delta$, while the $L_{n>0}$ play the role of lowering operators.  
If the original eigenstate $|\Delta \rangle$ is such that it is annihilated by all lowering operators, 
\begin{align}
L_n |\Delta \rangle = 0\  \forall n\geq 1
\end{align}
then $| \Delta \rangle$ is called a primary state. A highest weight representation is obtained by acting on a primary state with raising operators, and a Verma module $\mathcal{V}_{\Delta}$ is defined as the largest highest weight representation generated by the primary state $| \Delta \rangle$.
The states obtained by the action of raising operators on the primary state are called its descendants. They are organized into levels given by the value $N = -\sum_{ \{n_i\} }$ in \eqref{eq:deseigen}, which is the difference between the conformal weight of the descendent state and of the primary.  The first three levels of the Verma module $\mathcal{V}_{\Delta}$ are illustrated in the following diagram
\begin{align}
 \begin{tikzpicture}[scale = .3, baseline=(current  bounding  box.center)]
  \draw[-latex, very thick] (12, 0) -- (12, 19) node [left] {$N$};
  \foreach \x in {0, ..., 3}
  {
  \draw [dotted] (-15, {6*\x}) -- (12, {6*\x}) node [right] {${\x}$};
  }
  \node[fill = white] at (0, 0) (0) {$|\Delta\rangle$};
  \node[fill = white] at (-3.6,6) (1) {$L_{-1}|\Delta\rangle$};
  \node[fill = white] at (-7.2, 12) (11) {$L_{-1}^2|\Delta\rangle$};
  \node[fill = white] at (-10.8, 18) (111) {$L_{-1}^3|\Delta\rangle$};
  \node[fill = white] at (0,12) (2) {$L_{-2}|\Delta\rangle$};
  \node[fill = white] at (-4,18) (12) {$L_{-1}L_{-2}|\Delta\rangle$};
  \node[fill = white] at (6,18) (3) {$L_{-3}|\Delta\rangle$};
  \draw[-latex] (0) -- (1);
  \draw[-latex] (1) -- (11);
  \draw[-latex] (11) -- (111);
  \draw[-latex] (2) -- (12);
  \draw[-latex] (0) -- (2);
  \draw[-latex] (0) -- (3);
 \end{tikzpicture}
\end{align}
As is apparent, the number of linearly independent states at level $n \in \mathbb{N}$ is  the number of partitions of $p(n)$.

\subsection{Degenerate representations}\label{sec:nullvec}

Highest weight representations, and in particular Verma modules, are indecomposable representations of the Virasoro algebra. However, they need not be irreducible. Indeed, within the Verma module $\mathcal{V}_{\Delta}$ there could be a descendant state $|\chi \rangle$ which is itself a primary state, such that $|\chi \rangle$ and all its descendants form a subrepresentation of the Virasoro algebra. For each value of the central charge $c$, the existence of such a descendent is only possible if the primary state's conformal weight $\Delta$ takes certain specific values. A descendent state $|\chi\rangle$ that gives rise to a highest weight subrepresentation is called a null vector. 

The commutation relations \eqref{eq:Virasoro} allow us to write any of the lowering operators $L_{n\geq 1}$ in terms of commutators of $L_1$ and $L_2$, so that any state annihilated by $L_1$ and $L_2$ is a primary state. 
Let us find the conditions such that $\mathcal{V}_{\Delta}$ has a null vector at a certain level. 

At level $1$ there is only one descendent, $L_{-1} |\Delta\rangle$. Since $|\Delta \rangle$ is primary, it is only necessary to check that
\begin{align}
L_1 (L_{-1}|\Delta\rangle )&= 2 \Delta |\Delta\rangle = 0\, .
\end{align}
Then, we find that there is a null vector at level $1$ if and only if $\Delta = 0$. 

Level $2$ null vectors impose more interesting conditions, and will be important for the analytic conformal bootstrap method presented in \ref{sec:aboots}. We write a generic level $2$ descendant as $|\chi\rangle = (a_1 L^2_{-1}+ a_2 L_{-2}) |\Delta\rangle$, with $a_1, a_2$ some coefficients to be determined. The state $|\chi\rangle$ is a null vector if
\begin{align}
L_1 |\chi\rangle &=  \left[2(2\Delta +1)a_1 +3a_2\right] L_{-1}|\Delta \rangle = 0 \, , \\
L_2 |\chi\rangle &= \left[6\Delta a_1 + \left(4\Delta + \frac{c}{2}\right)a_2 \right] |\Delta\rangle = 0 \, ,  \label{eq:nulllev2}
\end{align}
which gives a system of equations for the coefficients $a_1, a_2$ that generate $|\chi \rangle$.This system admits a non-trivial solution only if
\begin{align}
\Delta &= \frac{5-c \pm \sqrt{(c-1)(c-25)}}{16}\, . \label{eq:nullD}
\end{align}

This means that, for a given value of $c$, there are two Verma modules with null vectors at level $2$, where the primary state has one of the weights given by \eqref{eq:nullD}. These two representations are in general distinct, but from \eqref{eq:nullD} we see that for the special values of $c = 1$ and $c = 25$ the corresponding values of $\Delta$ coincide. 

The search for null vectors can be continued at higher levels, and it is possible to derive a general condition. In order to express this result, it is useful to introduce new variables $\beta$ and $P$ as alternative parametrizations of the $c$ and $\Delta$, respectively. We write
\begin{align}
c &= 1-6\left( \beta - \frac{1}{\beta} \right)^2 \, ,\quad \Delta = \frac{c-1}{24} + P^2\, ,\label{eq:cdef}\\
\end{align}
where $P$ is called the momentum. Notice that the sign of $P$ in \eqref{eq:cdef} is irrelevant, and we have two options for labelling  the Verma module $\mathcal{V}_{\Delta}$, $\mathcal{V}_{\pm P}$. The transformation that changes the sign of $P$ is called  a reflection, and when applying the conformal bootstrap approach in chapter \ref{ch:cboots} we will look for physical quantities which are reflection invariant. For the moment, we ignore this ambiguity.  

The general result, due to Kac, is that at a given central charge $c$ any Verma module with dimension
\begin{align}
\Delta_{(r,s)} = \Delta(P_{(r,s)})\ ,\quad \text{ with}\  P_{(r,s)} = \frac{1}{2}\left(r \beta - \frac{s}{\beta}\right)\, ,\quad r,s \in \mathbb{N}_{\geq 1} \label{eq:PKac}
\end{align}
has a null vector at level $rs$. In other words, there exists a level $rs$ operator $L_{\langle r,s \rangle}$ such that 
\begin{align}
L_{n\geq 1} \left( L_{\langle r,s \rangle} |\Delta_{{(r,s)}} \rangle \right) = 0\ .
\end{align}
The number of (in principle different) Verma modules with null vectors at a given level is equal to the number of factorisations of the level number as a product of two natural numbers $rs$. 

The cases of level $1$ and $2$ can be expressed in these terms. For he level $1$ null vector we have
\begin{align}
P_{(1,1)} = \frac{1}{2}\left(\beta - \frac{1}{\beta}\right)\, , \quad L_{\langle 1,1 \rangle} = L_{-1}\, .
\end{align}

For the level $2$ null vectors, the two solutions \eqref{eq:nullD} correspond to 
\begin{align}
P = 
\begin{cases}
P_{(2,1)} &= \frac{1}{2}(2\beta - \frac{1}{\beta} )\, ,\\
P_{(1,2)} &= \frac{1}{2}(\beta - \frac{2}{\beta} ) \, ,.
\end{cases}\label{eq:nullP}
\end{align}
Using these values, it is possible to determine the operators $L_{\langle 2,1 \rangle}$ and $L_{\langle 1,2 \rangle}$ that generate the null vectors by computing the coefficients $a_1$ and $a_2$ of equation \eqref{eq:nulllev2}. We have
\begin{align}
L_{\langle 2,1 \rangle} &= \frac{-1}{\beta^2}L^2_{-1} + L_{-2}\, , \quad L_{\langle 1,2 \rangle} = -\beta^2L^2_{-1} + L_{-2}\, . \label{eq:nullop}
\end{align}

To summarize: For a given central charge $c$, the existence of null vectors at a certain level imposes constraints on the conformal weights of the primary states, which must take the values $\Delta_{(r,s)}$. Furthermore, for particular values of the central charge there may be pairs of natural numbers $(r,s)$ and $(r', s')$ such that $\Delta_{(r,s)} = \Delta_{(r',s')}$. This is the case for $c=1$ and $c = 25$ in equation \eqref{eq:nullD}, and it is also a fundamental feature of the Minimal Models, discussed in section \ref{sec:minmods}.

\subsection{Irreducible representations}

We have mentioned that highest weight representations are indecomposable, but could still be reducible. Let us now look what types of irreducible representations we can have. 

At a given $c$, Verma modules $\mathcal{V}_{\Delta}$ with generic values of $\Delta \neq \Delta_{(r,s)}$ are irreducible representations of the Virasoro algebra, since they do not contain any non-trivial sub-representations.

On the other hand, if $\Delta = \Delta_{(r,s)}$, then $\mathcal{V}_{\Delta_{(r,s)}}$ contains a sub-representation of the Virasoro algebra generated by its level $rs$ null vector, $|\chi_{\langle r,s \rangle}\rangle$, whose weight is $\Delta_{(r,s)} +rs$ and which we denote $\mathcal{V}_{|\chi_{(r,s)\rangle}}$. In order to build an irreducible representation we may choose to quotient out all the states in this subrepresentation, effectively setting the null vector (and its descendants) to $0$. In other words, we set
\begin{align}
|\chi_{\langle r,s \rangle}\rangle = L_{\langle r,s \rangle} |\Delta_{(r,s)}\rangle = 0\ . \label{eq:gnlv}
\end{align}
The resulting representation $R_{\langle r,s \rangle}$ is called a degenerate representation, and it can be defined as
\begin{align}
R_{\langle r,s \rangle} &= \frac{\mathcal{V}_{\Delta_{(r,s)}}}{\mathcal{V}_{|\chi_{\langle r,s \rangle}\rangle}}\, . \label{eq:degrep}
\end{align} 
If, in addition, the central charge takes certain rational values, corresponding to minimal models charges, the Verma module $\mathcal{V}_{\Delta_{(r,s)}}$ will contain many different null vectors. This means that there will be many different subrepresentations, and in order to obtain an irreducible representation they should all be quotiented out. For example, if there are two pairs of integers $(r,s)$ and $(r',s')$ such that $\Delta_{(r,s)} =\Delta_{(r',s')} $, then the degenerate representation should be
\begin{align}
R_{\langle r,s \rangle} &= \frac{\mathcal{V}_{\Delta_{(r,s)}}}{\mathcal{V}_{|\chi_{\langle r,s \rangle}\rangle} + \mathcal{V}_{|\chi_{\langle r',s' \rangle}\rangle} } \, , 
\end{align} 
where the sums in the denominator are not direct sums \cite{Ribault:2014hia}.

\section{Consequences of symmetry}\label{sec:FW}
In this section we explore the consequences of the symmetry algebra $\mathfrak{V}\otimes \bar{\mathfrak{V}}$ on a field theory. In particular, we will see how the symmetry algebra determines the structure of the fields, and how the requirement that the observables of the theory be invariant under symmetry transformations constraints correlations functions. We present this content in a way that is oriented towards the conformal bootstrap approach, taking \cite{Ribault:2014hia} as our main reference. 

\subsection{Fields}\label{ssec:FW}
An important axiom of conformal field theories is the existence of the Operator-State correspondence. It means that there is an injective, linear map from the states in the irreducible representations of the symmetry algebra $\mathfrak{V} \otimes \bar{\mathfrak{V}}$  to fields, objects depending on the coordinates $(z, \bar{z})$, and which will be used to construct correlation functions. The consequence is that there are primary and descendent fields, in a similar way as there are primary and descendent states, and the action of the symmetry generators on the fields can be obtained from their action on states\cite{Ribault:2014hia}. When a path integral description of the field theory is available, the operator-state correspondence can be made more explicit, see for example\cite{Simmons-Duffin:2016gjk}.

A primary field is denoted by $V_{\Delta, \bar{\Delta}}(z, \bar{z})$, where $\Delta$ and $\bar{\Delta}$ are the left-moving and right-moving conformal weights, respectively. The field  $V_{\Delta, \bar{\Delta}}(z, \bar{z})$ is defined to satisfy
\begin{align}
L^{(z)}_0 V_{\Delta, \bar{\Delta}}(z, \bar{z}) = \Delta V_{\Delta, \bar{\Delta}}(z, \bar{z}) \, , \quad
L^{(z)}_{n\geq 1} V_{\Delta, \bar{\Delta}}(z, \bar{z}) = 0\, , \label{eq:primfield}
\end{align} 
along with analogous relations for the antiholomorphic operators. 

Here, the Virasoro generators $L_n$ carry a superindex $(z)$ indicating the point on which they are defined. 
In order to determine the dependence of the fields on $(\zb{z})$, we identify the operator $L^{(z)}_{-1}$ with the generator of translations, by defining
\begin{align}
L_{-1}^{(z)}V_{\zb{\Delta}}(\zb{z}) = \partial_{z} V_{\zb{\Delta}}(\zb{z})\, , \quad \bar{L}_{-1}^{(\bar{z})}V_{\zb{\Delta}}(\zb{z}) = \partial_{\bar{z}} V_{\zb{\Delta}}(\zb{z}).\label{eq:Ldrv}
\end{align}

Fields are defined to be smooth everywhere on the Riemann sphere, except at points where another field is inserted. In that case, their behaviour will be controlled by the Operator product expansion, described in section \ref{sec:OPE}.

\subsubsection{Spectrum}
The collection of all fields belonging to a certain theory forms the theory's spectrum. This spectrum can be expressed as a sum of the representations of the symmetry algebra  $\mathfrak{V} \otimes \bar{\mathfrak{V}}$ associated with each field, i.e.
\begin{align}
\mathbb{S} = \bigoplus_{\Delta, \bar{\Delta}} M_{\Delta, \bar{\Delta}} \mathcal{R}_{\Delta}\otimes \bar{\mathcal{R}}_{\bar{\Delta}}\, \label{eq:genspec}
\end{align}
where the $\mathcal{R}_{\Delta}$ stands for some irreducible representation of $\mathfrak{V}$, (regardless if it is degenerate or not), and the numbers $M_{\Delta, \bar{\Delta}}$ are multiplicities determining how many copies of a given representation appear.  

Theories may be classified according to certain characteristics of their spectrums, as we have discussed while reviewing known theories in section \ref{sec:cmap}. If the spectrum is finite, the corresponding theory is called rational, and non-rational if it has an infinite spectrum. Non-rational theories spectrums can be continuous, as in Liouville theory, or discrete, as in the generalized minimal models. On the other hand, a theory is called diagonal if both the holomorphic and antiholomorphic representations appearing on each term of \eqref{eq:genspec} are the same, and non-diagonal in the opposite case. 

\subsection{Energy-momentum tensor} \label{emtensor}

The $(\zb{z})$-dependence of the Virasoro generators $L_n$ is also controlled by equation \eqref{eq:Ldrv}. Applying this equation to the field $L^{(z)}_{n} V_{\zb{\Delta}}(\zb{z})$, and using the commutation relations \eqref{eq:Virasoro}, we find
\begin{align}
\partial_z L^{(z)}_{n} = -(n+1) L^{(z)}_{n-1}\, . \label{eq:lprime}
\end{align}
This means that operators acting at different points are linearly related to one another or,  in other words, that generators defined at different points are different basis for the same symmetry algebra $\mathfrak{V}$. Of course, an analogous relation holds for the right-moving generators.

We may encode the action of the Virasoro generators into the fields $T(y)$ and $\bar{T}(\bar{y})$, defined by the mode decompositions 
\begin{align}
T(y) = \sum_{n \in \mathbb{Z}} \frac{L^{(z)}_n}{(y-z)^{n+2} }\, ,\quad  \bar{T}(\bar{y}) = \sum_{n \in \mathbb{Z}} \frac{\bar{L}^{(\bar{z})}_n}{(\bar{y}-\bar{z})^{n+2} }\, , \label{eq:Tmodes}
\end{align}
where the series is assumed to converge as $(y, \bar{y}) \to (z,\bar{z})$.
Notice that equation \eqref{eq:lprime} implies $\partial_{z}T(y) = \partial_{\bar{z}}\bar{T}(\bar{y}) = 0$, and we can write the expansions \eqref{eq:Tmodes} around any point of interest.  

Let us now focus on left-moving quantities, knowing that there are analogous properties for the right-moving ones. In the same way as the fields, $T(y)$ should be regular everywhere in the Riemann sphere where no other field is present. This includes the point at infinity, and means that $T(y)$ has the following behaviour \cite{Ribault:2014hia, fms97} :
\begin{align}
T(y)\underset{y\to \infty}{\to} O\left(\frac{1}{y^4}\right)\ . \label{eq:Tinf}
\end{align}

If $T(y)$ approaches a primary field $V_{\zb{\Delta}}(\zb{z})$ inserted at point $z$, the mode expansion \eqref{eq:Tmodes} gives
\begin{align}
T(y)V_{\zb{\Delta}}(\zb{z}) = \frac{\Delta V_{\zb{\Delta}}(\zb{z})}{(y-z)^2} +\frac{\partial_z V_{\zb{\Delta}}(\zb{z})}{(y-z)}+ \dots \, , \label{eq:TV}
\end{align}
where $\dots $ includes all the regular terms in the expansion. These terms are omitted because the expansion usually appears inside a contour integration around $z$, and only the singular terms contribute. 

The expansion \eqref{eq:TV} is an example of an Operator Product Expansion (OPE), which will be discussed in greater detail in section \ref{sec:OPE}. Equation \eqref{eq:TV} provides an alternative definition of a primary field,  as one for which the most singular term of its OPE with $T(y)$ is of order $(y-z)^{-2}$. The coefficient of this term is the conformal weight of the field. 

Another important example is the product $T(y)T(z)$: This case is equivalent to the Virasoro commutation relations \eqref{eq:Virasoro}, and it gives
\begin{align}
T(y)T(z) = \frac{c}{2}\frac{1}{(y-z)^4} + \frac{2T(z)}{(y-z)^2}  + \frac{\partial_z T(z)}{(y-z)}  + \dots \, ,  \label{eq:TT}
\end{align}
where $c$ is the Virasoro central charge. This equation shows that $T(y)$ is not a primary field, because of the non-zero coefficient at order $(y-z)^{-4}$.

$T(y)$ and $\bar{T}(y)$ are interpreted as the non-zero components of the stress-energy tensor\cite{fms97}, and from them we can recover the conserved charges (i.e. the symmetry generators) by means of Cauchy's integral formula. Integrating $T(y)$ around a closed contour we have
\begin{align}
L_n = \frac{1}{2i\pi } \oint_z dy (y-z)^{n+1}T(y) \, . \label{eq:modinv}
\end{align}
This identity, and its right-moving analogue, allow us to compute the effects of conformal transformations on correlation functions. in the following section we will use this identity to find the constraints that conformal symmetry imposes on correlation functions. 

\subsection{Ward Identities}\label{sec:ward}

One of the most important advantages of having conformal symmetry are the constraints it imposes on the observables of the theory, the correlation functions of fields. These constraints take the form of differential equations for the correlation functions, called Ward Identities.  

We write an $N$-point correlation function of (not necessarily primary) fields $V_i(\zb[i]{z})$ as
\begin{align}
\left\langle \prod_{i=1}^{N} V_{i}(\zb[i]{z}) \right\rangle\, , \quad z_i \neq z_j\ ,\ . \forall i\neq j \label{eq:genNpoint}
\end{align}
This a function of the fields' conformal weights, of the positions $(\zb[i]{z})$, and of the central charge $c$. The correlation function is linear on the fields, and its dependence on the fields ordering is given by their commutation relations, i.e. it is independent of the ordering if fields commute, but it can gain a sign if there are anti-commuting fields. 

We consider an infinitesimal conformal transformation given by a function $\epsilon(z)$, of the type of equation \eqref{eq:expeps}.  We can study its influence on a $N$-point correlation function $\langle \prod_{i=1}^{N} V_i(\zb[i]{z}) \rangle$ by inserting the energy momentum tensor at point $z\neq z_i$.  Since $T(z)$ is holomorphic, the $(N+1)$-point correlation function with $T(z)$ inserted is analytic in $\mathbb{C}-\{z_1, \dots, z_N\}$, and it's behaviour at infinity is controlled by equation \eqref{eq:Tinf}. 
Integration around a closed contour around $z= \infty$ gives the Ward identities
\begin{align}
\oint_{\mathcal{C}} dz \epsilon(z) \left\langle T(z) \prod_{i=1}^{N} V_i(\zb[i]{z}) \right\rangle = 0\ ,
\end{align}
provided the behaviour of $\epsilon(z)$ is of the type
\begin{align}
\epsilon(z) \underset{z\to \infty}{\sim } O(z^2)\ ,
\end{align}. The integral around the contour can be decomposed as a sum over smaller contours, each one going around one point $z_i$. Then, the general Ward identity reads
\begin{align}
\sum_{i = 1}^{N}\oint_{\mathcal{C}_i} dz  \left\langle V_1(\zb[1]{z})\dots \epsilon(z)T(z)V_{i}(\zb[i]{z})  \dots V_{N}(\zb[N]{z}) \right\rangle = 0  \, , \label{eq:Ward}
\end{align}
where the product $\epsilon(z)T(z)V_{i}(\zb[i]{z})$ is controlled by the expansion \eqref{eq:TV}. The identity \eqref{eq:Ward} has an antiholomorphic analogue, obtained by inserting $\bar{T}(\bar{z})$, and correlation functions are subject to both sets of constraints. 

Depending on the choice of $\epsilon(z)$ in \eqref{eq:Ward} we will obtain different types constraints. If $\epsilon(z) = \{1,\ z,\ z^2 \}$ the conformal transformation it generates is globally well defined, and we speak of global Ward Identities. If this is not the case $\epsilon(z)$ will have poles, and the transformation will be conformal only locally, giving rise to local Ward identities. 

The importance of local Ward identities is that they make it possible to compute correlation functions involving descendent fields in terms of correlation functions of their primaries. For example, consider an $N$-point correlation function of $(N-1)$ primary fields and a level $n\geq 2$ descendent of $V_{\zb[1]{\Delta}}(\zb[1]{z})$ generated by the operator $L^{(z_1)}_{-n}$. Making use of equation \eqref{eq:modinv}, we write this descendent by choosing $\epsilon(z) = (z-z_1)^{-(n-1)}$, 
\begin{align}
L^{(z_1)}_{-n}V_{\zb[1]{\Delta}}(\zb[1]{z}) = \frac{1}{2 \pi i} \oint_{z_1} dz \frac{T(z)V_{\zb[1]{\Delta}}(\zb[1]{z})}{(z-z_1)^{n-1}}\, .
\end{align}
Then, using the identity \eqref{eq:Ward} for this same $\epsilon(z)$ we obtain,
\begin{multline}
\left\langle \left( L^{(z_1)}_{-n} V_{\zb[1]{\Delta}}(\zb[1]{z}) \right) \prod_{i = 2}^{N} V_{\zb[i]{\Delta}}(\zb[i]{z}) \right\rangle =\\ \sum_{i=2}^{N} \left(\frac{-1}{(z_i-z_1)^{n-1}} \partial_{z_i} + \frac{(n-1)}{(z_i-z_1)^n}\Delta_i \right) \left\langle  \prod_{j = 1}^{N} V_{\zb[j]{\Delta}}(\zb[j]{z}) \right\rangle\, ,  \label{eq:locWard}
\end{multline}
where we explicitly see how the action of the creation operator $L^{(z_1)}_{-n}$ is expressed by a linear combination of differential operators acting on the correlation function of primary fields. 

Since it is possible to express any $N$-point function in terms of correlation functions of primary fields, we focus on these functions only, and study the constraints imposed by global Ward identities. 
For $\epsilon(z) = 1,\ z,\ z^2$, (corresponding to the generators $L_{-1}, L_0, L_1$),  a correlation function $Z = \left\langle \prod_{i=1}^{N} V_{\zb[i]{\Delta}}(\zb[i]{z}) \right\rangle $ of primary fields satisfies
\begin{equation}
\begin{aligned}
\epsilon(z) = 1 \to \quad &\sum^{N}_{i=1} \partial_{z_i} Z = 0\ , \\
\epsilon(z) = z \to \quad &\sum^{N}_{i=1} \left(z_i \partial_{z_i}+\Delta_i \right) Z = 0\, , \\
\epsilon(z) = z^2 \to \quad &\sum^{N}_{i=1} \left( z^2_i\partial_{z_i} + 2z_i\Delta_i \right) Z = 0, \label{eq:globWard}
\end{aligned}
\end{equation}
along with the corresponding antiholomorphic equations. 
These equations control how correlation functions behave under global conformal transformations. In section \ref{sec:2dsym} we mentioned that these transformations form the group $PSL(2,\mathbb{C})$. If we perform a transformation of this type on the holomorphic and anti-holomorphic sides, global Ward identities allow us to deduce the behaviour of the correlation function
\begin{multline}
\left\langle \prod_{i=1}^{N} V_{\zb[i]{\Delta}}(\zb[i]{z}) \right\rangle =
 \prod_{i=1}^N (cz_i + d)^{-2\Delta_i}(\bar{c}\bar{z}_i + \bar{d})^{-2\bar{\Delta}_i} \\ \times \left\langle \prod_{i=1}^{N} V_{\zb[i]{\Delta}}\left( \frac{az_i+b}{cz_i+d},\frac{\bar{a}\bar{z}_i+\bar{b}}{\bar{c}\bar{z}_i+\bar{d}} \right) \right\rangle\ . \label{eq:globT}
\end{multline}

This can also be thought of as a transformation law for the fields, and in particular we can study their behaviour under specific transformations. Let us discuss three particular cases:
\begin{itemize}
\item Dilatation, $(z, \bar{z}) \to (\lambda z, \lambda\bar{z} ) $.

This transformation is described by the $PSL(2, \mathbb{C})$ matrix $\left(\begin{smallmatrix}
\sqrt{\lambda} & 0 \\ 0 & \frac{1}{\sqrt{\lambda}}
\end{smallmatrix}\right) $, with $\lambda \in \mathbb{R}_{>0}$ . Equation  \eqref{eq:globT} gives
\begin{align}
V_{\zb{\Delta}}(\zb{z}) \to \lambda^{\Delta + \bar{\Delta}}  V_{\zb{\Delta}}(\lambda z, \lambda \bar{z})\, ,
\end{align}
The number $\Delta + \bar{\Delta}$ controls the behaviour of the field under dilatations, and it is called the conformal dimension of the field. 

\item Rotation, $(z, \bar{z}) \to (e^{i\theta} z, e^{-i\theta} \bar{z} ) $.

The left and right moving matrices giving the transformation are $\left( \begin{smallmatrix}
e^{i\frac{\theta}{2}} & 0 \\ 0 & e^{-i\frac{\theta}{2}}
\end{smallmatrix} \right) $ and $\left( \begin{smallmatrix}
e^{-i\frac{\theta}{2}} & 0 \\ 0 & e^{i\frac{\theta}{2}}
\end{smallmatrix} \right) $. Then
\begin{align}
V_{\zb{\Delta}}(\zb{z}) \to e^{-i\theta (\Delta- \bar{\Delta})}V_{\zb{\Delta}}(e^{i\theta} z, e^{-i\theta} \bar{z} )\, ,
\end{align}
In this case we find that the difference between the left-and-right conformal weights determines the transformation of the field. This quantity is called the conformal spin, \begin{align}
S= \Delta - \bar{\Delta}\, .
\end{align}

\item Inversion, $(z, \bar{z}) \to (\frac{-1}{z}, \frac{-1}{\bar{z}})$.

This transformation has the particularity of mapping the origin to the point at infinity, and it allows us to define a field at $z = \infty$. The matrix describing this inversion is $\left(\begin{smallmatrix}
0 & 1 \\ -1 & 0
\end{smallmatrix}\right) $, and we get
\begin{align}
V_{\zb{\Delta}}(\zb{z}) \to (-z)^{-2\Delta}(-\bar{z})^{-2\bar{\Delta}} V_{\zb{\Delta}}(\tfrac{-1}{z},\tfrac{-1}{\bar{z}}) \, .
\end{align}
Thus,in order to have a field well-behaved at $z = \infty$  we define 
\begin{align}
V_{\zb{\Delta}}(\infty) = \lim_{z\to\infty}z^{2\Delta}\bar{z}^{2\bar{\Delta}} V_{\zb{\Delta}}(\zb{z})\ . \label{eq:finf}
\end{align}
\end{itemize}

Next, we can solve the global Ward identities \eqref{eq:globWard} in order to find how they constrain the correlation functions. For a one-point function $\left\langle V_{\zb{\Delta}}(\zb{z}) \right\rangle$, global Ward identities imply that the function vanishes identically unless $\Delta = 0$, in which case it is constant. 

In the case of a a two-point function, the solution to equations \eqref{eq:globWard} is
\begin{align}
\left\langle V_{\zb[1]{\Delta}}(\zb[1]{z}) V_{\zb[2]{\Delta}}(\zb[2]{z}) \right\rangle = B(V_1) \frac{\delta_{\Delta_1, \Delta_2}\delta_{\bar{\Delta}_1, \bar{\Delta}_2}}{z_{12}^{2\Delta_1}\bar{z}_{12}^{2\bar{\Delta}_1}}\, , \label{eq:gen2point}
\end{align}
where $z_{ij}=(z_i - z_j)$ and we have introduced the $(\zb{z})$-independent \textbf{two-point structure constant} $B(V)$. The $\delta_{\Delta_1, \Delta_2}$ factors indicate that the two point function can only be non-zero if the conformal weights of both fields are equal. 

For three-point functions of primary fields we have
\begin{align}
\left\langle \prod_{i=1}^{3} V_{\zb[i]{\Delta}}(\zb[i]{z}) \right\rangle = C_{123}\left|\mathcal{F}^{(3)}(\Delta_1,\Delta_2,\Delta_3|z_1,z_2,z_3)\right|^2  \label{eq:gen3point}
\end{align}
where we introduced the \textbf{three-point structure constant} $C_{123}= C(V_1, V_2, V_3)$, which is independent of $(\zb[i]{z})$, and the function 
\begin{align}
\mathcal{F}^{(3)}(\Delta_1,\Delta_2,\Delta_3|z_1,z_2,z_3) = 
 z_{12}^{-\Delta_1 - \Delta_2 + \Delta_3}
z_{23}^{-\Delta_2 - \Delta_3 + \Delta_1}  z_{31}^{-\Delta_3 - \Delta_1 + \Delta_2} \, , \label{eq:3pbl}
\end{align}
which is called the three-point conformal block. 
The modulus squared notation in the three-point function \eqref{eq:gen3point} is short for the product of the holomorphic and anti-holomorphic parts, where the anti-holomorphc three-point conformal block is obtained by replacing $z_i \to \bar{z}_i$ and $\Delta_i \to \bar{\Delta_i}$ in the definition \eqref{eq:3pbl}.

We can use the explicit form \eqref{eq:gen3point} of the three-point function to determine the properties of the structure constant $C_{123}$ under permutations of the fields. We mentioned that correlation functions depend on the ordering through the fields commutation properties, such that under a permutation $\sigma$ we have,
\begin{align}
\left\langle \prod_{i=1}^{3} V_{\zb[i]{\Delta}}(\zb[i]{z}) \right\rangle = \eta_{123}(\sigma) \left\langle \prod_{i=1}^{3} V_{\zb[\sigma(i)]{\Delta}}(\zb[\sigma(i)]{z}) \right\rangle \label{eq:3pperm}
\end{align} 
where the factor $\eta_{123}(\sigma)$ is 
\begin{align}
\eta_{123}(\sigma) = \begin{cases}
-1 &\text{if } \sigma \text{ exchanges anticommuting fields, } \\
1 &\text{else.}
\end{cases}
\end{align}
The permutation properties of the three-point structure constant should ensure that equation \eqref{eq:3pperm} is satisfied. Using the explicit expression  \eqref{eq:gen3point} we can find the factor arising from the permutation of the arguments of the three-point conformal blocks, and find that $C_{123}$ behaves as
\begin{align}
\boxed{\frac{C_{\sigma(1)\sigma(2)\sigma(3)}}{C_{123}} = \eta_{123}(\sigma )\,\text{sgn}(\sigma)^{S_1 + S_2 + S_3}} \, .\label{permC}
\end{align}

The full $(\zb{z})$-dependence of two-and-three-point functions is completely fixed by global  conformal symmetry, and it is common to any CFT with symmetry algebra $\mathfrak{V} \otimes \bar{\mathfrak{V}}$ .
In this context, the structure constants $B(V)$ and $C_{123}$ appearing in the correlation functions \eqref{eq:gen2point} and \eqref{eq:gen3point} can be seen as fundamental theory-dependent objects that need to be determined in order to compute correlation functions. An important part of solving any CFT is determining its structure constants. 

The fact that two-and-three-point correlation functions are fixed by symmetry can be understood if we recall that global conformal transformations \eqref{eq:mob} have three degrees of freedom, allowing one to fix the position of any three points by applying a conformal mapping. Beyond three-point functions it is possible to construct conformally invariant cross-ratios, and thus the $(\zb{z})$ dependence of correlation functions can no longer be fixed. For an $N$-point function we can fix the positions $z_N$, $z_{N-1}$ and $z_{N-3}$, and be left out with $(N-3)$ invariant cross-ratios,
\begin{align}
x_i = \frac{(z_i-z_{N-2})(z_{N-1}-z_{N})}{(z_i-z_{N-1})(z_{N-2}-z_{N})}\ ,
\end{align} 
on which correlation functions can depend arbitrarily. The same is true for the right-moving coordinates $\bar{z}$. 

The four-point functions are the first ones whose $(\zb{z})$ dependence is not fixed by symmetry. Focusing on the left-moving side, there is only one invariant cross-ratio on each sector
\begin{align}
x = \frac{z_{12}z_{34}}{z_{13}z_{24}}\, ,\quad \bar{x} = \frac{\bar{z}_{12}\bar{z}_{34}}{\bar{z}_{13}\bar{z}_{24}} \label{eq:cratio}
\end{align} 
and global Ward identities give
\begin{multline}
\left\langle V_{\zb[1]{\Delta}}(\zb[1]{z})V_{\zb[2]{\Delta}}(\zb[2]{z})V_{\zb[3]{\Delta}}(\zb[3]{z})V_{\zb[4]{\Delta}}(\zb[4]{z}) \right\rangle \\= \left|\mathcal{F}^{(3)}(\Delta_1,\Delta_2,0|z_1,z_2,z_3) \mathcal{F}^{(3)}(\Delta_3,\Delta_4,0|z_3,z_4,z_2)\right|^2 F(x,\bar{x}) \, .
\end{multline}
Here $F(x, \bar{x})$ is a function of the invariant cross-ratios, and it is related to a four-point function with three points fixed by
\begin{align}
 F(x) =\left| x^{\Delta_1+\Delta_2} \right|^2 \left\langle V_{\Delta_1,\bar\Delta_1}(\zb{x})V_{\Delta_2,\bar\Delta_2}(0)V_{\Delta_3,\bar\Delta_3}(\infty) V_{\Delta_4,\bar\Delta_4}(1)\right\rangle \ , 
\label{fx}
\end{align}
where we have abbreviated the arguments of the fields with fixed points.
 
In order to compute the special functions $F(x, \bar{x})$ we will profit from the Operator Product Expansions. We have already encountered two examples of OPE, involving the stress-energy tensor, in expressions \eqref{eq:TV} and \eqref{eq:TT}. It is possible to build such an expansion between any two fields of a CFT, and inserting it into the four-point correlation functions will allow us to simplify their calculation. In sections \ref{sec:OPE} and \ref{sec:degfields} we focus on building the Operator Product Expansions, and in section \ref{sec:cblocks} we discuss how they can be used to compute four-point correlation functions. 

\subsection{Operator product expansion}\label{sec:OPE}

In conformal field theories, products of fields inserted at two nearby points can be expanded as a sum over a certain set of fields, both primary and descendants. This is called an Operator Product Expansion, or OPE, and we have encountered previous examples of them in equations \eqref{eq:TT} and \eqref{eq:TV}. In this section we will use Ward identities in order to constrain the general structure of these expansions. 

We begin by writting an OPE between two primary fields $V_{\zb[1]{\Delta}}(\zb[1]{z})$ and $V_{\zb[2]{\Delta}}(\zb[2]{z})$ in a generic way. Denoting by $V_{\sigma_3}(\zb[2]{z})$ a generic field, not necessarily primary, we have
\begin{align}
V_{\zb[1]{\Delta}}(\zb[1]{z})V_{\zb[1]{\Delta}}(\zb[2]{z}) = \sum_{\sigma_3 } C^{\sigma_3}_{12}(\zb[i]{z})  V_{\sigma_3}(\zb[2]{z})\, .\label{eq:genOPE1}
\end{align} 
where we introduced the OPE coefficients $C^{\sigma_3}_{12}(\zb[i]{z})$. The set of fields $V_{\sigma_3}$ over which the sum is performed is left undetermined, but knowing its precise form is not necessary for finding general constraints on the OPE structure. 

Expansions such as \eqref{eq:genOPE1} should be thought of as being valid only when inserted into a correlation function, which we omit in order to lighten the notation. For this reason, the dependence of the OPE on the ordering of the fields on the LHS of \eqref{eq:genOPE1} is the same as that of correlation functions, discussed below equation \eqref{eq:genNpoint}. Another important property of the OPEs is that they are associative. This will give rise to self-consistency conditions which, combined with the results of this section, can be used to determine the OPE coefficients. We discuss these constraints in section \ref{sec:crossing}. 

Let us find the constraints that symmetry imposes on the OPE structure. In way similar to the derivation of the local Ward identities \eqref{eq:locWard}, we can find the action of operators $L_{n}$ on the OPE \eqref{eq:genOPE1} by performing a contour integration and using the mode decomposition \eqref{eq:Tmodes} of the stress-energy tensor. We act with $\oint_{\mathcal{C}} dz (z-z_2)^{n+1}T(z)$ on both sides of equation \eqref{eq:genOPE1}, where we take $n\geq -1$ and the contour $\mathcal{C}$ goes around both $z_1$ and $z_2$. We write only the left-moving identities, knowing that all steps below can be performed also on the right-moving sector. 
The identities we obtain read
\begin{multline}
 \left( L_n^{(z_2)}+\sum_{m=-1}^{n}\binom{m+1}{n+1} z_{12}^{n-m}L_{m}^{(z_1)}\right)V_{1}(\zb[1]{z})V_{2}(\zb[2]{z}) =\\ \sum_{\sigma_3} C_{1,2}^{\sigma_3}(\zb[i]{z}) L^{(z_2)}_n V_{\sigma_3}(\zb[2]{z})\ . \label{eq:WardOPE}
\end{multline}
Inserting now the generic form \eqref{eq:genOPE1} of the OPE into the left-hand side of the identity \eqref{eq:WardOPE}, we obtain equations that determine the behavoiour of the OPE coefficients. For $n = -1, 0$, equation \eqref{eq:WardOPE} results in
\begin{multline}
n= -1:\ \left( \partial_{z_1} + \partial_{z_2} \right) C^{\sigma_3}_{12}\left( (\zb[1]{z}), (\zb[2]{z}) \right) = 0 \\ \Rightarrow C^{\sigma_3}_{12}\left( (\zb[1]{z}), (\zb[2]{z}) \right) = C^{\sigma_3}_{12}\left( \zb[12]{z} \right)\, ,
\end{multline}
\begin{multline}
n = 0: z_{12} \partial_{z_{12}} C^{\sigma_3}_{12}\left( \zb[12]{z} \right) = (\Delta_{\sigma_3} - \Delta_1 - \Delta_2 ) C^{\sigma_3}_{12}\left( \zb[12]{z} \right)\, . 
\end{multline}
Using also the right-moving identities, we find that the OPE coefficients satisfy
\begin{align}
C^{\sigma_3}_{12}\left( \zb[12]{z} \right) = C^{\sigma_3}_{12} \left|z^{\Delta_{\sigma_3} - \Delta_1 - \Delta_2 }_{12} \right|^{2}\, ,  \label{eq:coz}
\end{align}
where now $C^{\sigma_3}_{12}$ is independent of the positions, an the $(\zb[12]{z})$-dependence of the OPE terms has been determined. 

In section \ref{sec:ward} we showed how Ward identities can be used to relate correlation functions of descendent fields with correlation functions of primary field. let us now do the same with the OPE coefficients $C^{\sigma_3}_{12}$. First of all, since the fields of a CFT are organised into primaries and descendants, we can write the summation over the fields $V_{\sigma_3}$ as a double summation, first over primary fields, and then over all its descendants. Explicitly, we write
\begin{align}
\sum_{\sigma_3} = \sum_{\Delta_3} \sum_{L \in \mathcal{L}}\ ,
\end{align}
where $\mathcal{L}$ some basis for the creation operators $L$ generating the descendants of $V_{\sigma_3}$.
If we consider all terms originating from the primary field $V_{\zb[3]{\Delta}}$, we can write identities \eqref{eq:WardOPE} with $n \geq 1$ as
\begin{multline}
\sum_{L \in \mathcal{L}} C^{L|\Delta_3\rangle}_{12} \left(\Delta_3 + |L| + n\Delta_1 - \Delta_2  \right) z_{12}^{|L|+n} L^{(z_2)}V_{\zb[3]{\Delta}}(\zb[2]{z}) = \\ \sum_{L' \in \mathcal{L}} C^{L'|\Delta_3\rangle}_{12}  z^{|L'|}_{12} L^{(z_2)}_n L'^{(z_2)}V_{\zb[3]{\Delta}}(\zb[2]{z})\ , \label{eq:LWCds}
\end{multline}
where $|L|$ is the level of the operator $L$, and $C^{L|\Delta_3\rangle}_{12}$ is the OPE coefficient associated with the descendent $LV_{\zb[3]{\Delta}}$. We propose that these structure constants are related to the ones involving primaries by
\begin{align}
 C^{L|\Delta_3\rangle}_{12} 
 = 
 C^{3}_{12} f^{\Delta_3,L}_{\Delta_1,\Delta_2}\ ,\label{eq:COPEf}
\end{align}
where  $f^{\Delta_3,L}_{\Delta_1,\Delta_2}$ are certain universal coefficients. These coefficients can be determined from equations \eqref{eq:LWCds}, by comparing terms with the same powers of $z_{12}$. In order to make this comparison, we first note that the summation over the descendants can be organised level by level, i.e. 
\begin{align}
\sum_{L \in \mathcal{L}} = \sum_{N = 0}^{\infty}\; \sum_{L/ |L| = N }\, .
\end{align}
Then we note that in equation \eqref{eq:LWCds}, terms on the left-hand side at level $|L| = N-n$, with $N\geq n$, produce the same $z_{12}$ powers as terms at level $|L'| = N$ on the right-hand side.  Using equation \eqref{eq:COPEf} we get
\begin{multline}
\sum_{|L| = N-n} f^{\Delta_3,L}_{\Delta_1,\Delta_2} (N-n + \Delta_3+\Delta_1 - \Delta_2) L^{(z_2)}V_{\zb[3]{\Delta}}(\zb[2]{z}) = \\ \sum_{|L'| = N} f^{\Delta_3,L'}_{\Delta_1,\Delta_2} L^{(z_2)}_n L'^{(z_2)}V_{\zb[3]{\Delta}}(\zb[2]{z}) \, . \label{eq:feqs}
\end{multline}
By choosing different values of $(N, n)$, with $N\geq n$, we can use the system \eqref{eq:feqs} to determine the factors $f^{\Delta_3, L}_{\Delta_1, \Delta_2}$. 

Equation \eqref{eq:coz} determines the generic $(\zb{z})$-dependence of the terms in the OPE \eqref{eq:genOPE1}, and the system \eqref{eq:feqs} allows us to write all OPE coefficients in terms those involving only primary fields. We can take one further step and relate the OPE coefficients of primary fields, $C^3_{12}$, with the two-and-three-point structure constants introduced in section \ref{sec:ward}. With our results so far, along with their corresponding right-moving analogues, we can write an OPE of primary fields as
\begin{multline}
 V_{\Delta_1,\bar\Delta_1}(\zb[1]{z}) V_{\Delta_2,\bar\Delta_2}(\zb[2]{z}) 
 = \\
 \sum_{(\Delta_3,\bar\Delta_3)} C^{3}_{12}
 \left| z_{12}^{\Delta_3-\Delta_1-\Delta_2}\right|^2 \Big( V_{\Delta_3,\bar\Delta_3}(\zb[2]{z}) + O(z_{12})\Big)\ ,
\end{multline}
where we are keeping only the lowest powers of $z_{12}$. Inserting this expansion into a three-point function, and using the two-point function \eqref{eq:gen2point}, we find  
\begin{align}
C^3_{12}= \frac{C_{123}}{ B_3 }\ , 
 \label{cftt}
\end{align}
where $B_3 = B(V_3)$ is the two-point structure constant of equation \eqref{eq:gen2point}. We see then that OPE coefficients can be written in terms of structure constants, which emphasizes the role of these structure constants as key objects for any CFT. 
Combining the results of these section we can give a more precise expression for a general OPE of two primary fields, 
\begin{multline}
 V_{\Delta_1,\bar\Delta_1}(\zb[1]{z}) V_{\Delta_2,\bar\Delta_2}(\zb[2]{z}) 
 = \\
 \sum_{\Delta_3,\bar\Delta_3} \frac{C_{123}}{B_3} \left| z_{12}^{\Delta_3-\Delta_1-\Delta_2}\sum_{L\in\mathcal{L}} z_{12}^{|L|} f^{\Delta_3,L}_{\Delta_1,\Delta_2} L \right|^2 V_{\Delta_3,\bar\Delta_3}(\zb[2]{z})
 \ .
 \label{eq:vvs}
\end{multline}

While the general form of the OPE between two fields is determined by Ward identities, we have in general no information about which fields (or conformal families of fields) appear in the expansion.The rules that determine which fields appear in the OPE spectrum $\mathbb{S}_{12}$ are called the fusion rules of the theory, and they are often not easy to determine. However, if one of the fields involved in the OPE is a degenerate field (i.e. it is associated with the degenerate representations discussed in section \ref{sec:nullvec}), then it becomes possible to find the terms contributing to the OPE. We now discuss this  particular case.

\subsubsection{Degenerate fields and OPEs}\label{sec:degfields}

We define a degenerate field as the field corresponding to the primary state of one of the degenerate representations $R_{\langle r,s\rangle}$ defined in \eqref{eq:degrep}. Such a field is denoted by $V_{\langle r,s \rangle}(\zb{z})$, and its conformal weights are given by equation \eqref{eq:PKac} $\Delta = \bar{\Delta} = \Delta_{(r,s)}$. For the moment we focus on the left-moving properties of this field, which will serve as a starting point for treating the full picture in section \ref{sec:degfr}.

In addition to the primary field equations \eqref{eq:primfield}, there is an operator $L_{\langle r,s \rangle}$ at level $rs$ such that $V_{\langle r,s \rangle}(\zb{z})$ satisfies a relation analogous to condition \eqref{eq:gnlv}
\begin{align}
L^{(z)}_{\langle r,s \rangle} V_{\langle r,s \rangle}(\zb{z}) = 0\, . \label{eq:nullfield}
\end{align}

Equation \eqref{eq:nullfield} expresses that the field $V_{\langle r,s \rangle}$ has a null descendant, and this brings extra constraints on the correlation functions. Indeed, in addition to the Ward identities, correlation functions involving degenerate fields satisfy the so-called Belavin-Polyakov-Zamolodchikov equations \cite{Belavin:1984vu}, or BPZ equation, given by
\begin{align}
\left\langle  L^{(z)}_{\langle r,s \rangle}V_{\langle r,s \rangle}(\zb{z}) \prod_{i=1}^{N}V_{i}(\zb[i]{z}) \right\rangle = 0 \, . \label{eq:BPZL}
\end{align}
By means of the Local Ward identities, the operator $L^{(z)}_{\langle r,s \rangle}$ can be expressed as an order $rs$ differential operator, and the constraint \eqref{eq:BPZL} turns into a differential equation for the correlation functions. 
Let us think of a three-point function involving a degenerate field, of the type 
\begin{align}
\left \langle V_{\langle r,s \rangle}(\zb{z}) V_{\zb[1]{\Delta}}(\zb[1]{z}) V_{\zb[2]{\Delta}}(\zb[2]{z}) \right\rangle \, .
\end{align}
Solutions to the BPZ equations for this three-point functions will impose constraints on the dimensions $\Delta_1$, $\Delta_2$, constraining the values of the dimensions appearing in the OPE $V_{\langle r,s \rangle}(\zb{z}) V_{\zb[1]{\Delta}}(\zb[1]{z})$. Let us show how this works in the case of the fields $V_{\langle 1,1 \rangle}$, $V_{\langle 2,1 \rangle}$ and $V_{\langle 1,2 \rangle}$, corresponding to the level $1$ and $2$ null vectors found in section \ref{sec:nullvec}.

\begin{itemize}
\item $V_{\langle 1,1 \rangle}$

The BPZ operator is $L_{\langle 1,1 \rangle } = L_{-1} = \partial_{z}$, and the equation for the three-point function reads
\begin{align}
\partial_z \left\langle V_{\langle 1,1 \rangle }(\zb{z})V_{\zb[1]{\Delta}}(\zb[1]{z})V_{\zb[2]{\Delta}}(\zb[2]{z}) \right\rangle = 0\, . \label{eq:3bpz11}
\end{align}
Using the explicit expression \eqref{eq:gen3point} for the three point function, we have
\begin{multline}
\left\langle V_{\langle 1,1 \rangle }(\zb{z})V_{\zb[1]{\Delta}}(\zb[1]{z})V_{\zb[2]{\Delta}}(\zb[2]{z}) \right\rangle \propto \\ (z-z_1)^{\Delta_2-\Delta_1}(z_2-z)^{\Delta_1-\Delta_2}(z_1-z_2)^{-\Delta_1-\Delta_2} \, , 
\end{multline}
where we have used  $\Delta_{\langle 1,1 \rangle} = 0$. Applying the BPZ equation \eqref{eq:3bpz11} we arrive at the condition
\begin{align}
\left\langle V_{\langle 1,1 \rangle }(\zb{z})V_{\zb[1]{\Delta}}(\zb[1]{z})V_{\zb[2]{\Delta}}(\zb[2]{z}) \right\rangle \neq 0 \Rightarrow \Delta_1 = \Delta_2\, . \label{eq:lvl1bpz}
\end{align}
This condition means that the three-point function with the field $V_{\langle 1,1 \rangle}$ is proportional to the two-point function $\left\langle V_{\zb[1]{\Delta}}(\zb[1]{z})V_{\zb[2]{\Delta}}(\zb[2]{z}) \right\rangle$, so that the insertion of the field $V_{\langle 1,1 \rangle}$ does not affect correlation functions in general. We can express this by writing a so-called fusion rule between the degenerate representation $R_{\langle 1,1 \rangle}$, corresponding to the left-moving sector of the degenerate field, and the Verma module $\mathcal{V}_{\Delta}$ associated with the left-moving sector of a generic field. We write the rule as 
\begin{align}
R_{\langle 1,1\rangle}\times \mathcal{V}_{\Delta} = \mathcal{V}_{\Delta}\ .
\label{eq:11fr}
\end{align}
The BPZ equation \eqref{eq:3bpz11} has a right moving analogue, leading to a condition similar to \eqref{eq:lvl1bpz}. In this case it is straightforward to translate the fusion rule \eqref{eq:11fr} into an OPE that considers simultaneously left-moving and right-moving sectors, because there is only one solution in each sector, and thus only one way to combine them. This OPE can be written as 
\begin{align}
V_{\langle 1,1 \rangle}(\zb{y}) V_{\zb{\Delta}}(\zb{z}) = C_{\zb{\Delta}}  V_{\zb{\Delta}}(\zb{z})\, .
\end{align}
It can be shown that the constant $ C_{\zb{\Delta}}$ is independent of the conformal weights, and that the degenerate field $V_{\langle 1,1 \rangle}$ is the identity field \cite{rib16}.

\item $V_{\langle 2,1 \rangle }$ and $V_{\langle 1,2 \rangle }$

We begin by the degenerate field $V_{\langle 2,1 \rangle}$. The corresponding left-moving null vector equation takes the form
\begin{align}
\left( L^{2, (z)}_{-1} - \beta^2 L^{(z)}_{-2} \right) V_{\langle 2,1 \rangle}(\zb{z}) = 0\, ,
\end{align}
and local Ward identities lead to a differential equation for the three-point function
\begin{align}
 \left( \frac{-1}{\beta^2}\frac{\partial^2}{\partial z_1^2} + \sum_{i=2}^N\left(\frac{1}{z_{1i}}\frac{\partial}{\partial z_i} +\frac{\Delta_i}{z^2_{1i}}\right) \right)\left< V_{\langle 2, 1 \rangle}(\zb[1]{z}) \prod_{i=2}^N V_{\Delta_i}(\zb[i]{z}) \right>  = 0\ .
 \label{eq:bpz}
\end{align}
Again, using the explicit form of the three-point function \eqref{eq:gen3point} we can apply the differential operator in equation \eqref{eq:bpz} directly, and we obtain the condition
\begin{multline}
 \left< V_{\langle 2, 1 \rangle}(\zb[1]{z}) V_{\Delta_2}(\zb[2]{z}) V_{\Delta_3}(\zb[3]{z}) \right> \neq 0 \\ \Rightarrow  \quad 
 2(\Delta_2-\Delta_3)^2 -\beta^2(\Delta_2+\Delta_3) -2\Delta_{\langle 2, 1 \rangle}^2 +\beta^2\Delta_{\langle 2, 1 \rangle} = 0\ .
 \end{multline}
The solution to this condition becomes easier in terms of the momentums $P_i$. We obtain
\begin{align}
P_3 = P_2 \pm \frac{\beta}{2}\, . \label{eq:fr21s}
\end{align}

We can obtain the analogous results for the degenerate field $V_{\langle 1,2 \rangle}$ by performing the substitution $\beta \to -\beta^{-1}$ in equations \eqref{eq:bpz} and \eqref{eq:fr21s}. This substitution has the effect $P_{(2,1)} \to P_{(1,2)}$, as can be seen from the definition \eqref{eq:PKac}. Then, in the case of $V_{\langle 1,2 \rangle}$ we obtain  the condition
\begin{align}
P_3 = P_2 \pm \frac{1}{2\beta}\, . \label{eq:fr12s}
\end{align}

We can summarize conditions \eqref{eq:fr21s} and\eqref{eq:fr12s} by writing the following degenerate fusion rules,  
\begin{align}
  \mathcal{R}_{\langle 2,1\rangle}\times \mathcal V_P = \sum_\pm \mathcal V_{P\pm \frac{\beta}{2}}\quad , \quad  
  \mathcal{R}_{\langle 1,2\rangle}\times \mathcal V_P = \sum_\pm \mathcal V_{P\pm \frac{1}{2\beta}}\ .
  \label{eq:rv}
 \end{align} 
where we have chosen to label Verma modules by their momentums, instead of their conformal weights. 
As opposed to the level $1$ case, for the level $2$ degenerate fields there are two contributions to each fusion rule, and it is not trivial to decide how to combine them into an OPE that takes into account the left-moving and right-moving sectors simultaneously. This problem will be addressed in section \ref{sec:degfr}, under certain supplementary assumptions. 
\end{itemize}

The examples above show that when degenerate fields are involved, extra constraints arising from BPZ equations allow for the determination of the OPE spectrums, at least when dealing only with one sector, either left-or-right-moving. The fusion rules of equations \eqref{eq:11fr} and \eqref{eq:rv} can be thought of as particular examples of a bilinear, associative product of representations called the fusion product, which determines which representations appear in the OPE between two given fields. 

Degenerate fusion rules can be extended to higher level degenerate fields \cite{Ribault:2014hia}, giving
 \begin{align}
 \mathcal{R}_{\langle r,s \rangle}\times \mathcal{V}_P &= \sum_{i=0}^{r-1} \sum_{j=0}^{s-1} \mathcal{V}_{P - P_{\langle r,s \rangle}+ i\beta-j\beta^{-1}}\ .
\label{rtv}
\end{align}

\subsection{Conformal blocks} \label{sec:cblocks}

We now return to the problem posed at the end of section \ref{sec:ward}, and focus on the calculation of four-point functions by means of the OPE. The existence of the OPEs, and the fact that their general structure is fixed by conformal symmetry (see equation \eqref{eq:vvs}), brings a great advantage into the computation of correlation functions. Inserting an OPE inside an $N$-point correlation function allows to compute it as a combination of $(N-1)$-point correlation functions, and in this way any correlation function can be expressed in terms of the ones whose form is fixed by symmetry. The price to pay is the introduction of many structure constants, which need to be determined. 

Let us focus on the case of the four-point function of primary fields with three fixed points appearing in equation \eqref{fx}. We denote this function by $\left\langle V_{1}(x) V_2(0) V_3(\infty) V_4(1) \right\rangle$, where we have simplified notations by writing  $V_i(z) = V_{\zb[i]{\Delta}}(\zb{z})$. Taking the OPE between the first two fields as in equation \eqref{eq:vvs}, we obtain an expansion of the four-point function as a sum of three-point functions:
\begin{multline}
\left\langle V_{1}(x) V_2(0) V_3(\infty) V_4(1) \right\rangle = \\ \sum_{s} \frac{C_{12s}}{B_s} \left|x^{\Delta_s-\Delta_1-\Delta_2} \sum_{L \in \mathcal{L} } x^{|L|} f^{\Delta_s, L}_{\Delta_1, \Delta_2}\right|^2 \left\langle L^{(0)} \bar{L}^{(0)}\ V_s(0)V_3(\infty)V_4(1) \right\rangle\ . \label{eq:4ptOPE1}
\end{multline}

In equation \eqref{eq:4ptOPE1} the index $s$ identifies the primary fields produced by the OPE $V_1(x)V_2(0)$,  and the four-point functions is expressed a sum over the conformal families generated by each of these fields. Three-point correlation functions are fixed by symmetry, and we would like to use expression \eqref{eq:gen3point} to further simplify the expansion of the four-point function. Equation \eqref{eq:gen3point} gives correlation functions of primary fields, but fortunately local Ward identities like \eqref{eq:locWard} relate correlation functions of descendent fields to correlation functions of primary fields. The universal quantities 
\begin{align}
g^{L}_{\Delta_s,\Delta_3,\Delta_4} = 
 \frac{ \left< L V_{s}(0)V_{3}(\infty)V_{4}(1)\right> }{  \left<  V_{s}(0)V_{3}(\infty)V_{4}(1)\right>}\ ,
 \label{glvv}
\end{align}
along with the analogous right-moving ratios, can always be computed form Ward identities. Then, the four-point function becomes
\begin{multline}
\left\langle V_{1}(x) V_2(0) V_3(\infty) V_4(1) \right\rangle = \\ \sum_{s} \frac{C_{12s}}{B_s} \left|x^{\Delta_s-\Delta_1-\Delta_2} \sum_{L \in \mathcal{L} } x^{|L|} f^{\Delta_s, L}_{\Delta_1, \Delta_2} g^{L}_{\Delta_s,\Delta_3,\Delta_4} \right|^2 \left\langle V_s(0)V_3(\infty)V_4(1) \right\rangle\ . \label{eq:4ptOPE2}
\end{multline}
The final step is to replace the three-point function $\left\langle V_s(0)V_3(\infty)V_4(1) \right\rangle $ by an explicit expression. This can be done by combining the definition \eqref{eq:finf} of a field at infinity with the expression \eqref{eq:gen3point}. We find that only the three-point structure constant survives, and the final expression for the four-point correlation function is
\begin{align}
\boxed{\left\langle  \prod_{i=1}^{4}V_i(z_i) \right\rangle = \sum_{s} D_{s|1234} \mathcal{F}^{(s)}_{\Delta_s}(\Delta_i|x) \bar{\mathcal{F}}^{(s)}_{\bar{\Delta}_s}(\bar{\Delta}_i|\bar{x})}\, , \label{eq:s4pt}
\end{align}
where we have introduced the four-point structure constants 
\begin{align}
D_{1234}(V_s) = D_{s|1234} = \frac{C_{12s}C_{s34}}{B_s}\, ,\label{eq:4pc}
\end{align}
and the functions $\mathcal{F}^{(s)}_{\Delta_s}(\Delta_i|x) $ and $ \bar{\mathcal{F}}^{(s)}_{\bar{\Delta}_s}(\bar{\Delta}_i|\bar{x})$ which are, respectively, the left-moving and right-moving four-point $s$-channel conformal blocks; explicitly, 
\begin{align}
 \mathcal{F}^{(s)}_{\Delta_s}(\Delta_i|x) = x^{\Delta_s-\Delta_1-\Delta_2}\sum_{L\in\mathcal{L}} f_{\Delta_1,\Delta_2}^{\Delta_s,L} g^{L}_{\Delta_s,\Delta_3,\Delta_4}x^{|L|}\ . 
\label{gsd}
\end{align}

Conformal blocks are the fundamental constituents of the correlation functions, completely determined by the symmetry algebra. We consider them as known functions, because they can in principle be computed from expression \eqref{gsd}. The conformal blocks \eqref{gsd} we have shown correspond to a correlation function with three-points fixed at $(0, \infty, 1)$, but there are of course blocks corresponding to the more generic case, denoted by $\mathcal{F}^{(s)}_{\Delta_s}(\Delta_i|z_i)$. An expression for them can be obtained by following the same procedure as above, but keeping all points generic. 
The name $s$-channel is associated with taking the OPE when $x \to 0$. Other choices are possible and should be equivalent, as we will discuss in section \ref{sec:crossing}.

Finally, let us mention that although expression \eqref{gsd} for the conformal blocks is correct, it is not very practical for numerical implementation. We discuss an alternative representation of conformal blocks, due to Al. Zamolodchikov, in appendix \ref{sec:cbtrunc} of chapter \ref{ch:crsymfun}. 

\section{Summary}

Throughout this chapter we have discussed the constraints that conformal symmetry imposes on two-dimensional field theories. These constraints go from the structure of their spectrum, given by irreducible representations of the Virasoro algebra, to the Ward identities satisfied by the observables of the theory: its correlation functions. We saw that the $(\zb{z})$-dependence of 2-and-3-point correlation functions is fixed, up to certain structure constants, and that four-and-higher point correlation functions can be decomposed into combinations of structure constants and universal objects called conformal blocks. Equation \eqref{eq:s4pt} is an example of one of these decompositions, where the four-point structure constants $D_{s|1234}$ are combinations of two-point and three-point structure constants. 

Since we consider conformal blocks as known, the problem of computing correlation functions in CFTs is the problem of finding the OPE spectrums, which determine the terms contributing to the expansion of a correlation function, and  the appropriate two-and-three-point structure constants. Finding these elements in any given theory gives us the tools to, in principle, compute all correlation functions. The conformal bootstrap approach to be discussed in the next chapters aims at determining these elements, as a way of solving CFTs.

%% file: chapters/bootstrap.tex
\chapter{Conformal bootstrap} \label{ch:cboots}

The conformal bootstrap method consists of trying to solve and classify conformal quantum field theories by exploiting the consequences of symmetry and self-consistency. 

The previous chapter illustrated how conformal symmetry imposes a certain structure on correlation functions in two-dimensional CFTs, which can be expressed as combinations of structure constants and conformal blocks, summed over a certain spectrum.  

In this chapter we study the self-consistency conditions that determine the structure constants and spectrums of a large class of conformal field theories. In section \ref{sec:crossing} we describe the crossing-symmetry constraints, and their degenerate version. In section \ref{sec:aboots} we explicit the basic assumptions considered, and arrive at the equations that determine the structure constants. Finally, in sections \ref{sec:cstssol} we show explicit solutions to these equations and discuss a few of their properties. 

\section{Crossing symmetry} \label{sec:crossing}
In section \ref{sec:cblocks} we presented the s-channel decomposition of a four-point function of primary fields, equation \eqref{eq:s4pt}. This expansion originates from inserting the OPE $V_{\zb[1]{\Delta}}(\zb{z}) V_{\zb[2]{\Delta}}(0) $ into the correlation function, and each of its terms is a product between a four-point structure constant, $D_{s|1234}$, and the left-moving and right-moving $s$-channel conformal blocks, $\mathcal{F}^{(s)}_{\Delta_s}(\Delta_i|x)$ and  $\mathcal{\bar{F}}^{(s)}_{\bar{\Delta}_s}(\bar{\Delta}_i|\bar{x})$. However, we could have chosen to insert a different OPE. Following \cite{Ribault:2014hia}, choosing the OPE $V_{\zb[4]{\Delta}}(1) V_{\zb[1]{\Delta}}(\zb{z}) $ gives the $t$-channel decomposition, 
\begin{align}
\left\langle  \prod_{i=1}^{4}V_i(z_i) \right\rangle = \sum_{\zb[t]{\Delta}} D_{t|1234} \mathcal{F}^{(t)}_{\Delta_t}(\Delta_i|x) \bar{\mathcal{F}}^{(t)}_{\bar{\Delta}_t}(\bar{\Delta}_i|\bar{x})\, , 
\end{align}
where the $t$-channel four-point structure constants are
\begin{align}
D_{t|2341} = \frac{C_{23t}C_{t41}}{B_t}\, ,
\end{align}
and $\mathcal{F}^{(t)}_{\Delta_t}(\Delta_i|x)$ and $\mathcal{F}^{(t)}_{\bar{\Delta}_t}(\bar{\Delta}_i|\bar{x})$  are the  $t$-channel conformal blocks. These conformal blocks can be obtained by permuting the arguments of the $s$-channel conformal blocks, with the permutation $1234 \to 2341$ \cite{Ribault:2014hia}. For example, for the left-moving conformal blocks this amounts to
\begin{align}
\mathcal{F}^{(t)}_{\Delta_t}(\Delta_1,\Delta_2,\Delta_3,\Delta_4|x ) = (1-x)^{\Delta_2+\Delta_3 - \Delta_1-\Delta_4}\mathcal{F}^{(s)}_{\Delta_t}(\Delta_2,\Delta_3,\Delta_4,\Delta_1|1-x)\, ,\label{eq:stperm}
\end{align}
where we have used the definition of the cross-ratio \eqref{eq:cratio}. A similar relation holds for the right-moving conformal blocks. 
The $s$-and-$t$-channel conformal blocks are two bases of solutions for the Ward identities satisfied by the four-point correlation function. In any consistent CFT both expansions should coincide, because they represent the same function. This condition is called crossing-symmetry. Introducing a diagrammatic representation for the conformal blocks, 
\begin{align}
 \mathcal{F}^{(s)}_{\Delta_s}  =  
\begin{tikzpicture}[baseline=(current  bounding  box.center), very thick, scale = .3]
\draw (-1,2) node [left] {$2$} -- (0,0) -- node [above] {$s$} (4,0) -- (5,2) node [right] {$3$};
\draw (-1,-2) node [left] {$1$} -- (0,0);
\draw (4,0) -- (5,-2) node [right] {$4$};
\end{tikzpicture}
\quad\ ,\quad\
 \mathcal{F}^{(t)}_{\Delta_t}  =  
\begin{tikzpicture}[baseline=(current  bounding  box.center), very thick, scale = .3]
 \draw (-2,3) node [left] {$2$} -- (0,2) -- node [left] {$t$} (0,-2) -- (-2, -3) node [left] {$1$};
\draw (0,2) -- (2,3) node [right] {$3$};
\draw (0,-2) -- (2, -3) node [right] {$4$};
\end{tikzpicture}\, ,
\end{align}
crossing symmetry can be expressed as
\begin{align}
 \sum_{\Delta_s,\bar{\Delta}_s} \frac{C_{12s} C_{s34}}{B_s} \left| 
 \begin{tikzpicture}[baseline=(current  bounding  box.center), very thick, scale = .3]
\draw (-1,2) node [left] {$2$} -- (0,0) -- node [above] {$s$} (4,0) -- (5,2) node [right] {$3$};
\draw (-1,-2) node [left] {$1$} -- (0,0);
\draw (4,0) -- (5,-2) node [right] {$4$};
\end{tikzpicture} 
\right|^2 = \sum_{\Delta_t,\bar{\Delta}_t} \frac{C_{23t}C_{t41}}{B_t} \left|
\begin{tikzpicture}[baseline=(current  bounding  box.center), very thick, scale = .3]
 \draw (-2,3) node [left] {$2$} -- (0,2) -- node [left] {$t$} (0,-2) -- (-2, -3) node [left] {$1$};
\draw (0,2) -- (2,3) node [right] {$3$};
\draw (0,-2) -- (2, -3) node [right] {$4$};
\end{tikzpicture}
\right|^2\ ,
\label{eq:csd}
\end{align}
where we have used the explicit expression of the structure constants, and the modulus squared notation denotes the product between the left-moving and right-moving blocks. 

The crossing-symmetry constraints \eqref{eq:csd} can be written for different values of $(\zb{x})$, and they constitute a system of infinitely many equations for the structure constants and the spectrums of each channel. Supplemented by some extra hypothesis, for example assumptions on the analyticity properties of the structure constants, the system \eqref{eq:csd} could have the power to determine all the unknowns.
If it was possible to solve explicitly equations \eqref{eq:csd}, the conformal bootstrap program would be complete: Symmetry constraints, imposed by Ward identities, and self-consistency conditions, arising from crossing-symmetry, would determine all the quantities needed to compute any correlation function. However, equations \eqref{eq:csd} have too many degrees of freedom to be solved systematically, once and for all, and although they are in principle sufficient,they become impractical in the general case.


The analytic conformal bootstrap approach discussed in this work consists in solving a simpler system of crossing-symmetry equations, in which the spectrum of each channel is known and the conformal blocks take a simple form. Such a system is obtained by taking one of the fields in the four point function to be degenerate, and using the degenerate fusion rules \eqref{eq:rv} to write the $s$-and-$t$-channel expansions. 

Let us study the example of a four point function including the degenerate field $V_{\langle 2,1 \rangle}(\zb{z})$. For simplicity, we consider only the left-moving sector, and the general case will be treated in section \ref{sec:dcsy}. Consider the four point function $Z^{\langle 2,1 \rangle}_{123}(\zb{x}) = \left\langle V_{\langle 2,1 \rangle}(x) V_1(0)V_2(\infty)V_3(1) \right\rangle$, where we have ommited right-moving arguments to lighten the notation. Due to the presence of the degenerate field, this four-point function satisfies the BPZ equation \eqref{eq:bpz}, 
\begin{align}
  \left\{ \frac{x(x-1)}{\beta^2}\frac{\partial^2}{\partial x^2} + (2x-1){\frac{\partial}{\partial x}} +\Delta_{\langle 2,1 \rangle} +\frac{\Delta_1}{x}-\Delta_2 + \frac{\Delta_3}{1-x}\right\} Z^{\langle 2,1 \rangle}_{123}(x)=0\ ,
\label{eq:sode}
\end{align}
whose coefficients are singular at the points where the fields $V_1, V_2$ and $V_3$ are inserted. We can look for solutions having a specific behaviour near the singularities. For example, near the singular point $x = 0$ we want
\begin{align}
Z^{\langle 2,1 \rangle}_{123}(x) \underset{x \to 0}{\propto} x^{\lambda} (1 + O(x))\, ,\label{eq:exp0}
\end{align}
where $\lambda$ is called a characteristic exponent. Inserting this into the BPZ equation \eqref{eq:sode} we find that there are two possible values of the characteristic exponent $\lambda$, 
\begin{align}
\lambda_{\pm} =  \beta\left[\frac{-1}{2} \left(\beta - \frac{1}{\beta}\right) \pm P_1 \right]\ . \label{eq:lambdapm}
\end{align}
Each one of these exponents is related to one of the fields appearing in the OPE $V_{\langle 2,1 \rangle} V_1$. We have
\begin{align}
\lambda_{\pm} = \Delta(P_1 \pm \frac{\beta}{2})-\Delta_{\langle 2,1 \rangle}-\Delta(P_1)\, 
\end{align}
where on the RHS we have the power of $x$ corresponding to the terms of the OPE, as follows from \eqref{eq:vvs} and the fusion rule \eqref{eq:rv}.

The BPZ equation \eqref{eq:sode} has a two dimensional space of solutions, and the basis corresponding to the behaviour \eqref{eq:exp0} is given by the $s$-channel hypergeometric conformal blocks $F^{(s)}_{\pm}(z)$, 
\begin{align}
\left\{ 
\begin{array}{l}  \mathcal{F}^{(s)}_+(x) = x^{-\frac{\beta}{2} \left(\beta - \frac{1}{\beta}\right) + \beta P_1 } (1-x)^{-\frac{\beta}{2} \left(\beta - \frac{1}{\beta}\right) + \beta P_3 } F(A,B,C,x)\ ,
\\  
\begin{aligned}
\mathcal{F}^{(s)}_-(x)  &= \mathcal{F}^{(s)}_+(x) |_{P_1\to -P_1} \\ &= x^{-\frac{\beta}{2} \left(\beta - \frac{1}{\beta}\right) - \beta P_1 } (1-x)^{-\frac{\beta}{2} \left(\beta - \frac{1}{\beta}\right) + \beta P_3 } \\  &\quad \times F(A-C+1,B-C+1,2-C,x)   \ , 
\end{aligned}
\end{array}
\right.  
\label{gpm}
\end{align}
Here $F(A,B,C,x)$ is the hypergeometric function 
\begin{align}
 F(A,B,C,x) = \sum_{n=0}^\infty \frac{(A)_n(B)_n}{n!(C)_n}x^n\ ,\quad (A)_n = \frac{\Gamma(A+n)}{\Gamma(A)} \, ,
\label{fsn}
\end{align}
and the coefficients $A, B, C$ are given by
\begin{align}
\renewcommand{\arraystretch}{1.3}
\left\{\begin{array}{l}   A = \frac12 + \beta(P_1+P_2+P_3) \ , \\
      B = \frac12 + \beta(P_1-P_2+P_3)\ , \\
      C = 1 + 2\beta P_1 \ ,
\end{array}\right. 
\label{abc}
\end{align}


Alternatively, we can find the $t$-channel basis of solutions by looking for solutions with the behaviour, 
\begin{align}
Z^{\langle 2,1 \rangle}_{123}(x) \underset{x \to 1}{\propto} (x-1)^{\omega} \left( 1 + O(x-1)\right)\, .
\end{align}
Again there are two solutions $\omega_{\pm}$, corresponding to the OPE $V_{\langle 2,1 \rangle} V_3$, and their values are obtain by replacing $P_1 \to P_3$ in \eqref{eq:lambdapm}. 

Using the same definitions of $A,B$ and $C$, \eqref{abc}, the $t$-channel basis is
\begin{align}
 \left\{\begin{array}{l} 
\begin{aligned}
  \mathcal{F}^{(t)}_+(x) &= x^{-\frac{\beta}{2} \left(\beta - \frac{1}{\beta}\right) + \beta P_1} (1-x)^{-\frac{\beta}{2} \left(\beta - \frac{1}{\beta}\right) + \beta P_3 }\\ &\qquad \times F(A,B,A+B-C+1,1-x) \ ,
\end{aligned}\\
\begin{aligned}
   \mathcal{F}^{(t)}_-(x) &= x^{-\frac{\beta}{2} \left(\beta - \frac{1}{\beta}\right) + \beta P_1 } (1-x)^{-\frac{\beta}{2} \left(\beta - \frac{1}{\beta}\right) - \beta P_3 )} \\ &\qquad \times F(C-A,C-B,C-A-B+1,1-x)  \ ,
\end{aligned}  
\end{array}\right.   \label{gpmt}
\end{align}

We can use a diagrammatic representation of the hypergeometric conformal blocks, in which we denote the degenerate field by a dashed line, 
\begin{align}
 \mathcal{F}^{(s)}_\pm(x)  =  
 \begin{tikzpicture}[baseline=(current  bounding  box.center), very thick, scale = .4]
\draw (-1,2) node [left] {$P_1$} -- (0,0) -- node [above] {$P_1\pm \frac{\beta}{2}$} (4,0) -- (5,2) node [right] {$P_2$};
\draw[dashed] (-1,-2) node [left] {$\langle 2,1 \rangle$} -- (0,0);
\draw (4,0) -- (5,-2) node [right] {$P_3$};
\end{tikzpicture}
\ , \quad
 \mathcal{F}^{(t)}_\pm(x)  =  
 \begin{tikzpicture}[baseline=(current  bounding  box.center), very thick, scale = .4]
 \draw (-2,3) node [left] {$P_1$} -- (0,2) -- node [left] {$P_3\pm \frac{\beta}{2}$} (0,-2);
 \draw[dashed] (0, -2) -- (-2, -3) node [left] {$\langle 2,1 \rangle$};
\draw (0,2) -- (2,3) node [right] {$P_2$};
\draw (0,-2) -- (2, -3) node [right] {$P_3$};
\end{tikzpicture}
\ .
\label{tpic}
\end{align}
 
Both bases of solutions are related by 
\begin{align}
 \mathcal{F}^{(s)}_i(x) = \sum_{j=\pm} F_{ij} \mathcal{F}^{(t)}_j(x)\ ,
\label{gfg}
\end{align}
where $F_{ij}$ are the elements of the fusing matrix $F$, 
\begin{align}
 F = \left[\begin{array}{cc} F_{++} & F_{+-} \\ F_{-+} & F_{--} \end{array}\right] 
= \left[\renewcommand{\arraystretch}{1.3}\begin{array}{cc}
         \frac{\Gamma(C)\Gamma(C-A-B)}{\Gamma(C-A)\Gamma(C-B)} & \frac{\Gamma(C)\Gamma(A+B-C)}{\Gamma(A)\Gamma(B)} 
       \\   \frac{\Gamma(2-C)\Gamma(C-A-B)}{\Gamma(1-A)\Gamma(1-B)} & \frac{\Gamma(2-C)\Gamma(A+B-C)}{\Gamma(A-C+1)\Gamma(B-C+1)}
        \end{array}\right]\ ,
\label{fmd}
\end{align}
whose determinant is 
\begin{align}
 \det F = \frac{1-C}{A+B-C}\ .
\label{detf}
\end{align}

A similar analysis can be performed on the four-point function $Z^{\langle 1,2 \rangle}_{123}(\zb{x})$, involving the second level two degenerate field $V_{\langle 1,2 \rangle}$. The procedure is identical, and the solutions of this case can be found by replacing $\beta \to \frac{-1}{\beta}$ in the results for $V_{\langle 2,1 \rangle}$.

The results obtained from the BPZ equations for $Z^{\langle 2,1 \rangle}_{123}$, and their analogues for $Z^{\langle 1,2 \rangle}_{123}$, mean that the left-moving conformal blocks  of correlation functions involving degenerate fields are expressed in terms of hypergeometric functions, and that the terms in the $s$-and-$t$-channel decompositions are known, and finitely many. The same is true for the right-moving blocks, and the full solution can be built from a combination of left-moving and right-moving objects. 
The challenge of solving the degenerate crossing-symmetry equations is, then, twofold. First, it is necessary to construct simultaneous solutions to the holomorphic and antiholomorphic versions of equation \eqref{eq:sode}, by combining solutions of the type \eqref{gpm} or \eqref{gpmt} depending on the channel. Then, we must write and solve  equations for the structure constants in terms of the left-and-right moving fusing matrix elements. 

The method of using degenerate crossing-symmetry to find the structure constants even though degenerate fields may not be part of the spectrum was introduced in \cite{Teschner:1995yf}, and it is sometimes referred to as Teschner's trick. Once the structure constants are determined we must find a way to determine the $s$-and-$t$-channel spectrums in the generic case, i.e. without degenerate fields, and show that crossing-symmetry is satisfied in the general case \eqref{eq:csd}. In what comes next we will follow this strategy.

\section{Analytic conformal bootstrap}\label{sec:aboots}
This section focuses on stating the main assumptions we take in order to follow the analytic conformal bootstrap program, and on finding their most immediate consequences for the spectrum and structure constants. 

\subsection{Assumptions}\label{sec:assumptions}

In the following we will work under three main assumptions:

\begin{itemize}
\item Generic central charge.

We assume that the central charge is not subject to any particular constraints and takes values $c \in \mathbb{C}$. Furthermore, we expect to find structure constants which are analytic in $\beta$, at least in some region. The idea behind this assumption is to be able to perform analytic continuations of the solutions, finding a large class of theories related to one another. 

\item Existence of degenerate fields $V_{\langle 2,1\rangle}$ and $V_{\langle 1,2 \rangle}$.

As discussed in section \ref{sec:crossing}, the cornerstone of the  analytic conformal boostrap approach are the equations of degenerate crossing symmetry. We assume that the degenerate fields exist, in the sense that even if they do not appear in any OPE spectrums of other fields in the theory, the constraints coming from correlation functions involving degenerate fields are valid. This assumption is the first step to writing down the constraints, but we still need to check that the structure constants obtained satisfy the generic crossing symmetry equations \eqref{eq:csd}. 

\item Singled-valued correlation functions

We impose as a restriction that correlation functions of the theory should be single valued. This condition is not always necessary to define a consistent CFT (theories with parafermionic observables do not fulfil this condition), but here it will be adopted for simplicity. We will see how this condition affects the spectrum of fields that can be part of the theory, and what is its influence in correlation functions. 
We note that the single-valuedness condition is imposed on correlation functions that are physical, in the following sense: We've mentioned that degenerate fields can exist without being part of the spectrum of the theory. This would mean that the single-valuedness conditions will not necessarily be imposed on correlation functions involving degenerate fields.

\end{itemize}

Let us explore the immediate consequences of these assumptions.

\subsubsection{Single-valuedness}

Let us consider an $N$-point function of primary fields $V_{\zb[i]{\Delta}}(\zb[i]{z})$, with $i = 1,\dots N$. If we perform a rotation by an angle $\theta$, the transformation rule \eqref{eq:globT} gives 
\begin{align}
\left\langle \prod_{i=1}^{n}V_{\zb[i]{\Delta}}(\zb[i]{z}) \right\rangle = e^{-i\theta \sum_{j=1}^{N}S_j} \left\langle \prod_{i=1}^{n}V_{\zb[i]{\Delta}}(e^{i\theta}z_i,e^{-i\theta}\bar{z}_i ) \right\rangle\, .\label{rotn}
\end{align}
where 
\begin{align}
S_i = \Delta_i -\bar{\Delta}_i
\end{align}
are the conformal spins of each field. 
For a non-zero, single valued $N$-point function, the phase factor of a rotation by $\theta = 2\pi$ should be $1$. Thus, single-valuedness requires the condition
\begin{align}
\sum_{j=1}^{N} S_j \in \mathbb{Z}\, . \label{sumsp}
\end{align}

In particular, if we consider the case of a two-point function for a field $V_{\Delta, \bar{\Delta}}(z, \bar{z})$,  \eqref{eq:gen2point}, we obtain the condition
\begin{align}
S = \Delta - \bar{\Delta} \in \frac{\mathbb{Z}}{2}\, . \label{eq:halfspin}
\end{align}

We consider theories which include only primary fields satisfying the half-integer spin condition \eqref{eq:halfspin}. Fields with $S \in \mathbb{Z}$ are called bosons, while fields with $S \in \mathbb{Z}+\frac{1}{2}$ are fermions. 

\subsubsection{Degenerate fusion rules}\label{sec:degfr}

Let us study how the requirement that degenerate fields exist constrains the spectrum. As a consequence of imposing single-valuedness, all fields should satisfy the half-integer spin condition \eqref{eq:halfspin}. The assumption that degenerate fields exist implies that fields produced in the degenerate OPEs should also satisfy the condition \eqref{eq:halfspin}. The fusion rules \eqref{eq:rv} are written in terms of representations, and are valid only in the left-moving or right-moving sector. Here, we want to write fusion rules for field, and we need to build rules that take both the left-and-right-moving sectors into account. 

We take a primary field $V_{\zb{P}}$, labeled by its momentums. We assume that it satisfies the half-integer spin condition \eqref{eq:halfspin}, which in terms of the momentums reads
\begin{align}
S = P^2 - \bar{P}^2 \in \frac{\mathbb{Z}}{2}\, . \label{eq:halfP}
\end{align}
Applying the fusion rules \eqref{eq:rv} on both sectors, the most general form of the fusion $V_{\langle 2,1 \rangle} V_{\zb{P}}$ is
\begin{align}
V_{\langle 2,1 \rangle} \times V_{P,\bar{P}} \subseteq \sum_{\sigma = \pm} \left(\sum_{\epsilon = \pm} V_{P+ \tfrac{\epsilon \beta}{2},\bar{P}+ \tfrac{\sigma \epsilon\beta}{2}} \right)\, , \label{V214}
\end{align}
where there are four possible terms. Using \eqref{eq:halfP} and imposing the fields on the RHS of \eqref{V214} to have half-integer spins we find the condition
\begin{align}
P-\sigma \bar{P} \in \frac{1}{2\beta}\mathbb{Z}\, , \label{spinc21}
\end{align}
where each value of $\sigma$ identifies two of the four terms in \eqref{V214}. In order for all terms to be present, condition \eqref{spinc21} should hold for both values of $\sigma$. But if that is the case, the spin of the original field $V_{\zb{P}}$ satisfies
\begin{align}
S = \prod_{\pm} (P\pm\bar{P}) \in \frac{1}{4\beta^2}\mathbb{Z}\, . \label{eq:2sigmas}
\end{align}
If the original field $V_{\zb{P}}$ is such that $S \neq 0$,  compatibility between equations \eqref{eq:2sigmas} and \eqref{eq:halfP} restricts the values of the parameter $\beta$; in particular, it implies $\beta^2 \in \mathbb{Q}$. This restriction is incompatible with our assumption of \textbf{generic central charge}, and we conclude that \eqref{eq:2sigmas} cannot hold. This in turn means that the condition \eqref{spinc21} holds only for one values of $\sigma$, and that only the terms of \eqref{V214} associated to this value appear in the OPE. 

In the case of the OPE $V_{\langle 1,2 \rangle} V_{\zb{P}}$ we find an analogous situation. Defining the parameter $\tilde{\sigma} = \pm$ in the same way as $\sigma$ before, the condition to have half-integer spins in the OPE reads
\begin{align}
P-\tilde{\sigma} \bar{P} \in \frac{\beta}{2}\mathbb{Z}\, . \label{spinc12}
\end{align}
Again, due to compatibility with \eqref{eq:halfP} and with the generic central charge assumption, this can hold only for one value of $\tilde{\sigma}$. 

In order to define the fusion rules between primaries and the degenerate fields, we consider $\sigma$ and $\tilde{\sigma}$ as properties of each field. The fusion rules can be written as:
\begin{align}
V_{\langle 2,1 \rangle} \times V_{P,\bar{P}}^{\sigma, \tilde{\sigma}} = \sum_{\epsilon = \pm} V_{P+\epsilon \tfrac{\beta}{2},\bar{P}+\sigma\epsilon \tfrac{\beta}{2}}^{\sigma, \tilde{\sigma}}\quad ,
\quad
V_{\langle 1,2 \rangle} \times V_{P,\bar{P}}^{\sigma, \tilde{\sigma}} = \sum_{\epsilon = \pm} V_{P-\tfrac{\epsilon}{2\beta},\bar{P}- \tfrac{\tilde{\sigma}\epsilon}{2\beta}}^{\sigma, \tilde{\sigma}}\, ,\label{dfus}
\end{align}
where we have included the values of $\sigma$ and $\tilde{\sigma}$ as labels on the field. Notice that the fields on the right-hand sides have the same values of $\sigma,\tilde{\sigma}$ as the field $V_{P,\bar{P}}^{\sigma, \tilde{\sigma}}$, because their momentums satisfy the same relations \eqref{spinc21} and \eqref{spinc12}. 

\subsection{Diagonal and Non-Diagonal fields}\label{sec:dnd}

Equations \eqref{spinc21} and \eqref{spinc12} determine which values of the momentums (and hence on the conformal weights) the fields of the theory can have, while being consistent with our assumptions of single-valuedness, existence of degenerate fields, and generic values of the central charge. In this sense, our assumptions impose a strong constraint on the structure of the spectrum of the theory. 
Here, we solve these constraints in order to determine what are the allowed values for the momentums of the fields. 

The simultaneous solution to \eqref{spinc21} and \eqref{spinc12} strongly depends on the product $\sigma\tilde{\sigma} \in \{ 1, -1 \}$, and we will distinguish to classes of fields depending on this value. Once this product is fixed, the actual value of $\sigma$  is irrelevant: it amounts to changing the convention of naming fields by $(P, \bar{P}) \to (P, -\bar{P})$, a change that does not affect the conformal weights nor the representations associated to each field. This means that, when needed, we can set $\sigma = 1 $ without loss of generality. 

We find two types of solutions to equations \eqref{spinc21} and  \eqref{spinc12}:
\begin{itemize}
\item \textbf{Diagonal fields: $\tilde{\sigma} = \sigma$}

Diagonal fields must have $P - \sigma \bar{P}\in \frac{ \beta }{2}\mathbb{Z} \cap \frac{1}{2\beta}\mathbb{Z}$. For a generic central charge we have $\beta^2 \neq \mathbb{Q}$, and thus the only solution is $P=\bar{P}$, and hence $\Delta = \bar{\Delta} $. Introducing the notation
\begin{align}
\boxed{V^{D}_P = V^{+, +}_{P,P}} \, , \label{eq:dnot}
\end{align}
for a diagonal field, the fusion products with degenerate fields read
\begin{align}
V_{\langle 2,1 \rangle} \times V_{P}^D = \sum_{\epsilon = \pm} V_{P+\frac{\epsilon \beta}{2}}^{D}\quad ,\quad V_{\langle 1,2 \rangle} \times V_{P}^{D} = \sum_{\epsilon = \pm} V_{P-\frac{\epsilon}{2\beta}}^{D}\, .
\end{align}

Diagonal primary fields have $S = 0$. 

\item \textbf{Non-diagonal fields $\tilde{\sigma} = -\sigma$}

The solution to \eqref{spinc21} and \eqref{spinc12} in this case is given by momentums of the type
\begin{align}
\left\{\begin{array}{l}
P=P_{( r,s )}\, , \\ \bar{P}= \sigma P_{( r,-s )}\, , \end{array}\right. \quad 
\text{with}\quad  r,s,rs \in \frac{1}{2}\mathbb{Z}\, ,
\label{eq:rsval}
\end{align}
where we have also imposed the condtion \eqref{eq:halfP}, and $P_{(r,s)}$ was defined in \eqref{eq:PKac}.
The spin of a non-diagonal field is $S = -rs$. 

Introducing the notation
\begin{align}
\boxed{V^N_{( r,s )} = V^{+,-}_{P_{( r,s)},P_{( r,-s)}}} \, ,
\end{align}
the fusion products with degenerate fields read
\begin{align}
V_{\langle 2,1 \rangle}\times V_{( r,s)}^N = \sum_{\epsilon = \pm}  V_{( r+\epsilon,s)}^{N}\quad ,\quad V_{\langle 1,2 \rangle} \times V_{(r,s)}^{N} = \sum_{\epsilon = \pm}  V_{( r,s+\epsilon)}^N\, . 
\end{align}
\end{itemize}

Notice that what defines a field as diagonal or non-diagonal is the value $\sigma \tilde{\sigma}$, and not its momentums. For example, the non-diagonal field $V^N_{(r,0)}$ has spin zero, and the same left and right momentums as the diagonal field $V^{D}_{P_{(r,0)}}$. These two fields cannot be distinguished by their momentums alone, but by their fusion products with degenerate fields. In particular, the fusion product 
\begin{align}
V_{\langle 1,2 \rangle}\times V^{N}_{(r,0)} = \sum_{\epsilon = \pm} V^{N}_{(r, \epsilon)}\, ,
\end{align}
produces fields of spin $S_{\epsilon} = -\epsilon r$, while the fusion product  $V_{\langle 1,2 \rangle}\times V^{D}_{P_{(r,0)}}$ only produces spinless fields. 

This is a particular case in which we consider fields with spin $S=0$, but we are applying the fusion rules \eqref{dfus}, which were derived assuming $S\neq 0$. In principle, linear combinations of $V^D_{P_{(r,0)}}$ and $V^N_{(r,0)}$ are primary fields whose fusion products with $V_{\langle 2,1\rangle}$ and $V_{\langle 1,2\rangle}$ can involve four fields, rather than two fields as in the fusion products \eqref{dfus}.
Here we make the supplementary assumption that all fieds (even spinless ones) follow these fusion rules, which can be thought of as choosing an orthogonal basis of fields in which diagonal and non-diagonal fields do not mix. In other words, we are choosing that the two point functions between diagonal and non-diagonal fields with the same dimensions vanish. For example
\begin{align}
\left\langle V^D_{P_{(r,0)}}V^N_{(r,0)} \right\rangle = 0\ .
\end{align}
This constitutes an extra assumption because it is not a consequence of Ward Identities. 

A final remark about diagonal and non-diagonal fields: In what follows we will often refer to the indices $(r,s)$ of non-diagonal fields. For any field (diagonal or not), these indices are defined from the momentums, by 
\begin{align}
r= \frac{1}{\beta} (P-\tilde{\sigma}\bar{P})\, \quad s = \beta(\sigma\bar{P} - P) \, .\label{eq:indef}
\end{align}
In the case of diagonal fields, \eqref{eq:indef} implies that the indices are always $r,s = 0$, even though a diagonal field $V^{D}_{P_{(r,s)}}$ has a momentum of the form \eqref{eq:PKac}.

Knowing the full left-and-right moving fusion rules \eqref{dfus}, we can use them to construct solutions to the holomophic and antiholomorphic BPZ equations \eqref{eq:sode}, and solve degenerate crossing symmetry. This is the content of the next section. 

\subsection{Degenerate crossing symmetry} \label{sec:dcsy}
By using the degenerate fusion rules \eqref{dfus} we can write the OPEs involving degenerate fields. Omitting the $(\zb{z})$-dependence, which is determined by \eqref{eq:vvs}, and introducing the degenerate OPE coefficients $C_{\epsilon}(V)$ and $\tilde{C}_{\tilde{\epsilon}}(V)$, we have
\begin{align}
V_{\langle 2,1 \rangle} V
= \sum_{\epsilon = \pm} C_{\epsilon}(V) V^\epsilon
\quad ,\quad 
V_{\langle 1,2 \rangle} V
= \sum_{\epsilon = \pm} \tilde{C}_{\epsilon}(V) V^{\tilde{\epsilon}} \, ,\label{dOPE}
\end{align}
where we also introduced the notations 
\begin{align}
 V = V_{P,\bar{P}}^{\sigma, \tilde{\sigma}} 
 \quad \implies \quad V^\epsilon = V_{P+\epsilon \tfrac{\beta}{2},\bar{P}+\sigma\epsilon \tfrac{\beta}{2}}^{\sigma, \tilde{\sigma}}\quad ,\quad 
 V^{\tilde{\epsilon}} = V_{P-\tfrac{\epsilon}{2\beta},\bar{P}- \tfrac{\tilde{\sigma}\epsilon}{2\beta}}^{\sigma, \tilde{\sigma}}\ .
\end{align}

So the momentums of each term in the OPEs are known, and we have established in section \ref{sec:crossing} that each one of the basis elements of the BPZ equations corresponds to one of these fields. Using \eqref{dOPE}, we can write the degenerate crossing symmetry equations 

\begin{align}
 \Big\langle V_{\langle 2,1\rangle} V_1 V_2 V_3 \Big\rangle &= 
 \sum_{\epsilon_1 = \pm} d^{(s)}_{\epsilon_1} \mathcal{F}^{(s)}_{\epsilon_1} \bar{\mathcal{F}}^{(s)}_{\sigma_1\epsilon_1}
 = \sum_{\epsilon_3 = \pm} d^{(t)}_{\epsilon_3} \mathcal{F}^{(t)}_{\epsilon_3} \bar{\mathcal{F}}^{(t)}_{\sigma_3\epsilon_3}
 \ ,
 \label{eq:2dec}
\end{align}
where $\mathcal{F}_\epsilon^{(s)},\mathcal{F}^{(t)}_\epsilon$ are degenerate four-point conformal blocks of equations \eqref{gpm} and \eqref{gpmt}, and the degenerate four-point structure constants are  combinations of three-point strucure constants given by
\begin{align}
d^{(s)}_\epsilon = C_\epsilon(V_1)C(V_1^\epsilon,V_2,V_3) \, .
\end{align}
Notice that the structure constant $C(V_1^\epsilon,V_2,V_3)$ does not involve the degenerate field $V_{\langle 2,1 \rangle}$, and so the degenerate crossing symmetry equations impose constraints on generic three-point structure constants. Furthermore, the generic four-point structure constants \eqref{eq:4pc} are combinations of three-point and two-point structure constants, so that the constraints on three-point structure constants can be used to compute them. 

Since the $s$-and-$t$-channel conformal blocks are related by the fusing matrix \eqref{fmd}, their ratios can be computed explicitly. We define the ratio $\rho(V_1|V_2,V_3) = \frac{d^{(s)}_+}{d^{(s)}_-} $, whose expression in terms of structure constants is
\begin{align}
 \rho(V_1|V_2,V_3) = \frac{C_+(V_1) C(V_1^+,V_2,V_3)}{C_-(V_1) C(V_1^-,V_2,V_3)} \ . \label{eq:rcst}
\end{align}
We call expressions of this type a \textbf{shift equation} for $C(V_1, V_2, V_3)$, since it relates the values of the structure constants for values of $(\zb[1]{P})$ shifted by $\pm \frac{\beta}{2}$. 

In order to build equations for the four-point structure constants using equation \eqref{eq:rcst}, we will need to account for the degenerate OPE coefficients $C_{\pm}(V)$ and the two point structure constants $B(V)$. This can be done by looking at the coefficients from a four point functions involving two degenerate fields. For the function $
\left\langle V_{\langle 2,1 \rangle} V_1 V_{\langle 2,1 \rangle} V_1 \right\rangle $, the four-point structure constants are of the type 
\begin{align}
d^{(s)}_\epsilon = C_\epsilon(V_1)B(V_1^\epsilon)C_\epsilon(V_1)
\end{align}
and the corresponding ratio, depending only on $V_1$, is
\begin{align}
 \rho(V_1) = \frac{C_+(V_1)^2 B(V_1^+)}{C_-(V_1)^2 B(V_1^-)} \ .
 \label{eq:rvo}
\end{align}

We can build analogous equations by using the field $V_{\langle 1,2 \rangle}$, and in this case we write the shift equations as
\begin{align}
 \tilde{\rho}(V_1|V_2,V_3) = \frac{\tilde{C}_+(V_1) C(V_1^{\tilde{+}},V_2,V_3)}{\tilde{C}_-(V_1) C(V_1^{\tilde{-}},V_2,V_3)} \ , \label{eq:trcst}
\end{align}
and 
\begin{align}
 \tilde{\rho}(V_1) = \frac{\tilde{C}_+(V_1)^2 B(V_1^{\tilde{+}})}{\tilde{C}_-(V_1)^2 B(V_1^{\tilde{-}})} \ .
 \label{eq:trvo}
\end{align}

It follows that the dependence of the four-point structure constants $D_{s|1234} = D_{1234}(V_s)$ \eqref{eq:4pc} on the $s$-channel field $V_s$ is controlled by the shift equations
\begin{equation}
\begin{aligned}
&\boxed{\frac{D_{1234}(V_s^+)}{D_{1234}(V_s^-)} = \frac{\rho(V_s|V_1,V_2)\rho(V_s|V_3,V_4)}{\rho(V_s)}}\, , \\
&\boxed{\frac{D_{1234}(V_s^{\tilde{+}})}{D_{1234}(V_s^{\tilde{-}})} = \frac{\tilde{\rho}(V_s|V_1,V_2)\tilde{\rho}(V_s|V_3,V_4)}{\tilde{\rho}(V_s)}}\, ,
\end{aligned} \label{eq:drat}
\end{equation}
where the OPE coefficients are no longer involved.

The next step is to determine the ratios $\rho$ and $\tilde{\rho}$ explicitly, and obtain equations for the three-and-four-point structure constants. Using the relation \eqref{gfg} between $s$-and-$t$-channel blocks, and the analogous relation for the right-moving blocks, we can write
\begin{align}
\sum_{\epsilon_1 = \pm} d^{(s)}_{\epsilon_1} \mathcal{F}_{\epsilon_1}^{(s)} \bar{\mathcal{F}}^{(s)}_{\sigma_1\epsilon_1} &= \sum_{\epsilon_1 = \pm} d^{(s)}_{\epsilon_1} \left( \sum_{l = \pm} F_{\epsilon_1,l}\mathcal{F}^{(t)}_{l}  \sum_{m = \pm}\bar{F}_{\sigma_1\epsilon_1,m} \bar{\mathcal{F}}^{(t)}_{m} \right)\\
& = \sum_{l = \pm} \sum_{m = \pm} \left( \sum_{\epsilon_1= \pm } d^{(s)}_{\epsilon_1}F_{\epsilon_1,l} \bar{F}_{\sigma_1\epsilon_1,m} \right) \mathcal{F}^{(t)}_{l} \bar{\mathcal{F}}^{(t)}_{m}\, , \label{eq:blockst}
\end{align}
where the elements of the right-moving fusing matrix are $\bar{F}_{ij} = F_{ij}\vert_{P_i \to \bar{P}_i}$.

Comparing \eqref{eq:blockst} with the $t$-channel expansion of \eqref{eq:2dec}, we find  four equations relating the structure constants $d^{(s)}_\pm,d^{(t)}_\pm$ of both channels, two for each value of $\epsilon_3 = \pm$:
\begin{align}
 \sum_{\epsilon_1 = \pm} d^{(s)}_{\epsilon_1} F_{\epsilon_1,\epsilon_3} \bar F_{\sigma_1\epsilon_1,-\sigma_3\epsilon_3} 
 &= 0\, , \label{h21}
 \\
 \sum_{\epsilon_1 = \pm} d^{(s)}_{\epsilon_1} F_{\epsilon_1,\epsilon_3} \bar F_{\sigma_1\epsilon_1,\sigma_3\epsilon_3}
 &= d^{(t)}_{\epsilon_3}\label{ih21}\, .
\end{align}
In order for the four-point function $\langle V_{\langle 2,1 \rangle} V_1 V_2 V_3 \rangle$ to be nonzero, the homogeneous system \eqref{h21} must admit a non-trivial solution. This happens only if
\begin{align}
 \frac{F_{++}F_{--}}{F_{+-}F_{-+}} = \left(\frac{\bar{F}_{++}\bar{F}_{--}}{\bar{F}_{+-}\bar{F}_{-+}} \right)^{\sigma_1\sigma_3}\, , \label{cons21}
\end{align}
which we call a non-triviality condition. 
Explicitly, equation \eqref{cons21} reads
\begin{align}
 \prod_{\pm} \frac{\cos \pi \beta (P_1\pm P_2-P_3)}{\cos \pi \beta (P_1\pm P_2 + P_3)} = \prod_\pm \frac{\cos \pi \beta (\sigma_1\bar P_1\pm \bar P_2 - \sigma_3 \bar P_3)}{\cos \pi \beta (\sigma_1\bar P_1\pm \bar P_2 +\sigma_3\bar P_3)} \, . 
\end{align}
If the fields $V_1$, $V_2$, $V_3$, are diagonal or non-diagonal, as discussed in section \ref{sec:dnd}, 
the numbers $s_i =  \beta (\sigma_i \bar{P}_i-P_i)$ must be half-integer. Then, the non-triviality condition reads
\begin{align}
 \left(1-(-1)^{2\sum_{i=1}^3 s_i}\right)\sin(2\pi\beta P_1)\cos(2\pi \beta P_2)\sin(2\pi \beta P_3) = 0\ .
\end{align}
Assuming that we are in a generic situation where the trigonometric factors do not vanish, the four-point function admits a non-trivial solution provided
\begin{align}
 \boxed{\sum_{i=1}^3 s_i \in \mathbb{Z}}\, .
 \label{eq:sums}
\end{align}
Now, since fusion with $V_{\langle 2,1\rangle}$ \eqref{dfus} leaves
the number $s$ unchanged, this condition  can be interpreted as a constraint on the three-point functions that result from the fusion $V_{\langle 2,1\rangle} \times V_1$. 
It follows that any nonzero three-point function must obey this condition. 
The same analysis with the  correlation function $\left\langle V_{\langle 1,2 \rangle}V_1V_2V_3 \right\rangle $ leads to the analogous non-triviality condition
\begin{align}
\boxed{\sum_{i=1}^3 r_i \in \mathbb{Z}}\, .\label{eq:sumr}
\end{align}

The implications of the non-triviality conditions \eqref{eq:sums} and \eqref{eq:sumr} on three-point functions can be illustrated by two examples:
\begin{itemize}
\item In any  non-zero three-point function of the type $\left<V^DV^DV^N\right>$, the non-diagonal field must have integer indices $r,s\in\mathbb{Z}$. 
\item Any three-point function with an odd number of fermionic fields vanishes, because fermionic fields obey $r_i+s_i\in\mathbb{Z}+\frac12$ while $\sum_{i=1}^3 (r_i+s_i)\in\mathbb{Z}$. 
So our conditions imply the single-valuedness condition \eqref{sumsp} for three-point functions.
\end{itemize}

In the case where the non-triviality conditions are obeyed, we can write the ratios $\rho(V_1|V_2,V_3)$ in terms of fusing matrix elements:
\begin{align}
\rho(V_1|V_2,V_3) = \frac{d^{(s)}_{+}}{d^{(s)}_{-}} = -\frac{F_{-,\epsilon_3} \bar F_{-\sigma_1,-\sigma_3\epsilon_3}  }{ F_{+,\epsilon_3} \bar F_{\sigma_1,-\sigma_3\epsilon_3}}\, ,\quad  \forall \epsilon_3\in\{+,-\}\ .  \label{eq:rfus}
\end{align}
Inserting explicit expressions \eqref{fmd} for the fusing matrix elements, we obtain
\begin{multline}
 \rho(V_1|V_2,V_3) = -  \frac{\Gamma(-2\beta P_1)}{\Gamma(2\beta P_1)} \frac{\Gamma(-2\beta \sigma_1\bar P_1)}{\Gamma(2\beta \sigma_1\bar P_1)} 
 \\
 \times
 \prod_\pm \frac{\Gamma(\frac12+\beta P_1\pm \beta P_2 +\beta \epsilon_3P_3)}{\Gamma(\frac12 -\beta P_1 \pm \beta P_2 +\beta \epsilon_3 P_3)}
 \prod_\pm \frac{\Gamma(\frac12+\beta \sigma_1\bar P_1\pm \beta \bar P_2 -\beta \epsilon_3\sigma_3\bar P_3)}{\Gamma(\frac12 -\beta \sigma_1\bar P_1\pm \beta \bar P_2-\beta \epsilon_3\sigma_3\bar P_3)}\ .
\end{multline}
Using the non-triviality condition \eqref{eq:sums} we can write $\rho(V_1|V_2, V_3)$ in a $\epsilon_3,\sigma_3$ independent way:
\begin{align}
 \boxed{
 \begin{aligned}
 \rho(V_1|V_2,V_3) = -(-1)^{2s_2} &\frac{\Gamma(-2\beta P_1)}{\Gamma(2\beta P_1)} \frac{\Gamma(-2\beta \sigma_1\bar P_1)}{\Gamma(2\beta \sigma_1\bar P_1)} \\  &\times \frac{\prod_{\pm, \pm} \Gamma(\frac12 +\beta P_1 \pm \beta P_2\pm \beta P_3)}{\prod_{\pm, \pm} \Gamma(\frac12 -\beta \sigma_1\bar P_1 \pm \beta \bar P_2 \pm \beta \bar P_3)}
 \end{aligned}
 }\, .
 \label{eq:r}
\end{align}

Let us check that the behaviour of $\rho(V_1|V_2, V_3)$ under a permutation $V_2 \leftrightarrow V_3$ is in agreement with the behaviour \eqref{permC} of the three-point structure constants. Using eq. \eqref{eq:sums}, we find 
\begin{align}
 \rho(V_1|V_2,V_3)=(-1)^{2s_1}\rho(V_1|V_3,V_2)\ .
\end{align}
Noting that $(-1)^{2s_1} = (-1)^{S(V_1^+)-S(V_1^-)}$, and using the expression of $\rho(V_1|V_2, V_3)$ in terms of three-point structure constants, we conclude that this is the expected behaviour.

The analogous expression for the ratio $\tilde{\rho}=\frac{\tilde{d}^{(s)}_{+}}{\tilde{d}^{(s)}_{-}}$ of $s$-channel structure constants of the four-point function $\left\langle V_{\langle 1,2\rangle} V_1 V_2 V_3 \right\rangle $  is obtained by the substitutions $s\to r,\ \sigma\to \tilde{\sigma},\ \beta\to -\beta^{-1}$:
\begin{multline}
 \tilde{\rho}(V_1|V_2,V_3) = -(-1)^{2r_2} \frac{\Gamma(2\beta^{-1}P_1)}{\Gamma(-2\beta^{-1}P_1)} \frac{\Gamma(2\beta^{-1}\tilde{\sigma}_1\bar P_1)}{\Gamma(-2\beta^{-1}\tilde{\sigma}_1\bar P_1)} 
 \\
 \times
 \frac{ \prod_{\pm, \pm} \Gamma(\frac12 -\beta^{-1}P_1 \pm \beta^{-1}P_2\pm \beta^{-1}P_3)}{ \prod_{\pm, \pm} \Gamma(\frac12 +\beta^{-1}\tilde{\sigma}_1\bar P_1 \pm \beta^{-1}\bar P_2 \pm \beta^{-1}\bar P_3)}\ .
 \label{eq:rt}
\end{multline}

In order to build the shift equations \eqref{eq:drat} for the four-point structure constants we need explicit expressions for the ratios \eqref{eq:rvo} and \eqref{eq:trvo}. We can compute them from \eqref{eq:r} and \eqref{eq:rt} by choosing the appropriate values of the momentums. For the function $\left\langle V_{\langle 2,1 \rangle} V V_{\langle 2,1 \rangle} V \right\rangle $ we choose
\begin{align}
(P_1,\bar{P}_1) &= (P_3,\bar{P}_3)  = (P, \bar{P})\, ,\\
P_2 = \bar{P}_2 &= \beta-\tfrac{1}{2\beta}\, , 
\end{align} 
and we find 
\begin{align}
 \boxed{\rho(V) =  -\frac{\Gamma(-2\beta P)\Gamma(-2\beta \sigma \bar{P})}{\Gamma(2\beta P)\Gamma(2\beta \sigma \bar{P})}\frac{\Gamma(\beta^2 +2\beta P)\Gamma(1-\beta^2 +2\beta P)}{\Gamma(\beta^2 -2\beta\sigma\bar{P})\Gamma(1-\beta^2-2\beta\sigma\bar{P})}}\, . \label{OPEcg}
\end{align}
The ratio corresponding to the function $\left\langle V_{\langle 1,2 \rangle} V V_{\langle 1,2 \rangle} V \right\rangle $  can be obtained from \eqref{OPEcg} by switching $\beta \to -\beta^{-1}$, $\sigma \to \tilde{\sigma}$, giving
\begin{multline}
 \tilde{\rho}(V) =  -\frac{\Gamma(2\beta^{-1} P)\Gamma(2\beta^{-1} \tilde{\sigma} \bar{P})}{\Gamma(-2\beta^{-1} P)\Gamma(-2\beta^{-1} \tilde{\sigma} \bar{P})} \\ \times \frac{\Gamma(\beta^{-2} -2\beta^{-1} P)\Gamma(1-\beta^{-2} -2\beta^{-1} P)}{\Gamma(\beta^{-2} +2\beta^{-1}\tilde{\sigma}\bar{P})\Gamma(1-\beta^{-2}+2\beta^{-1}\tilde{\sigma}\bar{P})}\, . \label{OPEcgt}
\end{multline}
These ratios  simplify in the case of diagonal fields i.e. if $P=\sigma \bar P = \tilde{\sigma}\bar P$,
\begin{equation}
\begin{aligned}
 \rho(V^D_P) &= -\frac{\Gamma^2(-2\beta P)}{\Gamma^2(2\beta P)} \frac{\gamma(\beta^2+2\beta P)}{\gamma(\beta^2-2\beta P)} \, , \\
 \tilde{\rho}(V^D_P) &= -\frac{\Gamma^2(2\beta^{-1} P)}{\Gamma^2(-2\beta^{-1} P)} \frac{\gamma(\beta^{-2}-2\beta^{-1} P)}{\gamma(\beta^{-2}+2\beta^{-1} P)}\, , 
\end{aligned} \label{OPEcd}
\end{equation}
where we introduced $\gamma(x)=\frac{\Gamma(x)}{\Gamma(1-x)}$.

At this stage, we have enough elements to write explicitly the shift equations \eqref{eq:rcst}, \eqref{eq:trcst}, and \eqref{eq:drat} for the three-and-four-point structure constants. For some spectrums this is enough for checking crossing symmetry of four-point functions, since all the required structure constants can be computed by succesive application of the shift equations. We will offer some examples of this in  chapter \ref{ch:crsymfun}. In the next section we focus on finding explicit expressions for the structure constants.

\section{Structure constants}\label{sec:cstssol}
In this section we will show some solutions to the shift equations for the three-point structure constants. We begin by discussing field renormalizations and identifying quantities that are renormalization invariant. Next, we provide explicit expressions for some of these quantities, and discuss the methods to numerically compute the structure constants. 

\subsection{A comment on field normalization}

All fields and structure constants presented up to this point are defined only up to a field renormalization 
\begin{align}
 V_i(z) \to \lambda_i V_i(z)\, ,  \label{eq:renorm}
\end{align}
where the factor $\lambda_i$ is independent of $(\zb{z})$.
In certain cases it is customary fix this normalization by imposing extra constraints on the correlations functions. For example, in Liouville theory it is usual fix the values of the degenerate OPE coefficients $C_{+} = \tilde{C}_{+} = 1$, while in the minimal models normalization is fixed by setting two-point structure constants $B(V)=1$. Here, rather than imposing some of these constraints, we will introduce a reference normalization in which the three-point structure constants take a simple expression.

In a generic normalization we can write the three point structure constants as
\begin{align}
 C_{123} = \left(\prod_{i=1}^3 Y_i\right)C'_{123}\, , \label{3pnorm}
\end{align}
where $C'_{123}$ is normalization independent, and the factors $Y(V_i)$ account for the change in three-point correlation functions due field renormalizations. The change \eqref{eq:renorm} affects two-and-three-point correlation functions by modifying the structure constants as 
\begin{align}
B(V_i) \to \lambda^{2}_iB(V_i) \ ,\quad Y(V_i) \to \lambda_i Y(V_i)\, . 
\end{align}

As a consequence we find that the following quantities are renormalization independent,
\begin{align}
 C'_{123} \qquad \text{and} \qquad Y_i^2B_i^{-1}\ .\label{eq:rinv}
\end{align}
We will use the shift equations \eqref{eq:rcst} and \eqref{eq:trcst} to determine these quantities. This is enough for checking crossing symmetry, as can be seen by writing the four-point structure constants as
\begin{align}
 D_{s|1234} = \left(\prod_{i=1}^4 Y_i\right) C'_{12s} C'_{s34} \frac{Y^2_s}{B_s}\, , \label{eq:4pcp}
\end{align}
where the factor between brackets is common to both the $s$-and-$t$ channel decompositions.

\subsection{Explicit solutions}\label{sec:expsol}
Let us write the equations for the normalization-independent part of the three-point structure constant, $C'_{123}$. We define the normalization-independent ratios
\begin{align}
\rho'(V_1|V_2,V_3)= \frac{C'(V_1^{+},V_2,V_3)}{C'(V_1^{-},V_2,V_3)}\ ,\quad  \tilde{\rho}'(V_1|V_2,V_3)= \frac{C'(V_1^{\tilde{+}},V_2,V_3)}{C'(V_1^{\tilde{-}},V_2,V_3)}\, . \label{eq:rp}
\end{align}
Then, the combination
\begin{align}
\frac{\rho(V_1|V_2,V_3)}{\rho'(V_1|V_2,V_3)} = \frac{C_{+}(V_1)Y(V_1^{+})}{C_{-}(V_1)Y(V_1^{-})} \, , 
\label{eq:rrp}
\end{align}
depends on the field $V_1$. This means that $\rho'(V_1|V_2, V_3)$ accounts for the factor in \eqref{eq:r} that depends on combinations of the momentums $(P_i, \bar{P}_i)$ of the three fields, and the same is true for $\tilde{\rho}'(V_1|V_2, V_3)$. From the explicit expressions \eqref{eq:r} and \eqref{eq:rt} we can extract the expressions of the renormalization invariant ratios \eqref{eq:rp}.
\begin{align}
\rho'(V_1|V_2,V_3) &= (-1)^{2s_2} \beta^{-4\beta (P_1 + \sigma_1 \bar{P_1})}  \frac{\prod_{\pm, \pm} \Gamma\left(\frac{1}{2}+\beta (P_1\pm  P_2 \pm P_3 ) \right)}{\prod_{\pm, \pm}\Gamma\left(\frac{1}{2}-\beta (\sigma_1 \bar{P}_1\pm  \bar P_2 \pm  \bar P_3)\right)}\, , \label{eq:rop}\\
\tilde{\rho}'(V_1|V_2,V_3) &= (-1)^{2r_2} \beta^{-\frac{4}{\beta}(P_1+ \tilde{\sigma}_1 \bar{P_1})}  \frac{\prod_{\pm, \pm}\Gamma\left(\frac{1}{2}-\beta^{-1} (P_1\pm P_2 \pm P_3 ) \right)}{\prod_{\pm, \pm}\Gamma\left(\frac{1}{2}+ \beta^{-1}(\tilde{\sigma}_1 \bar{P}_1\pm \bar{P}_2 \pm \bar{P}_3) \right)}\, , \label{eq:trop}
\end{align}
where the powers of $\beta$ multiplying the $\Gamma$-functions have been conventionally chosen to avoid the appearance of similar factors in the solutions to these equations. 

Expressions \eqref{eq:rop} and \eqref{eq:trop} give the shift equations for the three-point struture constants $C'(V_1, V_2, V_3)$. We can write equations for the factor $Y^2B^{-1}(V)$ by using the ratios $\rho(V)$ and $\tilde{\rho}(V)$, \eqref{eq:rvo} and \eqref{eq:trvo}: 
\begin{equation}
\begin{aligned}
\frac{Y^{2}B^{-1}(V^{+}_{1})}{Y^{2}B^{-1}(V^{-}_{1})} &= \frac{\rho^2(V_1|V_2,V_3)}{\rho'^2(V_1|V_2,V_3)\rho(V_1)}\, ,  \\
\frac{Y^{2}B^{-1}(V^{+}_{1})}{Y^{2}B^{-1}(V^{-}_{1})} &= \frac{\tilde{\rho}^2(V_1|V_2,V_3)}{\tilde{\rho}'^2(V_1|V_2,V_3)\tilde{\rho}(V_1)}\, .  
\end{aligned}  \label{eq:YBr}
\end{equation}
Explicitly, the shift equations for $Y^2B^{-1}(V)$ are
\begin{multline}
\frac{(Y^2B^{-1})\left(V^{+}_{1} \right) }{(Y^2B^{-1})\left(V^{-}_{1} \right)}= -\beta^{8\beta(P_1+\sigma_1\bar{P}_1)} \frac{\Gamma(-2\beta P_1)\Gamma(-2\beta\sigma_1\bar{P}_1)}{\Gamma(2\beta P_1)\Gamma(2\beta\sigma_1\bar{P}_1)}
\\ 
\times
\frac{\Gamma(\beta^2-2\beta\sigma_1\bar{P}_1)\Gamma(1-\beta^2-2\beta\sigma_1\bar{P}_1)}{\Gamma(\beta^2+2\beta P_1)\Gamma(1-\beta^2+2\beta P_1)}\, ,\label{eq:sY2B21}
\end{multline}
\begin{multline}
\frac{(Y^2B^{-1})\left(V^{\tilde{+}}_{1} \right) }{(Y^2B^{-1})\left(V^{\tilde{-}}_{1} \right)}
= 
-\beta^{8\beta^{-1}(P_1+\tilde{\sigma}_1\bar{P}_1)} 
\frac{\Gamma(2\beta^{-1}P_1)\Gamma(2\beta^{-1}\tilde{\sigma}_1\bar{P}_1)}{\Gamma(-2\beta^{-1}P_1)\Gamma(-2\beta^{-1}\tilde{\sigma}_1\bar{P}_1)} 
\\ 
\times
\frac{\Gamma(\beta^{-2}+2\beta^{-1}\tilde{\sigma}_1\bar{P}_1)\Gamma(1-\beta^{-2}+2\beta^{-1}\tilde{\sigma}_1\bar{P}_1)}{\Gamma(\beta^{-2}-2\beta^{-1}P_1)\Gamma(1-\beta^{-2}-2\beta^{-1} P_1)} \, . \label{eq:sY2B12}
\end{multline}
The shift equations \eqref{eq:rop}, \eqref{eq:trop} for $C'(V_1, V_2, V_3)$, and  \eqref{eq:sY2B21} and \eqref{eq:sY2B12} for $Y^2B^{-1}(V)$, determine how the structure constants behave under shifts of the momentums by $\beta$ and $\frac{1}{\beta}$. Assuming the structure constants to be smooth functions of the momentums, the shift equations are enough to determine them up to a momentum independent constant provided $\beta^2$ is a real irrational number. The reason is that the ratio between any two solutions will be a doubly-periodic function of periods $\beta$ and $\frac{1}{\beta}$, and any smooth function with two incommensurable periods is constant. 

Let us denote the condition for having unique solutions as $\beta^2 \in \mathbb{R}$, and find solutions to the shift equations in this case. Notice that there are two different regimes where this condition holds: either $\beta \in \mathbb{R}$, which implies $c \leq 1$, or $\beta \in i \mathbb{R}$, in which case we have $c \geq 25$ and we reparametrize $\beta = i b$, with $b \in \mathbb{R}$. The case when solutions are not unique will be discussed at the end of this section. 

Before giving the solutions, one comment is in order. In the shift equations for the renormalization invariant quantities, equations \eqref{eq:rop}, \eqref{eq:trop}, \eqref{eq:sY2B21} and \eqref{eq:sY2B12}, we introduced certain powers of $\beta$ in order to simplify the form of the solutions. When writing the equations for $\beta = ib$, the factors introduced should be changed according to
\begin{align}
\beta^{\beta(P_1 + \sigma \bar{P}_1)} &\to b^{ib(P_1+\sigma_1 \bar{P}_1)} \ , \\
\beta^{\frac{1}{\beta}(P_1 + \tilde{\sigma} \bar{P}_1)} &\to b^{-\frac{i}{b}(P_1+\tilde{\sigma}_1 \bar{P}_1)}\, , 
\end{align}
and the rest of the equations are obtained by the replacement $\beta \to ib$. 

In order to write the solutions we introduce the Barnes double gamma function of period $\omega$, $\Gamma_{\omega}(x)$, which is invariant under $\omega \to \frac{1}{\omega}$ and satisfies the shift equation
\begin{align}
\frac{ \Gamma_\omega(x+\omega)}{\Gamma_\omega(x)} = \sqrt{2\pi}\frac{\omega^{\omega x-\frac12}}{\Gamma(\omega x)} \ . \label{eq:sgamma}
\end{align}
This function $\Gamma_{\omega}(x)$ is well defined for $\Re{\omega} > 0$, and it is a meromorphic function of $x$ with poles at 
\begin{align}
x = -m \omega - n \omega^{-1}\ , \quad m,n \in \mathbb{N}\, .
\end{align} 
When building solutions to the shift equations we will take $\omega = \beta$ or $\omega =  b$.

Let us begin by showing solutions for the factor $Y^2B^{-1}(V)$, and then move on to the three-point structure constants. We focus first on the case $\beta \in \mathbb{R}$. Using \eqref{eq:sgamma} we can directly check that a solution to the shift equations \eqref{eq:sY2B21} and \eqref{eq:sY2B12} can be written as
\begin{align}
\boxed{ (Y^{2}B^{-1})\left( V_{P,\bar P}^{\sigma,\tilde{\sigma}} \right) =(-1)^{P^2-\bar P^2} \prod_{\pm} \Gamma_\beta(\beta\pm 2P)\Gamma_\beta(\beta^{-1}\pm 2\bar{P})}\, ,  \label{Y2Bnd}
\end{align}
which is evidently invariant under changes in the sign of the momentums. 

If the field $ V_{P,\bar P}^{\sigma,\tilde{\sigma}}$ is diagonal, the expression becomes simpler. Due to the double sign in \eqref{Y2Bnd}, the diagonal case can be obtained by setting $P = \pm \bar{P}$. Then
\begin{align}
(Y^2B^{-1})(V^D_P) = \frac{1}{\prod_\pm \Upsilon_\beta(\beta\pm 2P)} \, ,\label{eq:YBUps}
\end{align}
where we've introduced another special function 
\begin{align}
\Upsilon_{\omega}(x) = \frac{1}{\Gamma_{\omega}(x) \Gamma_{\omega}(\omega + \omega^{-1} -x)}\, .\label{eq:Ups}
\end{align}

On the other hand, when $\beta = ib$, $b \in \mathbb{R}$, the solution for $Y^2B^{-1}(V)$ is
\begin{align}
\boxed{ (Y^{2}B^{-1})\left( V_{P,\bar P}^{\sigma,\tilde{\sigma}} \right) = \frac{(-1)^{P^2-\bar P^2}}{ \prod_{\pm} \Gamma_b(\pm 2iP)\Gamma_b\left( (b+\frac{1}{b})\pm 2i\bar{P}\right) } }\, , \label{Y2Bndb}
\end{align}
which for a diagonal field becomes
\begin{align}
(Y^2B^{-1})(V^D_P) = \prod_\pm \Upsilon_b(\pm2iP) \, .\label{eq:YBUpsb}
\end{align}
Notice that the solutions for $\beta \in \mathbb{R}$ and $\beta \in i\mathbb{R}$ are not the same, and they cannot be obtained as a continuation of one another. 

Let us now give a solutions to the shift equations \eqref{eq:rop} and \eqref{eq:trop} for the three-point structure constant $C'(V_1,V_2, V_3)$. The problem simplifies when we take the field $V_1$ to be diagonal, so we set $V_1 = V^D_{P_1}$ and restrict to this case.

We start by the case $\beta \in \mathbb{R}$. Using the properties of the special function $\Gamma_{\beta}(x)$ one can directly check that the ansatz
\begin{align}
\boxed{\begin{aligned} C'\left(V^{D}_{P_1}, V_2, V_3\right) =& \frac{ f_{2,3}(P_1) }{ \prod\limits_{\pm, \pm} 
\Gamma_\beta(\frac{\beta}{2}+\frac{1}{2\beta} +P_1\pm P_2 \pm P_3)
} \\ &\qquad \times \frac{1}{\prod\limits_{\pm, \pm} \Gamma_\beta(\frac{\beta}{2}+\frac{1}{2\beta}-P_1\pm \bar{P}_2 \pm \bar{P}_3)} 
\end{aligned} 
}\, , \label{cdnn}
\end{align}
satisfies the shift equations \eqref{eq:rop} and \eqref{eq:trop}, provided we make a suitable choice of the numerator. The factor $f_{2,3}(P_1)\in\{-1,+1\}$ should be determined by the equations
\begin{align}
\frac{ f_{2,3}(P_1 +\frac{\beta}{2})}{f_{2,3}(P_1 -\frac{\beta}{2})} = (-1)^{2s_2} \quad , \quad \frac{f_{2,3}(P_1 +\frac{1}{2\beta})}{ f_{2,3}(P_1 -\frac{1}{2\beta})} =(-1)^{2r_2}\, .\label{eq:f23}
\end{align}
In some cases this factor becomes simple: If $r_2,s_2\in \mathbb{Z}$, and in particular if either $V_2$ or $V_3$ are diagonal, we have $f_{2,3}(P_1) = 1$. 

If the three fields are diagonal, the solution \eqref{cdnn} simplifies, becoming
\begin{align}
 C'(V^D_1,V^D_2,V^D_3) &= \prod_{\pm, \pm}\Upsilon_\beta\left(\tfrac{\beta}{2}+\tfrac{1}{2\beta}+P_1\pm P_2\pm P_3\right) \label{eq:cpl} \\
 &= C'^{D}(P_1, P_2, P_3) \ , 
\end{align}
where in the second line we write the diagonal three-point structure constant $C'^D$, as a functions of the momentums. Up to normalization dependent factors, this expression coincides with the one derived in \cite{Zamolodchikov:2005fy}. 

Since the momentums $\pm P_1$ correspond to the same conformal weight (and hence to the same Virasoro representation),  let us see that the denominator of $C'(V^D_1, V_2, V_3)$ \eqref{cdnn} is invariant under this change of sign.
Neglecting the numerator, we have 
\begin{align}
\frac{C'(V^D_{P_1}, V_2, V_3)}{C'(V^D_{-P_1}, V_2, V_3)} \propto
 \frac{\prod\limits_{\pm, \pm} S_\beta(\frac{\beta}{2}+\frac{1}{2\beta}+P_1\pm \bar{P}_2 \pm \bar{P}_3)}{\prod\limits_{\pm, \pm} S_\beta(\frac{\beta}{2}+\frac{1}{2\beta} +P_1\pm P_2 \pm P_3)} \ , \label{eq:cpref}
\end{align}
where we've introduced another special function $S_\beta(x) = \frac{\Gamma_\beta(x)}{\Gamma_\beta(\beta+\frac{1}{\beta}-x)}$, satisfying the shift equation $\frac{S_\beta(x+\beta)}{S_\beta(x)} = 2 \sin(\pi\beta x)$. The function $S_{\beta}(x)$ satisfies the identity
\begin{align}
\prod_\pm \frac{S_\beta(x\pm P_{(r,-s)})}{S_\beta(x \pm P_{(r,s)})}=(-1)^{rs}, \quad \forall r,s \in \mathbb{Z} \, . \label{eq:IdS}
\end{align}
Making use of the non-triviality conditions \eqref{eq:sums} and \eqref{eq:sumr}, and the single-valuedness condition  \eqref{sumsp}, the ratio \eqref{eq:cpref} becomes
\begin{align}
 \frac{C'(V^D_{P_1}, V_2, V_3)}{C'(V^D_{-P_1}, V_2, V_3)} \propto (-1)^{2(r_2s_2 + r_3s_3)} = 1\ .
\end{align}
In particular, when the three fields are diagonal we can see that the expression \eqref{eq:cpl} is invariant under reflections of the momentums. This is thanks to the identity $\Upsilon_{\beta}(x) = \Upsilon_{\beta}\left(\beta + \beta^{-1} - x \right)$, which follows from \eqref{eq:Ups}.

In the case $\beta = ib$, the solution to the shift equations is
\begin{align}
\boxed{
\begin{aligned}
C'\left(V^{D}_{P_1}, V_2, V_3\right) =  f_{2,3}(P_1) \prod\limits_{\pm, \pm} &
\Gamma_b\left(\tfrac{b}{2}+\tfrac{1}{2b} + i (P_1\pm P_2 \pm P_3) \right)
\\ \times &\prod\limits_{\pm, \pm} \Gamma_b\left(\tfrac{b}{2}+\tfrac{1}{2b} - i (P_1\pm \bar{P}_2 \pm \bar{P}_3) \right)
\end{aligned} }\, , \label{cdnnb}
\end{align}
where the factor $f_{2,3}(P_1)$ obeys the shift equations \eqref{eq:f23} with $\beta \to ib$. 
In this case, when the three fields are diagonal we obtain
\begin{align}
 C'(V^D_1,V^D_2,V^D_3) &= \frac{1}{\prod_{\pm, \pm}\Upsilon_b\left(\tfrac{b}{2}+\tfrac{1}{2b}+ i (P_1\pm P_2\pm P_3) \right)} \label{eq:cplb} \\
 &= C'^{D}(P_1, P_2, P_3) \ , 
\end{align}
which gives the renormalization invariant part of the famous DOZZ solution for the three-point structure constants of Liouville theory, named after Dorn, Otto \cite{Dorn:1994xn} and  Zamoldchikov \& Zamolodchikov \cite{Zamolodchikov:1995aa}. 
Under reflections of $P_1$, expressions \eqref{cdnnb} and \eqref{eq:cplb} behave similarly to their counterparts with $\beta \in \mathbb{R}$. The proof of this statement follows the same steps as before, using analogous identities. 

Combining expressions \eqref{cdnn} or \eqref{cdnnb} with the permutation properties \eqref{permC} of structure constants makes it possible to determine any structure constant containing at least one diagonal field, regardless of the ordering of the arguments. So, we have found solutions to the shift equations for the normalization-independent quantities $C'$ and $Y^2B^{-1}$, with the exception of three-point structure constants that involve three non-diagonal fields. In that case, writing an explicit solution for $C'$ is not as straightforward, and we leave it for future work. Nonetheless, the shift equations \eqref{eq:rop} and \eqref{eq:trop} are not restricted to cases involving at least one diagonal field, and for certain spectrums they could be enough for computing structure constants involving non-diagonal fields only. 

Verifying crossing-symmetry requires four-point structure constants, which are controlled by equations \eqref{eq:drat}. The required renormalization invariant four-point structure constants can be built from our solutions by means of the relation \eqref{eq:4pcp}. Some examples are shown in chapter \ref{ch:crsymfun}. 

Let us comment about the uniqueness of the solutions of the shift equations, in the case where one of the momentums can take arbitrary values (i.e. it belongs to a diagonal field). The explicit solutions shown here are defined in terms of special functions, and they are unique when $\beta \in \mathbb{R}$ or $\beta \in i\mathbb{R}$. The solutions valid when $\beta \in \mathbb{R}$ can be extended to all values $\beta \notin i\mathbb{R}$, and the solutions for $\beta = ib$ can be continued to every $b \notin i\mathbb{R}$, provided the special functions remain well defined. However, these extensions are not unique: for $\beta^2 \notin \mathbb{R}$, solutions to the shift equations are defined up to multiplication by an arbitrary elliptic functions, and there are thus infinitely many different solutions. For a more detailed discussion in the diagonal case, see \cite{Zamolodchikov:2005fy}. 

Solutions to the shift equations can be interpreted as structure constants of consistent CFTs when they give rise to crossing-symmetric four-point functions, for a given spectrum. When there are infinitely many different solutions, it is unlikely that all of them are the structure constants of a consistent theory. Then, for a given theory, being able to analytically continue its structure constants to a different value of $c$ does not mean that the whole theory will have a continuation. This happens in the case of Liouville theory for $c \leq 1$\cite{rs15}, which will be discussed in section \ref{sec:liouville}.

\subsection{Relation between diagonal and non-diagonal solutions}

Our expressions for $Y^{2}B^{-1}$  and $C'$ are valid in cases when there are both diagonal and non-diagonal fields involved. We have seen that taking all fields to be diagonal usually simplifies some results, for example in equations \eqref{eq:YBUps} and \eqref{eq:cpl} or \eqref{eq:YBUpsb} and \eqref{eq:cplb}. It is then interesting to ask whether there exists a relation between generic solutions, and those involving only diagonal quantities. For concreteness we focus on the case $\beta \in \mathbb{R}$, knowing that all identities shown here translate to the case $\beta = ib$ in a straightforward way.  

Taking the square of \eqref{cdnn}  we can relate the general case with the diagonal one, 
\begin{align}
C'^2(V^D_1,V_2,V_3) = C'^{D}(P_1,P_2,P_3) C'^{D}(P_1,\bar{P}_2,\bar{P}_3)\, ,  \label{eq:cn2cd}
\end{align}
which suggest that quantities in the non-diagonal case can be computed as geometric means of the corresponding quantities in the diagonal case, i.e.
\begin{align}
C'(V_1, V_2, V_3) = \sqrt{C^D(P_1, P_2, P_3)C^D(\bar{P}_1, \bar{P}_2, \bar{P}_3)}\, .\label{eq:gmean}
\end{align}
Relations of this type appear when studying non-diagonal theories \cite{petkova1988two, petkova1989structure, Petkova:1994zs}, and in \cite{ei15} where used to compute ratios of certain structure constants. However, the appearance of the square root could spoil the analyticity of the diagonal structure constants, and it introduces a sign ambiguity that needs to be solved. Here, the explicit expression \eqref{cdnn}, derived as a solution of the shift equations, does not present these issues, or rather shows that there is a precise way to build the non-diagonal solutions. 

The factor $Y^2B^{-1}(V)$ in equation \eqref{Y2Bnd} can be shown to satisfy a relation of the same type. First, we check that it is invariant under an exchange of the left-and-right moving momentums, realised by switching $s \to -s$:
\begin{align}
\frac{(Y^2B^{-1})\left(V^N_{(r,s)}\right)}{(Y^2B^{-1})\left(V^N_{(r,-s)}\right)} =\prod_\pm \frac{S_\beta(\beta\pm 2P_{(r,-s)})}{S_\beta(\beta \pm 2P_{(r,s)})}
=(-1)^{4rs}=1 \, , \label{eq:Yrat}
\end{align}
where we used the identity \eqref{eq:IdS}.
Using this property we find
\begin{align}
 \left[(Y^2B^{-1})\left(V^N_{(r,s)}\right)\right]^2 
=  (Y^2B^{-1})\left(V^D_{P_{(r,s)}}\right) (Y^2B^{-1})\left(V^D_{\bar P_{(r,s)}}\right)\, , \label{eq:Y2Yd}
\end{align}
where we have used $\bar{P}_{(r,s)} = P_{(r,-s)}$.

These relations between diagonal and non-diagonal objects ultimately come from the shift equations, which satisfy analogous relationships. Indeed, we can write
\begin{align}
 \rho^2 = \prod_{\epsilon_3 = \pm} \frac{ F_{+,\epsilon_3} \bar F_{\sigma_1,-\sigma_3\epsilon_3}}{F_{-,\epsilon_3} \bar F_{-\sigma_1,-\sigma_3\epsilon_3}  } = \frac{F_{++}F_{+-}}{F_{-+}F_{--}} \left(\frac{\bar{F}_{++}\bar{F}_{+-}}{\bar{F}_{-+}\bar{F}_{--}} \right)^{\sigma_1} \ . \label{r2rdrd}
\end{align}
Equivalently
\begin{align}
 \rho^2(V_1|V_2, V_3) = \rho^D(P_1|P_2,P_3) \rho^D(\bar{P}_1|\bar{P}_2,\bar{P}_3)\, ,
 \label{rhrh}
\end{align}
where $\rho^D=\frac{F_{++}F_{+-}}{F_{-+}F_{--}}  $ is the expression for $\rho$ when all three fields are diagonal.

It turns out that relationships like \eqref{eq:gmean} between diagonal and non-diagonal solutions are common to a larger class of systems of differential equations, as discussed in the appendix of \cite{Migliaccio:2017dch}.

\subsection{Computing structure constants} \label{sec:strcalc}

Explicit expressions for structure constants, like \eqref{Y2Bnd}, \eqref{Y2Bndb}, \eqref{cdnn}, \eqref{eq:cpl} and their $\beta \in i\mathbb{R}$ counterparts,  are specially useful when it is not possible to relate all constants by using the shift-equations, for example in the case of continuous spectrums. In such a situation it can be useful to have the following integral expressions for the special functions, in order to compute them directly: 
\begin{multline}
\log\Gamma_{\omega}(x) =\\ \int_0^\infty\frac{dt}{t}\left[\frac{e^{-xt}-e^{-\frac{1}{2}(\omega + \omega^{-1})t}}{(1-e^{-\omega t})(1-e^{-\omega^{-1}t})} -\frac{\left(\frac{1}{2}(\omega + \omega^{-1})-x\right)^2}{2}e^{-t} -\frac{\frac{1}{2}(\omega + \omega^{-1})-x}{t}\right]\ , \label{eq:intG}  
\end{multline}
and
\begin{align}
\log\Upsilon_{\omega}(x) = \int_0^\infty\frac{dt}{t}\left[\left(\frac{1}{2}(\omega + \omega^{-1})-x\right)^2e^{-t} -\frac{\sinh^2\frac12\left(\frac{1}{2}(\omega + \omega^{-1})-x\right)t}{\sinh\left(\frac12 \omega t\right)\sinh\left(\frac12 \omega^{-1}t\right)}\right]\ \label{eq:intU}
\end{align}

These integral definitions hold for $\Re{x} > 0$, in the case of \eqref{eq:intG}, and $0< \Re{x} < \frac{1}{2}(\omega + \omega^{-1})$ for \eqref{eq:intU}. In order to extend the other values of the argument we can use the shift equations that each special function satisfies. 

When the spectrum of the conformal block decomposition is discrete, it can be more useful to use directly the shift equations. This approach can be simpler, since shift equations involve only $\Gamma(x)$ functions, and none of the other special functions that have been introduced. 

Computing structure constants by successive application of the shift equations \eqref{eq:drat} requires that we choose the starting values to which all constants are related, and the number of values to choose depends on the spectrum considered. For the purpose of checking crossing symmetry between the $s$-and-$t$-channels in a specific four-point function $\langle V_1 V_2 V_3 V_4 \rangle$ we may fix values on one channel, and then determine the other channel through shift equations. For example, identifying by $V^{(s)}_0$ the $s$-channel field with lowest conformal dimension, we may set
\begin{align}
D_{1234}(V^{(s)}_0) = 1\, ,
\end{align} 
and then use either the shift equations or the explicit expressions to compute the ratio
\begin{align}
\frac{D_{2341}(V^{(t)}_0)}{D_{1234}(V^{(s)}_0)} \label{eq:normrat}
\end{align} 
In this discussion we have tacitly assumed that all the structure constants within each channel are related among themselves through the shift equations.  If this is the case, once the ratio \eqref{eq:normrat} is determined, the shift equations can be used to compute all other structure constants. We have given here an illustration of the procedure, but we present more precise examples in appendix \ref{sec:cstsap}. 

%% file: chapters/mmlim.tex
\chapter{Crossing symmetric four-point functions}\label{ch:crsymfun}
In the previous chapter we have found solutions for the constraints that degenerate crossing symmetry imposes on the two-point and three-point structure constants. These solutions should be universal, in the sense that they apply to any theory satisfying the three main assumptions discussed in section \ref{sec:assumptions}. 

In order to show that a conformal field theory is consistent we need to show that crossing symmetry is satisfied in generic four-point functions, not necessarily involving degenerate fields. Showing that the structure constants we have found satisfy these constraints requires that we determine the  $s$-and-$t$-channel spectrums of four point functions.
In the first part of this chapter we will present some examples of known CFTs where our solutions apply. These theories are all either diagonal or rational, and in the second part of this chapter we will try to build a non-rational, non-diagonal CFT, and provide numerical evidence for its existence. 

\section{Examples of crossing-symmetric theories}\label{sec:theories}
In this section we present examples of conformal field theories to which the analytic conformal bootstrap solutions apply. We discuss each theory's spectrum and fusion rules and argue that the solutions of chapter \ref{ch:cboots}  apply. For Liouville theory and the generalized minimal models we give references where crossing symmetric functions have been computed. For the A-series and D-series minimal models we show examples of four-point functions. Finally, in the case of the Ashkin-Teller model we verify that its known structure constants satisfy the shift equations. 

\subsection{Liouville theory}\label{sec:liouville}

The name Liouville theory corresponds to a family of two-dimensional, non-rational conformal field theories parametrized by the central charge $c$. One of the most important characteristics of these theories is that they possess a diagonal, continuous spectrum, which can be formally written as
\begin{align}
\mathcal{S}_{L} = \int\limits_{P \in \mathbb{R}_{\geq 0}} dP\  \mathcal{V}_{P}\otimes \bar{\mathcal{V}}_{P}\, , \label{eq:Lspec}
\end{align}
where the integration is performed over the positive real line only, in order to avoid double counting of Verma modules with reflected momentums, $\mathcal{V}_P$ and $\mathcal{V}_{-P}$. 

Let us discuss OPEs and four-point correlation functions in Liouville theory, leaving aside technical details related to certain values of the central charge. 

In this theory, the spectrum of the OPE between two non-degenerate fields $V^D_{P_1}$ and $V^D_{P_2}$ is also given by equation \eqref{eq:Lspec}, such that the fusion rules are
\begin{align}
V^D_{P_1} \times V^D_{P_2} = \int\limits_{P_3 \in \mathbb{R}_{\geq 0}} dP_3\  V^{D}_{P_3}\, .
\end{align}

Applying this rule to a four-point correlation function 
\begin{align}
\left\langle V^{D}_{P_1}(\zb{x})V^{D}_{P_2}(0)V^{D}_{P_3}(\infty)V^{D}_{P_4}(1) \right\rangle \ ,
\end{align} 
crossing symmetry takes the form 
\begin{align}
\int\limits_{P_s \in \mathbb{R}_{\geq 0}} dP_s\ D_{s|1234} \left|\mathcal{F}^{(s)}_{\Delta_s}(\Delta_i|x)\right|^2 = \int\limits_{P_t \in \mathbb{R}_{\geq 0}} dP_t\ D_{t|2341} \left|\mathcal{F}^{(t)}_{\Delta_t}(\Delta_i|x)\right|^2\, . \label{eq:L4pt}
\end{align}

In order to compute four-point functions in Liouville theory one must be able to perform the integrations in \eqref{eq:L4pt}, which involves controlling possible singularities in the integrand. In particular, conformal blocks are singular whenever the internal momentum $P_s$ or $P_t$ coincides with one of the degenerate momentums $P_{(r,s)}$ with $r,s \in \mathbb{N}$, so that care must be taken when these momentums are real. Regarding the structure constants, it was established in section \ref{sec:cstssol} that the form of the solutions to the shift equations depends on the value of the central charge. The solutions are unique when $\beta \in \mathbb{R}$ or $\beta \in i\mathbb{R}$, and the structure constants in the four-point function \eqref{eq:L4pt} have to be chosen according to the value of $\beta$. Let us discuss the two cases where the solutions are unique:
\begin{itemize}
\item $\beta \in i\mathbb{R} \Rightarrow c \geq 25$.

In this case, setting $\beta = ib$, with $b \in \mathbb{R}_{\geq 0}$, the three-point structure constants and the factor $Y^2B^{-1}(V)$ are given by equations \eqref{eq:cplb} and \eqref{eq:YBUpsb}. We can combine these expressions to find the four-point structure constant
\begin{multline}
D_{1234}(P_s) = \frac{\prod_{\pm} \Upsilon_b(\pm 2i P_s)}{\prod_{\pm \pm }\Upsilon_b \left( \frac{1}{2}(b+\frac{1}{b}) +i( P_s \pm P_1 \pm P_2 )\right)  } \times \\ \frac{1}{\prod_{\pm \pm }\Upsilon_b \left( \frac{1}{2}(b+\frac{1}{b}) +i( P_s \pm P_3 \pm P_4) \right)}\, . \label{eq:L4p25}
\end{multline}
These structure constants have no singularities in the domain of integration $P_s \in \mathbb{R}_{\geq 0}$. Conformal blocks have poles for degenerate momentums $P_s =P_{(r,s) }$, but for $\beta  \in i \mathbb{R}$ we have,
\begin{align}
P_{(r,s) } \in i \mathbb{R}\, ,
\end{align}
so that the internal momentums $P_s, P_t \in \mathbb{R}_{\geq 0}$ never reach these values. Then, the poles of the integrands in equation \eqref{eq:L4pt} are away from the integration line, and with the help of the explicit expressions for the special functions $\Upsilon_b(x)$, equation \eqref{eq:intU}, the four-point functions can be computed. 

Four-point functions of the type of \eqref{eq:L4pt} have been computed in \cite{rs15} for different values of $c \geq 25$, and they were found to satisfy crossing symmetry. 

\item $\beta \in \mathbb{R} \Rightarrow c \leq 1$

In this case, the structure constants and the factor $Y^2B^{-1}(V)$ are given by equations \eqref{eq:cpl} and \eqref{eq:YBUps} respectively, so that the four-point structure constant takes the form
\begin{multline}
D_{1234}(P_s) = \frac{\prod_{\pm, \pm}\Upsilon_\beta\left(\tfrac{\beta}{2}+\tfrac{1}{2\beta}+P_s\pm P_1\pm P_2\right)}{\prod_{\pm} \Upsilon_\beta(\beta \pm 2P_s)} \\ 
\prod_{\pm, \pm}\Upsilon_\beta\left(\tfrac{\beta}{2}+\tfrac{1}{2\beta}+P_s\pm P_3\pm P_4\right)\, .\label{eq:L4p1}
\end{multline}

In this case, however, all the poles of the structure constants and of the conformal blocks take real values, and thus they are located on top of the integration line.
The presence of these poles implies that the integration in equation \eqref{eq:L4pt} cannot be performed as it is written. However, it was argued in \cite{rs15} that in order to avoid the poles it suffices to shift the integration in the imaginary direction by a small parameter $\epsilon$, thus integrating over 
\begin{align}
P \in \mathbb{R} + i \epsilon \, .
\end{align}
Numerical computations in \cite{rs15} show that Liouville theory four-point functions computed in this way are crossing-symmetric, and that the results do not depend on  the regularization parameter $\epsilon$. 
\end{itemize}

From this discussion we learn that solutions to the analytic conformal bootstrap of chapter \ref{ch:cboots} give rise to crossing-symmetric functions in Liouville theory for $c \in ]-\infty, 1] \cup [25, \infty[$. Furthermore, the structure constants and conformal blocks depend analytically on $\beta$ or $b$, and one may wonder whether it is possible to find a continuation of Liouville theory for other values of the central charge. Indeed, in \cite{Ribault:2014hia} and \cite{rs15} it is argued that the solution of Liouville theory for $c \geq 25$ can be extended to all complex values of $c$ such that $c \notin ]-\infty, 1]$, or $\beta \notin \mathbb{R}$, while the solution for $c \in ]-\infty, 1]$ cannot be extended beyond this line due to the impossibility of avoiding poles when performing the integration in the four-point function. Then, the analytic conformal bootstrap structure constants give solutions for Liouville theory for every $c \in \mathbb{C}$.

Liouville theory is an interesting example where all the assumptions from section \ref{sec:assumptions} hold: Every primary field is diagonal, so that the single-valuedness condition \eqref{sumsp} always holds. Moreover, the theory has a realisation for every value of $c \in \mathbb{C}$, and its structure constants depend analytically on $\beta$. Finally, structure constants determined by studying degenerate fields give rise to crossing-symmetric four-point functions, even though degenerate fields are not part of any OPE spectrums. Next, we discuss an example of a diagonal theory that does contain degenerate fields in its spectrum.

\subsection{Generalized minimal models}

Here we follow the definitions of \cite{Ribault:2014hia}, and define  generalized minimal models as a family of diagonal CFTs that exist for every $c \in \mathbb{C}$, and whose spectrum is made from all the degenerate representations of the type $R_{\langle r,s \rangle}$, $r,s \in \mathbb{N}_{\geq 1}$. We can write this spectrum as
\begin{align}
\mathcal{S}^{GMM} = \bigoplus_{r,s \in \mathbb{N}_{\geq 1}} R_{\langle r,s \rangle} \otimes \bar{R}_{\langle r,s \rangle}\, .
\end{align}
Here $\mathcal{S}^{GMM}$ is a discrete, infinite  spectrum, which in particular includes degenerate fields $V_{\langle 2,1 \rangle}$ and $V_{\langle 1,2 \rangle}$ we used to derive the degenerate crossing symmetry constraints. 

The fusion rules of these models arise form the degenerate fusion rules discussed in section \ref{sec:degfields}. Taking equation \eqref{rtv} and assuming that both representations are  degenerate: In terms of the fields, the fusion rules read
\begin{align}
V_{\langle r_1,s_1 \rangle} \times V_{\langle r_2,s_2 \rangle} &= \sum_{r_3\overset{2}{=}|r_1-r_2|+1}^{r_1+r_2-1}\ \sum_{s_3\overset{2}{=}|s_1-s_2|+1}^{s_1+s_2-1} V_{\langle r_3,s_3 \rangle}\ ,
\label{rrsr}
\end{align}
where the superscript in $\overset{2}{=}$ indicates that the corresponding sum runs by increments of $2$. 

The structure constants of the generalized minimal models should satisfy the same shift-equations as the structure constants of Liouville theory, and they can be obtained from them by taking a limit of \eqref{eq:L4p25} or \eqref{eq:L4p1}, depending on the value of $c$,  in which all momentums become degenerate. Explicit expressions can be found in   \cite{Ribault:2014hia}.

While taking limits of Liouville theory structure constants can lead to expressions for the generalized minimal model structure constants, fusion rules \eqref{rrsr} have to be imposed separately. Combining these elements it is possible to numerically verify crossing symmetry of four-point functions for different values of the central charge \cite{rs15}.

The examples of CFTs discussed so far have infinite diagonal spectrums, even though fusion rules of the generalized minimal models give rise to finite expansions. In the next section we discuss minimal models, which have finite spectrums, and in some cases involve non-diagonal fields.

\subsection{Minimal models}\label{sec:minmods}

The minimal models are theories whose spectrums are finite, and built from degenerate representations. They exist for rational values of the central charge, are given in terms of two coprime positive integers $p,q$ such that $\beta^2 = \frac{p}{q}$. The values of $c_{p,q}$ are \cite{fms97}
\begin{align}
c_{p,q} = 1 - 6 \frac{(q-p)^2}{pq} \, ,\label{eq:mmc}
\end{align}
where $c_{p,q} \leq 1 \forall p,q$.

When studying level two null vectors in section \ref{sec:nullvec}, we saw in equation \eqref{eq:nullD} that particular values of the central charge can cause degenerate conformal weights $\Delta_{(r,s)}$ and $\Delta_{(r', s')}$ to coincide. For example, we found that for $c = 1$ we have $\Delta_{(1,2)} = \Delta_{(2,1)}$. Let us see how the values $c = c_{p,q}$ also give rise to similar coincidences. Since $\beta^{2} = \frac{p}{q}$, we have
\begin{align}
P_{(q,p)} = \frac{1}{2} (q\beta-\frac{p}{\beta}) = 0\, .
\end{align}
This means that it is possible to shift the indices of a degenerate conformal weight  $\Delta_{(r,s)}$ without altering its value. For any $\lambda \in \mathbb{R}$ we have
\begin{align}
\Delta_{(r,s)} = \Delta_{(\lambda q + r, \lambda p +s)} = \Delta_{(\lambda q - r, \lambda p -s)}\, , \label{eq:eqlrs}
\end{align}
where for the second equality we have used the fact that $\Delta(P) = \Delta(-P)$, and $P_{(-r,-s)} = -P_{(r,s)}$. 

The equalities \eqref{eq:eqlrs} mean that the degenerate representations $\mathcal{R}_{\langle r,s \rangle}$ become multiply degenerate. This happens in two ways: on the one hand, $\mathcal{R}_{\langle r,s \rangle}$ now contains null vectors at levels $rs$, $(q-r)(p-s), (q + r)(q+s)$, etc, and on the other hand, each of these null vectors also has null descendants. For example, the null vector at level $rs$ has weight
\begin{align}
\Delta_{(r,s)} + rs = \Delta_{(r, -s)}\, .
\end{align} 
For generic values of $c$, $\Delta_{(r,-s)}$ does not give rise to a degenerate representation, because only one of its indices is negative (if they both were, a reflection would give two positive indices). However, for $c = c_{p,q}$ we can shift the indices by using equation \eqref{eq:eqlrs}, and choosing an appropriate $\lambda \in \mathbb{N}$ we can write $\Delta_{(r,-s)} $ as a degenerate conformal weight. For example, choosing $\lambda = 1$ we see that $\Delta_{(r,-s)} = \Delta_{(q+r, p-s)}$, which is evidently degenerate if $p > s$.

The coincidences between dimensions imply also coincidences between degenerate representations, 
\begin{align}
\mathcal{R}_{\langle r,s \rangle} = \mathcal{R}_{\langle q+r,p+s \rangle} = \mathcal{R}_{\langle q-r,p-s \rangle}\, . \label{eq:repeq}
\end{align}
This correspondence is a direct consequence of the special values of the central charge \eqref{eq:mmc}, and it gives a way to construct finite sets of degenerate fields that are closed under fusion. These sets form the spectrums of the rational conformal field theories called minimal models, and their fusion rules are obtained by combining the degenerate fusion rules \eqref{rrsr} with the equivalences \eqref{eq:repeq}. In what follows we will discuss two types of minimal models, known as the A-series and D-series. These models exist for infinitely many values of the central charge of the type $c_{p,q}$, which are dense in the line $c \leq 1$. They differ in that the D-series models spectrum contains a non-diagonal sector, absent in the A-series. We will leave aside the E-series models, because they only exist for a subset of $c_{p,q}$ that is not dense in $c \leq 1$.

A-series minimal models are diagonal theories which exists for all values of $c = c_{p,q}$  with $p,q$ two coprime positive integers. Their spectrum for a given $(p,q)$ are   
\begin{align}
 \mathcal{S}^{\text{A-series}}_{p,q} = \frac12 \bigoplus_{r=1}^{q-1}\bigoplus_{s=1}^{p-1} \mathcal{R}_{\langle r,s\rangle} \otimes \bar{\mathcal{R}}_{\langle r,s\rangle} \ , \label{eq:AMMspec}
\end{align}
where the factor $1/2$ accounts for the double counting introduced by the symmetry $\mathcal{R}_{\langle r,s\rangle} = \mathcal{R}_{\langle q- r, p-s \rangle}$.

The fusion rules for the A-series minimal models can be obtained by combining the degenerate fusion rules \eqref{rrsr}, and the equivalences between fields given by $\Delta_{(r,s)} = \Delta_{(q-r,p-s)}$. The way to incorporate this symmetry is by noting that for any two fields $V_{\langle r_1,s_1 \rangle},V_{\langle r_2,s_2 \rangle} \in \mathcal{S}^{\text{A-series}}_{p,q}$, the following OPEs should give the same result:
\begin{equation}
\begin{aligned}
V_{\langle r_1,s_1 \rangle}\times V_{\langle r_2,s_2 \rangle} &= V_{\langle q-r_1,p-s_1 \rangle}\times V_{\langle r_2,s_2 \rangle}\\ &=V_{\langle r_1,s_1 \rangle}\times V_{\langle q-r_2,p-s_2 \rangle} \\ &= V_{\langle q-r_1,p-s_1 \rangle}\times V_{\langle q-r_2,p-s_2 \rangle}\, .
\end{aligned}
\end{equation}
The resulting fusion rules are
\begin{align}
V_{\langle r_1,s_1 \rangle}\times V_{\langle r_2,s_2 \rangle} = \sum^{\min(r_1+r_2, 2q-r_1-r_2) -1 }_{r_3\overset{2}{=} |r_1 - r_2| +1} \sum^{\min(s_1+s_2, 2p-s_1-s_2) -1 }_{s_3\overset{2}{=} |s_1 - s_2| +1} V_{\langle r_3, s_3 \rangle} \, . \label{eq:afrule}
\end{align}

The indices in the fusion rules increase by steps of two. This will allow us to compute the structure constants of a particular OPE by repeated application of the shift equations \eqref{eq:drat}, avoiding the use of explicit expressions. We will use these fusion rules to compute the s-and-t-channel spectrums of four-point correlation functions, but let us first discuss the second kind of minimal models. 

D-series minimal models exist for certain values $c_{p,q}$ such that $p,q$ have different parity. For concreteness we will take $q$ even and $p$ odd, but equivalent models can be obtained by changing $p \leftrightarrow q$. With this convention, D-series minimal models exist for $q \geq 6$ even, and $p \geq 3$ and odd, with $q$ and $p$ coprime.  Their spectrum can be split into a diagonal and a non-diagonal sector, and is given by
\begin{align}
 \mathcal{S}_{p,q}^{\text{D-series}} \ =\  \frac12 \underbrace{\bigoplus_{r\overset{2}{=}1}^{q-1} \bigoplus_{s=1}^{p-1} \left|\mathcal{R}_{\langle r,s\rangle}\right|^2}_{\text{Diagonal}} \oplus \frac12 \ \underbrace{\bigoplus_{\substack{1\leq r\leq q-1 \\ r\equiv \frac{q}{2}\bmod 2}} \  \bigoplus_{s=1}^{p-1} \mathcal{R}_{\langle r,s\rangle} \otimes \bar{\mathcal{R}}_{\langle q-r,s\rangle} }_{\text{Non-diagonal}}\ . \label{eq:DMMspec}
\end{align}
Where again the factors of $\frac{1}{2}$ account for double counting of representations as in \eqref{eq:AMMspec}.

As opposed to the A-series spectrum \eqref{eq:AMMspec}, here the index $r$ only takes half the values between $1$ and $q-1$. In particular, the diagonal sector of the D-series spectrum has $r$ odd, while the non-diagonal sector has $r \equiv \frac{q}{2} \bmod 2$. The fields of the non-diagonal sector have spins
\begin{align}
S = - (r - \frac{q}{2})(s - \frac{p}{2})\, .
\end{align}

While we can choose $q$ even and $p$ odd without losing generality in the description of minimal models, the D-series spectrum \eqref{eq:DMMspec} shows qualitative differences depending on whether $q$ is a multiple of $4$ or not. If $q = 2 \bmod 4$, the indices in the  summations of the diagonal and the non-diagonal sectors take the same values. In particular the representation $\mathcal{R}_{\langle \frac{q}{2}, s \rangle} \otimes \bar{\mathcal{R}}_{\langle \frac{q}{2}, s \rangle}$ appears on both sectors, and we find a situation where we have diagonal and non-diagonal fields with the same momentums, as described at the end of section \ref{sec:dnd}. On the contrary, if $q = 0 \bmod  4$ the $r$ indices of each sectors take different values, and there are no representations appearing in both simultaneously. 

Before discussing D-series minimal models fusion rules, let us introduce a change in notation that will allow us to more easily relate the minimal model fields to the diagonal and non-diagonal fields of chapter \ref{ch:cboots}, and that will prove useful in  section \ref{sec:2zz12}:
For a minimal model of charge $c_{p,q}$ for $q$ even and $p$ odd, we redefine the indices of the fields by
\begin{align}
r \to r-\frac{q}{2}\ , \quad s \to s-\frac{p}{2}\, , 
\end{align}
a redefinition that is allowed by the symmetries \eqref{eq:eqlrs} with $\lambda = \frac{1}{2}$. Notice that now the indices can take negative values, and also that we now have $s \in \mathbb{Z}+\frac{1}{2}$. 

In this notation, the spectrums of the minimal models are, 
\begin{align}
 \mathcal{S}^{\text{A-series}}_{p,q} &= \frac12 \bigoplus_{r=1-\frac{q}{2}}^{\frac{q}{2}-1}\bigoplus_{s=1-\frac{p}{2}}^{\frac{p}{2}-1} \mathcal{R}_{\langle r,s\rangle} \otimes \bar{\mathcal{R}}_{\langle r,s\rangle} \ , \label{eq:AMMs2z}\\
 \mathcal{S}_{p,q}^{\text{D-series}} \ &=\  \frac12 \bigoplus_{r\overset{2}{=}1-\frac{q}{2}}^{\frac{q}{2}-1} \bigoplus_{s=1-\frac{p}{2}}^{\frac{p}{2}-1} \left|\mathcal{R}_{\langle r,s\rangle}\right|^2 \oplus \frac12 \ \bigoplus_{\substack{ |r| \leq \frac{q}{2}-1 \\ r \in 2\mathbb{Z} }} \  \bigoplus_{s=1-\frac{p}{2}}^{\frac{p}{2}-1} \mathcal{R}_{\langle r,s\rangle} \otimes \bar{\mathcal{R}}_{\langle r,-s\rangle}\ ,  \label{eq:DMMs2z}
\end{align}
where we have rewritten the indices in the non-diagonal sector of the D-series models by using the following series of transformations:
\begin{align}
\mathcal{R}_{\langle r,s \rangle} \otimes \mathcal{R}_{\langle q-r,s \rangle} \to \mathcal{R}_{\langle r,s \rangle} \otimes \mathcal{R}_{\langle r-q ,-s \rangle} \to \mathcal{R}_{\langle r-\frac{q}{2},s-\frac{p}{2} \rangle} \otimes \mathcal{R}_{\langle r-\frac{q}{2} ,\frac{p}{2}-s \rangle} \, .
\end{align}
This change in notation means that now fields in the non-diagonal sector of D-series minimal model have the form of the non-diagonal fields of section \ref{sec:dnd}, in the sense that their left-moving and right-moving momentums are, respectively
\begin{align}
P = P_{(r,s)}\ , \quad \bar{P} = P_{(r, -s)}\ , 
\end{align}
and its spin is now $S = -rs$. 
Furthermore, a degenerate representation $\mathcal{R}_{\langle r,s \rangle}$ appearing in the spectrums \eqref{eq:AMMs2z} or \eqref{eq:DMMs2z} has its first null vectors at levels $(r+\frac{q}{2}) (s+\frac{p}{2}) $ and $(\frac{q}{2}-r) (\frac{p}{2}-s) $ . 

Let us write the fusion rules of minimal models in these notations. We unify notations with chapter \ref{ch:cboots} by renaming the degenerate diagonal fields of minimal models as 
\begin{align}
V_{\langle r,s \rangle} = V^{D}_{P_{(r,s)}} = V^{D}_{(r,s)}\, ,
\end{align}
where the intermediate step is the notation defined in equation \eqref{eq:dnot}, and in the last step we have omitted the `$P$', in order to lighten the notation. 

The fusion rules of the D-series minimal models have been known for a long time, and in the basis we have chosen they satisfy a rule  we call conservation of diagonality \cite{petkova1988two,petkova1989structure, Petkova:1994zs,furlan1990fusion,run99}.  This rule can be summarized as 
\begin{align}
D \times D = N \times N = D, \quad D \times N = N \, ,  \label{eq:dcons}
\end{align}
meaning that the OPE between two diagonal ($D$) or two non-diagonal ($N$) fields has a diagonal spectrum, while the OPE between a  diagonal and a non-diagonal field has a non-diagonal spectrum. 

Then, the fusion rules of minimal models are, in the notation of equations \eqref{eq:AMMs2z} and \eqref{eq:DMMs2z}, 
\begin{align}
V^{A}_{(r_1, s_1)} \times V^{B}_{(r_2, s_2)} = \sum^{\frac{q}{2}-1 -|r_1 + r_2|}_{r_3 \overset{2}{=}|r_1 - r_2| +1 -\frac{q}{2}} \;  \sum^{\frac{p}{2}-1 - |s_1 + s_2|}_{s_3 \overset{2}{=}|s_1 - s_2| +1 -\frac{p}{2}} \; V^{(A \times B)}_{(r_3,s_3)}\, , \label{eq:MMfus2Z}
\end{align}
where $A,B \in \{D, N \}$ indicate if the fields are diagonal or not, and $A \times B$ should be computed according to \eqref{eq:dcons}.

The fusion rules \eqref{eq:MMfus2Z} can be used to determine the spectrums of four-point functions. We are interested in correlation functions of the type
\begin{align}
\left\langle   V^{A_1}_{(r_1,s_1)}(\zb{x}) V^{A_2}_{(r_2,s_2)}(0) V^{A_3}_{(r_3,s_3)}(\infty) V^{A_4}_{(r_4,s_4)}(1) \right\rangle \, , \label{eq:MM4pf}
\end{align}
where from now on we will omit the arguments of the fields. 

The $s$-channel decomposition arises from inserting the OPE $V^{A_1}_{(r_1,s_1)}\times V^{A_2}_{(r_2,s_2)}$ into the four-point function \eqref{eq:MM4pf}. The spectrum in this channel is obtained as the intersection between the fusion spectrums $V^{A_1}_{(r_1,s_1)}\times V^{A_2}_{(r_2,s_2)}$ and $V^{A_3}_{(r_3,s_3)}\times V^{A_4}_{(r_4,s_4)}$, i.e.
\begin{align}
\text{$s$-channel spectrum: } \left( V^{A_1}_{(r_1,s_1)}\times V^{A_2}_{(r_2,s_2)}\right) \cap \left( V^{A_3}_{(r_3,s_3)}\times V^{A_4}_{(r_4,s_4)}\right) \, . \label{eq:schinter}
\end{align}
When taking the intersection we consider $V^{A_s}_{(r_s, s_s)} = V^{A_s}_{(-r_s, -s_s)}$, due to reflection symmetry. 

In a similar way, the $t$-channel spectrum of a four-point function \eqref{eq:MM4pf} is 
\begin{align}
\text{$t$-channel spectrum: } \left( V^{A_2}_{(r_2,s_2)}\times V^{A_3}_{(r_3,s_3)}\right) \cap \left( V^{A_4}_{(r_4,s_4)} \times V^{A_1}_{(r_1,s_1)}\right) \, .\label{eq:tchinter}
\end{align}

Let us briefly discuss how we can compute four-point correlation functions in minimal models, in order to show a few examples. For more details on the numerical computations of structure constants and conformal blocks, see appendix \ref{sec:numap}. 

The spectrum of each channel can be determined by using the fusion rules \eqref{eq:MMfus2Z}, and following expressions  \eqref{eq:schinter} and \eqref{eq:tchinter}. 

Regarding the structure constants, since the fusion rules \eqref{eq:MMfus2Z} contain fields whose indices differ by $2$, we can compute all the structure constants of a given channel by repeatedly applying the shift equations \eqref{eq:drat}, instead of using explicit expressions. This method can be used to compute all the structure constants of a given four-point function, but it does not relate structure constants of different functions among themselves. However, it is enough for checking crossing symmetry.

Finally, we compute conformal blocks through Zamolodchikov's recursion representation, discussed in section \ref{sec:cblocks}. In order to compute this expansions numerically we need to introduce two parameters: $N_{max}$, which controls the truncation order of the recursion, and $\epsilon$, which introduces a small shift in the value of the central charge $c_{p,q}$, in order to avoid certain singularities of the conformal blocks. For more precisions on the role of these parameters see appendix \ref{sec:cbtrunc}.

We will now show examples of four-point correlation functions of minimal models, in order to show that the structure constants derived from the shift equations \eqref{eq:drat} give rise to crossing-symmetric functions. The Python code used to compute these functions is based on the 2-dimensional conformal bootstrap package available on \href{https://github.com/ribault/bootstrap-2d-Python}{GitHub}\cite{b2P}.


For each correlation function $Z(x, \bar{x}) = \left\langle V_1 V_2 V_3 V_4 \right\rangle$, we show:
\begin{itemize}
\item The minimal model to which the function belongs, along with the values $(p,q)$. 
\item An expression for $Z(x,\bar{x})$ in terms of the fields. We give two notations for the fields' indices: The more standard notation with $(r,s) \in [1, q-1]\times[1,p-1]$, and the one used in equations \eqref{eq:AMMs2z} and \eqref{eq:DMMs2z}, which was used in the computation.  
\item The primary fields of each channel's spectrum, derived from \eqref{eq:MMfus2Z} and equations \eqref{eq:schinter} and \eqref{eq:tchinter}.
\item A plot of the correlation function in both channels.
\item A plot of the relative difference $\eta_{(s,t)}(x)$ between the two channels, in logarithmic scale. The relative difference is computed as 
\begin{align}
\eta_{(s,t)}(x) = 2\left|\frac{Z^{(s)}(x) -Z^{(t)}(x)}{Z^{(s)}(x) + Z^{(t)}(x)} \right|\, ,
\end{align}
where $Z^{(s)}$ and $Z^{(t)}$ are the $s$-and-$t$-channel decompositions of the four-point function. 
\end{itemize}

We remark that for any $(p,q)$, the diagonal sector of the D-series minimal model spectrum \eqref{eq:DMMs2z} is contained in the spectrum of the A-series minimal model with the same $(p,q)$, \eqref{eq:AMMs2z}. This means that correlation functions between fields in the diagonal sector of a D-series minimal models also exist in the A-series models, and when presenting such examples we will list them as belonging to both. 

In each example below we plot the values of the four point functions for approximately $100$ evenly spaced points in the interval $]0,1[$. Correlation functions involving only two different fields, of the type $\langle V_1 V_2 V_1 V_2 \rangle$, have the same spectrum and structure constants in both channels, and they are therefore trivially symmetric under $x\to 1-x$. In this cases we take $100$ evenly spaced points for $x\in ]0, 1/2[$ only. 

All examples shown here were computed by setting the numerical parameters controlling the truncation of the conformal blocks and the shift of the central charge, $N_{max}$ and $\epsilon$, to:
\begin{align}
N_{max} = 40\, , \quad \epsilon = 10^{-12}\, .
\end{align}

\newpage
\subsubsection{Minimal models four-point functions}

\begin{itemize}

\item A-series minimal model with $(p,q)= (7, 8)$ .
\begin{align}
Z(x,\bar{x}) =\left\langle V^{D}_{\left(5,4 \right)}V^{D}_{\left(6,4\right)}V^{D}_{\left(5,5\right)}V^{D}_{\left(6,5\right)}\right\rangle  =\left\langle V^{D}_{\left(1,\frac{1}{2}\right)}V^{D}_{\left(2,\frac{1}{2}\right)}V^{D}_{\left(1,\frac{3}{2}\right)}V^{D}_{\left(2,\frac{3}{2}\right)}\right\rangle  .\label{eq:exMM1}
\end{align} 

Primary fields: 
\begin{flalign}
&\left[ \begin{array}{ll} 
s\text{-channel :} &V^{D}_{\left(-2, \frac{-5}{2}\right)} +V^{D}_{\left(-2, \frac{-1}{2}\right)} +V^{D}_{\left(0, \frac{-5}{2}\right)} +V^{D}_{\left(0, \frac{-1}{2}\right)} \\
t\text{-channel:} & V^{D}_{ \left(-2, \frac{-3}{2}\right)} +V^{D}_{ \left(-2, \frac{1}{2}\right)} +V^{D}_{ \left(0, \frac{-3}{2}\right)} +V^{D}_{ \left(0, \frac{1}{2}\right)} 
\end{array} \right] & \label{eq:exMM1S}
\end{flalign}
\vspace*{-15pt}
\begin{figure}[h]
\centering 
\includegraphics[scale=.7]{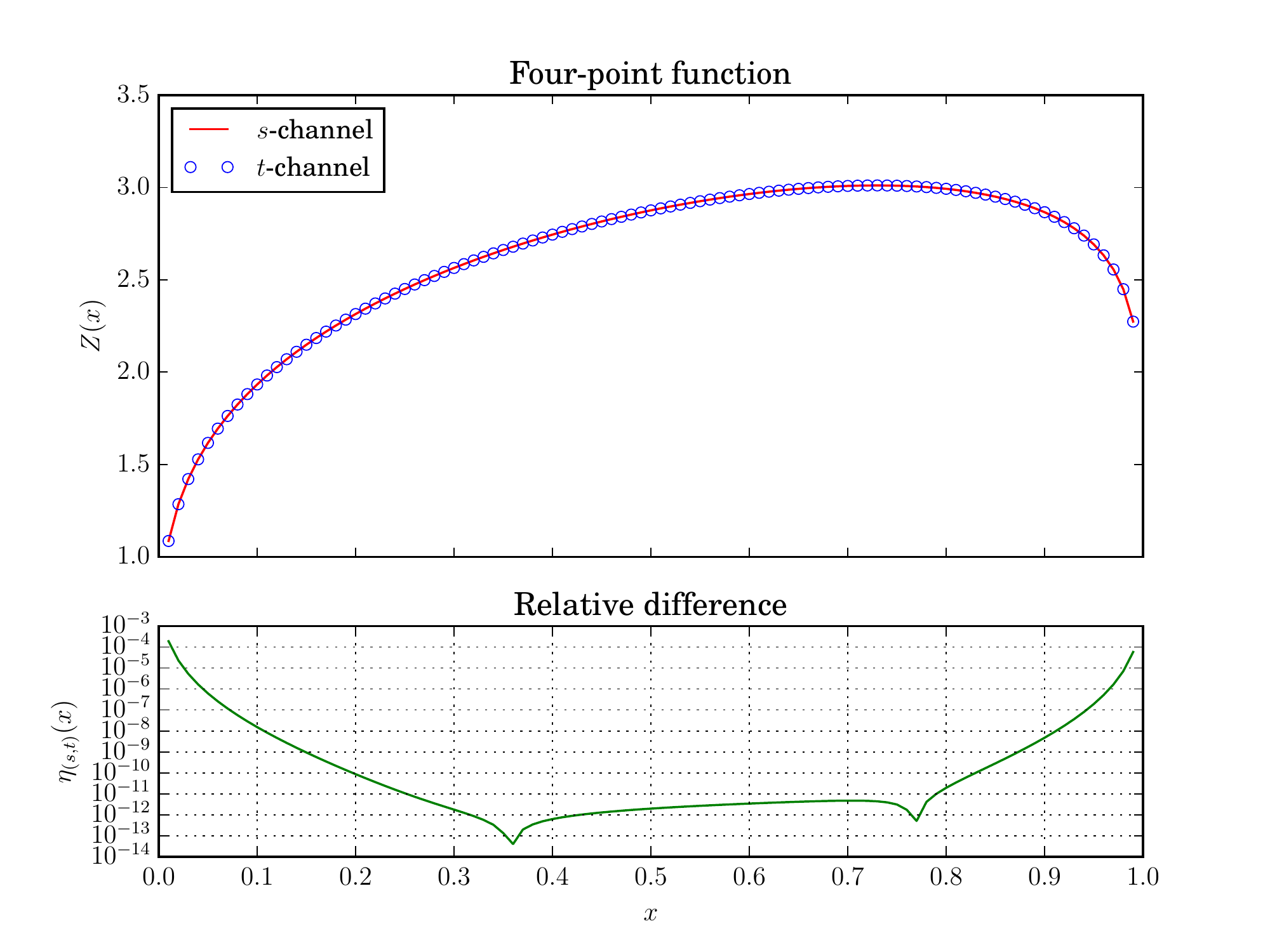}
\vspace*{-20pt}
\caption{Crossing symmetric four-point function in an A-series minimal model with $q = 0 \bmod 4$.} \label{fig:exMM1}
\end{figure}

The correlation function \eqref{eq:exMM1} involves four different diagonal fields, with both odd and even $r$ indices. We remark that the spectrums \eqref{eq:exMM1S} of each channel are different, so that there are different conformal blocks and structure constants in each channel. . 
Figure \ref{fig:exMM1} shows a good agreement between channels, with relative differences smaller than $10^{-7}$ for $x \in [0.1, 0.9]$, with differences several orders of magnitude smaller within this interval.  

\newpage

\item A-series minimal model with $(p,q)= (9, 10)$ .
\begin{align}
Z(x,\bar{x}) = \left\langle V^{D}_{\left(6,5 \right)}V^{D}_{\left(8,7\right)}V^{D}_{\left(6,3\right)}V^{D}_{\left(6,3\right)}\right\rangle  = \left\langle V^{D}_{\left(1,\frac{1}{2}\right)}V^{D}_{\left(3,\frac{5}{2}\right)}V^{D}_{\left(1,\frac{-3}{2}\right)}V^{D}_{\left(1,\frac{-3}{2}\right)}\right\rangle . \label{eq:exMM2}
\end{align}
Primary fields: 
\begin{flalign}
&\left[ \begin{array}{ll} 
s\text{-channel :} &V^{D}_{\left(-2, \frac{-3}{2}\right)} +V^{D}_{\left(-2, \frac{1}{2}\right)} +V^{D}_{\left(0, \frac{-3}{2}\right)} +V^{D}_{\left(0, \frac{1}{2}\right)} \\
t\text{-channel:} & V^{D}_{ \left(-2, \frac{1}{2}\right)} +V^{D}_{ \left(-2, \frac{5}{2}\right)} +V^{D}_{ \left(0, \frac{1}{2}\right)} +V^{D}_{ \left(0, \frac{5}{2}\right)} 
\end{array} \right] &  \label{eq:exMM2S}
\end{flalign}
\vspace*{-15pt}
\begin{figure}[h]
\centering 
\includegraphics[scale=.7]{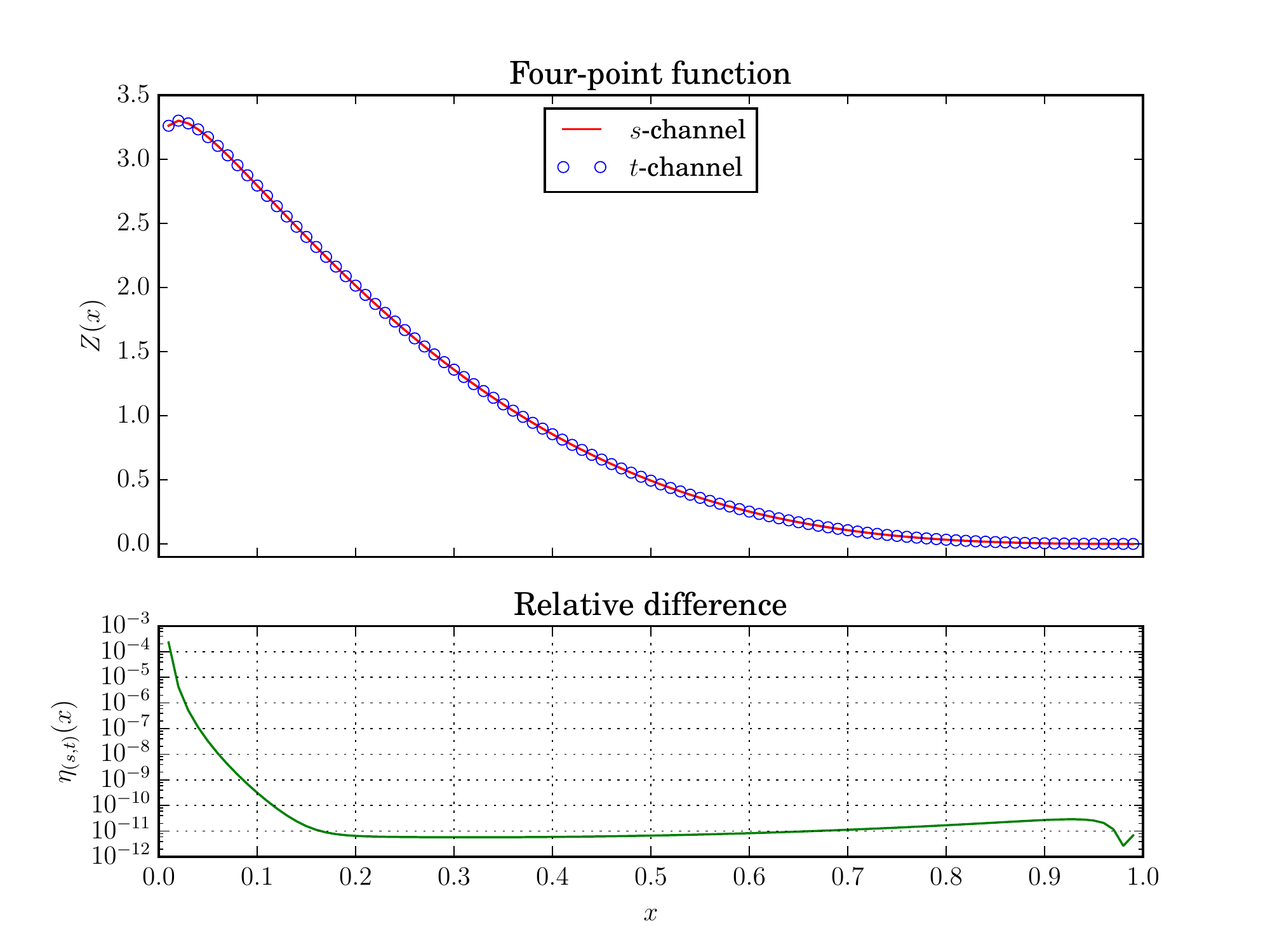}
\vspace*{-20pt}
\caption{Crossing symmetric four-point function in an A-series minimal model with $q = 2 \bmod 4$.} \label{fig:exMM2}
\end{figure}

The function \eqref{eq:exMM2} involves four different diagonal fields, all with $r$ odd. None of these fields belong to the $D$-series minimal models with the same $(p,q)$. Once again the spectrums of each channel \eqref{eq:exMM2S} are different, and we find a good agreement between both channels. 

\newpage

\item A-series \& D-series minimal model with $(p,q)= (5, 12)$ .
\begin{align}
Z(x,\bar{x}) = \left\langle V^{D}_{\left(7,3\right)}V^{D}_{\left(9,4\right)}V^{D}_{\left(7,3\right)}V^{D}_{\left(9,4\right)}\right\rangle = \left\langle V^{D}_{\left(1,\frac{1}{2}\right)}V^{D}_{\left(3,\frac{3}{2}\right)}V^{D}_{\left(1,\frac{1}{2}\right)}V^{D}_{\left(3,\frac{3}{2}\right)}\right\rangle . \label{eq:exMM3}
\end{align}
Primary fields: 
\begin{flalign}
&\left[ \begin{array}{ll} 
s\text{-channel :} &V^{D}_{\left(-3, \frac{-1}{2}\right)} +V^{D}_{\left(-1, \frac{-1}{2}\right)} +V^{D}_{\left(1, \frac{-1}{2}\right)} \\
t\text{-channel:} & V^{D}_{ \left(-3, \frac{-1}{2}\right)} +V^{D}_{ \left(-1, \frac{-1}{2}\right)} +V^{D}_{ \left(1, \frac{-1}{2}\right)} 
\end{array} \right] & \label{eq:exMM3S}
\end{flalign}
\vspace*{-15pt}
\begin{figure}[h]
\centering 
\includegraphics[scale=.7]{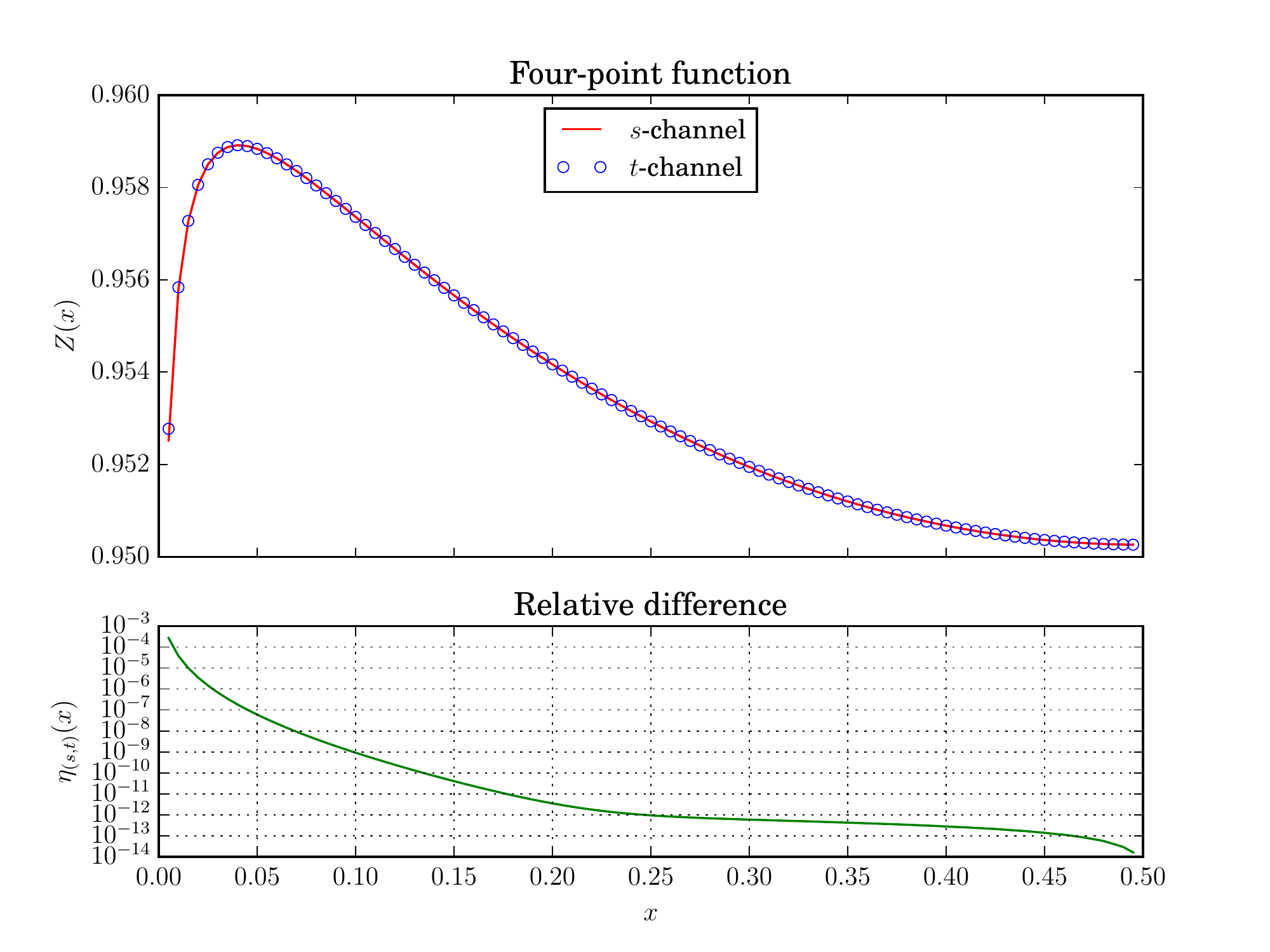}
\vspace*{-20pt}
\caption{Crossing symmetric four-point function in A-and-D-series minimal models with $q = 0 \bmod 4$.} \label{fig:exMM3}
\end{figure}

Equation \eqref{eq:exMM3} contains four fields on the diagonal sector of the D-series minimal model $(p,q) = (5,12)$, which also appear in the corresponding A-series model. Due to the symmetry of the function, which contains only two different fields, figure \ref{fig:exMM3} shows its values up to $x = 0.5$. We find a good agreement between both channels. 

\newpage
\item A-series \& D-series minimal model with $(p,q)= (11, 10)$ .
\begin{align}
Z(x,\bar{x}) = \left\langle V^{D}_{\left(3,7\right)}V^{D}_{\left(5,5\right)}V^{D}_{\left(7,5\right)}V^{D}_{\left(5,7\right)}\right\rangle  = \left\langle V^{D}_{\left(-2,\frac{3}{2}\right)}V^{D}_{\left(0,\frac{-1}{2}\right)}V^{D}_{\left(2,\frac{-1}{2}\right)}V^{D}_{\left(0,\frac{3}{2}\right)}\right\rangle . \label{eq:exMM4}
\end{align}
Primary fields: 
\begin{flalign}
&\left[ \begin{array}{ll} 
s\text{-channel :} &V^{D}_{\left(-2, \frac{-5}{2}\right)} +V^{D}_{\left(-2, \frac{-1}{2}\right)} +V^{D}_{\left(-2, \frac{3}{2}\right)} +V^{D}_{\left(-2, \frac{7}{2}\right)} +V^{D}_{\left(0, \frac{-5}{2}\right)} +\\ & V^{D}_{\left(0, \frac{-1}{2}\right)} +V^{D}_{\left(0, \frac{3}{2}\right)} +V^{D}_{\left(0, \frac{7}{2}\right)} +V^{D}_{\left(2, \frac{-5}{2}\right)} +V^{D}_{\left(2, \frac{-1}{2}\right)} +\\ & V^{D}_{\left(2, \frac{3}{2}\right)} +V^{D}_{\left(2, \frac{7}{2}\right)} \\
t\text{-channel:} & V^{D}_{ \left(-2, \frac{-9}{2}\right)} +V^{D}_{ \left(-2, \frac{-5}{2}\right)} +V^{D}_{ \left(-2, \frac{-1}{2}\right)} +V^{D}_{ \left(-2, \frac{3}{2}\right)} +V^{D}_{ \left(0, \frac{-9}{2}\right)} +\\ & V^{D}_{ \left(0, \frac{-5}{2}\right)} +V^{D}_{ \left(0, \frac{-1}{2}\right)} +V^{D}_{ \left(0, \frac{3}{2}\right)} +V^{D}_{ \left(2, \frac{-9}{2}\right)} +V^{D}_{ \left(2, \frac{-5}{2}\right)} +\\ & V^{D}_{ \left(2, \frac{-1}{2}\right)} +V^{D}_{ \left(2, \frac{3}{2}\right)} 
\end{array} \right] & \label{eq:exMM4S}
\end{flalign}
\vspace*{-20pt}
\begin{figure}[h]
\centering 
\includegraphics[scale=.7]{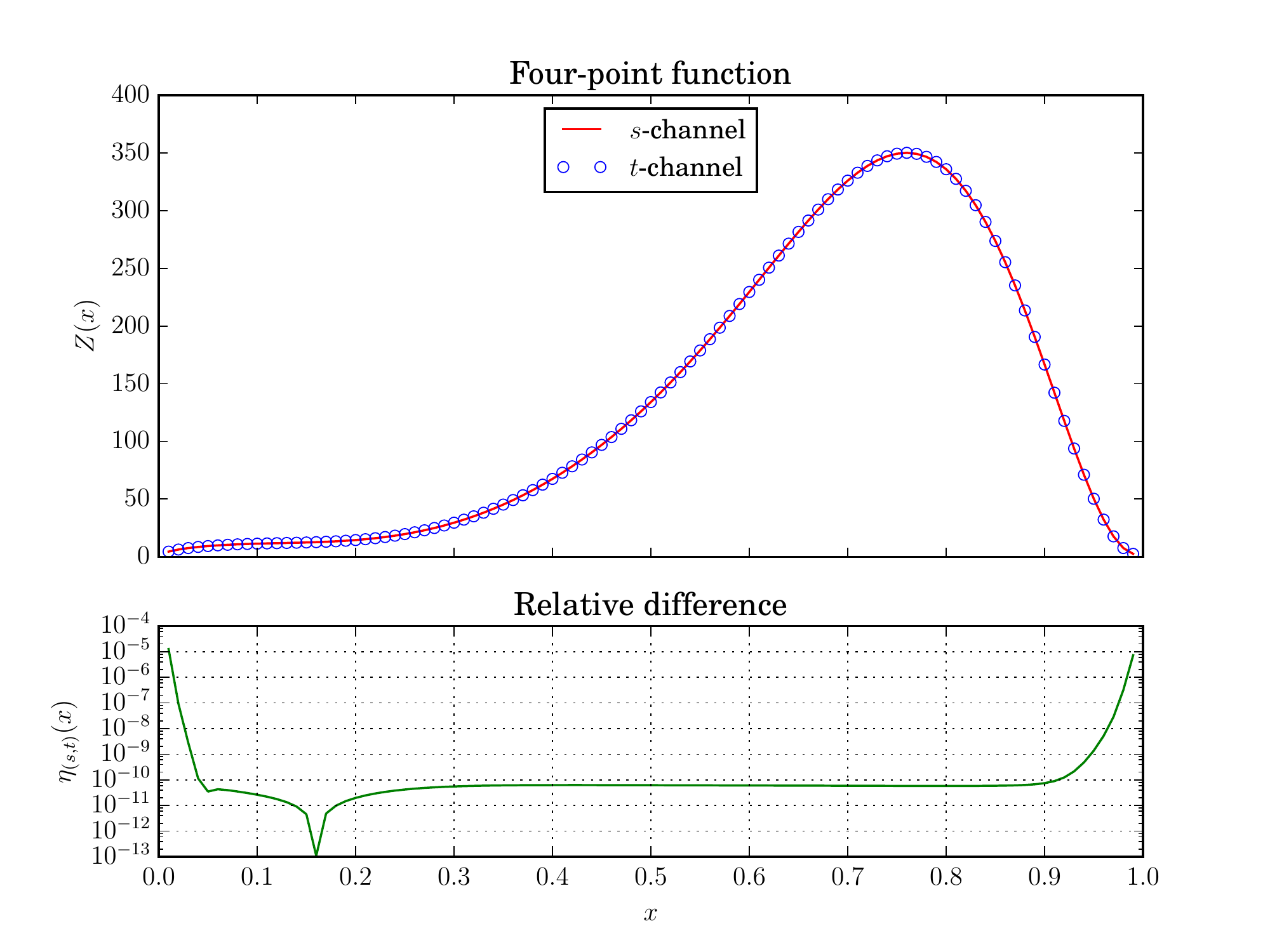}
\vspace*{-20pt}
\caption{Crossing symmetric four-point function in A-and-D-series minimal models with $q = 2 \bmod 4$.} \label{fig:exMM4}
\end{figure}


\newpage
\item D-series minimal model with $(p,q)= (7, 8)$ .
\begin{align}
Z(x,\bar{x}) = \left\langle V^{D}_{\left(5,4\right)}V^{N}_{\left(6,4\right)}V^{D}_{\left(3,2\right)}V^{N}_{\left(6,2\right)}\right\rangle  =  \left\langle V^{D}_{\left(1,\frac{1}{2}\right)}V^{N}_{\left(2,\frac{1}{2}\right)}V^{D}_{\left(-1,\frac{-3}{2}\right)}V^{N}_{\left(2,\frac{-3}{2}\right)}\right\rangle . \label{eq:exMM5}
\end{align}
Primary fields: 
\begin{flalign}
&\left[ \begin{array}{ll} 
s\text{-channel :} &V^{N}_{\left(0, \frac{-5}{2}\right)} +V^{N}_{\left(0, \frac{-1}{2}\right)} \\
t\text{-channel:} & V^{N}_{ \left(0, \frac{-1}{2}\right)} +V^{N}_{ \left(0, \frac{3}{2}\right)} 
\end{array} \right] & \label{eq:exMM5S}
\end{flalign}
\vspace*{-15pt}
\begin{figure}[h]
\centering 
\includegraphics[scale=.7]{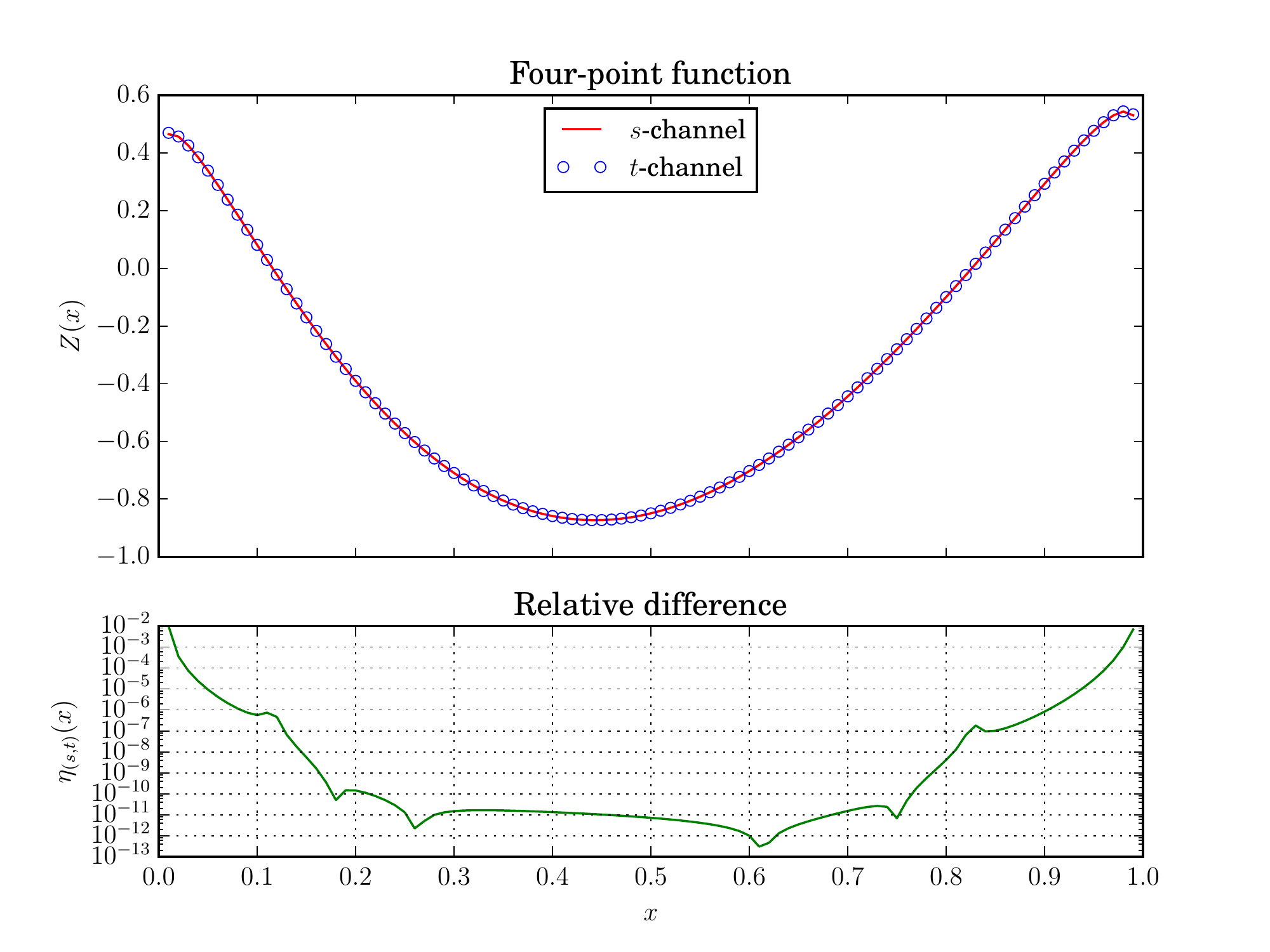}
\vspace*{-20pt}
\caption{Crossing symmetric four-point function in D-series minimal models with $q = 0 \bmod 4$.} \label{fig:exMM5}
\end{figure}

The function \eqref{eq:exMM5} is our first example of a function mixing diagonal and non-diagonal fields. Accordingly, the expansions \eqref{eq:exMM5S} run over  non-diagonal primary fields, and Figure \ref{fig:exMM5} shows that these spectrums produce agreement between both channels. 

\newpage
\item D-series minimal model with $(p,q)= (7, 10)$ .
\begin{align}
Z(x,\bar{x}) = \left\langle V^{D}_{\left(5,4\right)}V^{N}_{\left(5,4\right)}V^{D}_{\left(7,5\right)}V^{N}_{\left(7,2\right)}\right\rangle  =\left\langle V^{D}_{\left(0,\frac{1}{2}\right)}V^{N}_{\left(0,\frac{1}{2}\right)}V^{D}_{\left(2,\frac{3}{2}\right)}V^{N}_{\left(2,\frac{-3}{2}\right)}\right\rangle . \label{eq:exMM6}
\end{align}
Primary fields: 
\begin{flalign}
&\left[ \begin{array}{ll} 
s\text{-channel :} &V^{N}_{\left(0, \frac{-5}{2}\right)} +V^{N}_{\left(0, \frac{-1}{2}\right)} +V^{N}_{\left(2, \frac{-5}{2}\right)} +V^{N}_{\left(2, \frac{-1}{2}\right)} +V^{N}_{\left(4, \frac{-5}{2}\right)} +\\ & V^{N}_{\left(4, \frac{-1}{2}\right)} \\
t\text{-channel:} & V^{N}_{ \left(-2, \frac{-1}{2}\right)} +V^{N}_{ \left(-2, \frac{3}{2}\right)} +V^{N}_{ \left(0, \frac{-1}{2}\right)} +V^{N}_{ \left(0, \frac{3}{2}\right)} +V^{N}_{ \left(2, \frac{-1}{2}\right)} +\\ & V^{N}_{ \left(2, \frac{3}{2}\right)} 
\end{array} \right] & \label{eq:exMM6S}
\end{flalign}
\vspace*{-15pt}
\begin{figure}[h]
\centering 
\includegraphics[scale=.7]{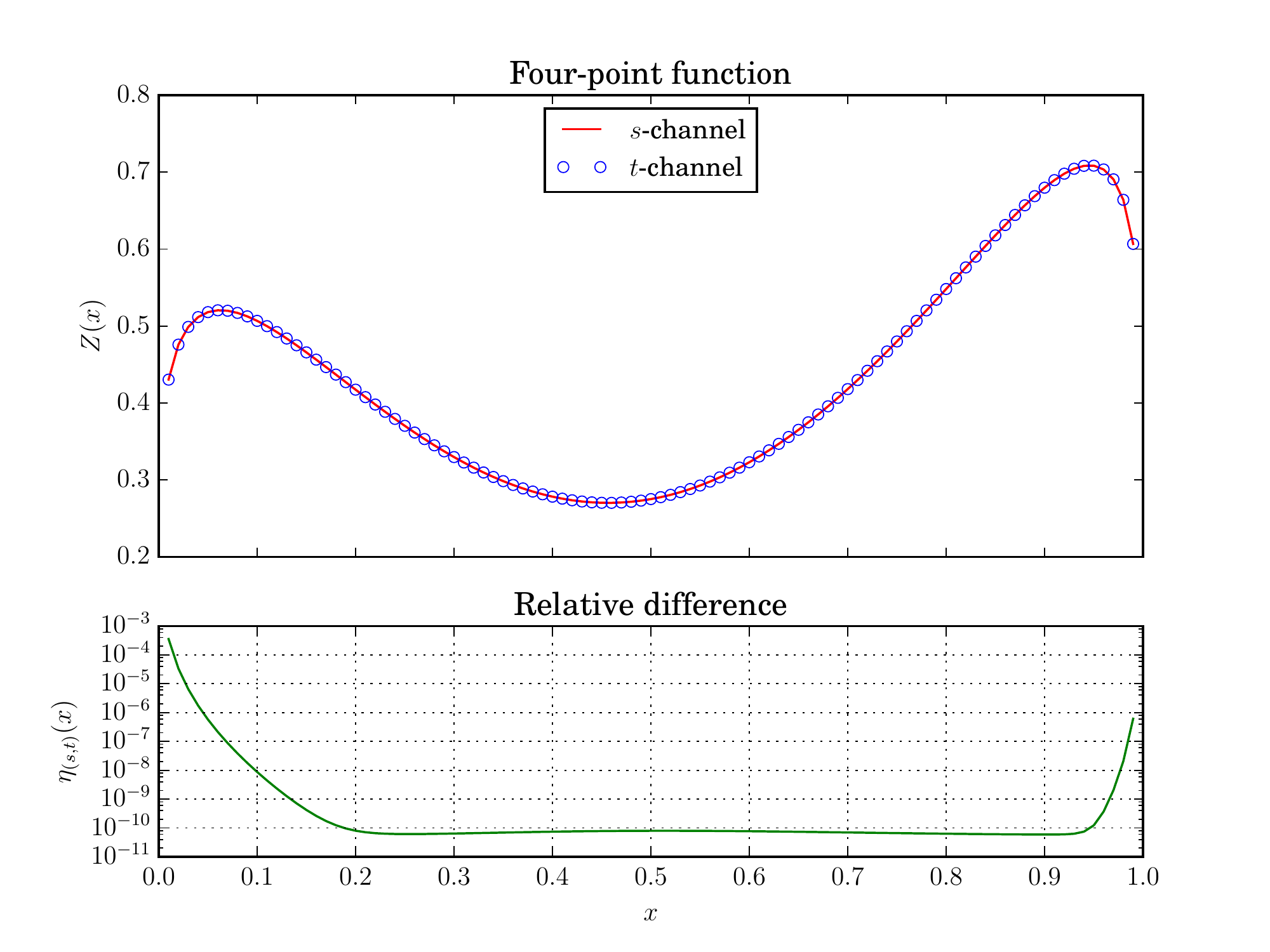}
\vspace*{-20pt}
\caption{Crossing symmetric four-point function in A-and-D-series minimal models with $q = 2 \bmod 4$.} \label{fig:exMM6}
\end{figure}

This example shows the expansion of the function \eqref{eq:exMM6} into its two non-diagonal channels. The spectrums \eqref{eq:exMM6S} are different, but figure \ref{fig:exMM6} shows that the computations of each channel agree. 
\end{itemize}

\newpage
Overall, we find a good agreement between channels in all examples.  For values of $x \in [0.1, 0.9]$ differences are  typically below $10^{-6}$, and below $10^{-10}$ for most of the interval. This is evidence that crossing symmetry is satisfied by four-point functions computed using the structure constants of the analytic conformal bootstrap. Typically, relative differences increase when approaching $x = 0$ or $x = 1$. This behaviour is expected: The $t$-channel expansion diverges when $x \to 0$, and the $s$-channel expansion, when $x \to 1$. Then, it is natural to find that agreement between channels worsens when approaching these singular points. 

In the previous examples we have chosen to take values of $x \in \mathbb{R}$, but correlation functions should also be crossing symmetric for complex values of $x$. Let us show that this is the case by computing the four-point function of equation \eqref{eq:exMM6} for some $x \in \mathbb{C}$. We introduce a parameter $\rho \in ]0,1[$, such that  
\begin{align}
x = \rho +\tfrac{i}{2}  \sin(2\pi \rho)\, . \label{eq:compex}
\end{align}
For these points, we compute the minimal model four-point function of equation \eqref{eq:exMM6}. The $s$-and-$t$-channel spectrums remain the same, so we omit them. Since 
the four-point function now takes complex values, we show the real and imaginary parts of each channel on the same plot :
\newpage
\begin{itemize}
\item D-series minimal model with $(p,q)= (7, 10)$ .
\begin{align}
Z(x,\bar{x}) = \left\langle V^{D}_{\left(5,4\right)}V^{N}_{\left(5,4\right)}V^{D}_{\left(7,5\right)}V^{N}_{\left(7,2\right)}\right\rangle  =\left\langle V^{D}_{\left(0,\frac{1}{2}\right)}V^{N}_{\left(0,\frac{1}{2}\right)}V^{D}_{\left(2,\frac{3}{2}\right)}V^{N}_{\left(2,\frac{-3}{2}\right)}\right\rangle . \label{eq:exMM7}
\end{align}
\vspace*{-20pt}
\begin{figure}[h]
\centering 
\includegraphics[scale=.7]{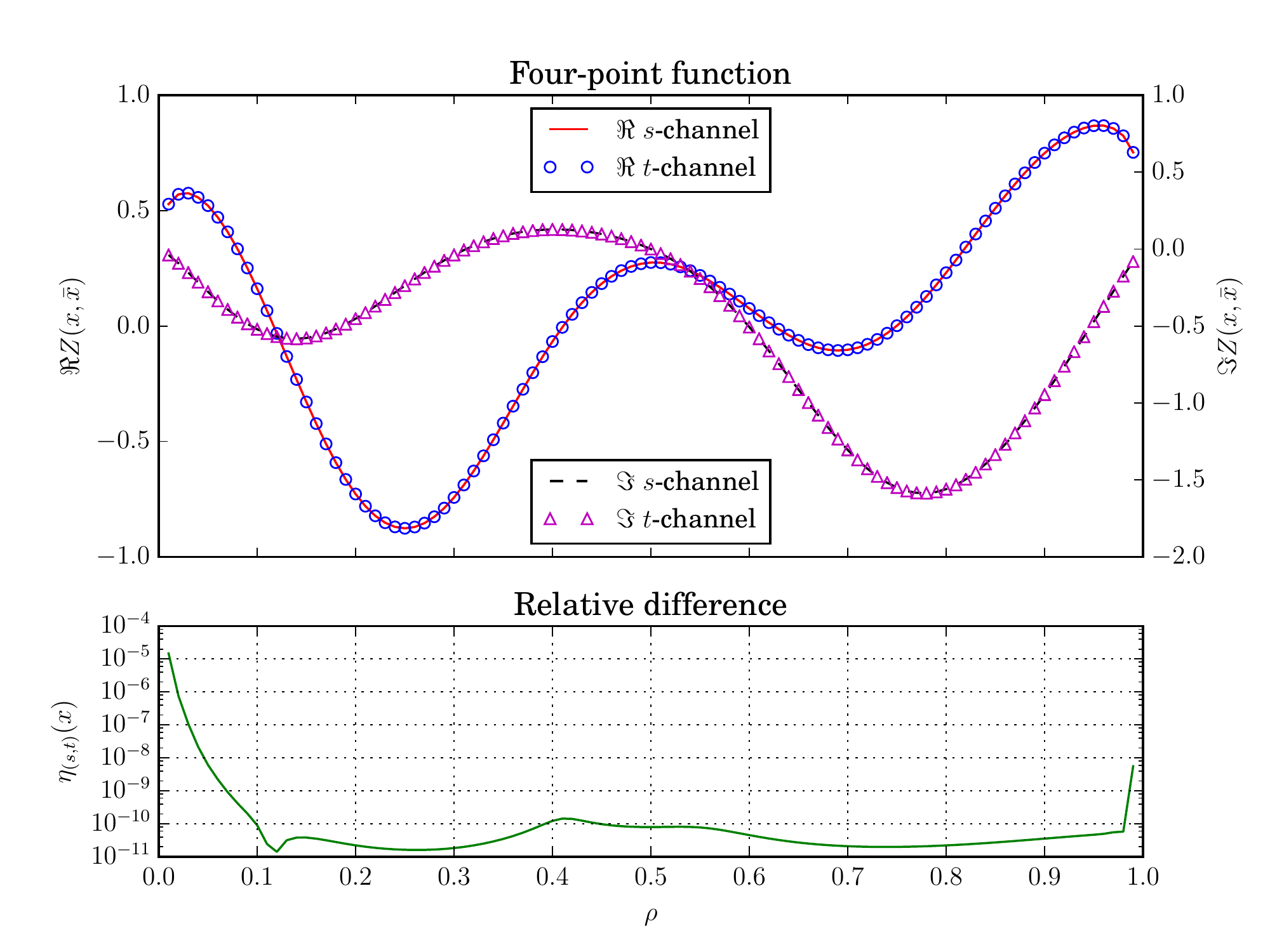}
\vspace*{-20pt}
\caption{Four-point correlation function of the D-series minimal model at $c = c_{7,10}$, for complex values of $x$ given by equation \eqref{eq:compex}. The real and imaginary parts of each channel are shown separately. } \label{fig:exMM7}
\end{figure}

\end{itemize}
This last example shows a crossing-symmetric four-point function for complex values of $x$, evidencing that the choice of $x \in \mathbb{R}$ in the previous examples does not play a key role in ensuring crossing symmetry.   

The examples of this section explore several cases of minimal model correlation functions. We have included models with $q =0 \bmod 4$ and $q =2 \bmod 4$, and correlation functions with both only diagonal fields and mixing diagonal and non-diagonal. Moreover, many of the examples contain different primary fields in each channel.

The conclusion we extract from our examples is that crossing symmetry is satisfied in the Minimal Models with the structure constants of chapter \ref{ch:cboots}. Studying A-and-D-series minimal models prevents us from computing four-point functions containing three-point structure constants of the type $C(V^N_1, V^N_2, V^N_3 )$, due to conservation of diagonality \eqref{eq:dcons}. In the next section we discuss the Ashkin-Teller model, which is not subject to this constraint.
 
\subsection{The Ashkin-Teller model}

The Ashkin-Teller models is a conformal field theory defined at $c=1$, which describes the critical point of a system of two different types of interacting spins $\sigma_1, \sigma_2 = \pm 1$ distributed  on a planar lattice. 
In this model we can find an example of a correlation function that does not satisfy conservation of diagonality \eqref{eq:dcons}, and thus includes structure constants involving three non-diagonal fields. Furthermore, these structure constants are known, and they take a particularly simple form \cite{zam85b}.

The four point function in question is
\begin{align}
\left\langle V^{D}_{P_{(0,\frac{1}{2})}} V^{D}_{P_{(0,\frac{1}{2})}} V^{N}_{P_{(0,\frac{1}{2})}}  V^{N}_{P_{(0,\frac{1}{2})}} \right\rangle \ , \label{eq:atfun}
\end{align}  
and its $s$-channel expansion runs over infinitely many non-diagonal primary fields of the type $V^{N}_{(r,s)}$, with $r \in 2\mathbb{Z}$ and $s \in \mathbb{Z}$. The structure constants $D_{(r,s)}$ of this expansion are given by
\begin{align}
 D_{(r,s)} = (-1)^{\frac{r}{2}} 16^{-\frac12 r^2 -\frac12 s^2}\ .
 \label{eq:drs}
\end{align}

Let us show that these structure constants satisfy the shift equations \eqref{eq:drat}. We compute the necessary ratios $\rho$ and $\tilde{\rho}$ from expressions \eqref{eq:r}, \eqref{eq:rt}, \eqref{OPEcgt} and \eqref{OPEcgt}. We set $\beta = 1$ (corresponding to $c=1$) and 
\begin{align}
P_{i = 1,2,3,4} &= -\frac{1}{4}, \quad \bar{P_1} = \bar{P_3} = -\frac{1}{4} \ ,\  \bar{P_2} = \bar{P_4} = \frac{1}{4}\, ,  \\
\sigma_{i = 1,2,3,4} &= -\frac{1}{4}, \quad \tilde{\sigma}_1 = \tilde{\sigma}_3 = 1 \ ,\  \tilde{\sigma}_2 = \tilde{\sigma}_4 = -1\, .
\end{align}
Using the Gamma function's duplication formula, and more specifically its consequence
\begin{align}
 \prod_{\pm, \pm}\Gamma(\tfrac12 \pm \tfrac14\pm\tfrac14 +x) = 2^{2-4x}\pi\Gamma(2x)\Gamma(2x+1)\ ,
\end{align}
we find that the shift equations are
\begin{align}
\frac{\rho \left(V^N_{(r,s)}|V^D_{(0,\frac{1}{2})}, V^D_{(0,\frac{1}{2})} \right)\rho \left(V^N_{(r,s)}|V^N_{(0,\frac{1}{2})}, V^N_{(0,\frac{1}{2})} \right)}{\rho\left( V^{N}_{(r,s)} \right)} = -16^{-2r}\, \\
\frac{\tilde{\rho} \left(V^N_{(r,s)}|V^D_{(0,\frac{1}{2})}, V^D_{(0,\frac{1}{2})} \right)\tilde{\rho} \left(V^N_{(r,s)}|V^N_{(0,\frac{1}{2})}, V^N_{(0,\frac{1}{2})} \right)}{\tilde{\rho}\left( V^{N}_{(r,s)} \right)} = 16^{-2s}\, . 
\end{align}
Computing the ratios $\frac{D_{(r+1, s)}}{D_{(r-1, s)}}$ and $\frac{D_{(r, s-1)}}{D_{(r, s-1)}}$, we can see that the constants \eqref{eq:drs} satisfy the conformal bootstrap shift equations \eqref{eq:drat}. 

If we wanted to compute the four-point function \eqref{eq:atfun} of the Ashkin-Teller model in the same way as we did for the minimal models, we would have to determine the structure constants by successive applications of the shift equations. Since in the $s$-channel expansion the index $s$ takes values $s\in \mathbb{Z}$, we could relate all structure constants to $D_{(0,0)} $ and $D_{(0,1)}$. However, we would then be unable to determine the relative normalization between the two families of structure constants, and this prevents us from using this method to compute the four-point function. Nevertheless, we take the fact that the structure constants \eqref{eq:drs} satisfy the shift equations as an indication that the analytic conformal bootstrap of chapter \ref{ch:cboots} can give solutions to this model also. 

This theory provides an example of a non-rational, non-diagonal spectrum, but it is limited to the value $c = 1$. In section \ref{sec:2zz12} we will attempt to build a family of non-rational, non-diagonal theories for generic values of $c$. 

\section{A non-rational, non-diagonal CFT}\label{sec:2zz12}

In this section, we will build a non-diagonal spectrum for generic values of $c$. and provide numerical evidence for the existence of a consistent theory with this spectrum. We will build the non-diagonal sector of such a theory by taking a generic central charge limit of the D-series minimal models. Then, we will compute four-point functions using this spectrum, and the shift equations \eqref{eq:drat}. We will find that crossing symmetry is satisfied to a great accuracy, and that the spectrum and structure constants found could belong to a consistent theory.  

\subsection{Limits of Minimal model spectrums} \label{sec:lmmspec}
The values of the minimal models central charge $c_{p,q}$ are dense on the line $c \leq 1$. Our strategy to build a spectrum for generic values of $c$ is to take a limit of the minimal models spectrum with $p,q \to \infty$ such that $c_{p,q} \to c_0 \in \mathbb{R}$, or equivalently $\frac{p}{q} \to \beta^2_0 \in \mathbb{R}$. Schematically the spectrum $\mathcal{S}_{c_0}$ we are looking for is, 
\begin{align}
\mathcal{S}_{c_0} = \lim_{\substack{p,q \to \infty \\ c_{p,q} \to c_0 } } \mathcal{S}^{\text{MM}}_{p,q}\ .
\end{align}

In order to take the limit of a minimal model spectrum we must specify how the indices $(r,s)$ of the representations in the spectrum change as we send $(p,q)$ to $\infty$. Let us begin with the case of a diagonal spectrum, such as the A-series spectrum \eqref{eq:AMMs2z}.

The A-series minimal model spectrum is built from representations $\mathcal{R}_{\langle r,s \rangle}$, with $(r,s) \in [1-\frac{q}{2}, 1-\frac{q}{2}] \times [1-\frac{p}{2},\frac{p}{2}-1] \subset \mathbb{Z} \times \left(\mathbb{Z}+\frac{1}{2}\right)$. Each representation has null vectors at levels $(r+\frac{q}{2})(s+\frac{p}{2})$ or $(\frac{q}{2}-r)(\frac{p}{2}-s)$, and higher. As we send $p, q \to \infty$, the bounds for $(r,s)$ go to $\pm \infty$, and in the limit the indices can take every value in $\mathbb{Z} \times \left(\mathbb{Z}+\frac{1}{2}\right)$. Furthermore, taking $p,q \to \infty$ sends the null vector's level to infinity, an indication that in the limit  representations lose their degeneracy and become Verma modules. Another way of seeing this is to note that we are taking the limit $\frac{p}{q} \to \beta^2_0$ for $ \beta_0 \in \mathbb{R}$ ,and that for $\beta_0 \notin \mathbb{Q}$, the momentums $P_{(r,s)}$  with $(r,s) \in \mathbb{Z} \times \left(\mathbb{Z}+\frac{1}{2}\right)$ do not correspond to degenerate representations because $s$ is a half-integer. Combining these arguments we can write
\begin{align}
\lim_{\substack{p,q \to \infty \\ \frac{p}{q}\to \beta_0^2}} \mathcal{S}^{\text{A-series}}_{p,q} = \frac{1}{2} \bigoplus_{r \in \mathbb{Z}} \bigoplus_{s \in \mathbb{Z}+\frac{1}{2} }\left|\mathcal{V}_{P_{(r,s)}} \right|^2\, . 
\end{align}
Finally, we note that for generic $\beta_0$, the momentums $P_{(r,s)}$ become dense in the real line and, in the limit, the summation over all values of $(r,s)$ amount to integrating over $P \in \mathbb{R}$. Then, assuming all representations appear with multiplicity $1$, we have
\begin{align}
\lim_{\substack{p,q \to \infty \\ \frac{p}{q}\to \beta_0^2}} \mathcal{S}^{\text{A-series}}_{p,q} = \int_{\mathbb{R}_+} dP\, \left|\mathcal{V}_P \right|^2 \, , \label{eq:AspeclimL}
\end{align}
so that we recover the spectrum \eqref{eq:Lspec} of Liouville theory. 
It is important to note that we have made a choice in the way of taking the limits of degenerate representations, by allowing the level of their null vectors to increase as we send $p,q \to \infty$. Had we chosen to keep these levels constant by working with the original indices of \eqref{eq:AMMspec} and keeping them fixed for each representation, we would have found the spectrum of generalized minimal models as a limit. In this section, and in the rest of this work, adopt the prescription that gives \eqref{eq:AspeclimL}.

Let us discuss the case of a non-diagonal spectrum. The non-diagonal sector of the D-series minimal models is, from \eqref{eq:DMMs2z}, 
\begin{align}
\mathcal{S}^{N, \text{D-series}}_{p,q} = \frac12 \ \bigoplus_{\substack{ |r| \leq \frac{q}{2}-1 \\ r\ \text{even} } } \  \bigoplus_{s=1-\frac{p}{2}}^{\frac{p}{2}-1} \mathcal{R}_{\langle r,s\rangle} \otimes \bar{\mathcal{R}}_{\langle r,-s\rangle}  \, ,
\end{align}
where $r$ take even integer values, and $s$ takes half-integer values. 
Each representation in this spectrum has integer spin, given by 
\begin{align}
S =- rs \in \mathbb{Z}\, .
\end{align} 
Since these values are discrete, we will keep the spins constants while taking the limit. As a consequence we expect that the limiting spectrum will remain discrete. Again, when $p,q \to \infty$ the bounds for $(r,s)$ also go to $\pm \infty$, and in the limit the indices take values $(r,s) \in 2\mathbb{Z}\times \left(\mathbb{Z}+ \frac{1}{2}\right) $. As in the diagonal case, taking the limit of the degenerate representations gives rise to Verma modules, and we find
\begin{align}
\lim_{\substack{p,q \to \infty \\ \frac{p}{q}\to \beta^2_0\in \mathbb{R}}} \mathcal{S}^{N, \text{D-series}}_{p,q} = \mathcal{S}_{2\mathbb{Z}, \mathbb{Z}+\frac{1}{2}}\, 
\end{align}
where $\mathcal{S}_{X,Y} = \otimes_{r \in X} \otimes_{s \in Y} \mathcal{V}_{P_{(r,s)}} \otimes \bar{\mathcal{V}}_{P_{(r,-s)}}$. Spectrums of this type, allowing even more general fractional values of the indices, were proposed in \cite{DiFrancesco:1987gwq} as corresponding to non-minimal CFTs with $c < 1$. 

Taking the limit of the full D-series minimal model spectrum, \eqref{eq:DMMs2z}, requires that we take also the limit of the diagonal sector. All the arguments used in taking the limit of A-series minimal models hold also in this case, and in the limit we get
\begin{align}
\lim_{\substack{p,q \to \infty \\ \frac{p}{q}\to \beta^2_0\in \mathbb{R}}} \mathcal{S}^{\text{D-series}}_{p,q} = \frac12 \int_{\mathbb{R}} dP\, \left|\mathcal{V}_P\right|^2 \oplus \mathcal{S}_{2\mathbb{Z},\mathbb{Z}+\frac12}\ ,
\label{eq:dlim}
\end{align}

A theory with the spectrum \eqref{eq:dlim} has diagonal fields $V^D_{P}$ of arbitrary momentums, similar to the fields of Liouville theory, and non-diagonal fields $V^{N}_{(r, s)}$ with $(r,s) \in 2\mathbb{Z} \times \left(\mathbb{Z}+\frac{1}{2}\right)$. 

In order to compute four point functions in this limit theory, we need to know its fusion rules. We expect that the conservation of diagonality \eqref{eq:dcons} that holds in the minimal models will also hold in their limit, and with these in mind we propose the fusion rule
\begin{align}
V^D_{P} \times V^N_{(r,s)} = \mathbb{S}_{2\mathbb{Z}, \mathbb{Z}+\frac{1}{2}}\, ,  \label{eq:dnr2z}
\end{align}
where we take the whole non-diagonal sector as the OPE spectrum, because in the limit there are no degenerate fields that could give rise to a finite spectrum. 
Notice that in the spectrum \eqref{eq:dlim}, where the $s$ indices of non-diagonal fields take half integer values, conservation of diagonality is enforced by the non-triviality conditions \eqref{eq:sums}. In other words, correlation functions of the type $\left\langle V^N_1 V^N_2 V^N_3 \right\rangle$ and $ \left\langle V^D_1 V^D_2 V^N_3 \right\rangle$ vanish, because they have
\begin{align}
\sum^3_{i=1} s_i \notin \mathbb{Z}\, .
\end{align}
Remember that when writing the non-triviality conditions \eqref{eq:sums} and \eqref{eq:sumr} diagonal fields are taken to have indices $(r,s) = 0$, as explained in equation \eqref{eq:indef}.

The proposed fusion rule \eqref{eq:dnr2z} would play a role in computation of the $s$-and-$t$-channel expansions of a correlation function $\left\langle V^D_{1} V^N_{2} V^D_{3} V^N_{4} \right\rangle$. In this case, all the structure constants in each channel can be computed directly from the shift equations \eqref{eq:drat}: For fields in $\mathcal{S}_{2\mathbb{Z}, \mathbb{Z} + \frac{1}{2}}$ all the indices $r$ differ by $2$, so we can reach all necessary values through shift equations. For the index $s$, we can relate all values to $s = \frac{1}{2}$ or $s=\frac{-1}{2}$. Then, all constants are related by the shift equations to $D_{1234}(V^N_{(0,\frac{1}{2})})$ or $D_{1234}(V^N_{(0,\frac{-1}{2})})$, and reflection symmetry implies 
\begin{align}
D_{1234}(V^N_{(0,\frac{1}{2})}) = D_{1234}(V^N_{(0,\frac{-1}{2})})\, , 
\end{align}
so that all structure constants are related to each other. We will later compute correlation functions using the fusion rule \eqref{eq:dnr2z}, in order to check numerically if our proposal is consistent. 

Following the same principle of conservation of diagonality, and based on the spectrum \eqref{eq:dlim}, we are tempted to write down the missing fusion rules 
\begin{align}
V^D_{P_1} \times V^D_{P_2} &= \int_{\mathbb{R}_{\geq 0}} dP_3\, V^D_{P_3}\, \label{eq:ddrulp}\\
V^N_{(r_1,s_1)} \times V^N_{(r_2,s_2)} &= \int_{\mathbb{R}_{\geq 0}} dP_3\, V^D_{P_3}\, . \label{eq:nnrlp}
\end{align}
The rule \eqref{eq:ddrulp} is the same as in Liouville theory. We have discussed in section \ref{sec:liouville} how this rule, combined with the solutions to the shift equations \eqref{eq:drat}, produces crossing symmetric four-point functions of the type $\left\langle V^D_1 V^D_2 V^D_3 V^D_4 \right\rangle$. Then, we may assume that the fusion rule \eqref{eq:ddrulp} also holds in the limit theory we are building. 

Let us discuss how these proposals would work for correlation functions of the type $\left\langle V^N_1 V^N_2 V^N_3 V^N_4 \right\rangle$ and  $\left\langle V^D_1 V^D_2 V^N_3 V^N_4 \right\rangle$, and for concreteness we focus on the $s$-channel expansions. Since the spectrums in \eqref{eq:ddrulp} and \eqref{eq:nnrlp} are continuous, we need an explicit expression for the four-point structure constants.  From the expression \eqref{eq:4pcp} of the four-point structure constants and the explicit solutions \eqref{cdnn} and \eqref{eq:YBUps}, the structure constants in the $s$-channel are 
\begin{multline}
D_{1234}(V^D_{P_s}) = \frac{1}{\prod_\pm \Upsilon_\beta(\beta\pm 2P)} \times \\  \frac{ f_{1,2}(P_s) }{ \prod\limits_{\pm, \pm} 
\Gamma_\beta(\frac{\beta}{2}+\frac{1}{2\beta} + P_s\pm P_1 \pm P_2)
\prod\limits_{\pm, \pm} \Gamma_\beta(\frac{\beta}{2}+\frac{1}{2\beta}-P_s\pm \bar{P}_1 \pm \bar{P}_2)} \times \\ \frac{ f_{3,4}(P_s) }{ \prod\limits_{\pm, \pm} 
\Gamma_\beta(\frac{\beta}{2}+\frac{1}{2\beta} +P_s\pm P_3 \pm P_4)
\prod\limits_{\pm, \pm} \Gamma_\beta(\frac{\beta}{2}+\frac{1}{2\beta}-P_s\pm \bar{P}_3 \pm \bar{P}_4)} \, . \label{eq:expD}
\end{multline} 
In order to use these structure constants it is necessary to determine sign factor $f_{1,2}(P_s)f_{3,4}(P_s)$. For a correlation function $\left\langle V^N_1 V^N_2 V^N_3 V^N_4 \right\rangle$  of four non-diagonal fields both $f_{1,2}$ and $f_{3,4}$ obey the same shift equations, because $(r_i, s_i) \in 2\mathbb{Z} \times \left( \mathbb{Z}+\frac{1}{2} \right)$ for all $i = 1,\dots,4$. Then, we can take them to be equal, such that $f_{1,2}(P_s)f_{3,4}(P_s) =1$. In principle, having determined the sign factor we could try to compute correlation functions of four non-diagonal fields in order to check whether our proposal for the fusion rules is compatible with crossing symmetry. However, at the moment we lack a numerical implementation of the special functions $\Gamma_{\beta}$, which will prevent us from preforming such tests.  This is however a minor technical issue that could be solved in the short term.

For correlation functions of the type $\left\langle V^{D}_{1}V^{D}_{2} V^{N}_{3}V^{N}_{4} \right\rangle$ the structure constants can be obtained from \eqref{eq:expD} by setting the $P_1 = \bar{P}_1$, $P_2 = \bar{P}_2$ and $f_{1,2}(P_s) = 1$. In this case, is there remains a factor $f_{3,4}(P_s)$, which can be non-analytic. This makes the computation of these correlation functions an intrinsically more complicated problem,  preventing us from testing correlation functions involving this structure constants, and therefore the compatibility between the proposed fusion rules \eqref{eq:ddrulp} \eqref{eq:nnrlp}. 

We can also evaluate whether the proposal \eqref{eq:dlim} could give rise to a consistent theory, by studying how taking the limit would work for a minimal model four-point correlation function. In minimal models, the $s$-or-$t$-channel expansions are sums over finite spectrums. As we take the limit, the spectrums become infinite, and the convergence of the correlation function limit depends on the behaviour of its terms as $(r,s) \to  \infty$. Each term is a product between structure constants and conformal blocks. According to reference \cite{rs15}, in Liouville theory the terms behave as decreasing exponentials in $\Delta + \bar{\Delta}$ as  $\Delta + \bar{\Delta} \to \infty$. The conformal blocks are universal, and the structure constants of D-series minimal models are related to the diagonal constants by a geometric mean relation of the type \eqref{eq:gmean}.  As a consequence, we expect to find the same exponential behaviour as in Liouville theory when taking the limit of minimal models, and the convergence of the limit depends on whether $\Delta + \bar{\Delta} \underset{(r,s)\to \infty}{\longrightarrow} \infty$. 
For fields in the non-diagonal sector of D-series minimal models, or in the spectrum $\mathcal{S}_{2\mathbb{Z}, \mathbb{Z} + \frac{1}{2}}$ we have 
\begin{align}
 \Delta_{(r,s)} + \Delta_{(r,-s)} = \frac{c-1}{12} + \frac12\left(r^2\beta^2 + \frac{s^2}{\beta^2}\right)\ , \label{eq:nddims}
\end{align}
which goes to infinity as $r,s\to \infty$ provided $\Re \beta^2>0$ i.e. $\Re c<13$. If this is the case, the infinite sum over $\mathcal{S}_{2\mathbb{Z}, \mathbb{Z} + \frac{1}{2}}$ converges, and D-series minimal model four-point functions could have a finite limit.

On the other hand, for diagonal states the  conformal dimensions $\Delta + \bar{\Delta}$ take the form
\begin{align}
2\Delta_{(r,s)} = \frac{c-1}{12} + \frac12\left(r\beta - \frac{s}{\beta}\right)^2\, ,
\end{align}
which does not go to infinity as $(r,s) \to \infty$. This means that the limit of the expansion of a minimal model four-point function over the diagonal spectrum may not converge. While we do expect certain four-point functions with diagonal spectrum to have a definite limit, in particular correlation functions of four diagonal fields, convergence problems may appear when the limit structure constants involved are of the type \eqref{eq:expD} and include non-trivial factors $f_{3,4}(P)$.

Since we cannot ensure that the limit of the diagonal spectrum will always exist, we only keep the non-diagonal spectrum $\mathcal{S}_{2\mathbb{Z},\mathbb{Z}+\frac12}$ and the fusion rule \eqref{eq:dnr2z} as a robust prediction from minimal models.

The following plot shows the $(P, \bar{P})$-plane for an arbitrarily chosen value of $\beta$, where the red dots correspond to primary fields of the spectrum $\mathcal{S}_{2\mathbb{Z},\mathbb{Z}+\frac12}$, and the thick blue line represents primary fields in the the spectrum of Liouville theory:
\begin{align}
\left[ \begin{array}{l}
\beta = 0.81 \\
c = -0.08
\end{array}\right]
\qquad 
\def\be{.81}
 \begin{tikzpicture}[scale = 1, baseline=(current  bounding  box.center)]
\draw [black, ->] (-2.2, 0) -- (2.2, 0) node[below]{$P$};
\draw [black, ->] (0, -2.2) -- (0, 2.2) node[left]{$\bar P$};
\draw [blue, very thick] (-2, -2) -- (2, 2);
\draw [blue, thin] (2, -2) -- (-2, 2);
\node[left] at (-2*\be/2 - 1/\be/4  , -2*\be/2 + 1/\be/4 ) {\tiny{$\left(-2, \frac{1}{2}\right)$}};
\node[left] at (-4*\be/2 - 1/\be/4  , -4*\be/2 + 1/\be/4 ) {\tiny{$\left(-4, \frac{1}{2}\right)$}};
\node[left] at (- 5/\be/4  , 5/\be/4) {\tiny{$\left(0, \frac{5}{2}\right)$}};
\node[left] at (- 3/\be/4  , 3/\be/4) {\tiny{$\left(0, \frac{3}{2}\right)$}};
\node[left] at (- 1/\be/4  , 1/\be/4) {\tiny{$\left(0, \frac{1}{2}\right)$}};
\clip (-2, 2 ) -- (-2, -2 ) -- (2, -2 ) -- (2, 2 ) -- cycle;
\foreach \x in {-4,-2,..., 4}{
  \foreach \y in {-7, -5,...,7}{
    \node[draw,circle,inner sep=1pt,fill,red] at (\x*\be/2 - \y/\be/4  , \x*\be/2 + \y/\be/4 ) {};
  }
}
 \end{tikzpicture} 
 \label{graph}
\end{align}
Primary fields along the diagonal (thick blue line) and antidiagonal (thin blue line) have spin zero, and the momentums in $\mathcal{S}_{2\mathbb{Z},\mathbb{Z}+\frac12}$ form a lattice  whose spacing is controlled by the value of $\beta$. 

The non-diagonal spectrum $\mathcal{S}_{2\mathbb{Z}, \mathbb{Z}+\frac{1}{2}}$ has been discussed in \cite{prs16}, where it was found that it appears in the $s$-and-$t$-channel expansions of the correlation function
\begin{align}
 Z_0 = \left< V^D_{P_{(0,\frac12)}} V^N_{(0,\frac12)} V^D_{P_{(0,\frac12)}} V^N_{(0,\frac12)} \right> \ .
 \label{eq:z0}
\end{align}
In that paper the structure constants where determined numerically, and it was shown that the function \eqref{eq:z0} is crossing  symmetric. Furthermore, it was argued that for $c = 0$, $Z_0$ reproduces cluster connectivities in critical percolation, and that for other values of $c$ it can be related to connectivities in the Potts model. Our analysis suggest that the spectrum $\mathcal{S}_{2\mathbb{Z}, \mathbb{Z}+\frac{1}{2}}$ should be common to many correlation functions, of which \eqref{eq:z0} is a particular case.

We conjecture that for any central charge $c$ such that $\Re{c} <13$ there exist a non-diagonal, non-rational conformal field theory whose non-diagonal spectrum is $\mathbb{S}_{2\mathbb{Z}, \mathbb{Z}+\frac{1}{2}}$, where the fusion rule \eqref{eq:dnr2z} holds, and whose   structure constants are solutions to the the shift equations of the analytic conformal bootstrap of chapter \ref{ch:cboots}. 

In order to support our conjecture, in the next section we will compute correlation functions of the conjectured theory for different values of $c$, and show that they are crossing-symmetric. 

\subsection{Numerical tests of crossing symmetry} \label{sec:ndnrex}

This section contains examples of correlation functions of our proposed non-diagonal, non-rational CFT. Computing functions in this theory is similar to the computations for minimal models, but there are two main differences:
On the one hand, we will compute correlation functions at generic central charges. Contrary to the case of minimal models, conformal blocks do not have singularities for these values of $c$, and it is no longer necessary compute them using a modified central charge.  
On the other hand, since the spectrums are now infinite, we need to truncate them in order to perform the calculation. We introduce a parameter $L$ that controls the order of truncation. Then, the numerical parameters we need to compute the correlation functions in this section are the spectrum truncation parameter $L$, and the conformal block truncation parameter $N_{max}$. In all the examples below the values of these parameter are set to 
\begin{align}
L = 5.5\ , \quad N_{max} = 20 \, . \label{eq:param2Z}
\end{align}
More details on the numerical computation of structure constants and conformal blocks can be found in appendix \ref{sec:numap}.

We compute correlation functions of the type 
\begin{align}
Z(x, \bar{x}) = \langle V^D_{P_1} V^N_{(r_2,s_2)} V^D_{P_3} V^N_{(r_4, s_4)} \rangle \, , \label{eq:fun2z}
\end{align}
where we take $ V^N_{(r_2,s_2)},V^N_{(r_4, s_4)} \in \mathcal{S}_{2\mathbb{Z},\mathbb{Z}+\frac{1}{2} }$, and we allow the momentums of the diagonal fields to take values $P_1, P_3 \in \mathbb{C}$. The reason for allowing values outside the diagonal sector $P \in \mathbb{R}$ of \eqref{eq:dlim} is the idea that correlation functions should allow for analytic continuations in the momentums. This can only be done for diagonal fields, because non-diagonal momentums remain discrete. 

The $s$-and-$t$-channel expansions of the function \eqref{eq:fun2z} are over the non-diagonal sector of \eqref{eq:dlim}, $\mathcal{S}_{2\mathbb{Z}, \mathbb{Z}+\frac{1}{2}}$. For this discrete spectrum the required structure constants can be computed directly from the shift equations, see section \ref{sec:cstsap} for more details.

In what follows we show examples of four-point correlation functions of the type \eqref{eq:fun2z}, for different values of the central charge $c$ such that $\Re{c} < 13$ and for different values of the fields momentums. The diagonal fields in \eqref{eq:fun2z} will sometimes be specified by their indices, and sometimes by their conformal weights $\Delta$. We also indicate the number of primary fields in the truncated spectrum, which depends on the parameter $L$ set in \eqref{eq:param2Z}, and on the value of the central charge.  

\newpage
\begin{itemize}
\item $c = 10^{-9}$
\begin{align}
Z(x,\bar{x}) = \left\langle V^D_{ P_{\left( 0, \tfrac{1}{2}\right)} } V^N_{ \left( 0, \tfrac{1}{2}\right) }V^D_{ P_{\left ( 0, \tfrac{1}{2}\right )} } V^N_{ \left ( 0, \tfrac{1}{2}\right ) }\right\rangle  \label{eq:exz0}
\end{align}
Number of primary fields in the spectrum: $24$.
\vspace*{-15pt}
\begin{figure}[h]
\centering 
\includegraphics[scale=.7]{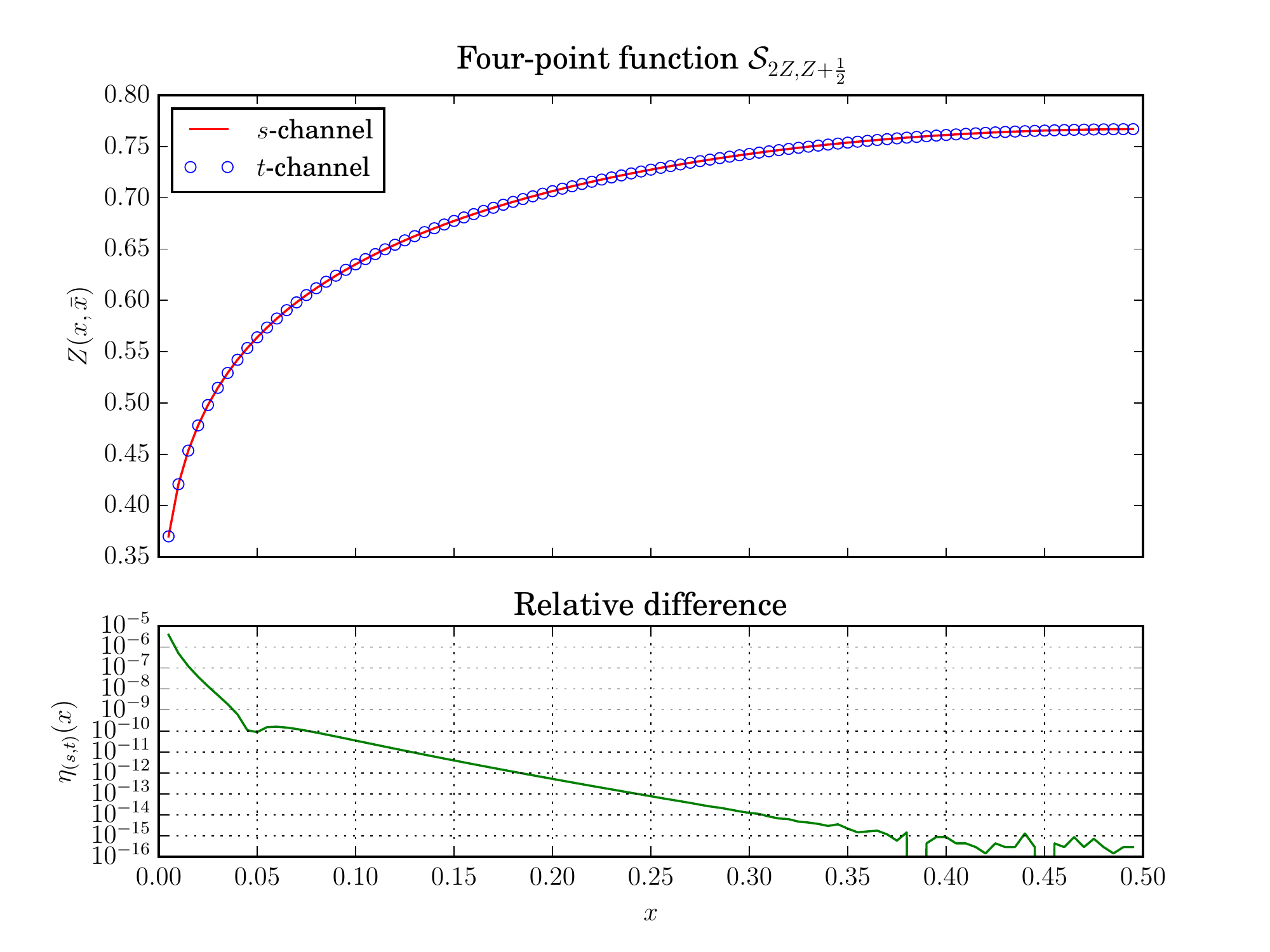}
\caption{$s$-and-$t$-channel decompositions of the four-point function \eqref{eq:z0} over the non-diagonal spectrum $\mathcal{S}_{2\mathbb{Z}, \mathbb{Z}+\frac{1}{2} }. $} \label{fig:z0}
\end{figure}

This example corresponds to the four-point function \eqref{eq:z0} discussed in \cite{prs16}. Since we cannot compute conformal blocks at $c=0$, we choose a value of $c$ that is near $0$. As expected form \cite{prs16}, this function shows good agreement between both channels. This correlation function is symmetric, and thus we give its values for $x < 0.5$. 

\newpage
\item $c = -0.54327$
\begin{align}
Z(x,\bar{x}) = \left\langle V^D_{  P_1 } V^N_{ \left ( 0,  - \tfrac{3}{2}\right ) } V^D_{  P_3 } V^N_{ \left ( 2, \tfrac{1}{2}\right ) } \right\rangle \label{eq:ex2Z1}
\end{align}
\begin{align}
 \Delta_1 = 2.341\ , \quad\Delta_3 = 1.546 
\end{align}
Number of primary fields in the spectrum: $24$.
\vspace*{-15pt}
\begin{figure}[h]
\centering 
\includegraphics[scale=.7]{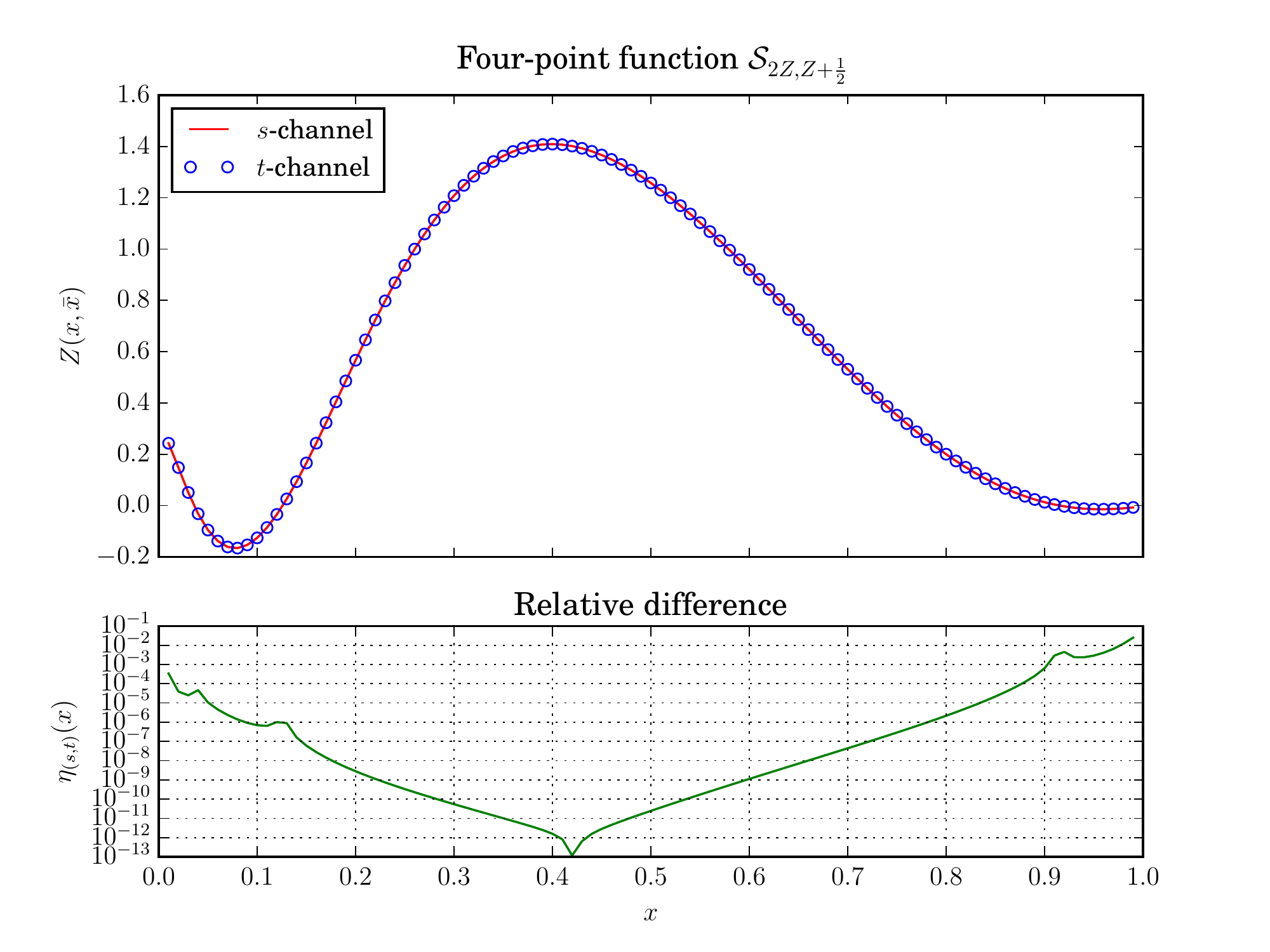}
\caption{Four-point function for arbitrary real values of the conformal weights of the diagonal fields, and negative central charge.  } \label{fig:ex2Z1}
\end{figure}

Here we see an example of a correlation function expanded over the spectrum $\mathcal{S}_{2\mathbb{Z}, \mathbb{Z}+\frac{1}{2}}$ for an arbitrary negative value of the central charge, arbitrary real values of the diagonal field's conformal weights.  We remark that the function \eqref{eq:ex2Z1} contains four different fields, such that not only the conformal blocks but also the structure constants are different in each channel. Figure \ref{fig:ex2Z1} shows good agreement between the $s$-and-$t$-channel expansions. 

\newpage
\item $c = (4.72+1.2i)$
\begin{align}
Z(x,\bar{x}) = \left\langle V^D_{  P_1 } V^N_{ \left ( 2,  \tfrac{3}{2}\right ) } V^D_{  P_3 } V^N_{ \left ( 0, \tfrac{1}{2}\right ) } \right\rangle \label{eq:ex2Z2}
\end{align}
\begin{align}
 \Delta_1 = (0.23+0.143i)\ , \quad\Delta_3 = (-0.546+2i)
\end{align}
Number of primary fields in the spectrum: $34$.
\vspace*{-15pt}
\begin{figure}[h]
\centering 
\includegraphics[scale=.7]{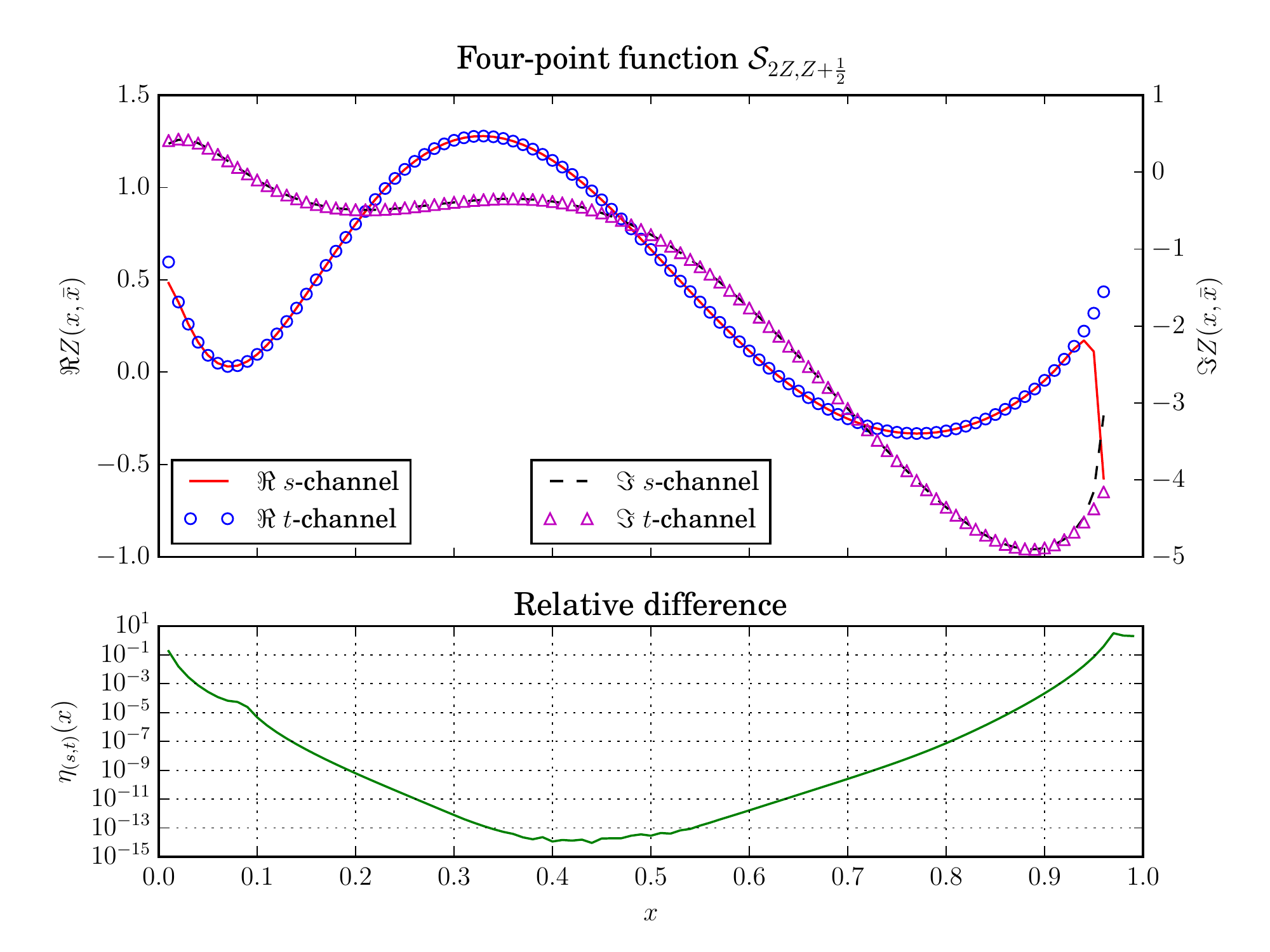}
\vspace*{-20pt}
\caption{Four-point function for arbitrary complex values of the conformal weights of the diagonal fields, at complex central charge.} \label{fig:ex2Z2}
\end{figure}

Due to the complex values of $c$, $\Delta_1$ and $\Delta_3$, both the $s$ and $t$ channel expansions of \eqref{eq:ex2Z2} take values in $\mathbb{C}$. In figure \ref{fig:ex2Z2} we plot the real and imaginary parts of each decompositions, finding that they both agree for $x$ sufficiently far from the extremes of the interval. 
\newpage

\item $c = -3.6721$
\begin{align}
Z(x,\bar{x}) = \left\langle V^D_{  P_1 } V^N_{ \left ( 0, - \frac{5}{2}\right ) } V^D_{  P_3 } V^N_{ \left ( 4, \frac{1}{2}\right ) } \right\rangle \label{eq:ex2Z3}
\end{align}
\begin{align}
 \Delta_1 = 0.567\ , \quad\Delta_3 = 1.982
\end{align}
Number of primary fields in the spectrum: $26$.
\vspace*{-15pt}
\begin{figure}[h]
\centering 
\includegraphics[scale=.7]{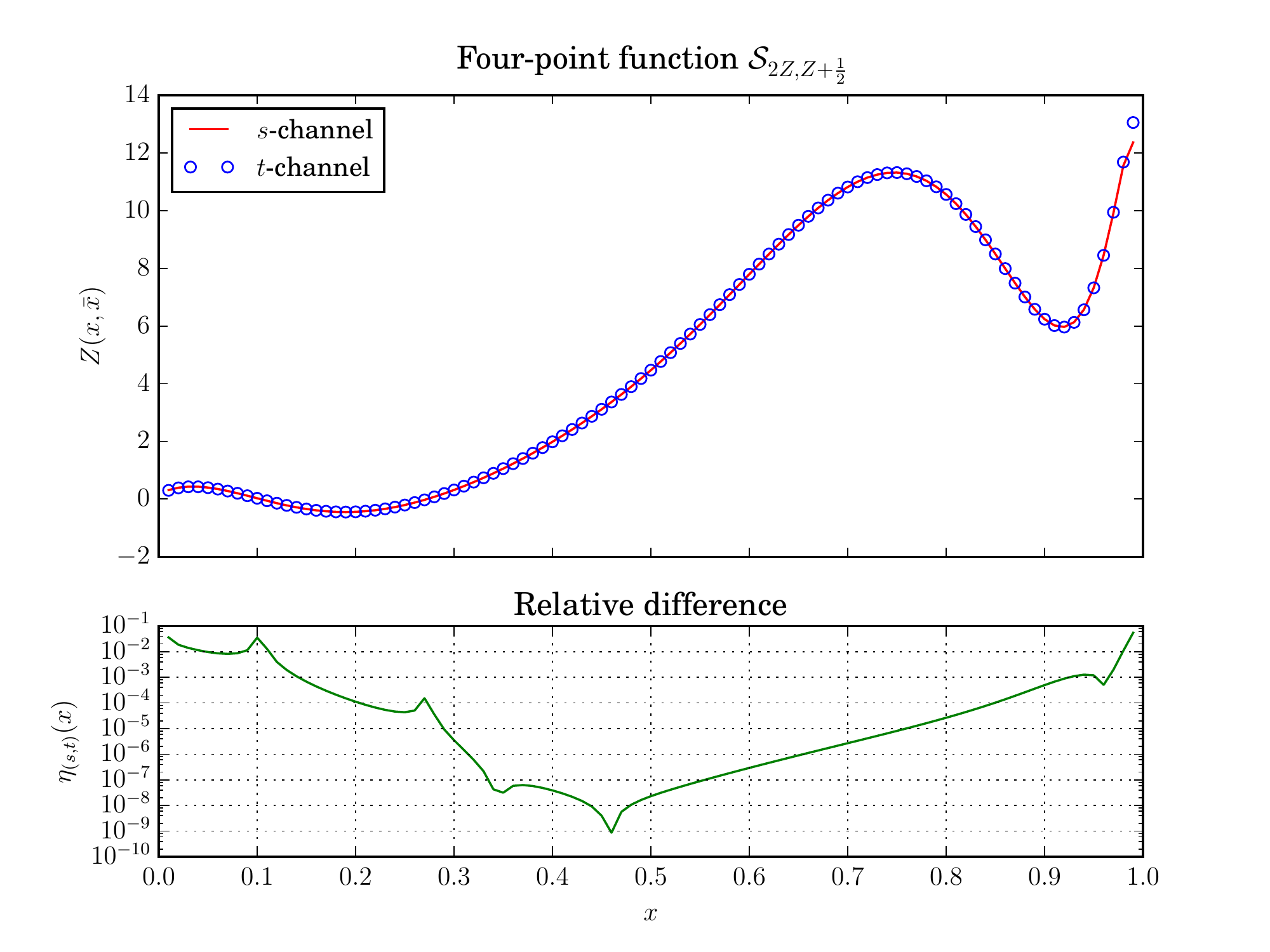}
\vspace*{-20pt}
\caption{Four-point function for arbitrary real values of the conformal weights of the diagonal fields, and central charge $c < -2$. } \label{fig:ex2Z3}
\end{figure}

\end{itemize}

Overall, these examples behave similarly to the ones discussed for Minimal Models. In all the cases studied we find a good agreement between channels, which worsens as we approach the points where the conformal blocks diverge. However, in the last example \eqref{eq:ex2Z3} the agreement between both channels seems not to be as strong, and one may worry that this is an indication that crossing symmetry is not satisfied. There is still some freedom to increase the values of the cutoffs $L$ and $N_{max}$, so let us dissipate these worries by showing that these relative differences can be drastically reduced by increasing the values of the cutoffs. Taking 
\begin{align}
L= 8 \ , \quad N_{max} = 40\, 
\end{align}
we find
\newpage
\begin{itemize}
\item c = -3.6721
\begin{align}
Z(x,\bar{x}) = \left\langle V^D_{  P_1 } V^N_{ \left ( 0, - \frac{5}{2}\right ) } V^D_{  P_3 } V^N_{ \left ( 4,  \frac{1}{2}\right ) } \right\rangle \label{eq:ex2Z4}
\end{align}
\begin{align}
 \Delta_1 = 0.567\ , \quad\Delta_3 = 1.982
\end{align}
Number of primary fields in the spectrum: $51$.
\vspace*{-15pt}
\begin{figure}[h]
\centering 
\includegraphics[scale=.7]{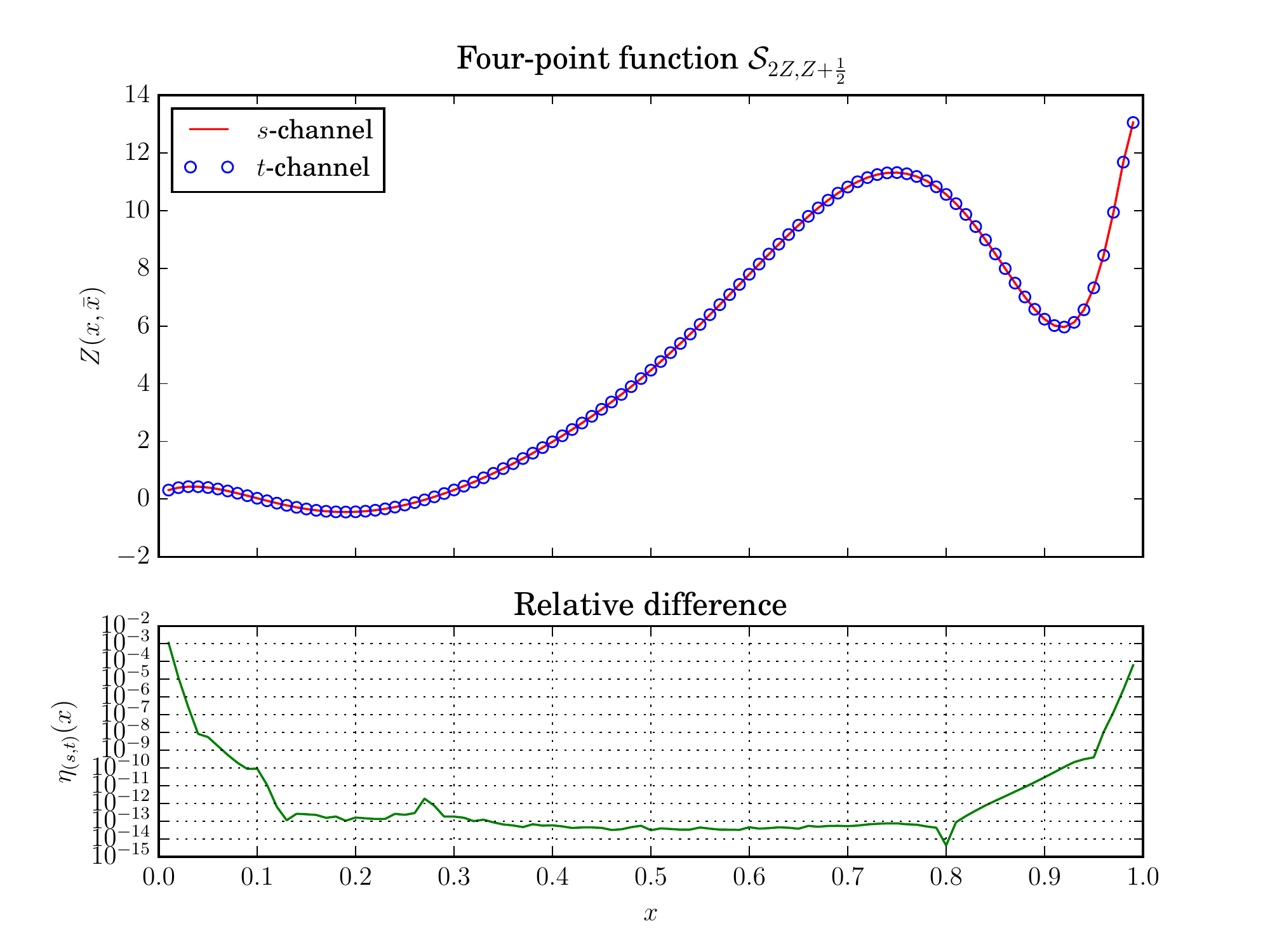}
\vspace*{-20pt}
\caption{A new computation of the four-point function \eqref{eq:ex2Z3}, with cutoff $L=8$ and $N_{max}=40$. } \label{fig:ex2Z4}
\end{figure}

This figure shows a new computation of the four-point function of figure \ref{fig:ex2Z3}, for increased values of the cutoffs. Comparing both figures we see that relative differences have decreased from $\sim 10^{-1}$ to $\sim 10^{-3}$ near $x=0$, and from $\sim 10^{-9}$ to $\sim 10^{-13}$ near the middle of the interval. Increasing the cutoffs has produces a significant decrease of relative differences across the whole interval. 
\end{itemize}

The example of figure \ref{fig:ex2Z4} shows that, indeed, increasing the cutoffs improves the agreement between channels, supporting the idea that the examples in this section correspond to crossing-symmetric four-point functions. 

Other aspects of the numerical computation can be addressed. On the one hand, since the calculation requires that we truncate the spectrum, we expect that as the values of the conformal weights and momentums become large and approach the truncation order, the agreement between channels reduces. On the other hand, values of $c$ near the minimal model values $c_{p,q}$ , like $c=0$ in the example \eqref{eq:z0}, suffer from the proximity of poles of the conformal blocks. However, we may compute correlation functions by taking $c$ slightly shifted from the problematic values, as was done in the minimal models examples.

We believe that the examples in this section provide strong evidence that crossing symmetry is satisfied for the spectrum $\mathcal{S}_{2\mathbb{Z}, \mathbb{Z} + \frac{1}{2}}$ and the structure constants coming from the analytic conformal bootstrap. Although some discrepancies between channels remain, they appear to be controlled by the truncation parameters necessary for the numerical implementation, and we find no hint that the spectrum should be modified. 
This supports our claim that there exist
consistent non-rational conformal field theories whose non-diagonal spectrum is $\mathbb{S}_{2\mathbb{Z}, \mathbb{Z}+\frac{1}{2}}$, and whose structure constants satisfy the analytic bootstrap equations of chapter \ref{ch:cboots}. Furthermore, the discussion in section \ref{sec:lmmspec} suggests that this non-rational, non-diagonal theory can be obtained as a limit of minimal models, and could be used to interpolate between them.


\begin{subappendices}
\section{On the computation of four-point functions} \label{sec:numap}

In this appendix we discuss more technical aspects related to the numerical computation of four-point functions. In particular, we focus on certain features of the calculation which are relevant for the four-point functions shown as examples for minimal models in section \ref{sec:minmods} and for the examples of section \ref{sec:lmmspec}.

In section \ref{sec:cstsap} we discuss the computation of structure constants by repeated application of the shift equations \eqref{eq:drat}, and the relative normalization between $s$-channel and $t$-channel structure constants. Section \ref{sec:cbtrunc} offers some details on the implementation of Zamolodchikov's recursion representation for the computation of conformal blocks. Finally, section \ref{sec:spectrunc} explains how non-rational spectrums were truncated to perform the computations in \ref{sec:ndnrex} .

\subsection{Structure constants and normalization} \label{sec:cstsap}

As discussed in \ref{sec:strcalc}, computing structure constants by recursive application of the shift equations requires that we fix the values of certain structure constants, in order to start the iteration. In a given four-point function, our choice is to fix the $s$-channel structure constant corresponding to the field of lowest conformal dimension to $1$. This method can be applied when the spectrum is discrete, and all the necessary structure constants can be related through shift equations. This is the case of the minimal models spectrums \eqref{eq:AMMs2z} and \eqref{eq:DMMs2z}, and for the non-diagonal sector of the limit spectrum \eqref{eq:dlim}.

In order to simplify our calculations we restrict ourselves to correlation functions such that their $s$-and-$t$-channel spectrums include the lowest dimension field of the sector of the spectrum \eqref{eq:AMMs2z}, \eqref{eq:DMMs2z}, or \eqref{eq:dlim} in which they are contained. There are two possible situations depending on the values taken by the $r$ index in the $s$-channel spectrum . Denoting the structure constants as $D_{1234}(V^{A_s}_{(r_s, s_s)})$ with $A_s \in \{D, N\}$ as in \eqref{eq:MMfus2Z}, we have
\begin{itemize}
\item $r$ even

In this case, the lowest dimension field is $ V^{A_s}_{(0, \frac{1}{2})}$. It can be diagonal in four-point functions of A-series minimal models or D-series minimal models with $q = 2 \bmod 4$, or non-diagonal in any D-series minimal model or in the non-rational theory with spectrum $\mathcal{S}_{2\mathbb{Z}, \mathbb{Z}+\frac{1}{2}}$. Then, we set
\begin{align}
D_{1234}(V^{A_s}_{(0,\frac{1}{2})}) = 1 = D_{1234}(V^{A_s}_{(0,\frac{-1}{2})})\, , \label{eq:gdstnorm}
\end{align}
where in the second equality we have used reflection symmetry. From these starting values we can compute all structure constants: The $V_{\langle 2,1 \rangle}$-shift equations in \eqref{eq:drat} will shift the index $r$ in steps of $2$, generating all values needed. On the other hand, using the $V_{\langle 2,1 \rangle}$-shift equation we can relate all required values of $s \in \mathbb{Z}+\frac{1}{2}$ to either  $s = \frac{1}{2}$ or $s = \frac{-1}{2}$, such that all constants are connected to one of the starting structure constants in \eqref{eq:gdstnorm}. 

\item $r$ odd

When $r$ takes odd values, the lowest dimension corresponds to the field  $ V^{A_s}_{(1, \frac{1}{2})}$. A field with these indices can appear in the spectrum of a four-point function in an A-series minimal models, or in a D-series minimal models with $q = 0 \bmod 4$, and it is always diagonal.

In this case, we begin by setting, 
\begin{align}
D_{1234}(V^{D}_{(1,\frac{1}{2})}) = 1 = D_{1234}(V^{D}_{(-1,\frac{-1}{2})})\, , \label{eq:rost}
\end{align}
where we have once more used reflection symmetry. With these constants fixed, we may use the $V_{\langle 2,1 \rangle}$-shift equation to obtain the constants
\begin{align}
D_{1234}(V^{D}_{(-1,\frac{1}{2})}) = D_{1234}(V^{D}_{(1,\frac{-1}{2})})\, .
\end{align}
Strictly speaking these are not starting values, because they are obtained by the applying the shift equations as every other constants. However, they are included in this analysis in order to illustrate how constants with indices of different signs can be obtained from the initial choice \eqref{eq:rost}, and also because the Jupyter notebooks used for the computation of structure constants take these four starting points. 

Starting from these structure constants, we can use the  $V_{\langle 2,1 \rangle}$-shift equation to generate constants with all the necessary odd values of $r$, and use the  $V_{\langle 1,2 \rangle}$-shift equation for obtaining constants with all necessary values of $s$. 
\end{itemize}

The above analysis explains how to compute all $s$-channel structure constants by choosing starting values, and the next step is to determine the $t$-channel structure constants. These constants can be determined by following the same scheme (choosing starting values and iteration the shift equations), but in order to compare both channels it is necessary to find their relative normalization. It suffices to know the ratio between the structure constants  of each channel corresponding to fields of lowest conformal dimensions, i.e. the ratio between the starting points. In the following, we restrict our study to four-point functions $\left\langle V^{A_1}_{1}V^{A_2}_{2}V^{A_3}_{3}V^{A_4}_{4} \right\rangle $  such that 
\begin{itemize}
\item The $s$-and-$t$-channel expansions include lowest dimension fields of one sector of the spectrum. More explicitly, they include either the fields $V^{D}_{(1,\frac{1}{2})}$,  $V^{D}_{(0,\frac{1}{2})}$ or $V^{N}_{(0,\frac{1}{2})}$.  

\item The field with lowest conformal dimension is the same on both channels.

\item The fields $V^{A_2}_{2}$ and $V^{A_4}_{4}$ are of the same type, i.e. $A_2 = A_4$. 
\end{itemize}
In this case, the ratio between the $s$-and-$t$-channel structure constants can be expressed through a simple formula which makes use of the shift equations \eqref{eq:drat}, 

\begin{multline}
\frac{D_{2341}\left( V^{A_0}_{\left( r_0,s_0 \right)} \right)}{D_{1234}\left( V^{A_0}_{\left( r_0,s_0 \right)} \right)} = (-1)^{\sum\limits_{i = 1}^{4}S_i } \prod_{j=0}^{m-1} \prod_{k=0}^{n-1} \left[ \frac{\tilde{\rho} \left( V^{A_2}_{\left( r_2, \, s_2 -\delta_s (2j+1) \right) }, V^{A_0}_{\left( r_0,s_0 \right)} , V_3 \right) } {\tilde{\rho}\left( V^{A_2}_{\left( r_2, \, s_2 -\delta_s (2j+1) \right) }, V^{A_0}_{\left( r_0,s_0 \right)}, V_1 \right) }\right]^{\delta_s}
\times\\
\left[ \frac{\rho \left( V^{A_2}_{\left( r_2 -\delta_r (2k+1) , \, s_4 \right) }, V^{A_0}_{\left( r_0,s_0 \right)}, V_3 \right) } {\rho\left( V^{A_2}_{\left( r_2 -\delta_r (2k+1) , \, s_4 \right) }, V^{A_0}_{\left( r_0,s_0 \right)}, V_1 \right) }\right]^{\delta_r} \, , \label{eq:stnorm}
\end{multline}
where $S_i$ are the spins of the fields, and we define $\delta_s, \delta_r \in \{1, -1\}$ by
\begin{align}
s_2 - s_4 &= 2 \delta_s m, \quad m\in \mathbb{Z}_{\geq 0}, \\
r_2 - r_4 &= 2 \delta_r n, \quad n\in \mathbb{Z}_{\geq 0}\, .
\end{align}

Equation \eqref{eq:stnorm} is obtained by writing the four-point structure constants in terms of three-point structure constants, and using their permutation properties \eqref{permC} along with the shift equations \eqref{eq:rcst} and \eqref{eq:trcst}. 

For certain symmetric correlation functions the relative normalization becomes trivial. For example, correlation functions with only two different fields of the type $ \left\langle V_1 V_2 V_1 V_2 \right\rangle$, or even in the less symmetric case $ \left\langle V_1 V_2 V_3 V_2 \right\rangle$.

In a four-point function where the lowest dimension fields of each channel are different, fixing relative normalizations might be more complicated. If the lowest dimension fields are connected by shift equations, or if there are fields that appear on both channels, it is possible to find some expression similar to \eqref{eq:stnorm}. On the contrary, if the fields are not related through shift equations and there are no fields common to both channels, explicit expressions have to be used to fix the relative normalization between channels.

Due to current limitations in the coding of the method to determine the relative normalization between channels, the correlation functions shown below are limited to functions satisfying the assumptions leading to equation \eqref{eq:stnorm}.

\subsection{Conformal blocks}\label{sec:cbtrunc}

In section \ref{sec:cblocks} we presented equation \eqref{gsd} as a way of computing conformal blocks. This expression, while accurate, is not very  efficient. On the one hand, the number of terms for each level $|L|$ is the number of partitions $p(|L|)$, which increases very rapidly with $|L|$. On the other hand, the factors $f_{\Delta_1,\Delta_2}^{\Delta_s,L_s} g^{L_s}_{\Delta_s,\Delta_3,\Delta_4}$ don not have a general expression, and their computation can be costly. 

An alternative to equation \eqref{gsd} is to use the recursion representation of conformal blocks, introduced by Al. Zamolodchikov in \cite{Zamolodchikov1987}. This representation, as opposed to expression \eqref{gsd}, does not have a known as a sum over descendants of some field, but it is more suitable for numerical implementation. Here we reproduce a variation of the recursion representation written in terms of the momentums defined in \eqref{eq:cdef}, instead of the conformal weights. This expression was used in reference \cite{rs15}, which uses the same numerical implementations as we do. 

In order to write the recursion representation we introduce the elliptic variable $q$ and the function $\theta_{3}(q)$, defined by
\begin{align}
 q = \exp -\pi \frac{F(\frac12,\frac12,1,1-x)}{F(\frac12,\frac12,1,x)}  \quad , \quad \theta_3(q) = \sum_{n\in{\mathbb{Z}}} q^{n^2}\ ,
\end{align}
where $F(\frac12,\frac12,1,x)$ is a special case of the hypergeometric function $_2F_1$, and $|q|<1$ for any $x\in\mathbb{C}$.

Then, for external left-moving momentums $P_i$, the the s-channel conformal block is given by
\begin{multline}
 \mathcal{F}^{(s)}_{P_s}(P_i|x) 
=  (16q)^{P^2_s } x^{-\frac{c-1}{24}-P^2_1-P^2_2} (1-x)^{-\frac{c-1}{24}-P^2_1-P^2_4}\\ \times \theta_3(q)^{-\frac{c-1}{6}-4(P^2_1+P^2_2+P^2_3+P^2_4)} \left(1+\sum_{N=1}^{\infty} \sum_{rs \leq N} A^N_{(r,s)} \frac{(16q)^N}{P^2_s - P^2_{(r,s)}} \right)\ , \label{eq:Zamrec}
\end{multline}
where $P_{(r,s)}$ are the momentums \eqref{eq:PKac}. The coefficients $A^N_{(r,s)}$ are determined by the recursion relation
\begin{align}
A^N_{(r,s)} = R_{(r,s)} \left( \delta_{N-rs, 0} + \sum_{r's'\leq N-rs} \frac{A^{N-rs}_{(r',s')}}{P^2_{(r,-s)} - P^2_{(r',s')}} \right) \, .
\end{align}
Here we have introduced a factor $R_{(r,s)}$, which is given by

\begin{multline}
 R_{m,n} = \frac{-2P_{(0,0)} P_{( m,n)}}{\prod_{r=1-m}^m \prod_{s=1-n}^n 2P_{(r,s)}}
\\ \times \prod_{r\overset{2}{=}1-m}^{m-1} \prod_{s\overset{2}{=}1-n}^{n-1}  \prod_\pm (P_2\pm P_1 + P_{( r,s)}) (P_3\pm P_4 +P_{(r,s)})\ ,
\end{multline}
where the factor $P_{(0,0)}$ in the numerator is included to cancel the same factor appearing in the denominator. 

The recursion representation \eqref{eq:Zamrec} converges very fast, which makes it suitable for numerical computations. It also has the advantage of showing explicitly the poles of the conformal block, which correspond to $P_s = P_{(r,s)}$ with $r,s \in \mathbb{N}$.

In generalized minimal models and minimal models, the fields in the $s$-and-$t$-channel spectrums have momentums of the type $P = P_{(r,s)}$. One may then worry about divergences of the conformal blocks, because these are precisely their poles. However, when the fusion rules involving degenerate fields are obeyed, equations \eqref{rrsr} for generalized minimal models and \eqref{eq:MMfus2Z} for minimal models, the residues $R_{m,n}$ in the recursion vanish. Then, the blocks computed through the recursion \eqref{eq:Zamrec} remain finite \cite{Ribault:2014hia}. 

In order to compute conformal blocks numerically we introduce a cutoff $N_{max}$, which serves as a truncation for the sum in \eqref{eq:Zamrec}. The truncated conformal block is a polynomial of degree $N_{max}$  in $q$, with the poles $P_{(r,s)}$ with $rs\leq N_{max}$.

\subsubsection{In minimal models} \label{sec:cbMM}

The recursion formula \eqref{eq:Zamrec} has extra singularities at the minimal values of the central charge $c = c_{p,q}$, due to coincidences between conformal weights produced by relations \eqref{eq:eqlrs}. Minimal models conformal blocks should be well defined, and the poles that appear for $c = c_{p,q}$ are an artefact of the recursion representation, rather than a physical property of the blocks themselves \cite{Javerzat:2018maf}. 

In order to compute conformal blocks for minimal models from the recursion relation \eqref{eq:Zamrec} it is necessary to introduce a second parameter $\epsilon$, which serves as a regularization parameter. When computing conformal blocks for a minimal model four-point function, we take a slightly shifted central charge $c_{\text{blocks}} = c_{p,q} + \epsilon$, for a small $\epsilon$, which allows us to avoid the extra singularities due to \eqref{eq:eqlrs}. 

Notice that this shift in the central charge only affects minimal models conformal blocks. When computing minimal models four-point functions, the structure constants are computed at $c = c_{p,q}$. 

\subsection{Truncation of Infinite spectrums} \label{sec:spectrunc}

The examples of section \ref{sec:ndnrex} require that we compute summations over the infinite spectrum $\mathcal{S}_{2\mathbb{Z}, \mathbb{Z}+\frac{1}{2} }$. In order to perform this we need to introduce a criterion to truncate this spectrum, keeping finitely many terms. Here we explain the method used.

The discussion leading to equation \eqref{eq:nddims} implies that terms in the conformal block expansion of a four-point function will contribute less as the conformal dimension $\Delta + \bar{\Delta}$ of the intermediate state increases. Then, it is natural to truncate the spectrum by imposing an upper bound to the conformal dimensions that contribute. 

The states in spectrum $\mathcal{S}_{2\mathbb{Z}, \mathbb{Z} + \frac{1}{2}}$ are identified by their indices $(r,s)$, and their conformal dimension is given by equation \eqref{eq:nddims}.We work in the case $\Re{\beta^2} > 0 $, so that the dimensions increase as $(r,s)$ increase. Then, a truncation in the dimensions implies a truncation of in the values of the indices. 

In order to truncate the spectrum choose a value $L$, and restrict the indices to the range 
\begin{align}
(r,s)\in \left[0, \frac{L}{\sqrt{\Re\beta^2}}\right]\times \left[0, \frac{L}{\sqrt{\Re\frac{1}{\beta^2}}}\right]\, . \label{eq:indrt}
\end{align}
Then, from this subset we keep only fields such that
\begin{align}
\Re{\left(r^2 \beta^2 + \frac{s^2}{\beta^2}\right)} < L^2 - \frac{1}{6}(\Re{c-1})\, . \label{eq:dimrt}
\end{align}

Finally, since the indices in equations \eqref{eq:indrt} are only positive, we need to incorporate fields with negative indices. In order to avoid repetition of states due to reflection, we keep the values of $r$ positive, and for each field obeying restrictions \eqref{eq:indrt} and \eqref{eq:dimrt} we add a a field of indices $(r,-s)$.

The following table shows an example of the fields of a truncated spectrum. For $\beta = 0.81$ and truncation parameter $L = 4$, the indices and dimensions of the fields are:
\begin{align}
\left[ \begin{array}{l} 
\beta = 0.81 \\
L = 4 
\end{array}
\right] \Rightarrow
\begin{array}{|c|c|c|}
\hline
(r,s) & \text{Spin} & \Delta+\bar{\Delta} \\
\hline
\left(0,\tfrac{1}{2}\right)&0&0.100390786\\
\left(2,\tfrac{1}{2}\right)&1&1.412590786\\
\left(2,\tfrac{-1}{2}\right)&-1&1.412590786\\
\left(0,\tfrac{3}{2}\right)&0&1.624548689\\
\left(2,\tfrac{3}{2}\right)&3&2.936748689\\
\left(2,\tfrac{-3}{2}\right)&-3&2.936748689\\
\left(0,\tfrac{5}{2}\right)&0&4.672864495\\
\left(4,\tfrac{1}{2}\right)&2&5.349190786\\
\left(4,\tfrac{-1}{2}\right)&-2&5.349190786\\
\left(2,\tfrac{5}{2}\right)&5&5.985064495\\
\left(2,\tfrac{-5}{2}\right)&-5&5.985064495\\
\left(4,\tfrac{3}{2}\right)&6&6.873348689\\
\left(4,\tfrac{-3}{2}\right)&-6&6.873348689\\ \hline
\end{array}
\end{align}

The number of allowed states increases rapidly as we increase $L$, as the upper bound in \eqref{eq:dimrt} is quadratic in $L$. 
\end{subappendices}

%% file: chapters/conclusions.tex
\chapter{Conclusions and outlook} \label{ch:concl}

In this thesis we have presented two main results regarding two-dimensional conformal field theories with symmetry algebra $\mathfrak{V} \otimes \bar{\mathfrak{V}}$. Firstly, we have shown an extension of the analytic conformal bootstrap approach to non-diagonal theories, giving explicit expressions for the structure constants. Furthermore, we have checked the validity of these results by numerically computing four-point functions and finding that crossing symmetry is satisfied to good degree. Secondly, we have proposed the existence of a non-diagonal, non-rational theory for generic values of $c$ such that $\Re{c}<13$. We have tested this proposal by computing correlation functions involving non diagonal fields, for different values of the central charge and the momentums of the diagonal fields. In this chapter we summarize the results of  our analysis, and suggest some directions for further work. 

\section{Analytic conformal bootstrap}

In chapter \ref{ch:cboots} we discussed how to apply the analytic conformal bootstrap method to non-diagonal theories. We began by establishing the three main assumptions of our approach: generic values of the central charge, single-valued correlation functions, and existence of two independent degenerate fields. Then we derived degenerate fusion rules \eqref{dfus} that are in agreement with these assumptions, and in section \ref{sec:dnd} we defined diagonal and non-diagonal fields based on their fusion properties with degenerate fields.

In section \ref{sec:dcsy}, using the degenerate fusion rules \eqref{dfus} we wrote the crossing symmetry equations \eqref{eq:2dec} for correlation functions involving degenerate fields. This led to the shift equations \eqref{eq:rvo} and \eqref{eq:trvo} for two-point structure constants, and equations \eqref{eq:rcst} and \eqref{eq:trcst} for three-point structure constants. Those expressions allowed us to write equations \eqref{eq:drat} for four-point structure constants, which have the advantage of not involving degenerate OPE coefficients. Writing these equations in terms of fusion matrix elements we found their explicit expressions, in terms of ratios of Gamma functions.

In section \ref{sec:expsol}  we looked for explicit solutions to the shift equations, and found expressions \eqref{Y2Bnd}, \eqref{Y2Bndb}, \eqref{cdnn} and \eqref{cdnnb} for the two-point and three-point structure constants. These expressions are written in terms of Barnes' double Gamma functions, and they apply for cases with diagonal and non-diagonal fields. We discussed how these solutions have the correct reflection and permutation properties. The solutions found in this section are meromorphic in the momentums of the fields, and hold for different values of $c$. 

With the explicit solutions in hand, we discussed the relation between the non-diagonal and diagonal solutions. We found that our expressions structure constants reduce to the known ones of the diagonal case when we take all arguments to be diagonal. Furthermore, we found that three-point structure constants satisfy a geometric-mean-like relationship \eqref{eq:gmean}. This relation had previously appeared in \cite{ei15}, and it ultimately originates from similar relations for the shift equations, \eqref{r2rdrd}. The geometric mean relation could arise as an educated guess for the non-diagonal structure constants, based on the holomorphic factorization of conformal blocks. However, having the explicit expressions shows that there is a consistent way to take the square root without spoiling analyticity of the structure constants, and eliminates the sign ambiguity associated with taking the square root. 

The results of the analytic conformal bootstrap were argued to apply to a large number of theories. For the cases restricted to diagonal fields, we recovered constants of Liouville theory and generalized minimal models. Then, we checked that our solutions also apply to A-series and D-series minimal models by using the shift equations \eqref{eq:drat} to compute four-point correlation functions in the $s$-and-$t$-channel decompositions, finding a good agreement between both channels. The case of D-series minimal models is particularly interesting, because it allowed us to test the non-diagonal shift equations. 

When looking for solutions to the analytic conformal bootstrap shift-equations we left aside the case of three-point structure constants with three non-diagonal fields, and it would be interesting to find an explicit expression for this case. However, the example of the Ashkin-Teller model showed that its four-point structure constants, which can involve three-point structure constants of this type, do obey the shift equations \eqref{eq:drat}. Moreover, for some discrete spectrums these structure constants could be computed directly form the shift equations. 

Finally, the three-point structure constants contain a sign factor $f_{2,3}(P_1)$. This factor becomes trivial in certain cases, for example for structure constants with three diagonal fields, or with non-diagonal fields of integer indices. The factor $f_{2,3}(P_1)$ is necessary for our expression to satisfy the shift equations, but outside the simple examples mentioned, determining it can be non-trivial. Difficulties in determining this factor can be a problem when trying to compute four-point correlation functions in certain channels involving these constants, as we will discuss in the next section. 

\section{Limits of minimal models }


In section \ref{sec:2zz12} we proposed a spectrum for a non-rational, non-diagonal theory. This proposal was obtained by taking a non-rational limit of the minimal models spectrums such that $p,q \to \infty$ and $c_{p,q} \to c_0 \in \mathbb{R} $. The diagonal sector of the proposed spectrum \eqref{eq:dlim} coincides with the continuous spectrum of Liouville theory, while the non-diagonal sector $\mathcal{S}_{2\mathbb{Z},\mathbb{Z} + \frac{1}{2}}$ is built from non-diagonal fields $V^{N}_{(r,s)}$ with $r \in 2\mathbb{Z} $ and $s \in \mathbb{Z} + \frac{1}{2}$. This non-diagonal spectrum appeared previously in \cite{prs16}, where it was used to compute the $s$-and-$t$-channels of the correlation function $Z_0$ \eqref{eq:z0} at $c = 0$. In that paper the structure constants were determined by a numerical estimation, without making use of the shift equations, and certain combinations of correlation functions $Z_0$ were shown to reproduce cluster connectivities in critical percolation. 

Furthermore, we proposed fusion rules for this non-rational theory, based on the conservation of diagonality \eqref{eq:dcons} of minimal models, or equivalently on the non-triviality conditions \eqref{eq:sums} and \eqref{eq:sumr} of the analytic conformal bootstrap. The fusion rule \eqref{eq:ddrulp} between two diagonal fields is the same rule as in Liouville theory, and we know from \cite{rs15} that, when combined with the structure constants from the analytic conformal bootstrap, it gives rise to crossing symmetric four-point functions of the type $\left\langle V^{D}_{P_1}V^{D}_{P_2}V^{D}_{P_3}V^{D}_{P_4} \right\rangle$. On the other hand, our proposal \eqref{eq:nnrlp} for the fusion rule between two non-diagonal fields could not be tested, and constitutes a weaker prediction. 

In order to support our proposal for fusion rule \eqref{eq:dnr2z} between diagonal and non-diagonal fields, we computed four-point correlation functions of the type
\begin{align}
\left\langle V^{D}_{P_1}V^{N}_{(r_2,s_2)} V^{D}_{P_3}V^{N}_{(r_4,s_4)} \right\rangle \ .
\end{align} 
The $s$-and-$t$-channel expansion of these functions are  infinite sums over the spectrum $\mathcal{S}_{2\mathbb{Z},\mathbb{Z} + \frac{1}{2}}$. We presented an argument, based on the convergence of these sums, indicating that such functions could be extended to complex values of $c$ such that $\Re{c} < 13$. For this non-diagonal spectrum the structure constants could be computed by successive application of the shift equations, and in this manner we computed several examples of four-point functions for different values of $c$, and for arbitrary values of the momentums $P_1$ and $P_3$. In all examples we found a good agreement between the $s$-and-$t$-channels, suggesting that these functions satisfy crossing symmetry. 

The examples of section \ref{sec:2zz12} provide support for our proposal of a non-rational, non-diagonal theory with spectrum \eqref{eq:dlim} for $\Re{c} < 13$. We can then include it into the map of known models \ref{fig:cftmap}, along with the rest of our examples. Identifying the proposed theory by it's non-diagonal spectrum, the map becomes,  
\begin{align}
\begin{tikzpicture}[scale = 4.5, baseline=(current  bounding  box.center)]
 \filldraw [red!50, opacity = .15] (2.15, -.3) -- (2.15, .75) -- (-.6, .75) -- (-.6, -.3) -- cycle;
 \filldraw [blue!50, opacity = .15] (2.15, -.3) -- (2.15, .75) -- (-.6, .75) -- (-.6, -.3) -- cycle;
\fill[pattern= north east lines,pattern color=gray] (1.6, -.3) -- (1.6, .75) -- (-.6, .75) -- (-.6, -.3);
\draw[gray, opacity =.6] (1.6, -.3) -- (1.6, .75);
\node [draw = red, fill = red!70, right] at (-.5, .65) {Liouville theory};
\node [draw = blue, fill = blue!15, right] at (-.5, .49) {Generalized minimal models};
\node [draw = gray, fill= gray!50, right] at (-.5, .30) {$\mathcal{S}_{2\mathbb{Z}, \mathbb{Z} + \frac{1}{2} }$};
\node [draw = gray, pattern= north east lines,pattern color=gray , right] at (-.5, .30) {$\mathcal{S}_{2\mathbb{Z}, \mathbb{Z} + \frac{1}{2} }$};
 \node[draw = green!70!black, fill = green!16!black!7, right] at (-.5, .12) {Minimal models};
  \filldraw [red, opacity = .5] (.98, -.02) -- (.98, .02) -- (2.15, .02) -- (2.15, -.02) -- cycle;
  \filldraw [green!70!black, opacity = .3] (-.6, -.02) -- (-.6, .02) -- (.99, .02) -- (.99, -.02) -- cycle;
 \foreach \p in {2,...,20}
  {
  \draw [green!70!black, thick, opacity = 1] ({1-6/(\p*(\p+1))}, -.03) -- ({1-6/(\p*(\p+1))}, .03);
  }
  \node [below] at (0, -.02) {$0$};  \node [below] at (.99, -.02) {$1$};
  \node [below] at (1.6, -.02) {$13$};
  \node[draw,circle,inner sep=2pt,yellow] at (.99, 0) {} ; \node [above] at (.99, .02) {Ashkin-Teller};
  \draw[-latex] (-.6, 0) -- (2.15, 0) node [below left] {$c$};
\end{tikzpicture} \label{fig:newcftmap}
\end{align}
where the dashed region to the left of $\Re{c} =13$ indicates where our proposal should exist. 

Even though the examples of section \ref{sec:2zz12} show non-trivial verifications of crossing symmetry, further checks would be useful to gain more confidence in our proposal. In particular, verifying  whether all the proposed fusion rules produce crossing symmetric four-point functions would be an important step. The remaining fusion rules can be tested by computing four-point functions of the type $\left\langle V^{N}_{(r_1, s_1)}V^{N}_{(r_2, s_2)}V^{N}_{(r_3, s_3)}V^{N}_{(r_4, s_4)} \right\rangle$, and and the $s$-channel decomposition of functions $\left\langle V^{D}_{P_1}V^{D}_{P_2} V^{N}_{(r_3, s_3)} V^{N}_{(r_4, s_4)} \right\rangle$ . This last calculation would also be useful to follow up on the results of \cite{prs16}, by determining  the remaining quantities needed to compute connectivities.  

The diagonal sector of the proposed theory reproduces correlation functions of Liouville theory. Then, it is natural to interpret our proposal as a non-diagonal extension of Liouville theory. This idea could be reinforced by verifying the remaining fusion rules, showing that computing the diagonal channel of a function with two diagonal and two non-diagonal fields gives consistent results. 

A complementary point of view is that the non-rational limits of minimal models can be used to `navigate' through the map \ref{fig:newcftmap}, interpolating between these models. From this point of view, the challenge is evaluating the limit theories at \textit{rational} values of $c$, and recovering the minimal models. 


Finally, let us mention that our proposal for a non-rational, non-diagonal theory is not unique. A second, similar proposal can be obtained by switching $\beta \to \frac{1}{\beta}$, obtaining the non-diagonal spectrum $\mathcal{S}_{\mathbb{Z}+\frac{1}{2}, 2\mathbb{Z}}$. Crossing symmetry should be satisfied in this theory in the same way as in our examples, and it could be obtained as a non-rational limit of D-series minimal models with $p$ even and $q$ odd. 

\section{Outlook}

The results presented in this thesis offer a few lines for further work. The most immediate step to take is to complete the analytic conformal bootstrap analysis by finding an explicit solution for three-point structure constants of three non-diagonal fields, and to build a numerical implementation for the four-point functions of section \ref{sec:2zz12} that remain to be tested. 

In this thesis we have given a proposal for a non-rational, non-diagonal theory, which could be interpreted as a non-diagonal extension of Liouville theory. We arrived at this proposal by taking non-rational limits of minimal models spectrums, but we mentioned that there are different ways of taking these limits. In principle, it should be possible to take a different non-rational limit of the minimal models, and obtain a non-diagonal extension of generalized minimal model. In order to do this, we would need to keep the degeneracy of the fields, and keep enforcing the fusion rules \eqref{rrsr} when taking the limit. The differences in the process of taking this limit could be more clearly controlled while taking limits of four-point functions, instead of spectrums.

In this sense, it is worth mentioning that the approach of taking limits of minimal models \textit{spectrums} is somewhat heuristic, because spectrums and OPEs acquire their meaning when they are part of correlation functions. For this reason, it would be important to show that the result is the same when taking the limit $p,q \to \infty$ on a minimal model four-point function, instead of only the spectrum. Doing this requires that we choose a sequence of values $p,q$ such that $\frac{p}{q} \to \beta_0$ for some non-rational $\beta_0$, which define a sequence of minimal models. Then, we should verify that correlation functions of these minimal models approach the result of our non-rational proposal. Different limits arise because it is necessary to make a choice on how to identify correlation functions of different theories at each step of the sequence. 

Conversely, an important question is how can we take the limit in the opposite sense, i.e. how can we recover the minimal models from the limit theory by going from irrational to rational values of the central charge. Certain aspects of this process can be non-trivial: Typically, a rational value of the central charge could produce divergences in the structure constants or the conformal blocks, which need to be kept under control. Furthermore, it would be necessary to determine whether the minimal models fusion rules emerge naturally, or have to be imposed as a supplementary constraint. These questions have been explored recently in \cite{Ribault:2018jdv}, suggesting that under certain conditions it is indeed possible to recover the minimal models in this way. 

Finally, perhaps the most interesting direction would be to extend the non-diagonal analytic conformal bootstrap analysis to other conformal theories. This can be done either by relaxing some of the main assumptions of section \ref{sec:assumptions}, or by studying conformal theories whose symmetry algebra is other than $\mathfrak{V} \otimes \bar{\mathfrak{V}}$, and in particular to $\mathcal{W}$-theories. 

Firstly, relaxing some of our assumptions could allow us to study theories with fractional indices, such as the ones discussed in \cite{ei15}. It would be important to find a way to apply the results discussed here to that case. 

Secondly, extending to $\mathcal{W}$-theories could produce interesting results. One of the main challenges in this direction is the definition of consistent fusion rules for the degenerate fields of $\mathcal{W}$-algebras. In the Virasoro algebra case discussed in this work, it was sufficient to define two different classes of fields, diagonal and non-diagonal, with different fusion rules with degenerate fields. In the case of $\mathcal{W}$-algebras the momentums have many components, and there exist different types of degenerate fields. The challenge would be to identify how many degenerate fields are needed to produce constraints analogous to the shift equations \eqref{eq:drat}, and how many different types of fields are needed in order to have a consistent set of degenerate fusion rules. This extension has been studied in the recent paper \cite{Dupic:2018pqr}. 

%% file: chapters/resume.tex
\chapter{Résumé}


Dans cette thèse nous nous concentrons sur la symétrie conforme, et ses conséquences sur les théories quantiques des champs bidimensionnelles. Les transformations conformes sont les transformations qui laissent les angles invariants, et elles forment une classe des transformations plus large que celles du groupe de Poincaré. En conséquence, les théories conformes des champs possèdent une symétrie augmentée qui les rende plus faciles à étudier que les théories des champs génériques. Parfois, ces symétries sont suffisantes pour rendre une 
théorie complètement résoluble, dans les sens où toutes ses fonctions de corrélation peuvent être, en principe, déterminées. 

Pour étudier les théories conformes des champs nous suivons l'approche connue sous le nom de bootstrap conforme, basée sur l'idée de construire, classifier et résoudre des théories en imposant des contraintes provenant des symétries et cohérence seulement. Une des avantages de cette approche est qu'elle permet d'arriver à des résultats applicables dans une grande classe de théories, car elle repose sur des principes très généraux. Par contre, le manque d'une description Lagrangienne des théories ainsi construites peut rendre leur interprétation physique difficile à identifier.  

Les résultats principaux de cette thèse ont été publiés dans \cite{Migliaccio:2017dch}.

\section{Théories conformes des champs bidimensionnelles}

Nous nous concentrons sur des théories des champs définis sur la sphère de Riemann, décrite par les coordonnés complexes $z$ et $\bar{z}$. En général ces coordonnés sont considérés comme indépendants, et les observables physique sont obtenus en choisissant $\bar{z} = z$. 
Une théorie quantique de champs bidimensionnelle est conforme quand l’algèbre de symétrie est ou contient l’algèbre de Virasoro $\mathfrak{V}$ \cite{Teschner:2017del}. Celle-ci est un algèbre de Lie avec une infinité des générateurs, caractérisée para un paramètre $c$ appelé la charge centrale. La charge centrale peut être écrite en termes d'un autre paramètre $\beta$ comme
\begin{align}
c = 1- 6 \left( \beta - \frac{1}{\beta}\right)^2\, .
\end{align}

Dans les théories conformes discutées ici les champs sont en correspondance avec les représentations de l’algèbre de symétrie \cite{Ribault:2014hia}. L'algèbre $\mathfrak{V}$ a des représentations de plus haut poids, où tous les états de la représentation s'obtiennent à partir de l'action d'un sous-ensemble des générateurs de l'algèbre, appelés opérateurs de création, sur un état primaire. Cet état primaire, et donc la représentation, est identifié par son poids conforme $\Delta$, qui est une valeur propre de l’opérateur qui génère les dilatations. Les états obtenus par l'action des opérateurs de création sur l'état primaire sont ses descendants. Ils sont aussi des états propres de l’opérateur de dilatation, et leurs dimensions diffèrent de celle de l'état primaire par des entiers. 

Le module de Verma $\mathcal{V}_{\Delta}$ est la représentation de plus haut poids la plus large généré à partir de l'état primaire de poids $\Delta$. Par contre, une représentation de plus haut poids se dit dégénéré quand un des descendants de l’état primaire s'annule. Les représentations dégénérées existent pour des valeurs particulières $\Delta_{\langle r,s \rangle}$ du poids conforme de l'état primaire, déterminés par deux nombres entiers $r$ et $s$. Ces valeurs sont
\begin{align}
\Delta_{\langle r,s \rangle} = \frac{c-1}{24} + \frac{1}{4}(r\beta - \frac{s}{\beta})^2\, ,
\end{align}
et nous écrivons une représentation dégénéré comme $\mathcal{R}_{\langle r,s \rangle}$.

Cette thèse étudie des théories conformes dont l'algèbre de symétrie est composée de deux copies de l'algèbre de Virasoro, avec la même charge centrale. L'algèbre complète est donc $\mathfrak{V} \otimes \bar{\mathfrak{V}}$, où $\mathfrak{V}$, dit le secteur gauche ou holomorphe, représente des transformations agissant sur la coordonné $z$, et $\bar{\mathfrak{V}}$ , dit le secteur droit ou antiholomorphe, fait de même avec la coordonné $\bar{z}$. Comme l’algèbre de symétrie contient deux secteurs, holomorphe et antiholomorphe, un champ portera deux poids conformes, un pour chaque secteur. Nous écrivons un champ primaire comme
\begin{align}
V_{\Delta, \bar{\Delta}}(z) \, ,
\end{align} 
où nous omettons la dépendance en $\bar{z}$ dans l'argument. En termes générales,   nous parlons d'une théorie non-diagonale quand elle contient des champs primaires dont $\Delta \neq \bar{\Delta}$. 

Plusieurs théories différentes possédant l'algèbre de symétrie  $\mathfrak{V} \otimes \bar{\mathfrak{V}}$ existent, et elles peuvent être représentes sur le plan complexe des valeurs de la charge centrale, comme le montre le diagramme suivante \cite{Ribault:2014hia}: 

\begin{align}
\begin{tikzpicture}[scale = 4.5, baseline=(current  bounding  box.center)]
 \filldraw [red!50, opacity = .15] (2.15, -.3) -- (2.15, .75) -- (-.6, .75) -- (-.6, -.3) -- cycle;
 \filldraw [blue!50, opacity = .15] (2.15, -.3) -- (2.15, .75) -- (-.6, .75) -- (-.6, -.3) -- cycle;
 \node [draw = red, fill = red!70, right] at (-.5, .65) {Liouville theory};
\node [draw = blue, fill = blue!15, right] at (-.5, .49) {Generalized minimal models};
 \node[draw = green!70!black, fill = green!16!black!7, right] at (-.5, .12) {Minimal models};
  \filldraw [red, opacity = .5] (.98, -.02) -- (.98, .02) -- (2.15, .02) -- (2.15, -.02) -- cycle;
  \filldraw [green!70!black, opacity = .3] (-.6, -.02) -- (-.6, .02) -- (.99, .02) -- (.99, -.02) -- cycle;
 \foreach \p in {2,...,20}
  {
  \draw [green!70!black, thick, opacity = 1] ({1-6/(\p*(\p+1))}, -.03) -- ({1-6/(\p*(\p+1))}, .03);
  }
  \node [below] at (0, -.02) {$0$};  \node [below] at (.99, -.02) {$1$};
  \node[draw,circle,inner sep=2pt,yellow] at (.99, 0) {} ; \node [above] at (.99, .02) {Ashkin-Teller};
  \draw[-latex] (-.6, 0) -- (2.15, 0) node [below left] {$c$};
\end{tikzpicture} \label{res-fig:cftmap}
\end{align}
Dans cette figure la couleur rouge correspond à la théorie de Liouville, une théorie avec un spectre diagonal, continu et infini, qui existe pour toute valeur de $c \in \mathbb{C}$. Cette théorie est unitaire pour $c \in \mathbb{R}_{\geq 1}$, et à partir de cette région elle peut être continué aux valeurs $c \notin \mathbb{R}_{<1}$. Pour $c \in \mathbb{R}_{<1}$ il existe une version diffèrent de la théorie de Liouville qui a le même spectre qu'avant, mais ne peut pas être obtenu comme une continuation analytique des théories existant pour $c \notin \mathbb{R}_{<1}$ \cite{rs15}. La couleur bleu montre les modèles minimaux généralises, qui existent pour toute $c \in \mathbb{C}$ et possèdent un spectre diagonal et infini mais discret. En verte nous identifions les modèles minimaux. Ces modèles existent seulement pour une série des valeurs rationnelles de $c$, qui sont cependant denses sur la demi-droite $c \in \mathbb{R}_{<1}$, et ils possèdent des spectres discrets et finis. Les modèles minimaux obéissent une classification A-D-E \cite{Cappelli:2010}, dont les différents modèles ont des spectres différents: Les modèles de la série A ont des spectres diagonaux, tant que les modèles des séries D et E  ont des spectres non-diagonaux. Les lignes vertes dans la figure correspondent aux modèles minimales unitaires. Finalement, le cercle jaune situé à $c = 1$ représente le modèle d'Ashkin-Teller, un exemple d'une théorie avec un spectre infini, discret et non diagonal. 

Malgré la variété des théories conformes représentes dans la figure \ref{res-fig:cftmap}, il n'y a pas d'exemples d'une théorie non-diagonal existant pour des valeurs génériques de la charge central. L'objective principal de cette thèse est d'obtenir une extension de la méthode connu comme bootstrap conforme pour des théories conformes non-diagonales, à fin de trouver des solutions générales pour ce type des théories. 

\section{Conséquences de la symétrie conforme}

La symétrie conforme impose des importantes contraintes sur la structure des théories des champs. Ici nous discutons les plus importantes d'entre elles, qui seront la base de la méthode du bootstrap conforme. 

Premièrement, la symétrie conforme est  suffisante pour fixer complètement la forme des fonctions de corrélation aux deux-et-trois points. Pour les fonctions à deux points nous avons
\begin{align}
\left\langle V_{\zb[1]{\Delta}}(z_1) V_{\zb[2]{\Delta}}(z_2) \right\rangle = B(V_1) \frac{\delta_{\Delta_1, \Delta_2}\delta_{\bar{\Delta}_1, \bar{\Delta}_2}}{z_{12}^{2\Delta_1}\bar{z}_{12}^{2\bar{\Delta}_1}}\, , \label{res-eq:gen2point}
\end{align}
où $z_{ij}=(z_i - z_j)$ et nous avons introduit la constante de structure à deux points $B(V)$, qui est indépendante des coordonnés. Le facteur $\delta_{\Delta_1, \Delta_2}$ indique que cette fonction peut être non-nul seulement si les poids conformes des deux champs coïncident. 

Dans le cas de la fonction à trois points nous trouvons
\begin{align}
\left\langle \prod_{i=1}^{3} V_{\zb[i]{\Delta}}(z_i) \right\rangle = C_{123}\left|\mathcal{F}^{(3)}(\Delta_1,\Delta_2,\Delta_3|z_1,z_2,z_3)\right|^2  \label{res-eq:gen3point}
\end{align}
où nous avons introduit la constante de structure à trois points  $C_{123}= C(V_1, V_2, V_3)$, indépendante des coordonnés, et la fonction 
\begin{align}
\mathcal{F}^{(3)}(\Delta_1,\Delta_2,\Delta_3|z_1,z_2,z_3) = 
 z_{12}^{-\Delta_1 - \Delta_2 + \Delta_3}
z_{23}^{-\Delta_2 - \Delta_3 + \Delta_1}  z_{31}^{-\Delta_3 - \Delta_1 + \Delta_2} \, , \label{res-eq:3pbl}
\end{align}
dit le bloc conforme à trois points. Dans la formule \eqref{res-eq:gen3point}, la notation module carré fait référence au produit des quantités holomorphes et antiholomorphes, qui sont reliés par les changements $z_i \to \bar{z}_i$ et $\Delta_i \to \bar{\Delta}_i$. 

À partir des expressions \eqref{res-eq:gen2point} et \eqref{res-eq:gen3point} nous voyons que les parties non-triviales des fonctions de  corrélation à deux et trois points sont les constantes de structure $B(V)$ et $C(V_1, V_2, V_3)$. Ainsi, pour calculer ces fonctions dans une théorie déterminée il suffit d'identifier ses constantes de structure.
 
Dans le cas des fonctions à quatre points, la symétrie conforme n'est plus suffisante pour les déterminer complètement. Pour calculer ces fonctions nous pouvons profiter d'une propriété des théories conformes appelé l'expansion de produit d'opérateurs ou OPE, d'après son sigle en anglais. Cet expansion permet d'écrire le produit de deux opérateurs $V_1(z_1)$ et $V_2(z_2)$ se rapprochant l'un de l'autre comme une somme, sur un certain spectre, des termes composés par un coefficient fixé par la symétrie conforme et un champ évalué dans le point $z_2$. La forme générale peut être écrite schématiquement comme
\begin{align}
 V_{\Delta_1,\bar\Delta_1}(z_1) V_{\Delta_2,\bar\Delta_2}(z_2) 
 =
 \sum_{\Delta_3,\bar\Delta_3} \frac{C_{123}}{B_3}  V_{\Delta_3,\bar\Delta_3}(z_2)
 \ .
 \label{res-eq:vvs}
\end{align}
où $V_{\Delta_3,\bar\Delta_3}(z_2)$ représente la contribution de toute la famille correspondant au champ , et nous voyons que les coefficients de chaque terme sont une combinaison des constantes de structure à deux et trois points. Pour une expression plus explicite, voir l'équation \eqref{eq:vvs}. L'ensemble des champs sur lequel la somme est réalisée s'appelle le spectre de l'OPE, et il est déterminé par un ensemble des règles connus sous le nom des règles de fusion. 

L'OPE permet d'obtenir une expansion similaire pour exprimer les fonctions à quatre points. Si l'on insère l'OPE \eqref{res-eq:vvs} dans une fonction à quatre points $ \left\langle \prod_{i=1}^{4} V_{\zb[i]{\Delta}}(z_i) \right\rangle $ nous obtenons l'expansion dit du canal $s$, qui a la forme suivante
\begin{align}
\left\langle  \prod_{i=1}^{4}V_i(z_i) \right\rangle = \sum_{s} D_{s|1234} \left| 
\begin{tikzpicture}[baseline=(current  bounding  box.center), very thick, scale = .3]
\draw (-1,2) node [left] {$2$} -- (0,0) -- node [above] {$s$} (4,0) -- (5,2) node [right] {$3$};
\draw (-1,-2) node [left] {$1$} -- (0,0);
\draw (4,0) -- (5,-2) node [right] {$4$};
\end{tikzpicture} \right| \, . \label{res-eq:s4pt}
\end{align}
Ici, les coefficients $D_{1234}(V_s)$ sont les constants de structure à quatre points, donnés par 
\begin{align}
D_{1234}(V_s) = D_{s|1234} = \frac{C_{12s}C_{s34}}{B_s}\, ,\label{res-eq:4pc}
\end{align}
et les diagrammes représentent les fonctions connus comme blocs conformes du canal $s$, qui sont entièrement fixes par l’algèbre de symétrie. 

L'expansion \eqref{res-eq:s4pt} montre que pour connaître les fonctions à quatre-point nous avons besoin des mêmes constants de structure que pour les fonctions à deux et trois points, et en plus il est nécessaire de déterminer les règles de fusion de la théorie. Avec ces ingrédients il devient possible de décomposer une fonction à $n$ points en terme des fonctions et coefficients connus, et dans ce sens toutes les fonctions de corrélation d'une théorie peuvent être, en principe, déterminées. En conséquence, pour résoudre une théorie conforme il faut être capable de déterminer le spectre de ses OPEs et ses constants de structure.  

Jusqu'à ce point nous avons discuté les contraintes que la symétrie conforme impose sur les champs et les fonctions de corrélations des théories bidimensionnelles. Pour trouver les éléments manquants, les constants de structure et le spectre des OPEs, nous pouvons imposer une condition de auto consistance connu comme symétrie de croisement. L'idée est la suivante: Pour obtenir l'expansion du canal $s$, expression \eqref{res-eq:s4pt}, nous avons inséré l'OPE entre les champs $V_1$ et $V_2$. Cependant, nous aurons pu utiliser l'OPE entre les champs $V_1$ et $V_4$, ce qui nous aurait conduits a une expansion différente, l'expansion du canal $t$, pour la même fonction à quatre points. Ces deux expansions doivent coïncider, et ainsi nous arrivons aux équations de la symétrie de croisement: 
\begin{align}
\sum_{s} \frac{C_{12s}C_{s34}}{B_s} \left| 
 \begin{tikzpicture}[baseline=(current  bounding  box.center), very thick, scale = .3]
\draw (-1,2) node [left] {$2$} -- (0,0) -- node [above] {$s$} (4,0) -- (5,2) node [right] {$3$};
\draw (-1,-2) node [left] {$1$} -- (0,0);
\draw (4,0) -- (5,-2) node [right] {$4$};
\end{tikzpicture} 
\right|^2 = \sum_{\Delta_t,\bar{\Delta}_t} \frac{C_{23t}C_{t41}}{B_t} \left|
\begin{tikzpicture}[baseline=(current  bounding  box.center), very thick, scale = .3]
 \draw (-2,3) node [left] {$2$} -- (0,2) -- node [left] {$t$} (0,-2) -- (-2, -3) node [left] {$1$};
\draw (0,2) -- (2,3) node [right] {$3$};
\draw (0,-2) -- (2, -3) node [right] {$4$};
\end{tikzpicture}
\right|^2\ \, .
\end{align}
Ces équations forment un système non-linéaire qui pourrait, accompagné des certaines hypothèses, être suffisant pour déterminer les constantes de structure et les spectres de chaque canal. La résolution de ce système est au cœur du bootstrap conforme, qui cherche  à résoudre les théories conformes à partir des conditions de symétrie et auto consistance. Néanmoins, ce système est trop complexe pour être attaqué en toute généralité, et il est nécessaire de trouver des stratégies pour le simplifier et trouver ses solutions. 

Dans ce travail nous allons suivre le bootstrap conforme analytique, dont l'idée principal est d'utiliser un type spécial de champ, connus comme champs dégénère, qui se caractérisent par la propriété que ses OPEs, et en conséquences les fonctions de corrélations qui ont des champs dégénères, ont des spectres finis et connus. En utilisant ces champs dégénères il devienne possible d'avoir un système d'équations de symétrie de croisement fini et plus facilement résoluble.

\section{Bootstrap conforme}
Notre étude des théories conformes non-diagonales repose sur trois hypothèses principales:
\begin{itemize}
\item Charge centrale générique:
Nous considérons que la charge centrale $c$ de l'algèbre de Virasoro peut prendre des valeurs complexes arbitraires. Cette hypothèse vise à éviter des particularités associées à des valeurs spécifiques de $c$, comme l'existence des modèles minimaux pour $c$ rationnelle.
\item Fonctions de corrélation univalués: Nous considérons que toutes les fonctions de corrélation entre les champs appartenant au spectre d'une théorie doivent être univalués. Cette restriction implique, comme il est décrit dans le chapitre \ref{ch:cboots}, que le spin des champs primaires du spectre doit obéir
\begin{align}
S = \Delta - \bar{\Delta} \in \frac{1}{2}\mathbb{Z}\, . \label{res-halfint}
\end{align}  
\item Existence des champs dégénérés: Notre analyse est basée sur l’étude des fonctions de corrélation qui incluent les champs dégénérés diagonaux $V_{\langle 2,1 \rangle}$ et $V_{\langle 1,2 \rangle}$, dont les poids conformes sont
\begin{align}
\Delta_{\langle 2,1 \rangle} = \frac{c-1}{24} + \frac{1}{4} \left(2\beta - \frac{1}{\beta}\right)^2\, , \\
\Delta_{\langle 1,2 \rangle} = \frac{c-1}{24} + \frac{1}{4} \left(\beta - \frac{2}{\beta}\right)^2\, .
\end{align}
Nous considérons que ces champs existent, dans le sens ou nous pouvons étudier des fonctions de corrélation qui les incluent même s'ils ne font pas partie du spectre d'une théorie.  
\end{itemize}

\subsection{Champs diagonaux et non-diagonaux} \label{res-sec:dndfields} 

À partir de nos assomptions nous pouvons déduire des contraintes sur le spectre d'une théorie non-diagonale. Pour que nos assomptions soient cohérentes entre elles il est nécessaire que les OPEs incluant au moins un champ dégénéré produisent des champs qui respectent les assomptions de charge centrale générique et spin demi-entier \eqref{res-halfint}. Les OPEs prennent une forme plus simple si nous exprimons les poids conformes comme
\begin{align}
\Delta = \frac{c-1}{24} + P^2\, ,
\end{align}
où $P$ s'appelle l'impulsion. En termes des impulsions, le spin d'un champ est
\begin{align}
S = \Delta - \bar{\Delta} = P^2 - \bar{P}^2\, .
\end{align}

Ainsi, nous trouvons dans le chapitre \ref{ch:cboots} que dans une théorie obéissant à nos assomptions les champs portent des étiquettes $\sigma, \tilde{\sigma} \in \{ 1, -1 \}$ qui contrôlent ses OPEs avec les champs dégénérés. Ces OPEs prennent la forme suivante
\begin{align}
V_{\langle 2,1 \rangle} \times V_{P,\bar{P}}^{\sigma, \tilde{\sigma}} = \sum_{\epsilon = \pm} V_{P+\epsilon \tfrac{\beta}{2},\bar{P}+\sigma\epsilon \tfrac{\beta}{2}}^{\sigma, \tilde{\sigma}}\quad ,
\quad
V_{\langle 1,2 \rangle} \times V_{P,\bar{P}}^{\sigma, \tilde{\sigma}} = \sum_{\epsilon = \pm} V_{P-\tfrac{\epsilon}{2\beta},\bar{P}- \tfrac{\tilde{\sigma}\epsilon}{2\beta}}^{\sigma, \tilde{\sigma}}\, .\label{res-dfus}
\end{align}

Les champs d'une telle théorie peuvent être classifiés en deux groupes, selon la relation entre $\sigma$ et $\tilde{\sigma}$. Afin de fixer les conventions, nous allons prendre $\sigma = 1$ sans perte de généralité. Les deux types de champs sont: 
\begin{itemize}
\item Champs diagonaux, $\tilde{\sigma} = \sigma$.
Les impulsions gauche et droite de ces champs satisfassent
\begin{align}
\bar{P} = P\, ,
\end{align} 
et elles ne sont pas soumises à d'autres contraintes, pouvant prendre en principe des valeurs arbitraires. Les champs diagonaux ont $S = 0$, et ils seront parfois désignés par
\begin{align}
V^D_{P} = V^{+,+}_{P, P}\, .
\end{align}
\item Champs non-diagonaux, $\tilde{\sigma} = -\sigma$.
Dans ce cas, les impulsions des champs sont donnés à partir de deux indices $r$ et $s$, telles que $r,s, rs \in \frac{1}{2}\mathbb{Z}$. Les impulsions gauche et droite sont
\begin{align}
P = P_{(r,s)} &= \frac{1}{2}\left(r\beta - \frac{s}{\beta} \right)\, ,\\
\bar{P} &= P_{(r,-s)} \, ,\\
\end{align}
et le spin est donné par
\begin{align}
S = P^2 - \bar{P}^2 = -rs\, .
\end{align}
Un champ non-diagonal peut être indiqué par 
\begin{align}
V^N_{(r,s)} = V^{+,-}_{P_{(r,s)}, P_{(r,-s)} }\, .
\end{align}
\end{itemize}
Il est important de signaler que les champs diagonaux et non-diagonaux se distinguent non seulement par les valeurs qui peuvent prendre ses impulsions, mais fondamentalement par ses OPEs \eqref{res-dfus} avec les champs dégénérés. Par exemple, les champs $V^{D}_{P_{r,0}}$ et $V^N_{(r,0)}$ ont les mêmes impulsions gauche et droite. Cependant, l'OPE avec le champ dégénéré $V_{\langle 1,2 \rangle}$ produit des champs de spin $0$ dans le cas du champ diagonal $V^{D}_{P_{r,0}}$, et des champs de spin $\pm r$ dans le cas du champ non-diagonal $V^N_{(r,0)}$. 

La forme des impulsions et les règles de fusion des champs non-diagonaux ont été dérivées dans le cas des champs de spin non-zéro. Cependant, dans l'exemple précédent nous avons appliqué ce résultat aux champs non-diagonaux de spin $0$. En faisant ceci nous avons pris une assomption supplémentaire que nous allons maintenir pendant le reste de ce travail: l'idée que nous avons choisi une base dont les champs diagonaux et non-diagonaux ne se mélangent pas, dans le sens, 
\begin{align}
\left\langle V^D_{P_{(r,0)}}V^N_{(r,0)} \right\rangle = 0\ . \label{res-dnd2}
\end{align}  
Celle-ci est une hypothèse supplémentaire et non une restriction imposée sur la fonction à deux points par la symétrie conforme. Certains modèles minimaux de la série D ont des champs diagonaux et non-diagonaux avec les mêmes impulsions qui respectent l'équation \eqref{res-dnd2}.

\subsection{Équations pour les constantes de structure}
Pour ces champs diagonaux et non-diagonaux, nous pouvons écrire les OPEs dégénérés de façon plus complète de la forme suivante
\begin{align}
V_{\langle 2,1 \rangle} V
= \sum_{\epsilon = \pm} C_{\epsilon}(V) V^\epsilon
\quad ,\quad 
V_{\langle 1,2 \rangle} V
= \sum_{\epsilon = \pm} \tilde{C}_{\epsilon}(V) V^{\tilde{\epsilon}} \, ,\label{res-dOPE}
\end{align}
où nous avons introduit les coefficients de l'OPE dégénéré $C_{\pm}(V)$ et $\tilde{C}_{\pm}(V)$, et les notations 
\begin{align}
 V = V_{P,\bar{P}}^{\sigma, \tilde{\sigma}} 
 \quad \implies \quad V^\epsilon = V_{P+\epsilon \tfrac{\beta}{2},\bar{P}+\sigma\epsilon \tfrac{\beta}{2}}^{\sigma, \tilde{\sigma}}\quad ,\quad 
 V^{\tilde{\epsilon}} = V_{P-\tfrac{\epsilon}{2\beta},\bar{P}- \tfrac{\tilde{\sigma}\epsilon}{2\beta}}^{\sigma, \tilde{\sigma}}\ .
\end{align}

À partir de ces OPEs nous pouvons écrire les équations de symétrie de croisement pour une fonction à quatre points qui inclue des champs dégénérés. Prenons par exemple la fonction $\langle V_{\langle 2,1 \rangle}(x) V_1(0) V_2(\infty) V_3(1) \rangle$, où les champs $V_1$, $V_2$ et $V_3$ peuvent être diagonaux ou non, et nous pouvons fixer leurs positions grâce aux transformations conformes. La symétrie de croisement s'exprime par 
\begin{align}
 \Big\langle V_{\langle 2,1\rangle} V_1 V_2 V_3 \Big\rangle &= 
 \sum_{\epsilon_1 = \pm} d^{(s)}_{\epsilon_1} \mathcal{F}^{(s)}_{\epsilon_1} \bar{\mathcal{F}}^{(s)}_{\sigma_1\epsilon_1}
 = \sum_{\epsilon_3 = \pm} d^{(t)}_{\epsilon_3} \mathcal{F}^{(t)}_{\epsilon_3} \bar{\mathcal{F}}^{(t)}_{\sigma_3\epsilon_3}
 \ ,
 \label{res-eq:2dec}
\end{align}
où $\mathcal{F}_\epsilon^{(s)},\mathcal{F}^{(t)}_\epsilon$ sont les blocs conformes dégénérés à quatre points (voir chapitre \ref{ch:cboots} pour son expression en termes de fonctions hypergéométriques), et les constantes de structure  quatre points dégénérés sont
\begin{align}
d^{(s)}_\epsilon = C_\epsilon(V_1)C(V_1^\epsilon,V_2,V_3) \, .
\end{align}

Les expansions du canal $s$ et $t$ expriment la fonction à quatre points dans deux bases différentes. Ces bases sont reliés par une matrice dit la matrice de fusion, et à partir de ces relations nous pouvons déterminer les rapports des coefficients de ces expansions.
La première conséquence que nous trouvons est que la fonction \eqref{res-eq:2dec} admet une solution non-trivial seulement si
\begin{align}
\sum_{i=1}^{3} s_i \in \mathbb{Z}\, , \label{res-nontrivs}
\end{align}
où les indices font référence aux indices des champs non-diagonaux, i.e. $s_i = 0 $ si $V_i$ est diagonal. 
Une condition similaire existe pour les fonctions à quatre points avec le champ dégénéré $V_{\langle 1,2 \rangle}$, avec les indices $r_i$ au lieu de $s_i$.
Ces conditions de non-trivialité peuvent être interprétées comme des contraintes sur les constantes de structure à trois points. Dans ce sens là, les constantes de structure d'une théorie qui obéisse à nos hypothèses doivent satisfaire
\begin{align}
C(V_1,V_2,V_3) \neq 0 \Rightarrow \sum_{i=1}^{3} s_i\, , \sum_{i=1}^{3} r_i \in \mathbb{Z}. \label{res-eq:nontriC}
\end{align}

Si les conditions de non-trivialité sont respectées, les rapports des coefficients peuvent être connus explicitement. Nous définissons le rapport $\rho(V_1|V_2,V_3) = \frac{d^{(s)}_+}{d^{(s)}_-} $, qui s'exprime en termes des constantes de structure comme
\begin{align}
 \rho(V_1|V_2,V_3) = \frac{C_+(V_1) C(V_1^+,V_2,V_3)}{C_-(V_1) C(V_1^-,V_2,V_3)} \ , \label{res-eq:rcst}
\end{align}
où 
\begin{align}
\rho(V_1|V_2,V_3) = -(-1)^{2s_2} &\frac{\Gamma(-2\beta P_1)}{\Gamma(2\beta P_1)} \frac{\Gamma(-2\beta \sigma_1\bar P_1)}{\Gamma(2\beta \sigma_1\bar P_1)} \\  &\times \frac{\prod_{\pm, \pm} \Gamma(\frac12 +\beta P_1 \pm \beta P_2\pm \beta P_3)}{\prod_{\pm, \pm} \Gamma(\frac12 -\beta \sigma_1\bar P_1 \pm \beta \bar P_2 \pm \beta \bar P_3)}\, .
\end{align}
Nous appelons les expressions comme \eqref{res-eq:rcst} une \textit{équation de shift}, car elle relie des valeurs des constantes de structure dont les impulsions diffèrent par $\pm \beta$. 

Si l'on regarde une fonction à quatre points avec deux champs dégénérés, nous pouvons trouver des relations similaires pour les coefficients des OPEs dégénérés et la constante de structure à deux points. À partir de la fonction $\left\langle V_{\langle 2,1 \rangle} V_1 V_{\langle 2,1 \rangle} V_1 \right\rangle$ nous trouvons un rapport qui dépend seulement du champ $V_1$,  
\begin{align}
 \rho(V_1) = \frac{C_+(V_1)^2 B(V_1^+)}{C_-(V_1)^2 B(V_1^-)} \ ,
 \label{res-eq:rvo}
\end{align}
avec
\begin{align}
\rho(V) =  -\frac{\Gamma(-2\beta P)\Gamma(-2\beta \sigma \bar{P})}{\Gamma(2\beta P)\Gamma(2\beta \sigma \bar{P})}\frac{\Gamma(\beta^2 +2\beta P)\Gamma(1-\beta^2 +2\beta P)}{\Gamma(\beta^2 -2\beta\sigma\bar{P})\Gamma(1-\beta^2-2\beta\sigma\bar{P})}\, .
\end{align}

Nous avons aussi des relations similaires qui correspondent aux fonctions à quatre points avec le champ dégénéré $V_{\langle 1,2 \rangle}$:
\begin{align}
 \tilde{\rho}(V_1|V_2,V_3) = \frac{\tilde{C}_+(V_1) C(V_1^{\tilde{+}},V_2,V_3)}{\tilde{C}_-(V_1) C(V_1^{\tilde{-}},V_2,V_3)} \ , \label{res-eq:trcst}
\end{align}
et
\begin{align}
 \tilde{\rho}(V_1) = \frac{\tilde{C}_+(V_1)^2 B(V_1^{\tilde{+}})}{\tilde{C}_-(V_1)^2 B(V_1^{\tilde{-}})} \ , 
 \label{res-eq:trvo}
\end{align}
où les rapports $\tilde{\rho}(V_1|V_2,V_3)$ et $\tilde{\rho}(V_1)$ sont
\begin{multline}
 \tilde{\rho}(V_1|V_2,V_3) = -(-1)^{2r_2} \frac{\Gamma(2\beta^{-1}P_1)}{\Gamma(-2\beta^{-1}P_1)} \frac{\Gamma(2\beta^{-1}\tilde{\sigma}_1\bar P_1)}{\Gamma(-2\beta^{-1}\tilde{\sigma}_1\bar P_1)} 
 \\
 \times
 \frac{ \prod_{\pm, \pm} \Gamma(\frac12 -\beta^{-1}P_1 \pm \beta^{-1}P_2\pm \beta^{-1}P_3)}{ \prod_{\pm, \pm} \Gamma(\frac12 +\beta^{-1}\tilde{\sigma}_1\bar P_1 \pm \beta^{-1}\bar P_2 \pm \beta^{-1}\bar P_3)}\ ,
 \label{res-eq:rt}
\end{multline}
et
\begin{multline}
\tilde{\rho}(V) =  -\frac{\Gamma(2\beta^{-1} P)\Gamma(2\beta^{-1} \tilde{\sigma} \bar{P})}{\Gamma(-2\beta^{-1} P)\Gamma(-2\beta^{-1} \tilde{\sigma} \bar{P})} \\ \times \frac{\Gamma(\beta^{-2} -2\beta^{-1} P)\Gamma(1-\beta^{-2} -2\beta^{-1} P)}{\Gamma(\beta^{-2} +2\beta^{-1}\tilde{\sigma}\bar{P})\Gamma(1-\beta^{-2}+2\beta^{-1}\tilde{\sigma}\bar{P})}\, . \
\end{multline}

En combinant les expressions \eqref{res-eq:rcst} et \eqref{res-eq:rvo}, et aussi \eqref{res-eq:trcst} et \eqref{res-eq:trvo} nous pouvons écrire des équations de shift pour les constantes de structure à quatre points, 
\begin{equation}
\begin{aligned}
&\boxed{\frac{D_{1234}(V_s^+)}{D_{1234}(V_s^-)} = \frac{\rho(V_s|V_1,V_2)\rho(V_s|V_3,V_4)}{\rho(V_s)}}\, , \\
&\boxed{\frac{D_{1234}(V_s^{\tilde{+}})}{D_{1234}(V_s^{\tilde{-}})} = \frac{\tilde{\rho}(V_s|V_1,V_2)\tilde{\rho}(V_s|V_3,V_4)}{\tilde{\rho}(V_s)}}\, .
\end{aligned} \label{res-eq:drat}
\end{equation}
Il est important de signaler que les coefficients des OPEs dégénérés n'apparaissent plus dans les expressions \eqref{res-eq:drat}, et alors ces équations peuvent être utilisées directement pour calculer des constantes de structure itérativement et vérifier la symétrie de croisement.  

À fin de résoudre les équations de shift pour les constantes de structure à trois points, il est nécessaire de discuter sur la normalisation des champs. Nous pouvons rénormaliser les champs en faisant une transformation 
\begin{align}
 V_i(z) \to \lambda_i V_i(z)\, ,  \label{res-eq:renorm}
\end{align}
où le facteur $\lambda_i$ est indépendant de $(z)$.
Dans une théorie particulière certains critères peuvent être pris pour fixer la normalisation. Par exemple, nous pouvons choisir $\lambda_i$ de sorte que la constante de structure à deux points ou les coefficients dégénérés des OPE soient fixés à $1$. Ici nous prenons une normalisation de référence, et nous allons résoudre les équations de shift pour les quantités qui sont invariantes par renormalisation. 
Nous écrivons la constante de structure à trois points comme
\begin{align}
 C_{123} = \left(\prod_{i=1}^3 Y_i\right)C'_{123}\, , \label{res-3pnorm}
\end{align}
où $C'_{123}$ est la constante dans notre normalisation de référence, et les facteurs $Y(V)$ prennent en compte les changements de normalisation. Suite à une transformation du type \eqref{res-eq:renorm}, les constantes de structure se transforment selon
\begin{align}
B(V_i) \to \lambda^{2}_iB(V_i) \ ,\quad Y(V_i) \to \lambda_i Y(V_i)\, . 
\end{align}

Ainsi, nous allons déterminer les facteurs qui sont invariantes par renormalisation des champs, i.e.
\begin{align}
 C'_{123} \qquad \text{et} \qquad Y_i^2B_i^{-1}\ .\label{res-eq:rinv}
\end{align}
Ces facteurs sont les seuls dont nous avons besoin pour vérifier la symétrie de croisement, et donc la cohérence de la théorie. Ceci peut se voir en écrivant la constante de structure à quatre points comme
\begin{align}
 D_{s|1234} = \left(\prod_{i=1}^4 Y_i\right) C'_{12s} C'_{s34} \frac{Y^2_s}{B_s}\, , \label{res-eq:4pcp}
\end{align}
où nous voyons que le facteur qui dépend de la normalisation sera le même dans les deux canaux. 

Pour trouver la constante à trois points $C'_{123}$, nous introduisons des nouveaux rapports $\rho'(V_1|V_2, V_3)$ et $\tilde{\rho}'(V_1|V_2, V_3)$, 
\begin{align}
\rho'(V_1|V_2,V_3)= \frac{C'(V_1^{+},V_2,V_3)}{C'(V_1^{-},V_2,V_3)}\ ,\quad  \tilde{\rho}'(V_1|V_2,V_3)= \frac{C'(V_1^{\tilde{+}},V_2,V_3)}{C'(V_1^{\tilde{-}},V_2,V_3)}\ , . \label{res-eq:rp}
\end{align}
En combinant ces expressions avec \eqref{res-eq:rcst} nous voyons que la combinaison
\begin{align}
\frac{\rho(V_1|V_2,V_3)}{\rho'(V_1|V_2,V_3)} = \frac{C_{+}(V_1)Y(V_1^{+})}{C_{-}(V_1)Y(V_1^{-})} \, , 
\label{res-eq:rrp}
\end{align}
dépend seulement du champ $V_1$. Ceci veut dire que $\rho'(V_1|V_2, V_3)$ peut absorber toute la partie de $\rho(V_1|V_2, V_3)$ qui dépends des combinaisons des trois impulsions. Un analyse similaire peut être suivi avec $\tilde{\rho}'(V_1|V_2, V_3)$, et de cette manière nous trouvons que $C'_{123}$ obéisse
\begin{align}
\frac{C'(V_1^{+},V_2,V_3)}{C'(V_1^{-},V_2,V_3)} &= (-1)^{2s_2} \beta^{-4\beta (P_1 + \sigma_1 \bar{P_1})}  \frac{\prod_{\pm, \pm} \Gamma\left(\frac{1}{2}+\beta (P_1\pm  P_2 \pm P_3 ) \right)}{\prod_{\pm, \pm}\Gamma\left(\frac{1}{2}-\beta (\sigma_1 \bar{P}_1\pm  \bar P_2 \pm  \bar P_3)\right)}\, , \label{res-eq:rop}\\
\frac{C'(V_1^{\tilde{+}},V_2,V_3)}{C'(V_1^{\tilde{-}},V_2,V_3)} &= (-1)^{2r_2} \beta^{-\frac{4}{\beta}(P_1+ \tilde{\sigma}_1 \bar{P_1})}  \frac{\prod_{\pm, \pm}\Gamma\left(\frac{1}{2}-\beta^{-1} (P_1\pm P_2 \pm P_3 ) \right)}{\prod_{\pm, \pm}\Gamma\left(\frac{1}{2}+ \beta^{-1}(\tilde{\sigma}_1 \bar{P}_1\pm \bar{P}_2 \pm \bar{P}_3) \right)}\, , \label{res-eq:trop}
\end{align}
où les puissances de $\beta$ dans les préfacteurs ont été choisies de manière conventionnelle pour éviter l'apparition des facteurs similaires dans la constante de structure. 

Une fois que nous avons identifié la partie des équations de shift \eqref{res-eq:rcst} qui corresponds  $C'_{123}$, nous pouvons déduire les équations pour l'autre facteur invariant par renormalisation $Y^2B^{-1}(V)$. Ces équations sont
\begin{multline}
\frac{(Y^2B^{-1})\left(V^{+}_{1} \right) }{(Y^2B^{-1})\left(V^{-}_{1} \right)}= -\beta^{8\beta(P_1+\sigma_1\bar{P}_1)} \frac{\Gamma(-2\beta P_1)\Gamma(-2\beta\sigma_1\bar{P}_1)}{\Gamma(2\beta P_1)\Gamma(2\beta\sigma_1\bar{P}_1)}
\\ 
\times
\frac{\Gamma(\beta^2-2\beta\sigma_1\bar{P}_1)\Gamma(1-\beta^2-2\beta\sigma_1\bar{P}_1)}{\Gamma(\beta^2+2\beta P_1)\Gamma(1-\beta^2+2\beta P_1)}\, ,\label{res-eq:sY2B21}
\end{multline}
\begin{multline}
\frac{(Y^2B^{-1})\left(V^{\tilde{+}}_{1} \right) }{(Y^2B^{-1})\left(V^{\tilde{-}}_{1} \right)}
= 
-\beta^{8\beta^{-1}(P_1+\tilde{\sigma}_1\bar{P}_1)} 
\frac{\Gamma(2\beta^{-1}P_1)\Gamma(2\beta^{-1}\tilde{\sigma}_1\bar{P}_1)}{\Gamma(-2\beta^{-1}P_1)\Gamma(-2\beta^{-1}\tilde{\sigma}_1\bar{P}_1)} 
\\ 
\times
\frac{\Gamma(\beta^{-2}+2\beta^{-1}\tilde{\sigma}_1\bar{P}_1)\Gamma(1-\beta^{-2}+2\beta^{-1}\tilde{\sigma}_1\bar{P}_1)}{\Gamma(\beta^{-2}-2\beta^{-1}P_1)\Gamma(1-\beta^{-2}-2\beta^{-1} P_1)} \, . \label{res-eq:sY2B12}
\end{multline}

Les équations de shift \eqref{res-eq:rop}, \eqref{res-eq:trop} pour $C'(V_1, V_2, V_3)$, et \eqref{res-eq:sY2B21} et \eqref{res-eq:sY2B12} pour $Y^2B^{-1}(V)$, déterminent comment les constantes de structure se comportent face aux changements des impulsions par $\pm \beta$ et $\pm \frac{1}{\beta}$. Si l'on considère que les constantes de structure sont des fonctions \texttt{SMOOTH} des impulsions, les équations de shift sont suffisantes pour déterminer les constantes de structure à un facteur prés si $\beta^2$ est un nombre réel irrationnel. 
Nous écrivons cette condition comme $\beta^2 \in \mathbb{R}$, et cherchons de résoudre les équations de shift dans ce cas-ci. Il y a deux régimes possibles: soit $\beta \in \mathbb{R}$, et donc $c<1$, ou $\beta \in i \mathbb{R}$, $c \geq 25$ et nous prenons la paramétrisation $\beta = ib$ avec $b \in \mathbb{R}$. Les cas hors de ce régime sont discutés dans le chapitre \ref{ch:cboots}.

Une dernière remarque: pour le cas $\beta = i b$ les équations de shift peuvent être réécrites directement à exception des prefacteurs en puissances de $\beta$, qui doivent être modifiés selon
\begin{align}
\beta^{\beta(P_1 + \sigma \bar{P}_1)} &\to b^{ib(P_1+\sigma_1 \bar{P}_1)} \ , \\
\beta^{\frac{1}{\beta}(P_1 + \tilde{\sigma} \bar{P}_1)} &\to b^{-\frac{i}{b}(P_1+\tilde{\sigma}_1 \bar{P}_1)}\, .
\end{align}

Nous pouvons maintenant écrire des expressions explicites pour les constantes de structure. Ces solutions seront exprimées en termes de la fonction double gamma de Barnes, $\Gamma_{\omega}(x)$,  qui est invariant par $\omega \to \frac{1}{\omega}$ et obéisse
\begin{align}
\frac{ \Gamma_\omega(x+\omega)}{\Gamma_\omega(x)} = \sqrt{2\pi}\frac{\omega^{\omega x-\frac12}}{\Gamma(\omega x)} \ . \label{res-eq:sgamma}
\end{align}
La fonction $\Gamma_{\omega}(x)$ est bien défini pour $\Re{\omega} > 0$ et elle est une fonction méromorphe de $x$ avec des pôles pour
\begin{align}
x = -m \omega - n \omega^{-1}\ , \quad m,n \in \mathbb{N}\, .
\end{align}  
Pour écrire les solutions nous prenons $\omega = \beta$ ou $\omega  = b$, selon le cas. 

Nous commençons par le facteur $Y^2B^{-1}(V)$ dans le cas $\beta \in \mathbb{R}$. À partir des équations \eqref{res-eq:sgamma} nous pouvons vérifier que l'ansatz 
\begin{align}
\boxed{ (Y^{2}B^{-1})\left( V_{P,\bar P}^{\sigma,\tilde{\sigma}} \right) =(-1)^{P^2-\bar P^2} \prod_{\pm} \Gamma_\beta(\beta\pm 2P)\Gamma_\beta(\beta^{-1}\pm 2\bar{P})}\, ,  \label{res-Y2Bnd}
\end{align}
obéisse aux équations de shift. 

Quand le champ $V_{P,\bar P}^{\sigma,\tilde{\sigma}}$ est diagonal la solution \eqref{res-Y2Bnd} se simplifie et nous obtenons 
\begin{align}
(Y^2B^{-1})(V^D_P) = \frac{1}{\prod_\pm \Upsilon_\beta(\beta\pm 2P)} \, ,\label{res-eq:YBUps}
\end{align}
où nous avons introduit une autre fonction spéciale 
\begin{align}
\Upsilon_{\omega}(x) = \frac{1}{\Gamma_{\omega}(x) \Gamma_{\omega}(\omega + \omega^{-1} -x)}\, .\label{res-eq:Ups}
\end{align}

Dans le cas $\beta = ib$, $b \in \mathbb{R}$, la solution prends la forme 
\begin{align}
\boxed{ (Y^{2}B^{-1})\left( V_{P,\bar P}^{\sigma,\tilde{\sigma}} \right) = \frac{(-1)^{P^2-\bar P^2}}{ \prod_{\pm} \Gamma_b(\pm 2iP)\Gamma_b\left( (b+\frac{1}{b})\pm 2i\bar{P}\right) } }\, , \label{res-Y2Bndb}
\end{align}
et si le champ est diagonal nous avons
\begin{align}
(Y^2B^{-1})(V^D_P) = \prod_\pm \Upsilon_b(\pm2iP) \, .\label{res-eq:YBUpsb}
\end{align}

Nous nous concentrons maintenant sur les solutions pour la constante de structure à trois points $C'(V_1, V_2, V_3)$. Ces solutions sont plus simples quand le champ $V_1$ est diagonal, i.e. $V_1 =V^D_1$, et donc nous nous concentrons sur ce cas-ci. 
Pour $\beta \in \mathbb{R}$ nous avons, 
\begin{align}
\boxed{\begin{aligned} C'\left(V^{D}_{P_1}, V_2, V_3\right) =& \frac{ f_{2,3}(P_1) }{ \prod\limits_{\pm, \pm} 
\Gamma_\beta(\frac{\beta}{2}+\frac{1}{2\beta} +P_1\pm P_2 \pm P_3)
} \\ &\qquad \times \frac{1}{\prod\limits_{\pm, \pm} \Gamma_\beta(\frac{\beta}{2}+\frac{1}{2\beta}-P_1\pm \bar{P}_2 \pm \bar{P}_3)} 
\end{aligned} 
}\, . \label{res-cdnn}
\end{align}
Où le facteur $f_{2,3}(P_1)\in\{-1,+1\}$ doit être déterminé à partir des équations de shift suivantes
\begin{align}
\frac{ f_{2,3}(P_1 +\frac{\beta}{2})}{f_{2,3}(P_1 -\frac{\beta}{2})} = (-1)^{2s_2} \quad , \quad \frac{f_{2,3}(P_1 +\frac{1}{2\beta})}{ f_{2,3}(P_1 -\frac{1}{2\beta})} =(-1)^{2r_2}\, .\label{res-eq:f23}
\end{align}
La détermination de ce facteur est simple dans certains cas. Par exemple, quand il y a un autre champ diagonal, ou quand les indices des champs diagonaux sont des nombres entiers, nous avons $f_{2,3}(P_1) = 1$. 

Quand les trois champs sont diagonaux, la solution peut être exprimée en termes des fonctions $\Upsilon_{\beta}(x)$, 
\begin{align}
 C'(V^D_1,V^D_2,V^D_3) &= \prod_{\pm, \pm}\Upsilon_\beta\left(\tfrac{\beta}{2}+\tfrac{1}{2\beta}+P_1\pm P_2\pm P_3\right) \label{res-eq:cpl} \\
 &= C'^{D}(P_1, P_2, P_3) \ . 
\end{align}
Dans la deuxième ligne nous avons écrit la constante de structure diagonal $C'^D$ comme une fonction des impulsions seulement. Ignorant des facteurs dépendants de la normalisation, cette expression coïncide avec celle qui a été dérivée dans \cite{Zamolodchikov:2005fy}. 

Pour le cas $\beta = ib$ la solution pour la constante de structure prends la forme
\begin{align}
\boxed{
\begin{aligned}
C'\left(V^{D}_{P_1}, V_2, V_3\right) =  f_{2,3}(P_1) \prod\limits_{\pm, \pm} &
\Gamma_b\left(\tfrac{b}{2}+\tfrac{1}{2b} + i (P_1\pm P_2 \pm P_3) \right)
\\ \times &\prod\limits_{\pm, \pm} \Gamma_b\left(\tfrac{b}{2}+\tfrac{1}{2b} - i (P_1\pm \bar{P}_2 \pm \bar{P}_3) \right)
\end{aligned} }\, , \label{res-cdnnb}
\end{align}
où le facteur $f_{2,3}(P_1)$ obéisse les mêmes équations qu'avant.  
Dans ce cas, la solution pour trois champs diagonaux est
\begin{align}
 C'(V^D_1,V^D_2,V^D_3) &= \frac{1}{\prod_{\pm, \pm}\Upsilon_b\left(\tfrac{b}{2}+\tfrac{1}{2b}+ i (P_1\pm P_2\pm P_3) \right)} \label{res-eq:cplb} \\
 &= C'^{D}(P_1, P_2, P_3) \ , 
\end{align}
qui donne la partie invariante par renormalisation de la célèbre solution DOZZ pour les constantes de structure de la théorie de Liouville, nommé après Dorn, Otto \cite{Dorn:1994xn} et  Zamoldchikov \& Zamolodchikov \cite{Zamolodchikov:1995aa}. 

Il est intéressant de noter la relation qui existe entre les constantes de structure des théories diagonales et non-diagonales. Les équations \eqref{res-eq:cpl} et \eqref{res-eq:cplb} montrent que les solutions diagonales peuvent être obtenues comme un cas particulier des solutions non-diagonales. En plus, à partir des expressions explicites \eqref{res-cdnn} et \eqref{res-cdnnb} nous pouvons écrire 
\begin{align}
C'^2(V^D_1,V_2,V_3) = C'^{D}(P_1,P_2,P_3) C'^{D}(P_1,\bar{P}_2,\bar{P}_3)\, ,  \label{res-eq:cn2cd}
\end{align}
ce qui nous amène à exprimer les constantes de structure non-diagonales comme une moyenne géométrique des constantes de structure diagonales, i.e.
\begin{align}
C'(V_1, V_2, V_3) = \sqrt{C^D(P_1, P_2, P_3)C^D(\bar{P}_1, \bar{P}_2, \bar{P}_3)}\, .\label{res-eq:gmean}
\end{align}
Des relations de ce type apparaissent dans l'étude des théories non-diagonales, \cite{petkova1988two, petkova1989structure, Petkova:1994zs}, et dans \cite{ei15} elles ont été utilisées pour calculer des rapports entre certaines constantes de structure. Cependant, la présence de la racine carrée pourrait poser certains problèmes en termes d'analyticité des solutions, et de la détermination du signe des constantes. Les solutions explicités présentées ici n'ont pas ces problèmes, et montrent qu'il est possible de prendre la racine carrée d'une manière précise pour construire des solutions non-diagonales. 

Les relations du type \eqref{res-eq:gmean} apparaissent au niveau des équations de shift, et dans l'appendice de \cite{Migliaccio:2017dch} il est discuté qu'elles sont une propriété d'un type particulier des systèmes d'équations.
\section{Fonctions à quatre points}
Dans cette section nous discutons des exemples des théories où nos solutions sont valables.   Nous présentons aussi une proposition pour une théorie non-diagonal et non-rational dont certaines les fonctions à quatre points obéissent la symétrie de croisement, et qui peut être obtenu comme un limite des modèles minimaux. 

\subsection{Exemples des théories connus}
Nous avons montré que les constantes de structure des théories diagonales peuvent être obtenues comme des cas particuliers des constantes de structure des théories non-diagonales. Ceci veut dire que nos solutions son directement applicables aux théories diagonales telles que la théorie de Liouville, les modèles minimaux de la série $A$ ou les modèles minimaux généralisés. Une discussion plus détaillé de ces modèles se trouve dans le chapitre \ref{ch:crsymfun}.

Nous allons discuter ici les cas des modèles minimaux de la série $D$, car ils nous permettent de tester les constantes de structure non-diagonales. Ces modèles minimaux existent pour des valeurs rationnelles de la charge centrale, donnés en termes de deux entiers positifs copremiers $p$ et $q$, dont un est pair. Ces valeurs de $c$ correspondent à prendre $\beta^2 = \frac{p}{q}$, et sont
\begin{align}
c_{pq} = 1- 6 \frac{(p-q)^2}{pq}\, . \label{res-cpq}
\end{align}
Les valeurs $c_{pq}$ sont denses dans la demi-droite $c < 1$. 

Le spectre des modèles minimaux de la série $D$ est composé d'une quantité finie des champs dégénérés, et peut être divisé en un secteur diagonal et un secteur non-diagonal. Prenant $q$ pair afin de fixer les conventions, nous pouvons exprimer le spectre en termes des représentations associés à chaque champ. Si $\mathcal{R}_{\langle r,s \rangle}$ est la représentation associé au champ $V_{\langle r,s \rangle}$ nous avons
\begin{align}
 \mathcal{S}_{p,q}^{\text{D-series}} \ =\  \frac12 \underbrace{\bigoplus_{r\overset{2}{=}1}^{q-1} \bigoplus_{s=1}^{p-1} \left|\mathcal{R}_{\langle r,s\rangle}\right|^2}_{\text{Diagonal}} \oplus \frac12 \ \underbrace{\bigoplus_{\substack{1\leq r\leq q-1 \\ r\equiv \frac{q}{2}\bmod 2}} \  \bigoplus_{s=1}^{p-1} \mathcal{R}_{\langle r,s\rangle} \otimes \bar{\mathcal{R}}_{\langle q-r,s\rangle} }_{\text{Non-diagonal}}\ . \label{res-eq:DMMspec}
\end{align}
Ces spectres ont des propriétés différentes selon $q = 0 \bmod 4$ ou $q = 2 \bmod 4$. Si $q=2 \bmod 4$, les représentations  $\mathcal{R}_{\langle \frac{q}{2},s \rangle} \otimes \mathcal{R}_{\langle \frac{q}{2},s\rangle} $ apparaît deux fois dans le spectre: une fois dans le secteur diagonal et une fois dans le secteur non-diagonal. Dans ce cas, nous avons des champs diagonaux et non-diagonaux qui ont les mêmes dimensions, comme nous avons discuté dans la section \ref{res-sec:dndfields}. Si $q$ est multiple de $4$ ce phénomène ne se produit pas. 

Les règles de fusion de ces modèles sont connues \cite{petkova1988two,petkova1989structure, Petkova:1994zs,furlan1990fusion,run99}, et dans une certaine base elles obéissent à une règle que nous appelons la conservation de la diagonalité. Ceci veut dire que l'OPE entre deux champs diagonaux ou deux champs non-diagonaux produit des champs diagonaux, tandis qu'un OPE entre en champ diagonal et un champ non-diagonal produit des champs non-diagonaux. Nous pouvons exprimer cette propriété comme
\begin{align}
D \times D = N \times N = D, \quad D \times N = N \, .  \label{res-eq:dcons}
\end{align} 

Pour vérifier que les constantes de structure de la section précédente produisent des fonctions à quatre points qui respectent la symétrie de croisement dans ces modèles minimaux, nous calculons numériquement des fonctions de corrélation et comparons les deux canaux. Nous nous concentrons sur des fonctions du type $Z(x)=\left\langle V^D_{\langle r_1, s_1 \rangle}(x) V^N_{\langle r_2, s_2 \rangle}(0) V^D_{\langle r_3, s_3 \rangle}(\infty) V^N_{\langle r_4, s_4\rangle}(1) \right\rangle$ où, d'après les règles \eqref{res-eq:dcons}, les expansions du canal $s$ et $t$ sont des sommes sur le spectre non-diagonal. Dans ce cas-ci les constantes de structure à quatre points peuvent être calculées directement par itération des équations de shift, et nous suivons cette méthode. 

Les détails de calculs numériques peuvent être consultés dans l'appendice \ref{sec:numap}. Les calculs présentés ici sont basés sur le package de Python de bootstrap conforme à deux dimensions disponible sur \href{https://github.com/ribault/bootstrap-2d-Python}{GitHub}\cite{b2P}.

\begin{itemize}
\newpage
\item Modèle minimal de la série D avec $(p,q)= (7, 8)$ .
\begin{align}
Z(x) = \left\langle V^{D}_{\left(5,4\right)}V^{N}_{\left(6,4\right)}V^{D}_{\left(3,2\right)}V^{N}_{\left(6,2\right)}\right\rangle  . \label{res-eq:exMM5}
\end{align}
Champs primaires: 
\begin{flalign}
&\left[ \begin{array}{ll} 
s\text{-channel :} &V^{N}_{\left(4, 6\right)} +V^{N}_{\left(4, 3 \right)} \\
t\text{-channel:} & V^{N}_{ \left(4, 3 \right)} +V^{N}_{ \left(4, 5\right)} 
\end{array} \right] & \label{res-eq:exMM5S}
\end{flalign}
\vspace*{-15pt}
\begin{figure}[h]
\centering 
\includegraphics[scale=.7]{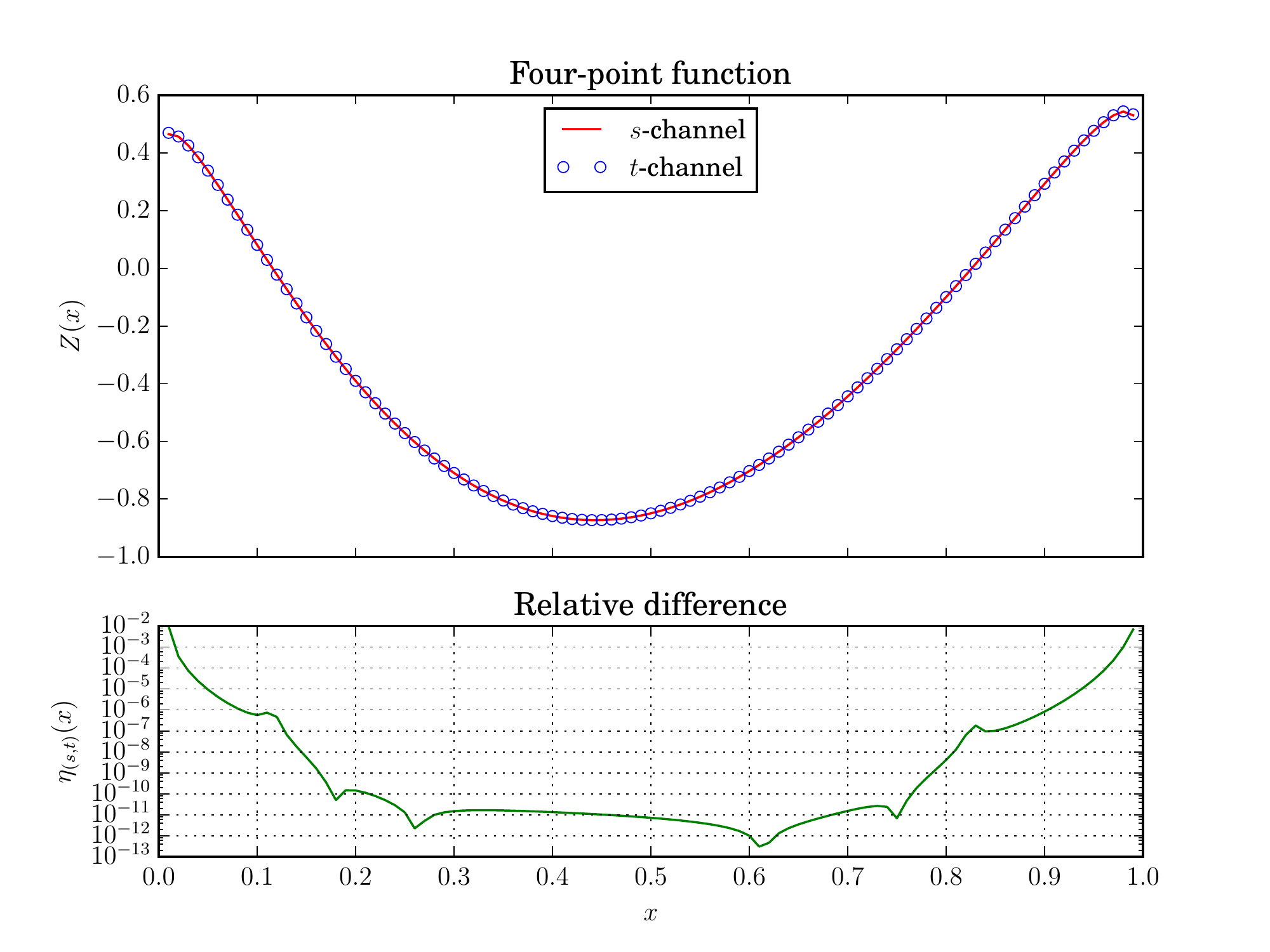}
\vspace*{-20pt}
\caption{Fonction à quatre points dans un modèle minimal de la série $D$ avec $q = 0 \bmod 4$.} \label{res-fig:exMM5}
\end{figure}

\newpage
\item Modèle minimal de la série D avec $(p,q)= (7, 10)$ .
\begin{align}
Z(x) = \left\langle V^{D}_{\left(5,4\right)}V^{N}_{\left(5,4\right)}V^{D}_{\left(7,5\right)}V^{N}_{\left(7,2\right)}\right\rangle   . \label{res-eq:exMM6}
\end{align}
Champs primaires: 
\begin{flalign}
&\left[ \begin{array}{ll} 
s\text{-channel :} &V^{N}_{\left(5, 6\right)} +V^{N}_{\left(5, 3\right)} +V^{N}_{\left(7, 6\right)} +V^{N}_{\left(7, 3\right)} +V^{N}_{\left(9, 6\right)} +\\ & V^{N}_{\left(9, 3\right)} \\
t\text{-channel:} & V^{N}_{ \left(3, 3\right)} +V^{N}_{ \left(3, 5\right)} +V^{N}_{ \left(5, 3\right)} +V^{N}_{ \left(5, 5\right)} +V^{N}_{ \left(7, 6\right)} +\\ & V^{N}_{ \left(7, 5\right)} 
\end{array} \right] & \label{res -eq:exMM6S}
\end{flalign}
\vspace*{-15pt}
\begin{figure}[h]
\centering 
\includegraphics[scale=.7]{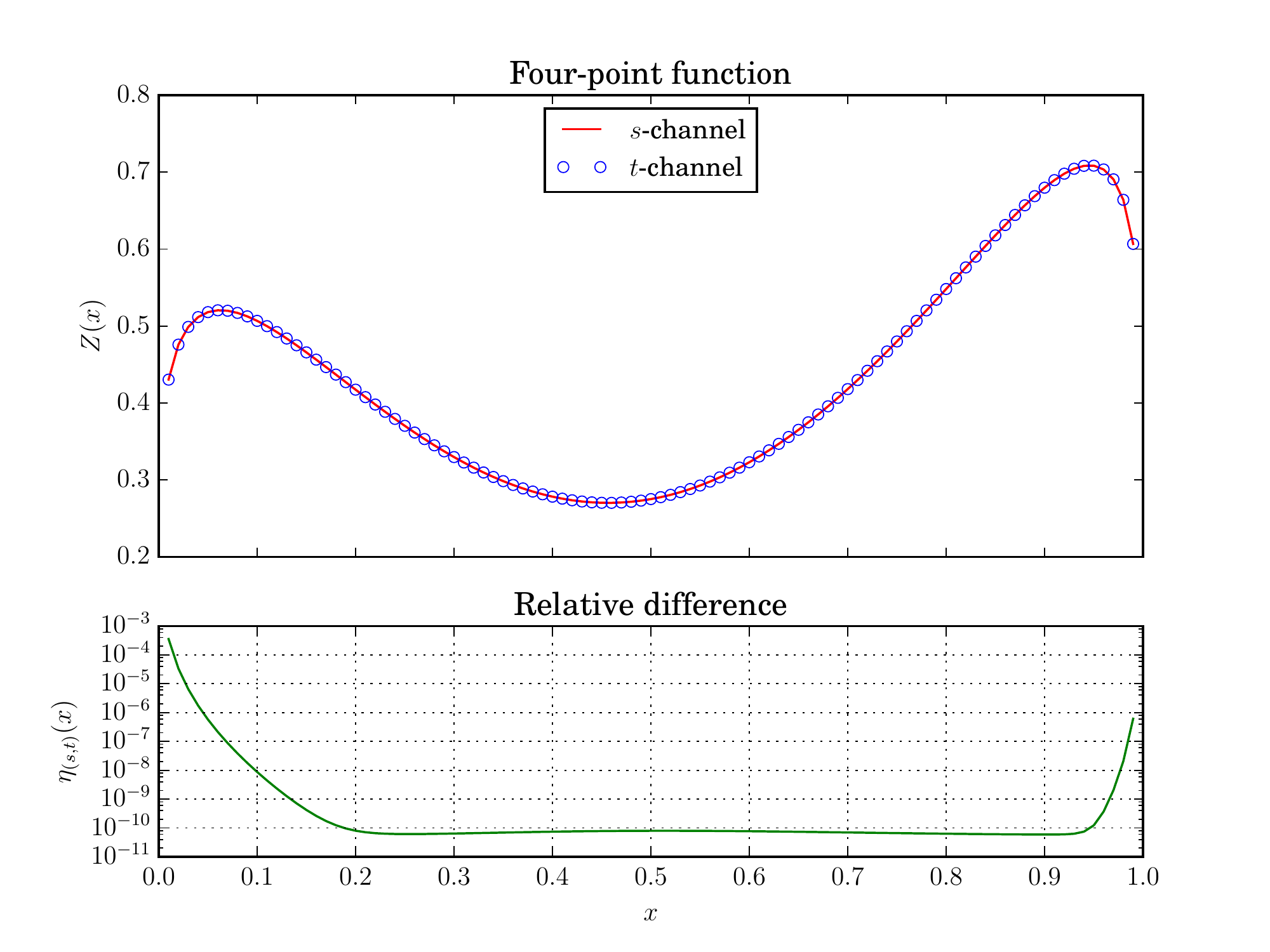}
\vspace*{-20pt}
\caption{Fonction à quatre points dans un modèle minimal de la série $D$ avec $q = 2 \bmod 4$.} \label{res-fig:exMM6}
\end{figure}

\end{itemize}

Ces exemples montrent que les canaux $s$ et $t$ des fonctions à quatre points calculés à partir des équations de shift de la coïncident avec une précision supérieure à $10^{-8}$ pendant la majorité de intervalle. Les différences plus importantes se trouvent prés des extrêmes $x = 0$ et $x=1$, ce qui est attendu d'après les propriétés des blocs conformes de chaque canal: Les blocs du canal $s$ divergent près de $x = 1$, tandis que les blocs du canal $t$ divergent près de $x= 0$. Nous prenons ces exemples et ceux que se trouvent dans le chapitre \ref{ch:crsymfun} comme évidence que la symétrie de croisement est respectée par les fonctions à quatre points de ces modèles calculés avec nos constantes de structure. 

\subsection{Limite des modèles minimaux}
Si nos constantes de structure non-diagonales donnent lieu à des fonctions à quatre points consistantes dans les modèles minimaux de la série $D$, ces théories sont assez particulières: elles existent pour des valeurs rationnelles de $c$, et leurs spectres sont finis. Nous allons construire dans cette section une proposition pour une famille des théories non-diagonales et non-rationnelles qui existent pour des valeurs plus génériques de $c$. 

Pour construire notre proposition, nous prenons une limite non-rationnelle des modèles minimaux: Comme les modèles minimaux de la série $D$ existent pour des valeurs $c = c_{p,q}$ qui sont denses dans la demi-droite $c<1$, nous pouvons rapprocher n'importe quelle valeur irrationnelle $c = c_0 \in \mathbb{R}_{<1}$ en choisissant $p,q$ suffisamment grandes. Notre proposition est qu'une théorie non-diagonale existe à ces valeurs $c_0$, et son spectre $\mathcal{S}_{c_0}$ peut être obtenu comme
\begin{align}
\mathcal{S}_{c_0} = \lim_{\substack{p,q \to \infty \\ c_{p,q} \to c_0 } } \mathcal{S}^{\text{MM}}_{p,q}\ ,
\end{align}
où $\mathcal{S}^{\text{MM}}_{p,q}$ est le spectre d'un modèle minimal. 

Nous prenons maintenant la limite du spectre des modèles minimaux de la série $D$, \eqref{res-eq:DMMspec}. Pour prendre cette limite nous devons choisir le comportement des indices $r,s$ de chaque représentation quand $p,q \to \infty$. Les champs du secteur non-diagonal ont un spin
\begin{align}
S_{(r,s)} = - (r - \frac{q}{2}) (s-\frac{p}{2}) \in \mathbb{Z}\, .
\end{align}
Comme les spins sont discrets, nous voulons envoyer $p,q \to \infty$ en les laissant constants. 
Nous pouvons faire une réparamétrisation du spectre afin que ce choix soit plus évident. Quand $c = c_{p,q}$ nous avons
\begin{align}
\Delta_{\langle r, s \rangle} = \Delta_{\langle r-\lambda q, s-\lambda p \rangle}\, , 
\end{align}  
pour toute $\lambda \in \mathbb{R}$. En choisissant $\lambda = \frac{1}{2}$ nous pouvons renommer les indices comme
\begin{align}
r \to r-\frac{q}{2}\, , \\
s \to s-\frac{p}{2}\, , 
\end{align}
et le spectre \eqref{res-eq:DMMspec} est réécrit comme 
\begin{align}
 \mathcal{S}_{p,q}^{\text{D-series}} \ &=\  \frac12 \bigoplus_{r\overset{2}{=}1-\frac{q}{2}}^{\frac{q}{2}-1} \bigoplus_{s=1-\frac{p}{2}}^{\frac{p}{2}-1} \left|\mathcal{R}_{\langle r,s\rangle}\right|^2 \oplus \frac12 \ \bigoplus_{\substack{ |r| \leq \frac{q}{2}-1 \\ r \in 2\mathbb{Z} }} \  \bigoplus_{s=1-\frac{p}{2}}^{\frac{p}{2}-1} \mathcal{R}_{\langle r,s\rangle} \otimes \bar{\mathcal{R}}_{\langle r,-s\rangle}\ ,  \label{res-eq:DMMs2z}
\end{align}
où maintenant nous avons valeurs négatives des indices, les indices $ \in \mathbb{Z}+\frac{1}{2}$, et les spins des champs sont maintenant
\begin{align}
S=-rs\, , 
\end{align}
comme pour les champs diagonales de la section \ref{res-sec:dndfields}
Dans cette notation, prendre la limite $p,q \to \infty$ en laissant les spins constants veut dire que les indices $r,s$ restent inchangés. Dans le secteur diagonal $\mathcal{S}^{\text{D,D-series}}_{p,q}$, quand nous envoyons $p,q \to \infty$ les bornes supérieure et inférieure pour $r,s$ vont aussi vers $\infty$, de sorte que les indices prennent tous les valeurs possibles. Dans la limite, $\beta^2 = \frac{p}{q}\to \beta_0^2 \in \mathbb{R}$, et les impulsions des champs diagonaux avec indices $r,s \in \mathbb Z$ couvrent toute la droite réel. Alors, en prenant la limite nous trouvons un spectre continu qui coïncide avec celui de la théorie de Liouville (voir chapitre \ref{ch:crsymfun}),
\begin{align}
\lim_{\substack{p,q \to \infty \\ \frac{p}{q}\to \beta^2_0\in \mathbb{R}}} \mathcal{S}^{\text{D,D-series}}_{p,q} = \frac12 \int_{\mathbb{R}} dP\, \left|\mathcal{V}_P\right|^2 \oplus \mathcal{S}_{2\mathbb{Z},\mathbb{Z}+\frac12}\ ,
\label{res-eq:dlim}
\end{align}

Pour le secteur non-diagonal $\mathcal{S}^{\text{N,D-series}}_{p,q}$ nous suivons la même procédure: en envoyant $p,q \to \infty$ les bornes pour les indices $r,s$ vont vers $\infty$, et les indices prennent des valeurs $(r,s) \in 2\mathbb{Z}, \mathbb{Z}+\frac{1}{2} $. Dans ce cas ci le spectre reste discret, car les spins prennent des valeurs discrètes. Nous avons
\begin{align}
\lim_{\substack{p,q \to \infty \\ \frac{p}{q}\to \beta^2_0\in \mathbb{R}}} \mathcal{S}^{\text{N,D-series}}_{p,q} = \frac12 \mathcal{S}_{2\mathbb{Z},\mathbb{Z}+\frac12}\  , \label{res-eq:ndlim}
\end{align}
où $\mathcal{S}_{X,Y} = \otimes_{r \in X} \otimes_{s \in Y} \mathcal{V}_{P_{(r,s)}} \otimes \bar{\mathcal{V}}_{P_{(r,-s)}}$. Des spectres de ce type ont été proposés dans  \cite{DiFrancesco:1987gwq} comme des spectres des théories conformes non-minimales avec $c<1$.

En prenant la limite nous arrivons à un spectre dont le secteur diagonal est donné par \eqref{res-eq:dlim}, et le secteur non-diagonal par \eqref{res-eq:ndlim}. Pour calculer des fonctions à quatre points nous avons besoin des règles de fusion pour des champs appartenant à ce spectre. Nous proposons des règles de fusion qui obéissent à la conservation de la diagonalité \eqref{res-eq:dcons}, pour deux raisons: D'un coté, ces règles seraient une extension de celles du modèle minimal, qui obéissent  cette propriété. De l'autre, 
dans une théorie avec spectre non-diagonal $\mathcal{S}_{2\mathbb{Z}, \mathbb{Z}+\frac{1}{2}}$ la conservation de la diagonalité est garanti par les conditions de non-trivialité \eqref{res-eq:nontriC}. Ainsi, la règle qui donne une expansion sur le secteur non-diagonal serait
\begin{align}
V^D_{P_1} \times V^N_{(r_2, s_2)} = \sum_{r \in 2\mathbb{Z}} \sum_{s \in \mathbb{Z}+\frac{1}{2}} V^{N}_{(r,s)}\, . \label{res-eq:dnnrule}
\end{align}
Dans le chapitre \ref{ch:crsymfun} nous discutons les autres règles de fusions proposées.

A fin d'utiliser cette règle pour calculer des fonctions de corrélation, nous devons déterminer quelle est la condition pour que une expansion sur le spectre $\mathcal{S}_{2\mathbb{Z}, \mathbb{Z}+\frac{1}{2}}$  converge. Chaque terme de l'expansion se comporte, à premier ordre, comme une exponentiel décroissant dans la dimension conforme $\Delta + \bar{\Delta}$, et donc la condition de convergence est
\begin{align}
\Delta + \bar{\Delta} \underset{r,s \to \infty}{\to} 0\, .
\end{align}
Pour les champs non-diagonaux du spectre \eqref{res-eq:ndlim} la dimension conforme est
\begin{align}
\Delta_{(r,s)} + \bar{\Delta}_{(r,-s)} = \frac{c-1}{12} + \frac12\left(r^2\beta^2 + \frac{s^2}{\beta^2}\right)\ , \label{res-eq:nddims}
\end{align}
qui tends vers $\infty$ quand $r,s\to \infty$ si $\Re \beta^2>0$ i.e. $\Re c<13$. 

À partir de la règle \eqref{res-eq:dnnrule} nous pouvons calculer des fonctions à quatre points pour vérifier la symétrie de croisement. Nous nous concentrons sur des fonctions du type $Z(x) = \left\langle  V^D_{P_1}(x) V^N_{(r_2, s_2)}(0) V^D_{P_3}(\infty) V^N_{(r_4, s_4)}(1) \right\rangle$ où $P_1, P_3$ peuvent prendre des valeurs arbitraires, $(r_i, s_i) \in 2\mathbb{Z} \times \mathbb{Z}+\frac{1}{2}$, $i=2,4$, et avec une charge centrale tel que $\Re c < 13$. Nous montrons à continuation quelques exemples, où les constantes de structure ont été calculées par application récurrente des équations de shift. Pour calculer ces fonctions de corrélation il est nécessaire de faire une troncation du spectre non-diagonal, et nous spécifions dans chaque cas la quantité des champs primaires inclus. 

Ces calculs numériques sont basés sur le package \cite{b2P} disponible sur \href{https://github.com/ribault/bootstrap-2d-Python}{GitHub}.

\begin{itemize}
\newpage
\item $c = -0.54327$
\begin{align}
Z(x) = \left\langle V^D_{  P_1 } V^N_{ \left ( 0,  - \tfrac{3}{2}\right ) } V^D_{  P_3 } V^N_{ \left ( 2, \tfrac{1}{2}\right ) } \right\rangle \label{res-eq:ex2Z1}
\end{align}
\begin{align}
 \Delta_1 = 2.341\ , \quad\Delta_3 = 1.546 
\end{align}
Quantité des champs primaires dans le spectre: $24$.
\vspace*{-15pt}
\begin{figure}[h]
\centering 
\includegraphics[scale=.7]{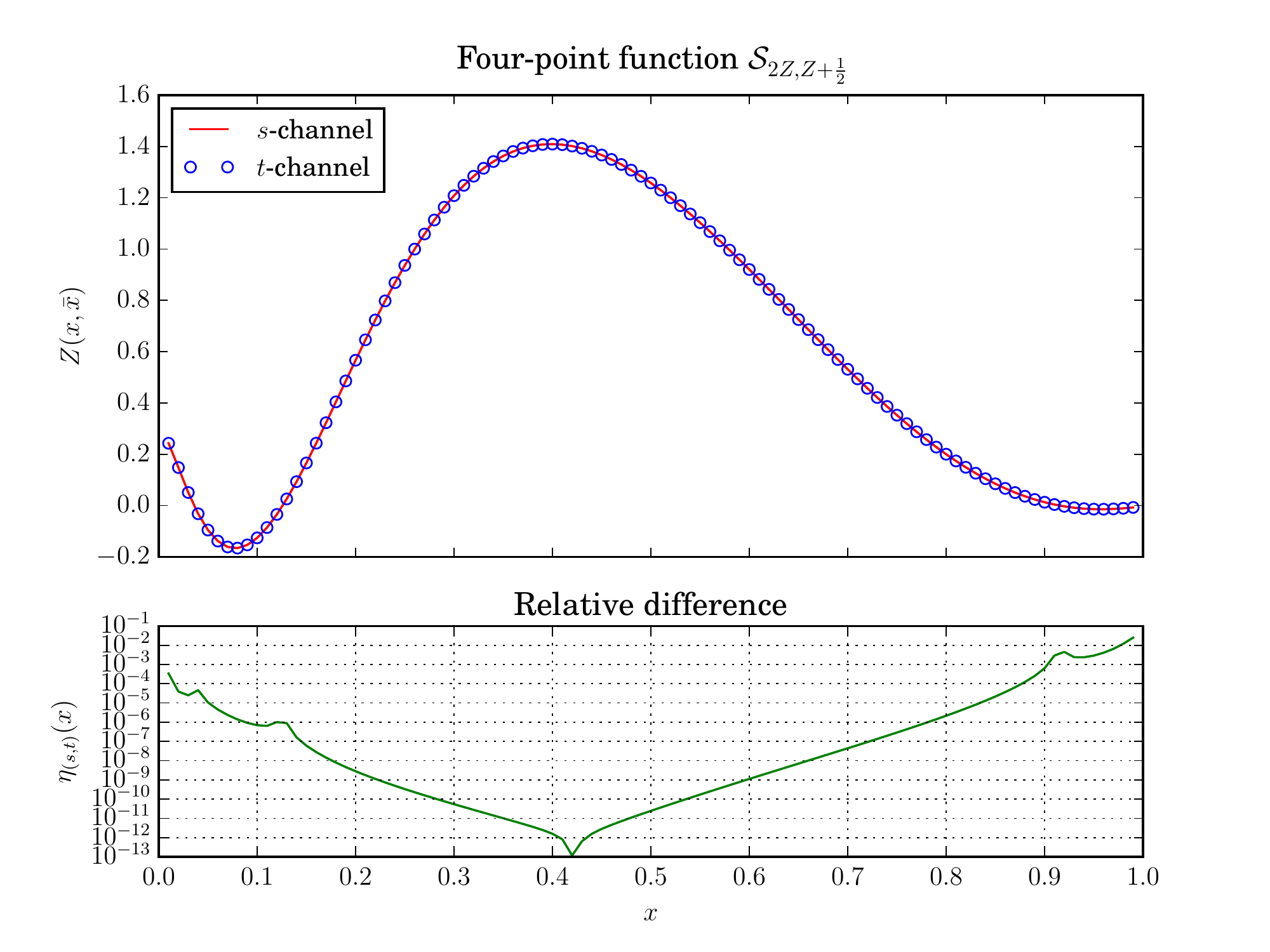}
\caption{Fonction à quatre points pour des valeurs réels arbitraires des poids conformes des champs diagonaux, avec charge centrale négative. } \label{res-fig:ex2Z1}
\end{figure}

\newpage
\item $c = (4.72+1.2i)$
\begin{align}
Z(x) = \left\langle V^D_{  P_1 } V^N_{ \left ( 2,  \tfrac{3}{2}\right ) } V^D_{  P_3 } V^N_{ \left ( 0, \tfrac{1}{2}\right ) } \right\rangle \label{res-eq:ex2Z2}
\end{align}
\begin{align}
 \Delta_1 = (0.23+0.143i)\ , \quad\Delta_3 = (-0.546+2i)
\end{align}
Quantité des champs primaires dans le spectre:  $34$.
\vspace*{-15pt}
\begin{figure}[h]
\centering 
\includegraphics[scale=.7]{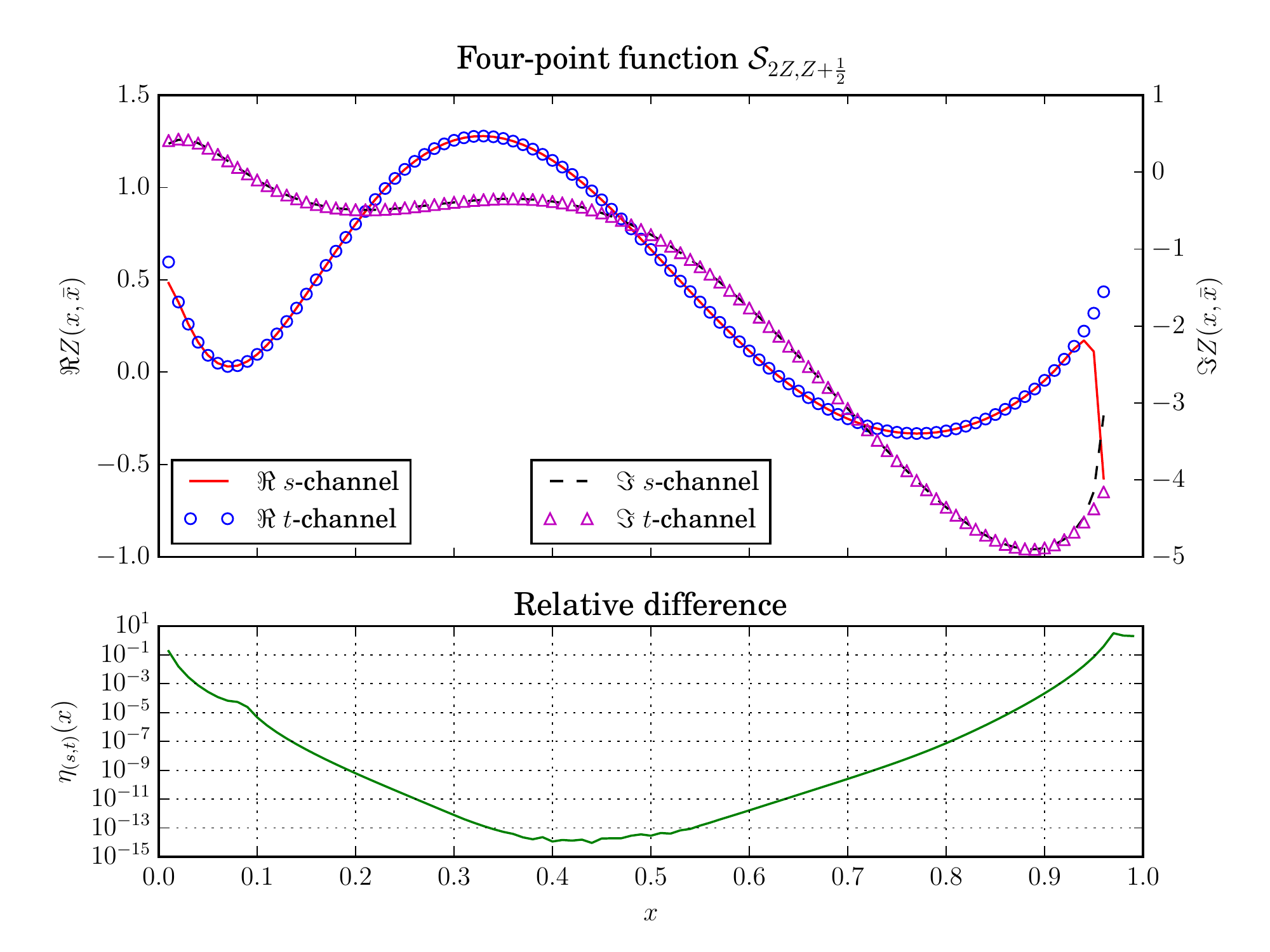}
\vspace*{-20pt}
\caption{Fonction à quatre points pour des valeurs complexes arbitraires des poids conformes des champs diagonaux et de la charge centrale.} \label{res-fig:ex2Z2}
\end{figure}

Ici les valeurs complexes de $c$, $\Delta_1$ et $\Delta_3$ font que les expansions des canaux $s$ et $t$ prennent des valeurs complexes. Nous montrons séparément les parties réel et imaginaire de chaque canal, et nous voyons une bonne coïncidence entre elles. 
\newpage
\item c = -3.6721
\begin{align}
Z(x) = \left\langle V^D_{  P_1 } V^N_{ \left ( 0, - \frac{5}{2}\right ) } V^D_{  P_3 } V^N_{ \left ( 4,  \frac{1}{2}\right ) } \right\rangle \label{res-eq:ex2Z4}
\end{align}
\begin{align}
 \Delta_1 = 0.567\ , \quad\Delta_3 = 1.982
\end{align}
Quantité des champs primaires dans le spectre:  $51$.
\vspace*{-15pt}
\begin{figure}[h]
\centering 
\includegraphics[scale=.7]{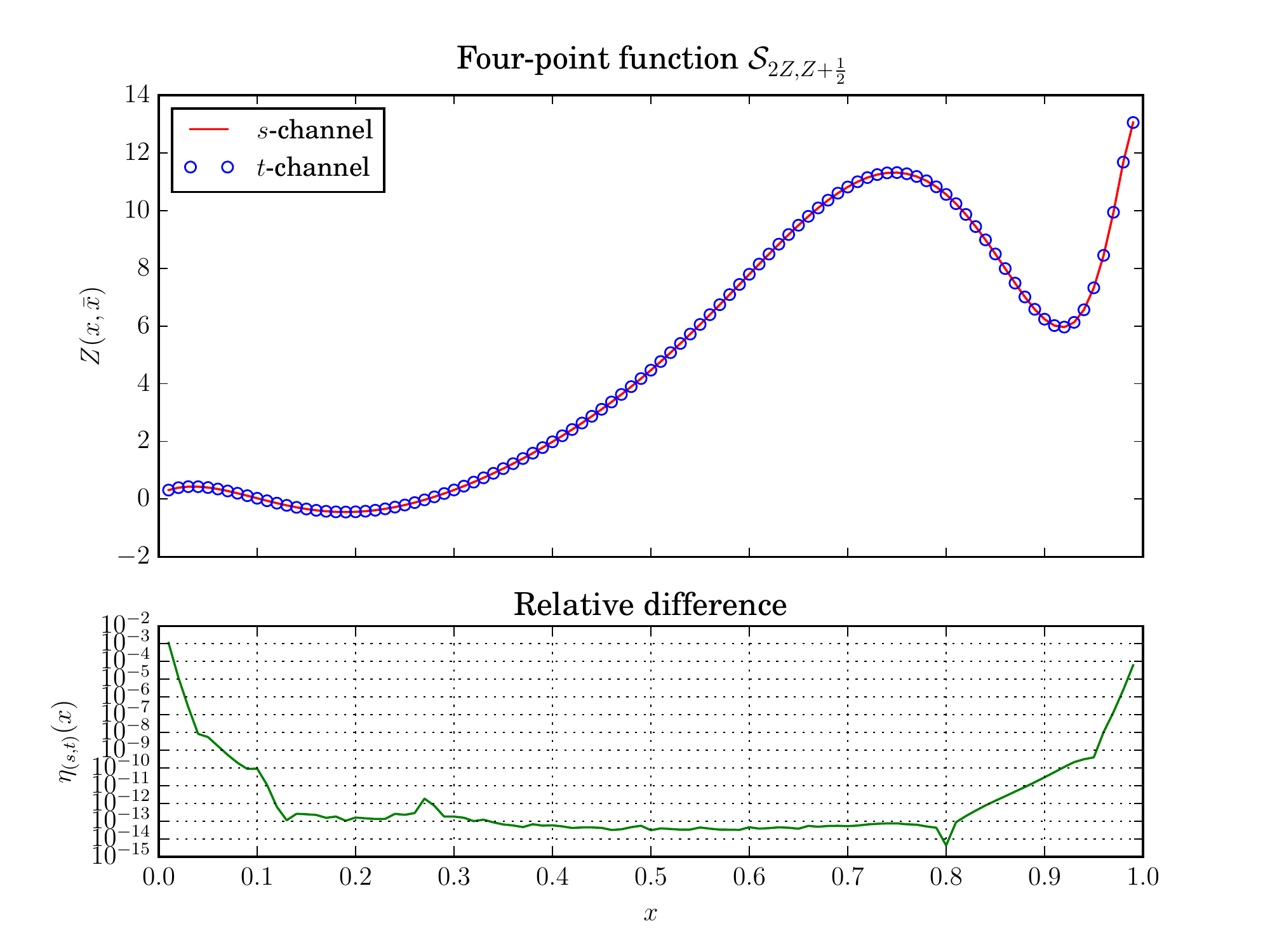}
\vspace*{-20pt}
\caption{Fonction à quatre points pour des valeurs réels arbitraires des poids conformes des champs diagonaux et charge centrale $c < -2$. } \label{res-fig:ex2Z4}
\end{figure}
\end{itemize}

Ces exemples montrent que les fonctions à quatre points entre des champs de notre spectre proposé, calculés avec les constates de structure issus du bootstrap conforme non-diagonal, obéissent avec une très bonne précision à la symétrie de croisement. Nous interprétons ceci comme indication de l'existence d'une théorie non-diagonal et non-rational pour des valeurs de $c$ tels que $\Re{c}<13$, qui peut être interprété comme une limite des modèles minimaux. 

\section{Conclusions}

Dans cette thèse nous avons présenté une extension de la méthode de bootstrap conforme analytique applicable aux théories non-diagonales, et nous avons introduit une proposition pour une théorie conforme non-diagonal et non-rationnelle pour des valeurs de $c$ telles que $\Re{c}<13$. 

En ce qui concerne le bootstrap conforme analytique, nous avons présenté trois hypothèses principales et nous avons vu quelles contraintes imposent-elles sur le spectre des théories non-diagonales. Ainsi, nous avons trouvé qu'il y a deux types des champs différents, diagonaux et non-diagonaux, qui se distinguent par ses règles de fusion avec les champs dégénérés. En plus, nous avons vu que les impulsions des champs non-diagonaux sont contrôles par deux indices discrètes $r$ et $s$. Ensuite, en étudiant des fonctions à quatre points avec des champs dégénérés nous avons trouvé des équations de shift qui contrôlent les constantes de structure, et nous avons trouvé des solutions pour les constantes à deux, trois et quatre points (voir chapitre \ref{ch:cboots} pour plus de détails), et nous avons discuté la relation entre les constantes de structure des théories diagonales et non-diagonales. Nos résultats ont été validés grâce aux calculs numériques des fonctions à quatre points dans des théories connus, et notamment dans les modèles minimaux de la série $D$, qui ont montre que nos constants donnent lieu aux fonctions qui respectent la symétrie de croisement avec une très haute précision.  

D'autre part, nous avons présenté une proposition pour une famille des théories conformes non-diagonales et non-rationnelles dont le spectre a été trouvé en prenant une limite des spectres des modèles minimaux. Des calculs numériques ont montré que certaines fonctions à quatre points de ces théories obéissent à la symétrie de croisement, ce qui constitue un argument non-trivial à faveur de leur existence. Nous pouvons ajouter ces théories dans la carte des théories conformes que nous avons montré au début:
\begin{align}
\begin{tikzpicture}[scale = 4.5, baseline=(current  bounding  box.center)]
 \filldraw [red!50, opacity = .15] (2.15, -.3) -- (2.15, .75) -- (-.6, .75) -- (-.6, -.3) -- cycle;
 \filldraw [blue!50, opacity = .15] (2.15, -.3) -- (2.15, .75) -- (-.6, .75) -- (-.6, -.3) -- cycle;
\fill[pattern= north east lines,pattern color=gray] (1.6, -.3) -- (1.6, .75) -- (-.6, .75) -- (-.6, -.3);
\draw[gray, opacity =.6] (1.6, -.3) -- (1.6, .75);
\node [draw = red, fill = red!70, right] at (-.5, .65) {Liouville theory};
\node [draw = blue, fill = blue!15, right] at (-.5, .49) {Generalized minimal models};
\node [draw = gray, fill= gray!50, right] at (-.5, .30) {$\mathcal{S}_{2\mathbb{Z}, \mathbb{Z} + \frac{1}{2} }$};
\node [draw = gray, pattern= north east lines,pattern color=gray , right] at (-.5, .30) {$\mathcal{S}_{2\mathbb{Z}, \mathbb{Z} + \frac{1}{2} }$};
 \node[draw = green!70!black, fill = green!16!black!7, right] at (-.5, .12) {Minimal models};
  \filldraw [red, opacity = .5] (.98, -.02) -- (.98, .02) -- (2.15, .02) -- (2.15, -.02) -- cycle;
  \filldraw [green!70!black, opacity = .3] (-.6, -.02) -- (-.6, .02) -- (.99, .02) -- (.99, -.02) -- cycle;
 \foreach \p in {2,...,20}
  {
  \draw [green!70!black, thick, opacity = 1] ({1-6/(\p*(\p+1))}, -.03) -- ({1-6/(\p*(\p+1))}, .03);
  }
  \node [below] at (0, -.02) {$0$};  \node [below] at (.99, -.02) {$1$};
  \node [below] at (1.6, -.02) {$13$};
  \node[draw,circle,inner sep=2pt,yellow] at (.99, 0) {} ; \node [above] at (.99, .02) {Ashkin-Teller};
  \draw[-latex] (-.6, 0) -- (2.15, 0) node [below left] {$c$};
\end{tikzpicture} \label{res-fig:newcftmap}
\end{align}
Ici, la région couverte des lignes grises montré les valeurs de $c$ pour lesquelles nos théories proposées devraient exister. Il est intéressant de remarquer qu'il y a une deuxième famille des théories conformes non-diagonales pour ces valeurs de $c$ qui peuvent être obtenus comme une limite des modèles minimaux de la série $D$ avec $p$ pair, au lieu de $q$. Le spectre non-diagonal de ces théories serait $\mathcal{S}_{\mathbb{Z}+\frac{1}{2}, \mathbb{Z}}$.

Les résultats de ce travail suggèrent différents possibilités pour des recherches futures. 
Premièrement, il serait intéressant d'étudier des généralisations du bootstrap analytique non-diagonal. Une manière de faire ceci est d'affaiblir certaines de nos assomptions, donnant lieu à d'autres théories. Dans \cite{ei15}, par exemple, les auteurs ont supposé l'existence d'un seul champ dégénéré, ce qui permet l'existence des champs non-diagonaux avec des indices fractionnaires. D'autre part, les principes du bootstrap analytique peuvent être appliqués aux théories avec des algèbres de symétrie plus larges que l’algèbre de Virasoro, comme les théories de symétrie $\mathcal{W}$. Cette généralisation a été récemment étudiée dans \cite{Dupic:2018pqr}.

Deuxièmement, notre proposition pour des théories non-diagonales à $c$ générique peut être plus profondément étudiée. D'abord, il y a des propositions des règles de fusion qui restent à vérifier, ce qui demande l'étude des fonctions de corrélation du type et $\langle V^N_1 V^N_2 V^N_3 V^N_4 \rangle$  $\langle V^D_1 V^D_2 V^N_3 V^N_4 \rangle$. D'autre part, nous avons pris des limites des modèles minimales d'une manière particulière, qui fait que dans la limite le spectre n'aille plus des représentations dégénérés. Cependant, il est en principe possible de prendre une limite diffèrent en trouvant, par exemple, une version non-diagonal des modèles minimaux généralisés. En autre, la méthode de prendre des limites des modèles minimaux à partir de leurs spectres est un peut heuristique, et il serait plus rigoureux d'étudier des limites des fonctions à quatre points quand la charge centrale approche des valeurs non-rationnelles. Finalement, une question intéressante est la prise des limites dans le sens opposé, i.e. en commençant par une théorie non-diagonale et non-rationnelle et en allant vers les modèles minimaux quand la charge centrale devient rationnelle. Ce problème a été traité dans \cite{Ribault:2018jdv}, où il est décrit comment dans certaines situations il serait effectivement possible de reconstruire les modèles minimaux en suivant cette procédure.  

%% file: refs.tex
\providecommand{\href}[2]{#2}\begingroup\raggedright\endgroup

%% file: backcover.tex
\newgeometry{
left=14mm,
top=20mm,
right=10mm,
bottom=30mm}

\pagestyle{empty}

\begin{textblock*}{61mm}(3mm,3mm)
	\noindent\includegraphics[height=24mm]{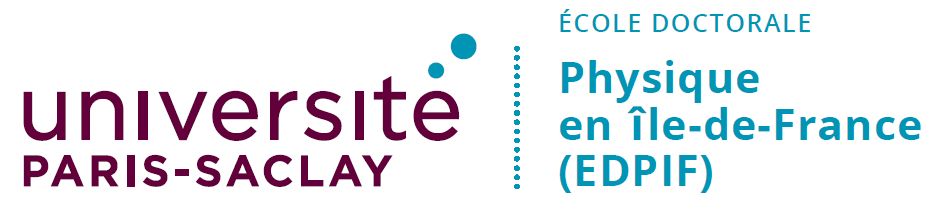} 
\end{textblock*}

\phantom{Adding text so the page is not perceived as empty.}
\begin{center}
\fbox{\parbox{\textwidth}{
{\bf Titre:}\PhDTitleFR 
\medskip

{\bf Mots clés:} \keywordsFR 
\vspace{-2mm}

\begin{multicols}{2}
{\bf Résumé:} 
\abstractFR 
\end{multicols}
}}
\end{center}

\vspace*{0mm}

\begin{center}
\fbox{\parbox{\textwidth}{
{\bf Title:} \PhDTitleEN 

\medskip

{\bf Keywords:}  \keywordsEN 
\vspace{-2mm}
\begin{multicols}{2}
	
{\bf Abstract:} 
\abstractEN
\end{multicols}
}}
\end{center}

\begin{textblock*}{161mm}(10mm,270mm)
\textblockcolor{white}
\color{bordeau}
{\bf\noindent Université Paris-Saclay	         }

\noindent Espace Technologique / Immeuble Discovery 

\noindent Route de l’Orme aux Merisiers RD 128 / 91190 Saint-Aubin, France 
\end{textblock*}

\begin{textblock*}{20mm}(182mm,255mm)
\textblockcolor{}
\includegraphics[width=20mm]{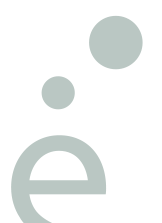}
\end{textblock*}

\restoregeometry